\pdfoutput=1
\documentclass[aip,jmp,amsmath,amssymb,reprint]{revtex4-2}
\usepackage{amsthm,mathtools}
\usepackage{booktabs}
\usepackage{graphicx}
\graphicspath{{figures/}}
\usepackage{tikz}
\usepackage{tikz-cd}
\usetikzlibrary{arrows.meta}
\usepackage[colorlinks=true,linkcolor=blue!60!black,citecolor=blue!60!black,urlcolor=blue!60!black]{hyperref}

\numberwithin{equation}{section}

\theoremstyle{plain}
\newtheorem{theorem}{Theorem}[section]
\newtheorem{proposition}[theorem]{Proposition}
\newtheorem{lemma}[theorem]{Lemma}
\newtheorem{corollary}[theorem]{Corollary}
\newtheorem{maintheorem}{Theorem}

\theoremstyle{definition}
\newtheorem{definition}[theorem]{Definition}
\newtheorem{hypothesis}[theorem]{Hypothesis}
\theoremstyle{remark}
\newtheorem{remark}[theorem]{Remark}

\newcommand{\R}{\mathbb{R}}
\newcommand{\Z}{\mathbb{Z}}
\newcommand{\Q}{\mathbb{Q}}
\newcommand{\F}{\mathbb{F}_2}
\newcommand{\RP}{\mathbb{RP}}
\newcommand{\CH}{\mathrm{CH}}
\newcommand{\Inv}{\operatorname{Inv}}
\newcommand{\cl}{\operatorname{cl}}
\newcommand{\pib}{\bar\pi}
\newcommand{\Ss}{\mathsf{E}}   % compactified extended phase space
\newcommand{\Dd}{\mathsf{D}}   % double
\newcommand{\osx}{\natural}
\newenvironment{notationlist}{\par\bigskip\small\setlength{\parindent}{0pt}\setlength{\parskip}{2pt}}{\par}
\newcommand{\ntrow}[3]{\par\noindent\parbox[t]{0.2\textwidth}{\raggedright #1}\hspace{0.02\textwidth}%
  \parbox[t]{0.56\textwidth}{\raggedright #2}\hspace{0.02\textwidth}%
  \parbox[t]{0.2\textwidth}{\raggedright #3}\par}
\newcommand{\ntrule}{\par\noindent\rule[0.5ex]{\textwidth}{0.4pt}\par}
\newcommand{\ntgroup}[1]{\par\medskip\noindent\emph{#1}\par\nopagebreak}
\newcommand{\Sig}{\Sigma}

\begin{document}

\title{Conley index theory for asymptotically autonomous systems and UV\texorpdfstring{$\to$}{->}IR connections}

\author{Alireza Behtash}
\email{ali@stacksai.com}
\affiliation{Stacks, 6841 Middleboro Dr, Raleigh, North Carolina 27612, USA}
\author{Syo Kamata}
\email{skamata11phys@gmail.com}
\affiliation{Department of Physics, The University of Tokyo, 7-3-1 Hongo, Bunkyo-ku, Tokyo 113-0033, Japan}
\author{Mauricio M. Martinez}
\email{m.martinezguerrero@montana.edu}
\affiliation{Department of Chemistry and Biochemistry, Montana State University, 103 Chemistry and Biochemistry
Building, Bozeman, Montana 59717, USA}

\date{\today}

\begin{abstract}
Many non-equilibrium systems, such as expanding plasmas, evolve by equations that depend explicitly on
time and interpolate between autonomous early-time (UV) and late-time (IR) limits. After time is
compactified, fixed points lie on invariant boundary fibers, where the Conley index is one-sided and
blind to the UV. For time dependence that admits a $C^1$ compactification, we reflect the extended
phase space across these fibers. Both sheets of the double run from the UV to the IR, and the
UV index reappears, suspended by one degree. Extending one sheet slightly beyond both fibers gives an explicit UV$\to$IR
transport map between the Conley indices of the limit systems, and its non-vanishing forces
connecting solutions. If one block with a fixed strict entrance/exit decomposition serves every
frozen system, this map is an isomorphism at every admissible rate. If the block also has
nonzero relative homology, the isolated invariant sets of the two limits are connected at every
rate. In the Gubser flow of Israel--Stewart theory, the separatrix between two classes of initial data
is a transversal connection at every finite relaxation scale $c$, with splitting
$\Delta(c)\propto1/c$. In the Bjorken flow at
the lowest moment truncation, we prove existence and uniqueness of the attractor solution without
regularization. For the three-dimensional Gubser system we construct the global Morse
decomposition. In the one-loop flow of the $(4+\epsilon)$-dimensional $O(N)$ model with cubic anisotropy,
a shift of $N$ through $N=4$ has vanishing transport map, but a driven connection and a nonzero
connecting homomorphism detect it at every rate.
\end{abstract}

\maketitle

%======================================================================
\section{Introduction}\label{sec:intro}
%======================================================================

\subsection{Explicit time dependence and UV/IR connections}\label{subsec:questions}

Many evolution equations of physics depend explicitly on time. Examples are the moment equations of
kinetic theory for an expanding plasma, in which the expansion rate decreases in time, and
renormalization group (RG) flows whose control parameters run with the scale. In both cases the
equations interpolate between autonomous limits at early and late times, which we call the
ultraviolet (UV) and infrared (IR) limits. A non-autonomous system with a $d$-dimensional phase space
is naturally regarded as an autonomous system in $d+1$ dimensions, with the time as an additional
variable. This enlarged system has no compact invariant sets, so the time direction has to be
compactified. Once this is done, all fixed points of the enlarged system lie on the two boundary fibers
of the compactified time direction. This is a major deviation from the autonomous case, where the fate
of the flow is not tied to how the physical time varies, and it requires careful treatment. As we shall
see, the Conley index of the compactified system is one-sided at the boundary, and the remedies we
develop lead to a modified description of the UV/IR structure.

Prototypes of such systems arise in the kinetic theory of expanding plasmas, for instance for the
Boltzmann equation in the boost-invariant Bjorken flow~\cite{Bjorken:1982qr} or in the conformal Gubser
flow~\cite{Gubser:2010ze,Gubser:2010ui}, where the dynamics is cast into evolution equations for the
moments of the distribution function~\cite{Behtash:2017wqg,Behtash:2018moe}, and for the Israel--Stewart
theory of second-order hydrodynamics~\cite{Israel:1979wp}. In earlier work the solutions of these models
were interpreted as RG flows parametrized by the moments and the temperature, with the transport
coefficients as the renormalized parameters~\cite{Behtash:2018moe,Behtash:2019}. Independently of this
interpretation, they are explicit non-autonomous systems whose global phase portraits can be studied
rigorously, and they teach us lessons about the non-perturbative nature of non-equilibrium physics.

When the RG equations are viewed as a gradient flow~\cite{Wallace:1974dx,Friedan:2009ik}, the UV and IR
fixed points are critical points of a function $\mathcal L$ that decreases along the flow, so that
$\mathcal L_{\rm UV}>\mathcal L_{\rm IR}$. Even when the flow is not known to be a gradient, such a
function often exists, namely the $c$-function in two dimensions~\cite{Zamolodchikov:1986gt}, the
$F$-function in three~\cite{Jafferis:2011zi,Casini:2012ei} and the $a$-function in
four~\cite{Jack:1990eb,Komargodski:2011vj}. A monotone function orders the fixed points, but it does not
say which of them are connected. Three questions remain. The first is which UV fixed points are connected
to a given IR fixed point. The second is whether these connections are robust when a control parameter,
such as the number of fermion flavors in quantum field theories, is tuned. A familiar example is the loss of conformality at the edge
of the conformal window, where an IR fixed point merges with a UV fixed point and both
disappear~\cite{Banks:1981nn,Kaplan:2009kr}. The third question is specific to the non-autonomous
setting. It asks whether a connection survives when the system is \emph{driven} in time at a finite
rate. In dynamical systems this is the question of rate-induced tipping~\cite{Ashwin:2012,Ashwin:2017},
and in statistical physics it is the question of a finite-rate quench through a critical
point~\cite{Kibble:1976sj,Zurek:1985qw}.

Conley index theory~\cite{Conley:1978,Mischaikow:2002} gives partial but deformation-invariant answers to
all three questions. It assigns to every isolated invariant set of a flow a pointed homotopy type,
together with algebraic data (connection matrices) that constrain the connecting orbits between its
pieces. It does not require non-degeneracy, it is invariant under continuation, and it can be computed
rigorously by computer~\cite{Arai:2009,Bush:2012}. For gradient flows of Morse functions on closed
manifolds it reduces to Morse theory, whose Morse--Smale--Witten complex counts gradient lines between
critical points of adjacent index~\cite{Smale:1960,Smale:1961,Witten:1982im,Schwarz:1993}, and whose
holomorphic version, Picard--Lefschetz theory, underlies the Lefschetz-thimble decomposition of path
integrals (see Ref.~\onlinecite{Fujimori:2018kqp} and references therein). In mathematical physics the Conley index
was first used to construct shock and traveling-wave profiles as connecting
orbits~\cite{Conley:1971cs,Smoller:1983}, and it is a natural tool for detecting global
bifurcations~\cite{Gukov:2016tnp}. A change of an index or of a connecting homomorphism along a family
forces a loss of isolation, and hence a transition. The converse fails in general, since connecting
orbits can appear or disappear in algebraically canceling pairs without changing any index. The index
cannot detect every connection, but the connections it does detect are robust. The aim of this paper is
to formulate these answers for asymptotically autonomous systems.

\subsection{Main results}\label{subsec:main}

We assume that the time dependence admits a $C^1$ compactification (Hypothesis~\ref{hyp:H}). The
compactified extended phase space is then $\Ss=X\times[-1,1]$, where $X$ is the phase space of the
non-autonomous system, and its boundary fibers $X\times\{\mp1\}$ are invariant and carry the UV and IR
limit flows $f_\mp$. For an isolated invariant set $S\subset\Ss$ we write $K_\mp$ for its parts in the
two boundary fibers.

The mechanism behind our constructions is simple. Once time is a coordinate on a compact interval, every
solution moves away from the UV end and toward the IR end. Near an IR fixed point the time direction is
therefore attracting, and the IR index is unchanged. Near a UV fixed point the time direction is
repelling, but the boundary leaves no room to exit, so the UV fixed point is invisible to the index of
the extended phase space. Reflecting the extended phase space across its boundary fibers removes this
asymmetry. In the resulting double $\Dd$ the time direction at the UV fiber becomes a genuine unstable
direction, and the UV index reappears, suspended by one degree. Keeping only one sheet of the double,
extended slightly beyond the UV and IR fibers, gives the one-sided extension $S^\osx$ of $S$, in which
the UV and IR data form an attractor--repeller pair. Its connecting homomorphism is the transport map,
which plays the role of the continuation map of Morse and Floer homology. The two main results are the
following.

\begin{maintheorem}[Transport map]\label{thmA}
Let $S\subset\Ss$ be an isolated invariant set. The connecting homomorphism of the UV--IR
attractor--repeller pair of the one-sided extension $S^\osx$ defines degree-preserving homomorphisms
$\Psi_k:\CH_k(K_-;f_-)\to\CH_k(K_+;f_+)$, which fit into a long exact sequence with $\CH_*(S^\osx)$. If
$\Psi\neq0$, then the non-autonomous equation has a solution in $S$ that converges to $K_-$ as
$t\to-\infty$ and to $K_+$ as $t\to+\infty$. With field coefficients, $\Psi$ is determined up to
isomorphism by the indices of $K_\pm$ and $S^\osx$, and it is invariant under continuation.
\end{maintheorem}

\begin{maintheorem}[Common blocks and rate independence]\label{thmB}
Let $N\subset X$ be a compact submanifold with a fixed decomposition $\partial N=\partial^{\rm
in}\sqcup\partial^{\rm out}$ such that every frozen vector field $F(\tau,\cdot)$, $\tau\in[-1,1]$, points
strictly into $N$ on $\partial^{\rm in}$ and strictly out of $N$ on $\partial^{\rm out}$. Then, for every
rate function $g$ as in Hypothesis~\ref{hyp:H}, the sets $K_\pm=\Inv(N,f_\pm)$ satisfy
$\CH_*(K_\pm;f_\pm)\cong H_*(N,\partial^{\rm out})$, $\CH_*(S^\osx)=0$, and $\Psi$ is an isomorphism. In
particular, if $H_*(N,\partial^{\rm out})\neq0$, there is a solution in $N$ from $K_-$ to $K_+$ at every
rate. The same holds for a block $N_\tau=\{x:\phi_\tau(x)\in N\}$ defined through time-dependent coordinates
$\xi=\phi_\tau(x)$, provided the entrance and exit conditions hold for the full compactified flow, including the term
$g(\tau)\,\partial_\tau\phi_\tau$ that comes from the time dependence of the block
(Corollary~\ref{cor:moving}).
\end{maintheorem}

\begin{figure}[t]
\centering
\begin{tikzpicture}[>=Stealth,font=\small,line width=1pt,
  box/.style={draw,rounded corners=3pt,align=center,inner sep=4pt,minimum height=1.0cm,text width=4.8cm},
  core/.style={box,fill=gray!12},
  main/.style={box,line width=2pt,fill=blue!14},
  side/.style={box,dashed,text width=3.4cm},
  app/.style={box,fill=red!10,text width=2.6cm}]
 \node[core] (H) at (0,0) {Hypothesis~\ref{hyp:H}\\ $C^1$ compactification of time};
 \node[core] (D) at (0,-1.6) {double $\Dd$ and extension $S^\osx$\\ Lemmas~\ref{lem:G}--\ref{lem:isolation}};
 \node[core] (B) at (0,-3.2) {boundary indices\\ Proposition~\ref{prop:boundary}};
 \node[side] (slab) at (-4.8,-3.2) {index of $\Ss$\\ is the IR index\\ Theorem~\ref{thm:slab}};
 \node[side] (eq) at (4.8,-3.2) {equivariant structure\\ and duality\\ Theorems~\ref{thm:equivariant}, \ref{thm:duality}};
 \node[main] (A) at (0,-4.8) {Theorem~\ref{thmA}, transport map $\Psi$\\ Theorem~\ref{thm:transport}};
 \node[main] (R) at (0,-6.4) {Theorem~\ref{thmB}, rate independence\\ Theorem~\ref{thm:rate}, Corollary~\ref{cor:moving}};
 \node[side] (F) at (-4.8,-6.4) {Fredholm index\\ and transversality\\ Proposition~\ref{prop:palmer}};
 \node[app] (G) at (-4.95,-8.6) {Gubser separatrix\\ Sec.~\ref{sec:gubser}};
 \node[app] (Bj) at (-1.65,-8.6) {Bjorken attractor\\ Sec.~\ref{sec:bjorken}};
 \node[app] (ON) at (1.65,-8.6) {$O(N)$ shift\\ Sec.~\ref{sec:ON}};
 \node[app] (G3) at (4.95,-8.6) {3D Gubser flow\\ Sec.~\ref{sec:full3d}};
 \draw[->] (H) -- (D);
 \draw[->] (D) -- (B);
 \draw[->] (B) -- (A);
 \draw[->] (A) -- (R);
 \draw[->,dashed] (B) -- (slab);
 \draw[->,dashed] (B) -- (eq);
 \draw[->] (R.south) -- ([xshift=-6mm]G.north east);
 \draw[->] (R.south) -- ([xshift=4mm]Bj.north);
 \draw[->] (R.south) -- (ON.north);
 \draw[->,dashed] ([xshift=-4mm]G.north |- F.south) -- ([xshift=-4mm]G.north);
 \draw[->,dashed] ([xshift=-2mm]F.south east) -- ([xshift=-4mm]Bj.north);
 \draw[->,dashed] (G3.north |- eq.south) -- (G3.north);
\end{tikzpicture}
\caption{Logical structure of the paper. Solid arrows mark the chain of results that leads to the two main
theorems (thick boxes) and their applications, and dashed boxes mark results that explain the
construction or add structure to it. The applications use Theorem~\ref{thmB} to establish isolation and
connections, and Proposition~\ref{prop:palmer} to establish transversality. The analysis of the full
three-dimensional Gubser flow uses the double and its equivariant structure, but not the transport map.}
\label{fig:roadmap}
\end{figure}
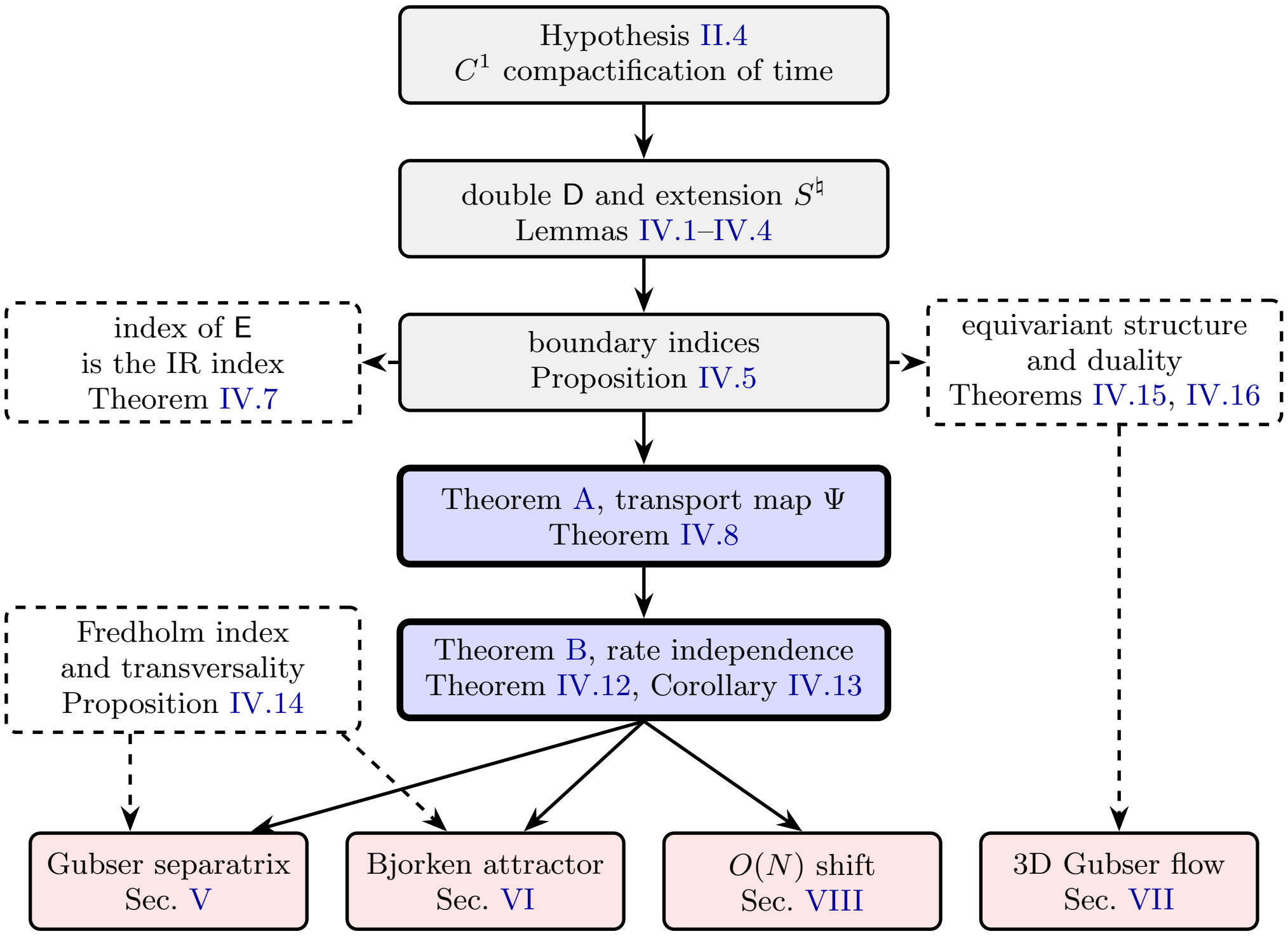

Theorem~\ref{thmA} is Theorem~\ref{thm:transport}, and Theorem~\ref{thmB} combines Theorem~\ref{thm:rate}
with Corollary~\ref{cor:moving}, which allows the block to depend on time. The common strict
entrance/exit block is the essential hypothesis of Theorem~\ref{thmB}. A common isolating neighborhood
alone gives continuation in $g$ but not the vanishing of $\CH_*(S^\osx)$. The conclusion concerns
the isolated invariant sets $K_\pm$ of the frozen limits. Theorem~\ref{thmB} does not show that a
solution tracks a particular branch of equilibria or attractors of the frozen systems, and it does not
exclude rate-induced tipping between Morse sets inside $N$. Figure~\ref{fig:roadmap} shows
how the other results depend on each other and on these two theorems. The minimal construction behind
them consists of the double and the one-sided extension (Sec.~\ref{subsec:double}), together with the
indices of sets that lie in a boundary fiber (Proposition~\ref{prop:boundary}). An IR set keeps its
autonomous index, whereas a UV set has trivial index in $\Ss$ and the suspended index in the double.

The remaining results of Sec.~\ref{sec:theory} either explain why this construction is needed or add
structure to it. Theorem~\ref{thm:slab} shows that the index in $\Ss$ by itself equals the IR index and
contains no information about connections. Proposition~\ref{prop:palmer} relates the doubled Morse
indices to the Fredholm index of the linearized operator along a connecting solution, which counts the
free parameters of a family of connections and gives a criterion for transversality.
Theorems~\ref{thm:equivariant} and~\ref{thm:duality} describe the $\Z_2$-equivariant structure of the full
double, whose connecting homomorphism vanishes over every coefficient ring, and the constraint that time reversal
imposes on indices. Throughout, $\Psi$ gives a sufficient condition for connections. Its non-vanishing
forces connecting solutions, and a change of $\Psi$ along a family signals a global transition, but $\Psi$
is not a complete invariant of the UV/IR connectivity. Algebraically, $\Psi$ is the connecting
homomorphism of an attractor--repeller pair of an autonomous flow, composed with two explicit boundary
isomorphisms. It is thus an explicit geometric realization of a connecting homomorphism, graded by the
indices of the UV and IR limit systems, and not a new kind of invariant. Section~\ref{subsec:previous}
compares it with the nonautonomous index of J\"anig.

\subsection{Relation to previous work}\label{subsec:previous}

Asymptotically autonomous equations go back to Markus~\cite{Markus:1956}, who showed that the
$\omega$-limit sets of their bounded solutions are invariant under the limit flow. Thieme~\cite{Thieme:1992}
showed that the long-time behavior can nevertheless differ from that of the limit flow, and Mischaikow,
Smith and Thieme~\cite{Mischaikow:1995} identified the correct replacement, namely that such
$\omega$-limit sets are internally chain transitive for the limit flow.

The theory of non-autonomous dynamical systems has developed several notions of attractors and Morse
decompositions for processes and skew-product flows~\cite{Kloeden:2011}. Pullback attractors describe the
state of the system at a fixed time after an infinitely long transient, and forward attractors describe
its behavior in the far future. Rasmussen~\cite{Rasmussen:2007tams,Rasmussen:2007lnm} introduced past and
future attractors, which are uniform with respect to the initial time, and the corresponding
non-autonomous Morse decompositions. These notions are formulated directly for the process and apply to
arbitrary, not necessarily asymptotically autonomous, time dependence. They describe where solutions go,
but they do not carry an algebraic invariant that forces connections, and for an asymptotically
autonomous system the past and future notions probe only the UV and the IR limit flow, respectively
(Appendix~\ref{app:appendix}). The present paper restricts the time dependence to the class of
Hypothesis~\ref{hyp:H} and, in exchange, attaches to every isolated invariant set of $\Ss$ a homological
invariant, the transport map. Its non-vanishing forces solutions that join a UV set to an IR set, it is
invariant under deformations of the equation, and under the conditions of Theorem~\ref{thmB} it is
an isomorphism at every rate. The two approaches are therefore complementary. Non-autonomous Morse
decompositions organize the asymptotic behavior, whereas the index theory tests whether a given UV set
and a given IR set are connected, and how robustly.

The closest index theory is the nonautonomous Conley index of J\"anig~\cite{Janig:2017}, defined for
skew-product semiflows, including those of semilinear parabolic equations. Their base is the hull of
the equation, the closure of the set of its time translates (Sec.~\ref{sec:nonaut}). The theory
includes a homology index with attractor--repeller sequences~\cite{Janig:2019hom} and a connecting
homomorphism, and a nonzero connecting homomorphism forces connections between the attractor and the
repeller~\cite{Janig:2019conn}. The index needs an initial element of the base whose $\omega$-limit set is
the whole base. For an asymptotically autonomous equation with different limits $f^{-\infty}\neq
f^{\infty}$, the hull of $f$ alone contains no such element. J\"anig~\cite{Janig:2020} therefore closes the
base into a circle. He adjoins a second asymptotically autonomous equation $g$ that runs from
$f^{\infty}$ back to $f^{-\infty}$, typically the reversed history $g(t,x)=f(-t,x)$, and works over the
cycle $(f,g)$. The double $\Dd$ of the present paper also has a circle as its base, but it contains no
reversed history. It consists of two copies of the compactified extended phase space glued along the
UV and IR fibers. Both sheets run from the UV to the IR, and the reflection $R$ exchanges them. The
one-sided extension $S^\osx$ selects one sheet and extends it slightly beyond the boundary
fibers.

The hypotheses and conclusions differ as well. J\"anig assumes only that $f(t,\cdot)\to f^{\pm\infty}$
uniformly on compact sets, at no prescribed rate, and he allows infinite-dimensional semiflows. His
index and its attractor--repeller theory apply to general compact isolated invariant sets, without
hyperbolicity or gradient-like conditions on the limit systems~\cite{Janig:2019hom,Janig:2019conn}. In
this generality he bounds the Betti numbers of the index over the cycle, with field coefficients, by
those of the index for the limit $f^{\infty}$~\cite[Theorem~6.1]{Janig:2020}. His results on specific
solutions concern good equations, whose limit systems have finitely many hyperbolic equilibria and
gradient-like dynamics. If the index has rank $n_0$ in one degree, he obtains at least $n_0$ solutions
connecting equilibria of equal Morse index~\cite[Theorem~7.1 and Corollary~7.2]{Janig:2020}. He also
shows that a generic asymptotically linear parabolic equation has a hyperbolic solution between
equilibria of the limit equations~\cite[Theorem~7.3]{Janig:2020}. We assume instead the $C^1$
compactification of Hypothesis~\ref{hyp:H}, which prescribes how the limits are approached, and we work
in finite dimensions. In this setting, Theorem~\ref{thmA} relates the indices of $K_-$ and $K_+$
directly, in the grading of the UV and IR limit systems, without passing to an index over a cycle, and
Theorem~\ref{thmB} gives a criterion for every admissible rate. The Fredholm index of
Proposition~\ref{prop:palmer} and the applications use this explicit form. Whether, under hypotheses common to both settings, $\Psi$ agrees
with the connecting homomorphism of an attractor--repeller decomposition in J\"anig's index over a
suitable cycle is an open question. We do not use such an identification.

The compactification of time is that of Wieczorek, Xie and Jones~\cite{Wieczorek:2021}, from which we
take Hypothesis~\ref{hyp:H}. It was developed for rate-induced tipping~\cite{Ashwin:2012,Ashwin:2017}, in
which a parameter of an autonomous system is shifted at a finite rate and one asks whether a solution
that tracks a branch of attractors of the frozen systems survives the shift. In that setting the
compactification describes tipping through the connecting orbits and invariant manifolds of the
compactified system. Theorem~\ref{thmB} answers a different question. It gives conditions under which
the isolated invariant sets $K_\pm$ are connected at every rate, namely a common block with nonzero
relative homology $H_*(N,\partial^{\rm out})$, and it does not decide whether a
solution tracks a particular branch. For families of autonomous flows, transition
matrices~\cite{Reineck:1988,Franzosa:1998,McCord:1992tm} detect connecting orbits that are forced when a
parameter is varied infinitely slowly. On the Floer-theoretic side, Salamon~\cite{Salamon:1990}
explained how the Conley index underlies Floer homology, and the continuation maps of Morse and Floer
homology count solutions of gradient equations with an explicitly $s$-dependent vector field, whose
Fredholm index is a difference of Morse indices or, more generally, a spectral
flow~\cite{Schwarz:1993,Robbin:1995}.

Relative to these works, the theory of Sec.~\ref{sec:theory} adds three things. First,
Theorem~\ref{thmA} realizes UV$\to$IR transport in the classical Conley index of an
autonomous flow. The boundary isomorphisms of Proposition~\ref{prop:boundary} grade it by the indices of
the UV and IR limit systems, and Proposition~\ref{prop:palmer} relates this grading to the Fredholm
index of the linearized operator along a connecting solution. Second, Theorem~\ref{thmB} gives a common-block criterion, valid for every admissible rate, under
which $\Psi$ is an isomorphism and which forces a connection when $H_*(N,\partial^{\rm out})\neq0$,
whereas transition matrices refer to the slow limit and the work of Wieczorek, Xie and Jones treats a
given rate through invariant manifolds. Third, the full double carries a $\Z_2$ action, and
Theorem~\ref{thm:equivariant} shows that its connecting homomorphism vanishes over every coefficient
ring. This
explains why the one-sided extension, rather than the full double, detects connections.

On the physics side, the attractor of boost-invariant expanding plasmas was found by Heller and
Spali\'nski~\cite{Heller:2015dha} and has since been established in hydrodynamics, kinetic theory and
holography (see Refs.~\onlinecite{Romatschke:2017vte,Blaizot:2017ucy} and the reviews in
Refs.~\onlinecite{Florkowski:2017olj,Jankowski:2023fdz}). Its existence has been argued numerically, by
slow-roll and resummation methods, and by phase-space analyses of the moment
equations~\cite{Behtash:2017wqg,Behtash:2018moe}. In Ref.~\onlinecite{Behtash:2019} the existence of a flow line from the far-from-equilibrium UV fixed
point to the hydrodynamic fixed point of the moment equations was raised as a question, and
Ref.~\onlinecite{Behtash:2020jhep} described this attractor solution through a Morse decomposition of
the truncated moment equations. At the truncation $L=1$, Proposition~\ref{prop:bjorken} proves existence
and uniqueness of this solution at every rate, together with the invertibility of its linearized
operator, by an index computation that requires no regularization of the late-time end of the
evolution. For the Gubser flow of IS theory, the fixed points, their linearizations and numerical flow
diagrams were obtained in Ref.~\onlinecite{Behtash:2017wqg}. Reference~\onlinecite{Behtash:2020jhep}
used skew-product flows and attractor--repeller pairs to classify the invariant submanifolds of the
same system and to determine the possible flow lines between its UV and IR fixed points, including the
constant solution from $\Sig_U(B)$ to $\Sig_I(B)$. Relative to that work, the present paper proves the
transversality of this connection at every finite relaxation scale, with an explicit splitting
$\Delta(c)$ between the slopes of its unstable and stable subspaces (Proposition~\ref{prop:transversal} and
Remark~\ref{rem:gap}). It also isolates the connection by an explicit block with quantitative
perturbation bounds (Corollary~\ref{cor:gubserB} and Lemma~\ref{lem:gubser-block}), and it computes the
Morse decomposition and the indices of the system compactified at infinity (Lemma~\ref{lem:TC} and
Theorem~\ref{thm:M}).
The model of Sec.~\ref{sec:ON} is the
one-loop flow of the $O(N)$ model with cubic anisotropy~\cite{Aharony:1973,Gukov:2016tnp}, the simplest setting in which two fixed points exchange stability as a
parameter is tuned. It serves as a controlled analog of the merger of fixed points discussed
above~\cite{Kaplan:2009kr}, and it lets us compute the transition data of a parameter shift explicitly.

\subsection{Applications and outline}

The applications show the main results at work in settings of increasing complexity, and their detailed
proofs are collected in the appendixes. In the Gubser flow of Israel--Stewart theory
(Sec.~\ref{sec:gubser}), the separatrix between two classes of initial data has $\Psi=0$ in the
collisionless limit. At every finite relaxation scale $c$ it is a transversal connection
(Proposition~\ref{prop:transversal}) with an explicit splitting $\Delta(c)\propto1/c$ of its stable and
unstable directions (Remark~\ref{rem:gap}), and an explicit moving block isolates it with $\Psi=\pm1$
(Corollary~\ref{cor:gubserB} and Lemma~\ref{lem:gubser-block}). In the Bjorken flow at $L=1$
(Sec.~\ref{sec:bjorken}), Theorem~\ref{thmB} gives a regulator-free proof of existence and uniqueness of
the attractor solution (Proposition~\ref{prop:bjorken}). For the full three-dimensional Gubser system
(Sec.~\ref{sec:full3d}), we construct the global Morse decomposition of the compactified and doubled
flow by explicit isolating blocks (Theorem~\ref{thm:M}). Since the base flow freezes at infinity after
compactification, the results of Sec.~\ref{sec:theory} enter there only through sets in the finite
part. Finally, in the one-loop RG flow of the $(4+\epsilon)$-dimensional $O(N)$ model
(Sec.~\ref{sec:ON}), the exchange of stability at $N=4$ leaves the unlabeled Morse polynomial
unchanged, while the Morse indices of the labeled fixed points $\Sig_C$ and $\Sig_H$ are exchanged. For
a parameter shift of $N$ through $4$ the transport map vanishes at every rate. The shift has an explicit
driven connection along the diagonal, of Fredholm index $-1$, and the grouped connecting homomorphism
from the UV point $\Sig_H$ to the Morse set of the IR points $\Sig_I$, $\Sig_C$ and $\Sig_H$ is an
isomorphism (Proposition~\ref{prop:ONramp}).

Section~\ref{sec:nonaut} recalls non-autonomous systems, compactifies time and discusses two warm-up
examples, and Sec.~\ref{sec:background} reviews Conley index theory. Section~\ref{sec:theory} develops the
theory, Secs.~\ref{sec:gubser}--\ref{sec:ON} contain the applications, and Sec.~\ref{sec:concl}
concludes. Appendixes~\ref{app:theory}--\ref{app:ON} contain the proofs deferred from the main text, and
Appendix~\ref{app:appendix} collects non-autonomous notions and lists the notation.

%======================================================================
\section{Non-autonomous systems}\label{sec:nonaut}
%======================================================================

\subsection{Processes and skew-product flows}

A non-autonomous ordinary differential equation is
\begin{equation}\label{eq:nonaut}
\dot x = f(t,x),\qquad x\in X,\quad t\in\R,
\end{equation}
where $X\subset\R^n$ is open (more generally, a smooth manifold) and $f:\R\times X\to\R^n$ is $C^1$.
Its solutions define a \emph{process} $\varphi(t,s,x)$, the value at time $t$ of the solution with
$x(s)=x$, which satisfies
\begin{equation}\label{eq:process}
\varphi(s,s,x)=x,\qquad \varphi\big(t,r,\varphi(r,s,x)\big)=\varphi(t,s,x),
\end{equation}
whenever defined. A process is a two-parameter family of maps, and it depends on $t-s$ only when $f$ is
autonomous. Treating time as a variable, $\frac{d}{dt}(x,s)=(f(s,x),1)$, gives an autonomous system on
$\R\times X$ without equilibria and without compact invariant sets. It is therefore of no direct use
for index theory, and the time direction has to be compactified.

\begin{definition}[Skew-product flow]
Let $P$ be a metric space with a flow $\theta:\R\times P\to P$ (the \emph{base flow}), and let
$\phi:\R\times P\times X\to X$ be a \emph{cocycle}, i.e.\ $\phi(0,p,\cdot)=\mathrm{id}_X$ and
$\phi(t+s,p,x)=\phi\big(t,\theta_s p,\phi(s,p,x)\big)$. The \emph{skew-product flow} on $P\times X$ is
$\Pi_t(p,x)=\big(\theta_t p,\phi(t,p,x)\big)$.
\end{definition}

The canonical choice of base is the \emph{hull}~\cite{Sell:1971} $\mathcal H(f)$, the closure of the set
of translates $f_s\coloneqq f(s+\cdot,\cdot)$ in $C(\R\times X,\R^n)$ with the compact-open topology.
The base flow is translation, $\theta_t(h)=h(t+\cdot,\cdot)$, and $\phi(t,h,x)$ is the time-$t$ value
of the solution of $\dot x=h(t,x)$, $x(0)=x$. In particular $\phi(t,f_s,x)=\varphi(s+t,s,x)$.

\begin{definition}
$f$ is \emph{asymptotically autonomous} if there are $C^1$ vector fields $f_\pm$ on $X$ such that
$f(t,\cdot)\to f_\pm$ uniformly on compact subsets of $X$ as $t\to\pm\infty$. We call $f_-$ the UV limit
and $f_+$ the IR limit.
\end{definition}

\begin{lemma}\label{lem:hull}
Let $f$ be asymptotically autonomous and not autonomous. Then $\mathcal H(f)=\{f_s:s\in\R\}\cup\{f_-,f_+\}$,
with $f_\pm$ regarded as time-independent, and $s\mapsto f_s$ extends to a continuous surjection
$[-\infty,+\infty]\to\mathcal H(f)$. It is a homeomorphism if $f_-\neq f_+$.
\end{lemma}

\begin{proof}
Uniform convergence on compacta gives $f_{s}\to f_\pm$ in the compact-open topology as $s\to\pm\infty$,
and $s\mapsto f_s$ is continuous, so the map extends continuously to $[-\infty,\infty]$. Its image is
compact, hence closed, and contains all translates, so it is the hull. If $f_s=f_{s'}$ with $s\neq s'$,
then $f$ is periodic in $t$, and since it converges as $t\to\infty$, it is then independent of $t$,
contradicting the assumption. Likewise $f_s=f_\pm$ would force $f$ to be autonomous. Hence the map is
injective when $f_-\neq f_+$, and a continuous bijection from a compact space to a Hausdorff space is a
homeomorphism.
\end{proof}

\subsection{The compactified extended phase space}\label{subsec:slab}

Lemma~\ref{lem:hull} identifies the base of the skew-product flow with a compact interval whose two
end points are fixed by the base flow. In practice we work with an explicit coordinate on this
interval.

\begin{hypothesis}\label{hyp:H}
There are $g\in C^3([-1,1])$ with $g(\pm1)=0$ and $g>0$ on $(-1,1)$, and $F\in C^1([-1,1]\times X,\R^n)$,
such that $f(t,x)=F(\sigma(t),x)$, where $\sigma$ solves $\sigma'=g(\sigma)$, $\sigma(0)=0$.
\end{hypothesis}

Under Hypothesis~\ref{hyp:H}, $\sigma$ is an increasing diffeomorphism $\R\to(-1,1)$ and
$f_\pm=F(\pm1,\cdot)$. The \emph{compactified system} on the \emph{compactified extended phase space}
$\Ss\coloneqq X\times[-1,1]$ is
\begin{equation}\label{eq:slab}
\dot x = F(\tau,x),\qquad \dot\tau = g(\tau).
\end{equation}
It is the skew-product flow of~\eqref{eq:nonaut} over (a resolution of) the hull. Orbits with
$\tau\in(-1,1)$ are, up to a shift of time, the solutions of~\eqref{eq:nonaut}, and the boundary
fibers $X_\pm\coloneqq X\times\{\pm1\}$ are invariant and carry the autonomous flows $\dot x=f_\pm(x)$.

Hypothesis~\ref{hyp:H} is an assumption on the class of systems, not a property of all asymptotically
autonomous ones. The following criterion makes precise when it holds.

\begin{proposition}[When Hypothesis~\ref{hyp:H} holds]\label{prop:Hcrit}
Let $f\in C^1(\R\times X,\R^n)$.
\begin{enumerate}
\item[(i)] Suppose there is $c>0$ such that, for every compact $K\subset X$,
$\sup_{x\in K}|\partial_tf(t,x)|=O(e^{-c|t|})$ as $|t|\to\infty$, and $D_xf(t,\cdot)$ converges uniformly on
$K$ as $t\to\pm\infty$. Then $f$ is asymptotically autonomous, and Hypothesis~\ref{hyp:H} holds with
$g(\tau)=\kappa(1-\tau^2)$ and $\sigma(t)=\tanh(\kappa t)$, for every $\kappa\in(0,c/2)$.
\item[(ii)] Conversely, if Hypothesis~\ref{hyp:H} holds, then for every compact $K\subset X$ there is
$C_K$ with $\sup_{x\in K}|\partial_tf(t,x)|\le C_K\,\sigma'(t)$ for all $t$. In particular
$t\mapsto\sup_K|\partial_tf(t,\cdot)|$ is integrable over $\R$, and it decays exponentially as
$t\to\pm\infty$ if $g'(\pm1)\neq0$.
\end{enumerate}
\end{proposition}

\begin{proof}
(i) Let $K\subset X$ be compact. Since $\sup_K|\partial_tf(t,\cdot)|$ is integrable, $f(t,\cdot)$
converges uniformly on $K$ to limits $f_\pm$ as $t\to\pm\infty$, and the uniform convergence of $D_xf(t,\cdot)$
shows that $f_\pm$ are $C^1$ with $Df_\pm=\lim D_xf(t,\cdot)$. Hence $f$ is asymptotically autonomous. The
function $\sigma(t)=\tanh(\kappa t)$ solves $\sigma'=g(\sigma)$ with $\sigma(0)=0$, and $g$ is a polynomial
with $g(\pm1)=0$ and $g>0$ on $(-1,1)$. Put $F(\tau,x)=f(\kappa^{-1}\operatorname{artanh}\tau,x)$ for
$|\tau|<1$ and $F(\pm1,\cdot)=f_\pm$. Then $F$ is $C^1$ on $(-1,1)\times X$ and continuous on
$[-1,1]\times X$, and $D_xF$ extends continuously to $\tau=\pm1$ by the convergence of $D_xf$. With
$t=\kappa^{-1}\operatorname{artanh}\tau$,
\begin{equation}
\partial_\tau F(\tau,x)=\frac{\partial_tf(t,x)}{\kappa(1-\tau^2)}=\frac{\cosh^2(\kappa t)}{\kappa}\,\partial_tf(t,x),
\end{equation}
and therefore $\sup_K|\partial_\tau F(\tau,\cdot)|\le\kappa^{-1}e^{2\kappa|t|}\sup_K|\partial_tf(t,\cdot)|\to0$
as $|\tau|\to1$, because $2\kappa<c$. By the mean value theorem, $F$ has the one-sided partial derivatives
$\partial_\tau F(\pm1,x)=0$. Both partial derivatives of $F$ are therefore continuous on $[-1,1]\times X$,
and $F\in C^1([-1,1]\times X,\R^n)$.

(ii) By the chain rule, $\partial_tf(t,x)=\partial_\tau F(\sigma(t),x)\,\sigma'(t)$, and $\partial_\tau F$ is
bounded on the compact set $[-1,1]\times K$. Moreover $\sigma'>0$ and $\int_\R\sigma'\,dt=2$. If $g'(1)\neq0$,
then $g'(1)<0$, $1-\sigma(t)$ decays like $e^{g'(1)t}$, and $\sigma'=g(\sigma)=O(1-\sigma)$. The end
$t\to-\infty$ is treated in the same way.
\end{proof}

Part (ii) shows that Hypothesis~\ref{hyp:H} is a genuine restriction. For instance,
$f(t,x)=\big(1+\sin(t^2)/(1+t^2)\big)\,x$ converges to $x$, together with its $x$-derivative, uniformly on
compact sets as $t\to\pm\infty$, but $|\partial_tf(t,x)|$ behaves like $2|x\cos(t^2)|/|t|$, which is not
integrable. Hence no choice of $g$ and $F$ realizes this system. Part (i) covers exponentially fast
approach to the limits. Slower, algebraic approach requires a $g$ with a degenerate zero, as at the IR
end of the Bjorken flow of Sec.~\ref{sec:bjorken}, where $g(\tau)=\frac14(1+\tau)(1-\tau)^2$.
Wieczorek, Xie and Jones~\cite{Wieczorek:2021} treat general decay rates. In all examples below $F$ and
$g$ are explicit.

Equilibria of~\eqref{eq:slab} are the points $(e,\pm1)$ with $f_\pm(e)=0$. Their linearization is block
triangular, with eigenvalues those of $Df_\pm(e)$ together with $g'(\pm1)$. We write
\begin{equation}
\mu(e)\coloneqq\#\{\text{eigenvalues of }Df_\pm(e)\text{ with positive real part}\}
\end{equation}
for the \emph{fiber index}. If $g'(-1)>0>g'(1)$, the Morse index of $(e,-1)$ in the ambient space
$X\times\R$ is $\mu(e)+1$ and that of $(e,+1)$ is $\mu(e)$. In $\Ss$, however, the extra repelling
direction at $\tau=-1$ points into a half-space only, that is, the equilibria are hyperbolic, but one-sided.
This is the source of all subtleties below.

\subsection{A warm-up example}\label{subsec:warmup}

\begin{figure}[t]
\centering
\includegraphics[width=0.62\textwidth]{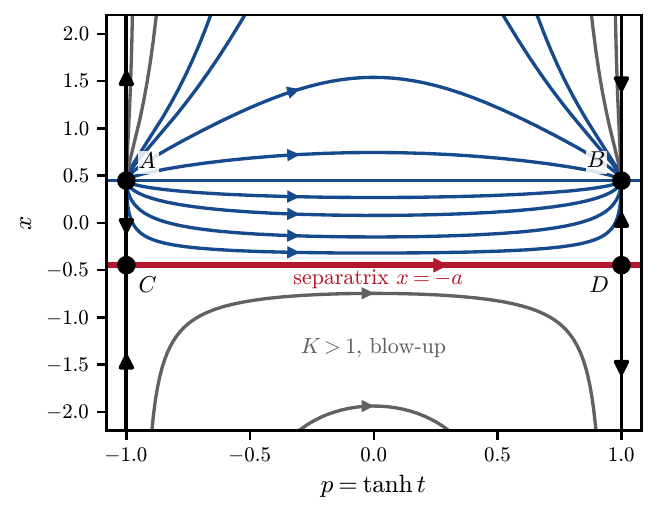}
\caption{Phase portrait of the warm-up system~\eqref{eq:warmup} in the compactified extended phase space
$(p,x)$, $p=\tanh t$. Blue curves are the bounded solutions $A\to B$ ($K<1$), the red line is the
separatrix $x=-a$ from $C$ to $D$, and gray curves are solutions with $K>1$, which blow up in finite time.
Black arrows on the invariant boundary fibers $p=\pm1$ show the UV and IR limit flows $f_\mp$. The UV
fiber repels in the normal direction, so $A$ and $C$ have Morse indices $2$ and $1$ in the ambient space,
whereas $D$ and $B$ have indices $1$ and $0$.}
\label{fig:warmup}
\end{figure}

Consider
\begin{equation}\label{eq:warmup}
\dot x = -\frac{4p}{3}\Big(x^2-\frac15\Big),\qquad \dot p = 1-p^2,\qquad p\in[-1,1],
\end{equation}
which satisfies Hypothesis~\ref{hyp:H} with $F(p,x)=-\frac{4p}{3}(x^2-a^2)$, $a\coloneqq1/\sqrt5$,
$g(p)=1-p^2$ and $\sigma=\tanh$. It is the restriction of the Gubser flow of Sec.~\ref{sec:gubser} to
its invariant plane $T=0$. The equilibria and their Morse indices in $X\times\R$ are
\begin{equation}\label{eq:warmup-fp}
A=(a,-1):\ 2,\qquad B=(a,+1):\ 0,\qquad C=(-a,-1):\ 1,\qquad D=(-a,+1):\ 1 .
\end{equation}
With $p(t)=\tanh t$, the new time $s=\log\cosh t$ turns the $x$-equation into the autonomous equation
$dx/ds=-\frac43(x^2-a^2)$. With $\kappa\coloneqq 8/(3\sqrt5)$, the general solution is
\begin{equation}\label{eq:warmup-sol}
x(t)=a\,\frac{\cosh^{\kappa}t+K}{\cosh^{\kappa}t-K},\qquad K=\frac{x(0)-a}{x(0)+a}\in\R\cup\{\infty\},
\end{equation}
with $K=\infty$ corresponding to $x\equiv-a$. Equivalently,
\begin{equation}\label{eq:warmup-K}
K=\frac{x-a}{x+a}\,\big(1-p^2\big)^{-\kappa/2}
\end{equation}
is a constant of motion of~\eqref{eq:warmup} on $\{|p|<1\}$. Consequently, the solutions fall into three classes.
\begin{itemize}
\item for $K<1$, i.e.\ $x(0)>-a$, the solution is bounded and $x(t)\to a$ as $t\to\pm\infty$, and these
solutions run from $A$ to $B$ (with $x\in(-a,a)$ for $K<0$, $x\equiv a$ for $K=0$, and $x>a$ for
$0<K<1$),
\item for $K=\infty$, $x\equiv-a$ runs from $C$ to $D$,
\item for $K>1$, i.e.\ $x(0)<-a$, the solution blows up in finite time in both directions, at
$|t|=t_*$ with $\cosh^{\kappa}t_*=K$.
\end{itemize}
Every solution is even in $t$. This reflects the reversibility of~\eqref{eq:warmup} under
$(x,p,t)\mapsto(x,-p,-t)$, discussed in Sec.~\ref{subsec:symmetries}. Supplementing these orbits
with the boundary flows $A\to C$ along $p=-1$ and $D\to B$ along $p=+1$, the flow-defined order on the
equilibria is the total order $A>C>D>B$. Figure~\ref{fig:warmup} shows the phase portrait.

The IR limit of a solution jumps discontinuously from $B$ to $D$ as the initial condition crosses the
separatrix $x=-a$. Similar phenomena occur in many physical contexts, e.g.\ RG flows, and they are the
kind of global information the Conley index is designed to capture. Note that the connections $A\to B$
form a continuous one-parameter family, and we return to the consequences of this degeneracy in
Sec.~\ref{subsec:reduced-index}.

\subsection{A phase transition and its time-dependent version}\label{subsec:phase}

The next example is the simplest system with an exchange of stability, and it is a useful caricature of
an RG flow. Read $z$ as a scale and $X$ as a running coupling. The right-hand side of~\eqref{eq:phase}
below is then a beta function with two zeros, and the control parameter $\alpha$ plays the role of
$\epsilon=4-d$ in the $\phi^4$ theory, whose sign decides whether the Gaussian or the Wilson--Fisher
point is the infrared one. Consider, for $0<z<\infty$ and a constant control parameter $\alpha\in\R$,
\begin{equation}\label{eq:phase}
\frac{dX}{dz}=-\frac{\alpha\,(X^2-1)}{2z}.
\end{equation}
The time $t=\log z$ makes it autonomous. It has equilibria $X_c^\pm=\pm1$, and the solutions are
\begin{equation}\label{eq:phase-sol}
X(z,\alpha)=\frac{z^\alpha-C_1}{z^\alpha+C_1},\qquad 0<C_1<\infty,
\end{equation}
with the two convergent expansions
\begin{equation}\label{eq:phase-ts}
X(z,\alpha)=\begin{cases}
-1-2\sum_{n\ge1}\big(-z^\alpha/C_1\big)^n, & z^\alpha\ll C_1,\\[0.5ex]
+1+2\sum_{n\ge1}\big(-C_1/z^\alpha\big)^n, & z^\alpha\gg C_1 .
\end{cases}
\end{equation}
The UV and IR fixed points are determined by the sign of $\alpha$, namely $X_c^-\to X_c^+$ for $\alpha>0$ and
$X_c^+\to X_c^-$ for $\alpha<0$, so the critical value is $\alpha_c=0$.

In Conley-index terms, for $\alpha\neq0$ the interval $N=[-2,2]$ is an isolating block with a single
exit point, so its index is trivial for both signs of $\alpha$. At $\alpha_c$ the Morse decomposition
changes, namely the Morse indices of $X_c^\pm$ are exchanged ($0\leftrightarrow1$). At $\alpha=0$ every
point is an equilibrium and $N$ is no longer isolating. This loss of isolation allows the Morse indices
of the labeled fixed points to jump, since continuation invariance holds only while an isolating neighborhood persists. An observer far
from the fixed points can measure only the total index, and the total index does not see the transition. Only the labeled Morse indices do, and this pattern recurs in every example
below.

\paragraph{Time-dependent parameter.}
In many physical situations the control parameter itself runs. In a film of finite thickness the
effective dimension, and hence the analog of $\epsilon$, changes from three to two as the scale passes
the thickness (dimensional crossover), and in an expanding plasma the relevant couplings depend on the
temperature, which decreases in time. To model this, let $\alpha=\alpha(z)$ be time dependent, for instance through $d\alpha/dz=\beta(\alpha)$, with
limits $\alpha_\pm$ as $z\to\infty$ and $z\to0$ respectively. In the time $t=\log z$ the general
solution follows by quadrature,
\begin{equation}
\frac{X-1}{X+1}=K\,e^{-A(t)},\qquad A(t)=\int_0^t\alpha(t')\,dt',
\end{equation}
and $X\equiv\pm1$ are solutions for every $\alpha(\cdot)$. Suppose $\alpha$ is asymptotically constant
with $\alpha_{\rm UV}>0$. Then generic solutions start at the UV equilibrium that is repelling for the
UV limit ($X=-1$), and they end at the equilibrium that is attracting for the IR limit. The IR fixed
point is therefore determined by $\operatorname{sign}\lim_{z\to\infty}\alpha(z)$ even when the flow
starts from the same UV fixed point, while the universality class at intermediate $z$ can change as
$\alpha$ crosses $\alpha_c$. The same discussion applies to the UV end.

This is a parameter shift in the sense of Ref.~\onlinecite{Ashwin:2017}. The vector field is
$f(t,X)=-\frac12\alpha(t)(X^2-1)$, so $\partial_tf=-\frac12\alpha'(t)(X^2-1)$, and by
Proposition~\ref{prop:Hcrit}(i) the system satisfies Hypothesis~\ref{hyp:H} whenever $\alpha'(t)$ decays
exponentially as $t\to\pm\infty$. For $d\alpha/dz=\beta(\alpha)$ this holds if $\alpha_+$ is a zero of
finite order of $\beta$, since $d\alpha/dt=z\beta(\alpha)$ then decays like a power of $z=e^t$ as
$z\to\infty$, while $d\alpha/dt=O(e^t)$ as $t\to-\infty$. The dimension of the family of generic solutions,
$1=1-0$, is the Fredholm index of Proposition~\ref{prop:palmer}, and Theorems~\ref{thm:transport}
and~\ref{thm:rate} describe when UV$\to$IR connections are forced independently of how fast $\alpha$
varies. In the language of the crossover, the infrared universality class is fixed by the asymptotic
value of the running parameter alone, while the route taken at intermediate scales depends on the
whole history $\alpha(\cdot)$. The theory of Sec.~\ref{sec:theory} is designed to separate these two
kinds of information.

%======================================================================
\section{Background on the Conley index}\label{sec:background}
%======================================================================

We collect the standard facts used below (see Refs.~\onlinecite{Conley:1978,Salamon:1985,Mischaikow:2002}).
Throughout, $\psi$ is a (local) flow on a locally compact metric space $Y$.

\subsection{Isolated invariant sets and index pairs}

For compact $N\subset Y$ let $\Inv(N)\coloneqq\{y\in N:\psi(\R,y)\subset N\}$. $N$ is an
\emph{isolating neighborhood} if $\Inv(N)\subset\operatorname{int}N$, and $S$ is an \emph{isolated
invariant set} if $S=\Inv(N)$ for such an $N$. An \emph{index pair} for $S$ is a compact pair
$L\subset N$ such that (i) $\cl(N\setminus L)$ is an isolating neighborhood of $S$ and $N\setminus L$
is a neighborhood of $S$, (ii) $L$ is positively invariant relative to $N$, i.e.\ if $y\in L$ and
$\psi([0,t],y)\subset N$ then $\psi([0,t],y)\subset L$, and (iii) $L$ is an exit set, i.e.\ if $y\in N$ and
$\psi(t,y)\notin N$ for some $t>0$, then there is $s\in[0,t]$ with $\psi([0,s],y)\subset N$ and
$\psi(s,y)\in L$. Index pairs exist, and the pointed homotopy type
\begin{equation}
h(S)=[N/L,[L]]
\end{equation}
is independent of the choice~\cite{Conley:1978,Salamon:1985}. The homology index is
$\CH_*(S)=\tilde H_*(N/L)$, and the cohomology index $\CH^*(S)$ is defined with Alexander--Spanier
cohomology. A hyperbolic equilibrium with $k$-dimensional unstable manifold has $h=S^k$ and
$\CH_*(\,\cdot\,;\Lambda)=\Lambda[k]$ for every coefficient ring $\Lambda$, i.e.\ $\Lambda$ in degree $k$ and $0$ otherwise. We write $\bar0$ for the homotopy type of a pointed one-point space.

Index pairs are conveniently produced from \emph{isolating blocks}~\cite{Conley:1971}, namely compact sets $N$ whose
boundary decomposes into entrance and exit points, meeting along tangency loci at which grazing orbits
touch $\partial N$ from outside (external tangencies). For a block, $(N,N^-)$ with $N^-$ the exit set is
an index pair. Internal tangencies, at which an orbit inside $N$ grazes $\partial N$ and returns,
violate condition~(ii).

The reduced Euler characteristic $\tilde\chi(h(S))\coloneqq\sum_k(-1)^k\operatorname{rank}\CH_k(S)$
is additive over Morse decompositions. For a set whose Morse sets are hyperbolic equilibria with Morse
indices $m_i$, it equals $\sum_i(-1)^{m_i}$.

\subsection{Morse decompositions, connection matrices and continuation}

A \emph{Morse decomposition} of $S$ is a finite family of disjoint isolated invariant subsets
$\{M_i\}$, indexed by a partial order, such that every $y\in S\setminus\bigcup M_i$ has $\alpha(y)\subset
M_i$ and $\omega(y)\subset M_j$ with $i>j$. For an \emph{attractor--repeller pair} $(A,R)$ there is a
long exact sequence
\begin{equation}\label{eq:AR}
\cdots\to\CH_k(A)\to\CH_k(S)\to\CH_k(R)\xrightarrow{\ \partial\ }\CH_{k-1}(A)\to\cdots.
\end{equation}
If there are no connecting orbits from $R$ to $A$ in $S$, then $S=A\sqcup R$, $h(S)=h(A)\vee h(R)$
and $\partial=0$. The connecting homomorphism is natural under flow-equivalences and invariant under
continuation~\cite{Franzosa:1989,Salamon:1985}. More generally a \emph{connection
matrix}~\cite{Franzosa:1989} is a degree $-1$, strictly upper triangular boundary operator $\Delta$ on
$\bigoplus_i\CH_*(M_i)$ whose homology is $\CH_*(S)$ and whose restriction to any interval $I$ of the
order is a connection matrix for the corresponding set $M(I)$. A nonzero entry between Morse sets that
are adjacent in the flow-defined order forces a connecting orbit, while entries between Morse sets of equal
index vanish for degree reasons. For hyperbolic equilibria of indices $k+1$ and $k$ whose invariant
manifolds intersect transversally, $\partial$ counts the connecting orbits with orientation
signs~\cite{McCord:1988}.

For RG flows these notions have a direct meaning. When the flow is gradient-like, a Lyapunov function
such as a $c$- or $F$-function orders the fixed points, and the Morse decomposition by fixed points with
this order is the familiar hierarchy UV $>$ IR. The Morse index of a fixed point is the number of
relevant directions, i.e., the number of couplings that must be tuned to reach it. A nonzero entry of the
connection matrix then guarantees an RG trajectory between two fixed points without solving the flow,
and the attractor--repeller sequence~\eqref{eq:AR} says how the index of a whole region of coupling space
is assembled from its UV and IR parts.

\paragraph{Continuation, products and duality.}
If $\psi^\lambda$, $\lambda\in[0,1]$, is a continuous family of flows and $N$ is an isolating
neighborhood for every $\psi^\lambda$, then $h(\Inv(N,\psi^\lambda))$ is independent of $\lambda$, and
the same holds for attractor--repeller pairs and their connecting
homomorphisms~\cite{Conley:1978,Salamon:1985,Franzosa:1989}. For a product flow~\cite{Conley:1978},
$h(S_1\times S_2)=h(S_1)\wedge h(S_2)$. For a flow on an $n$-dimensional manifold,
McCord's duality theorem~\cite{McCord:1992dual} gives, with $\F$ coefficients (or $\Z$ coefficients
if the manifold is orientable),
\begin{equation}\label{eq:duality}
\CH_k(S,\psi)\cong\CH^{n-k}(S,\psi^{-1}).
\end{equation}
Continuation makes these invariants robust. A change of scheme, a small change of a parameter,
or the addition of higher-order terms to a beta function does not change the index as long as the
region of coupling space under consideration stays isolated, i.e., as long as no fixed point or
connecting orbit enters or leaves through its boundary.

\subsection{Symmetries and equivariant cohomology}\label{subsec:equiv}

Let a finite group $G$ act on $Y$ by homeomorphisms commuting with $\psi$, and let $S$ be a
$G$-invariant isolated invariant set. Then $S$ admits $G$-invariant index pairs, the $G$-homotopy type
of $N/L$ is an invariant of $S$, and it is invariant under $G$-equivariant continuation. This is the
equivariant refinement of the Conley index due to Floer~\cite{Floer:1987} (see
also Ref.~\onlinecite{Bartsch:1993}). For a subgroup $H$ the fixed-point set $(N/L)^H$ is the index of $S\cap Y^H$
for the restricted flow on $Y^H$, and when $G$ acts freely off its fixed points the orbit space
$(N/L)/G$ is the index of $S/G$ for the induced flow on $Y/G$. For the involutions used in this paper
both facts are verified directly in the proof of Theorem~\ref{thm:equivariant}.

Finally, a \emph{time-reversal symmetry}~\cite{Lamb:1998} is an involution $\mathcal S$ with
$\mathcal S\circ\psi_t=\psi_{-t}\circ\mathcal S$.

%======================================================================
\section{Conley index theory for asymptotically autonomous systems}\label{sec:theory}
%======================================================================

Throughout this section Hypothesis~\ref{hyp:H} holds, $\Ss=X\times[-1,1]$ carries the flow
\eqref{eq:slab}, and $X_\pm=X\times\{\pm1\}$.

\subsection{The double and the one-sided extension}\label{subsec:double}

\begin{lemma}\label{lem:G}
The function $G(\theta)\coloneqq-g(\cos\theta)/\sin\theta$, $\theta\notin\pi\Z$, extends to an odd
$C^1$ function on $S^1=\R/2\pi\Z$. It satisfies $G<0$ on $(0,\pi)$, $G>0$ on $(-\pi,0)$,
$G(0)=G(\pi)=0$, $G'(0)=g'(1)/2$ and $G'(\pi)=g'(-1)/2$.
\end{lemma}

\begin{proof}
$h(\theta)=g(\cos\theta)$ is even, of class $C^3$, and satisfies $h(0)=h'(0)=0$. By Taylor's formula
with integral remainder, $h(\theta)=\theta^2k(\theta)$ with $k\in C^1$ and $k(0)=h''(0)/2=-g'(1)/2$.
Hence $G(\theta)=-(\theta/\sin\theta)\,\theta\,k(\theta)$ is $C^1$ near $0$ with $G'(0)=g'(1)/2$. The
point $\theta=\pi$ is treated in the same way. Oddness and the signs are immediate.
\end{proof}

\begin{definition}[Double]
The \emph{double} is $\Dd\coloneqq X\times S^1$ with the flow
\begin{equation}\label{eq:double}
\dot x = F(\cos\theta,x),\qquad \dot\theta = G(\theta),
\end{equation}
together with the \emph{reflection} $R(x,\theta)=(x,-\theta)$. The open \emph{upper} and
\emph{lower sheets} are $X\times(0,\pi)$ and $X\times(-\pi,0)$, and $R$ exchanges them.
\end{definition}

For $g(\tau)=1-\tau^2$ one simply has $G(\theta)=-\sin\theta$. By Lemma~\ref{lem:G} and
Hypothesis~\ref{hyp:H}, the vector field~\eqref{eq:double} is of class $C^1$, so the double is a $C^1$
version of gluing two copies of $\Ss$ along $X_\pm$. This is all the regularity used below. If
$F\in C^r$ and $g\in C^{r+2}$ for some $r\ge1$, the same argument gives $k\in C^r$ in the proof of
Lemma~\ref{lem:G}, and the double is of class $C^r$. The normal eigenvalues at the boundary fibers are halved,
$g'(\pm1)\mapsto g'(\pm1)/2$, which does not affect any index.

\begin{lemma}\label{lem:double}
(a) The map $q:\Dd\to\Ss$, $q(x,\theta)=(x,\cos\theta)$, is a semiconjugacy of~\eqref{eq:double}
onto~\eqref{eq:slab}, and its restriction to $X\times[0,\pi]$ is a homeomorphism that conjugates the flows.
(b) $R$ commutes with the flow, $\operatorname{Fix}(R)=X\times\{0\}\sqcup X\times\{\pi\}$ is mapped by
$q$ onto $X_+\sqcup X_-$, where the flows are $f_+$ and $f_-$, $R$ acts freely off $\operatorname{Fix}(R)$,
and $q$ induces a flow-equivalence $\Dd/R\cong\Ss$.
\end{lemma}

\begin{proof}
On $(0,\pi)$ one has $\frac{d}{dt}\cos\theta=-\sin\theta\,G(\theta)=g(\cos\theta)$. Both sides vanish
at $\theta\in\{0,\pi\}$, and $\cos:[0,\pi]\to[-1,1]$ is a homeomorphism. Hence the base flow of the double
is determined by that of $\dot\tau=g(\tau)$. Along an orbit, $\cos\theta(t)$ is the solution of
$\dot\tau=g(\tau)$ through $\cos\theta(0)$, and the sign of $\sin\theta$ is constant. Part (b) follows from $G$
being odd and $\cos$ being even. Two points have the same image under $q$ if and only if they are equal or
exchanged by $R$.
\end{proof}

\begin{figure}[t]
\centering
\begin{tikzpicture}[>=Stealth,scale=1.0,line width=1.3pt]
\begin{scope}
 \draw (0,0) -- (3,0);
 \draw[->,line width=1.6pt] (1.3,0) -- (1.7,0);
 \fill (0,0) circle (2.8pt); \fill (3,0) circle (2.8pt);
 \node[above] at (0,0.1) {UV}; \node[below] at (0,-0.1) {$\tau=-1$};
 \node[above] at (3,0.1) {IR}; \node[below] at (3,-0.1) {$\tau=+1$};
 \node at (1.5,-1.6) {(a) base of $\Ss$};
\end{scope}
\begin{scope}[xshift=6cm]
 \draw (0,0) circle (1);
 \fill (-1,0) circle (2.8pt) node[left=3pt] {$\theta=\pi$};
 \fill (1,0) circle (2.8pt) node[right=3pt] {$\theta=0$};
 \draw[->,line width=1.6pt] (100:1) arc (100:80:1);
 \draw[->,line width=1.6pt] (260:1) arc (260:280:1);
 \draw[dashed,black!55,line width=1pt] (-1,0) -- (1,0);
 \node at (0,1.35) {\small upper sheet};
 \node at (0,-1.35) {\small lower sheet};
 \node at (0,-1.9) {(b) base of the double $\Dd$};
\end{scope}
\begin{scope}[xshift=11.5cm]
 \draw[black!45,line width=1pt] (0,0) circle (1);
 \draw[red!70!black,line width=3pt] (-15:1) arc (-15:195:1);
 \fill (-1,0) circle (2.8pt); \fill (1,0) circle (2.8pt);
 \node at (0,1.35) {$A_\eta=[-\eta,\pi+\eta]$};
 \node at (0,-1.9) {(c) one-sided extension};
\end{scope}
\end{tikzpicture}
\caption{(a) The base of the extended phase space $\Ss$. The UV and IR fibers are invariant and the base flow runs from UV to
IR. (b) The base of the double, $\dot\theta=G(\theta)$. The reflection $R$ is the mirror image in the
dashed line and fixes the UV and IR fibers. (c) The arc used to isolate the one-sided extension $S^\osx$.}
\label{fig:double}
\end{figure}
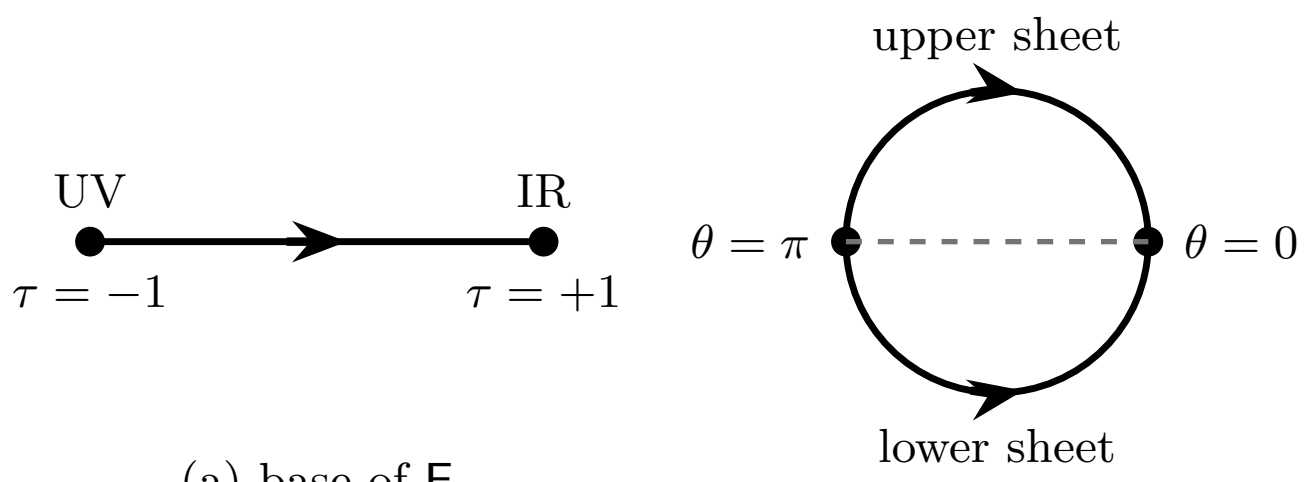

Let $S\subset\Ss$ be an isolated invariant set, and write $S\cap X_\pm=K_\pm\times\{\pm1\}$. Put
\begin{equation}
\tilde S\coloneqq q^{-1}(S)\subset\Dd,\qquad S^\osx\coloneqq\tilde S\cap\big(X\times[0,\pi]\big).
\end{equation}
We call $S^\osx$ the \emph{one-sided extension} of $S$. It is a copy of $S$ in which the UV fiber has
become two-sided (Figs.~\ref{fig:double} and~\ref{fig:onesided}).

\begin{lemma}\label{lem:isolation}
(a) $K_\pm$ is an isolated invariant set of $f_\pm$ in $X$.
(b) $\tilde S$ and $S^\osx$ are isolated invariant sets of $\Dd$, and $\tilde S$ is $R$-invariant.
(c) $(K_+\times\{0\},K_-\times\{\pi\})$ is an attractor--repeller pair for both $S^\osx$ and $\tilde S$.
Every orbit of $S^\osx$ with $\theta\in(0,\pi)$ has $\alpha$-limit in $K_-\times\{\pi\}$ and
$\omega$-limit in $K_+\times\{0\}$, and it projects under $q$ to a solution of~\eqref{eq:nonaut}
defined for all $t\in\R$, with $\operatorname{dist}(x(t),K_\mp)\to0$ as $t\to\mp\infty$.
\end{lemma}

\begin{proof}
(a) Let $N$ isolate $S$ and $N_\pm\coloneqq\{x:(x,\pm1)\in N\}$. Since $X_\pm$ is invariant,
$\Inv_{f_\pm}(N_\pm)\times\{\pm1\}=S\cap X_\pm$, and $\operatorname{int}_{\Ss}N\cap X_\pm\subset
\operatorname{int}_X(N_\pm)\times\{\pm1\}$.
(b) $q$ is proper and maps orbits onto orbits, so $\Inv(q^{-1}N)=q^{-1}(\Inv N)=\tilde S\subset
q^{-1}(\operatorname{int}N)\subset\operatorname{int}q^{-1}(N)$. For $S^\osx$, fix $\eta\in(0,\pi/2)$,
let $A_\eta=[-\eta,\pi+\eta]\subset S^1$ be the arc through $\pi/2$, and put $N^\osx\coloneqq q^{-1}(N)
\cap(X\times A_\eta)$. A full orbit in $N^\osx$ lies in $\tilde S$. If it met the open lower sheet,
its $\theta$-coordinate would sweep the whole interval $(-\pi,0)$, which is not contained in $A_\eta$.
Hence $\Inv(N^\osx)=S^\osx$, and $S^\osx\subset\operatorname{int}N^\osx$ because $0$ and $\pi$ are
interior points of $A_\eta$.
(c) On the open sheets $\theta$ is strictly monotone and tends to $\pi$ and $0$ as $t\to-\infty$ and
$t\to+\infty$ respectively. Limit sets of orbits in a compact invariant set are nonempty and
invariant, so they lie in the fixed fibers. The last claim follows from Lemma~\ref{lem:double}(a).
\end{proof}

\subsection{Boundary indices and the infrared}

We first compute the indices of sets that lie entirely in a boundary fiber, i.e., of the frozen UV and IR
fixed points and their invariant sets, and then show that the index of the extended phase space by itself
retains only the infrared data.

\begin{proposition}[Boundary indices]\label{prop:boundary}
Let $K\subset X$ be an isolated invariant set of $f_+$, respectively $f_-$. Then
\begin{equation}
\begin{aligned}
h_{\Ss}(K\times\{+1\})&\cong h_{f_+}(K), & h_{\Dd}(K\times\{0\})&\cong h_{f_+}(K),\\
h_{\Ss}(K\times\{-1\})&\cong\bar0, & h_{\Dd}(K\times\{\pi\})&\cong\Sigma\,h_{f_-}(K),
\end{aligned}
\end{equation}
where $\Sigma$ denotes the reduced suspension. Moreover $R_*=+1$ on $\CH_*(K\times\{0\})$ and $R_*=-1$ on
$\CH_*(K\times\{\pi\})\cong\CH_{*-1}(K;f_-)$, with coefficients in any ring.
\end{proposition}

\begin{proof}
Consider the IR fiber of the double, the other cases being identical. Let $N_K$ isolate $K$ for $f_+$,
and consider the family of flows $\dot x=F(\cos\lambda\theta,x)$, $\dot\theta=G(\theta)$,
$\lambda\in[0,1]$, on $N_K\times[-\eta,\eta]$. Since the $\theta$-equation does not involve $x$ and $0$
is its only zero in $[-\eta,\eta]$, a full orbit in $N_K\times[-\eta,\eta]$ has $\theta\equiv0$, where
the $x$-flow is $f_+$ for every $\lambda$. Hence $N_K\times[-\eta,\eta]$ isolates $K\times\{0\}$
throughout the family, and by continuation the index equals that of the product flow at $\lambda=0$.
By the product formula this is $h_{f_+}(K)\wedge h_G(\{0\})=h_{f_+}(K)\wedge S^0=h_{f_+}(K)$, since
$0$ is an attractor of $\dot\theta=G$. At $\theta=\pi$ the point $\pi$ is a repeller, with index $S^1$,
so one obtains $h_{f_-}(K)\wedge S^1=\Sigma h_{f_-}(K)$. In $\Ss$, $\tau=+1$ is an attractor of
$\dot\tau=g$ on $[1-\eta,1]$ (index $S^0$). The point $\tau=-1$ in $[-1,-1+\eta]$ has index pair
$([-1,-1+\eta],\{-1+\eta\})$, which is contractible, so the index is $\bar0$.

The family is $R$-equivariant because $\cos\lambda\theta$ is even in $\theta$, and so is the family
used near $\pi$, which is built from $\cos(\pi+\lambda(\theta-\pi))$. By naturality of continuation it
suffices to compute $R_*$ for the product flow. Near $\theta=0$, $R$ reflects the attracting interval,
which is homotopic to the identity as a map of index pairs. Near $\theta=\pi$ the index pair is
$(N_K\times J,\,L_K\times J\cup N_K\times\partial J)$ with $J=[\pi-\eta,\pi+\eta]$. Here $R$ acts by
the reflection of $(J,\partial J)$, which has degree $-1$, and the claim follows from the K\"unneth
formula.
\end{proof}

For a hyperbolic equilibrium $e$ with fiber index $\mu(e)$ this gives $S^{\mu(e)}$ in the IR, in both
$\Ss$ and the double, and in the UV it gives $\bar0$ in $\Ss$ and $S^{\mu(e)+1}$ in the double.

The extra unstable direction of a UV point in the double has a simple physical meaning. A solution that
leaves a UV fixed point is specified not only by the direction in which it leaves, i.e., by the values of
the $\mu(e)$ relevant couplings, but also by the moment at which it leaves, since the equations depend
explicitly on time and there is no time-translation symmetry to remove this parameter. The suspension
counts this departure time as one more relevant direction. In $\Ss$, by contrast, the UV fiber can only
be left toward later times, and its one-sided index is trivial.

\begin{remark}[The boundary isomorphisms]\label{rem:boundary-iso}
At the UV fiber, the proof of Proposition~\ref{prop:boundary} gives the commutative diagram
\begin{equation}\label{eq:UVsusp}
\begin{tikzcd}[column sep=large]
h_{f_-}(K_-)\arrow[r,"-\wedge S^1"]\arrow[d,"\tilde H_k"'] &
\Sigma h_{f_-}(K_-)\arrow[r,"\simeq"]\arrow[d,"\tilde H_{k+1}"'] &
h_{\Dd}(K_-\times\{\pi\})\arrow[d,"\tilde H_{k+1}"]\\
\CH_k(K_-;f_-)\arrow[r,"\cong"] &
\tilde H_{k+1}(\Sigma h_{f_-}(K_-))\arrow[r,"\cong"] &
\CH_{k+1}(K_-\times\{\pi\})
\end{tikzcd}
\end{equation}
The first arrow is the product with the index $S^1$ of the repeller $\pi$ of $\dot\theta=G(\theta)$, and
the second is the continuation from the product flow on $N_K\times[\pi-\eta,\pi+\eta]$ to the
flow~\eqref{eq:double} of the double. At the IR fiber the attractor $0$ of $\dot\theta=G(\theta)$ has
index $S^0$, and the same construction gives $h_{f_+}(K_+)\wedge S^0=h_{f_+}(K_+)\simeq
h_{\Dd}(K_+\times\{0\})$ without a shift of degree. We denote the resulting isomorphisms by
\[
s_k:\CH_k(K_-;f_-)\xrightarrow{\cong}\CH_{k+1}(K_-\times\{\pi\}),\qquad
a_k:\CH_k(K_+;f_+)\xrightarrow{\cong}\CH_k(K_+\times\{0\}).
\]
\end{remark}

\begin{theorem}[Index in the extended phase space]\label{thm:slab}
Let $S\subset\Ss$ be an isolated invariant set. Then $(K_+\times\{1\},K_-\times\{-1\})$ is an
attractor--repeller pair of $S$, and the inclusion of the attractor induces an isomorphism
\begin{equation}
\CH_*(K_+;f_+)\xrightarrow{\ \cong\ }\CH_*(S).
\end{equation}
In particular $\tilde\chi(h_{\Ss}(S))=\tilde\chi(h_{f_+}(K_+))$, i.e.\ the index in the extended phase space
contains no information about $K_-$ or about connecting solutions.
\end{theorem}

\begin{proof}
The function $\tau$ is strictly increasing along orbits with $|\tau|<1$, so the argument of
Lemma~\ref{lem:isolation}(c) shows that the pair is an attractor--repeller decomposition. By
Proposition~\ref{prop:boundary}, $\CH_*(K_-\times\{-1\})=0$ and $\CH_*(K_+\times\{1\})\cong
\CH_*(K_+;f_+)$, and the exact sequence~\eqref{eq:AR} gives the claim.
\end{proof}

Theorem~\ref{thm:slab} is a topological form of infrared universality. The naive enlarged phase space,
with time adjoined as a coordinate, carries exactly the index of the late-time limit flow,
independently of the UV data and of the dynamics at intermediate times. This is the situation familiar from hydrodynamic
attractors, where the late-time behavior is independent of the initial conditions. The index of the extended phase
space is therefore insensitive to the questions raised in Sec.~\ref{subsec:questions}, which concern the connections
between UV and IR fixed points, and the double and the one-sided extension are needed to restore the UV information.

\subsection{The UV\texorpdfstring{$\to$}{->}IR transport map}

\begin{theorem}[Transport map]\label{thm:transport}
Let $S\subset\Ss$ be an isolated invariant set. The attractor--repeller pair of
Lemma~\ref{lem:isolation}(c) in the one-sided extension $S^\osx$, combined with
Proposition~\ref{prop:boundary}, defines homomorphisms $\Psi_k:\CH_k(K_-;f_-)\to\CH_k(K_+;f_+)$ by the
commutative square
\begin{equation}\label{eq:transport-square}
\begin{tikzcd}[column sep=huge]
\CH_k(K_-;f_-)\arrow[r,"\Psi_k"]\arrow[d,"s_k"',"\cong"] &
\CH_k(K_+;f_+)\arrow[d,"a_k","\cong"']\\
\CH_{k+1}(K_-\times\{\pi\})\arrow[r,"\partial_{k+1}"'] & \CH_k(K_+\times\{0\})
\end{tikzcd}
\end{equation}
i.e.\ $\Psi_k=a_k^{-1}\circ\partial_{k+1}\circ s_k$, where $s_k$ and $a_k$ are the boundary isomorphisms
of Remark~\ref{rem:boundary-iso} and $\partial_{k+1}$ is the connecting homomorphism of~\eqref{eq:AR}.
The connecting homomorphism has degree $-1$ in the grading of the indices of $S^\osx$, but since the
UV index is suspended once, $\Psi$ has degree $0$ in the grading of the frozen systems. These maps fit into
the exact sequence
\begin{multline}\label{eq:transport-les}
\cdots\to\CH_k(K_-;f_-)\xrightarrow{\Psi_k}\CH_k(K_+;f_+)\to\CH_k(S^\osx)\\
\to\CH_{k-1}(K_-;f_-)\xrightarrow{\Psi_{k-1}}\CH_{k-1}(K_+;f_+)\to\cdots.
\end{multline}
Moreover, the following hold.
\begin{enumerate}
\item[(i)] with field coefficients, $\CH_k(S^\osx)\cong\operatorname{coker}\Psi_k\oplus\ker\Psi_{k-1}$,
\item[(ii)] if $\Psi\neq0$, then~\eqref{eq:nonaut} has a solution $x:\R\to X$ with
$\{(x(t),\sigma(t))\}\subset S$ and $\operatorname{dist}(x(t),K_\mp)\to0$ as $t\to\mp\infty$, whichever index triple is
used to define $\Psi$,
\item[(iii)] with field coefficients and finite dimensional $\CH_*(K_\pm;f_\pm)$, one has
\begin{equation}
\begin{aligned}
\operatorname{rk}\Psi_0&=\dim\CH_0(K_+;f_+)-\dim\CH_0(S^\osx),\\
\operatorname{rk}\Psi_k&=\dim\CH_k(K_+;f_+)+\dim\CH_{k-1}(K_-;f_-)-\operatorname{rk}\Psi_{k-1}
-\dim\CH_k(S^\osx),
\end{aligned}
\end{equation}
where the second identity holds for $k\geq1$.
Hence $\Psi$ is determined up to isomorphism by the indices of $K_\pm$ and $S^\osx$, and it is invariant
under continuation of $(F,g)$ as long as $S^\osx$ and $K_\pm$ continue, i.e.\ as long as there are
common isolating neighborhoods.
\end{enumerate}
\end{theorem}

\begin{proof}
Lemma~\ref{lem:isolation} provides the attractor--repeller pair, and Proposition~\ref{prop:boundary}
identifies the indices as $\CH_{k+1}(K_-\times\{\pi\})\cong\CH_k(K_-;f_-)$ and
$\CH_k(K_+\times\{0\})\cong\CH_k(K_+;f_+)$. Substituting these into~\eqref{eq:AR}
gives~\eqref{eq:transport-les}, and (i) is linear algebra.

Call an index pair $(N,L)$ regular if $L$ has a flow collar in $N$, i.e.\ a neighborhood in $N$ whose
points reach $L$ within a time $\leq1$ that depends continuously on the point. Every index pair $(N,L)$
can be enlarged to a regular index pair $(N,L')$ with $L\subset L'$. For (ii), let
$N_0\subset N_1\subset N_2$ be an index triple for the pair, with $(N_1,N_0)$ regular, and let
$\partial:H_{k+1}(N_2,N_1)\to H_k(N_1,N_0)$ be its connecting map. Suppose that there are no connecting
orbits, so that $S^\osx=A\sqcup U$ with $A=K_+\times\{0\}$ and $U=K_-\times\{\pi\}$. Then $(U,A)$ is also an attractor--repeller pair, and refining the index pair $(N_2,N_0)$ for it gives
$N_0\subset M_1\subset N_2$ such that $(N_2,M_1)$ is an index pair for $A$. Enlarging it to a regular
pair $(N_2,M_1')$ gives an index pair for $A$ that contains $(N_1,N_0)$. For nested index pairs
$(N,L)\subset(N',L')$ of the same isolated invariant set, the map $N/L\to N'/L'$ induced by the
inclusion is a pointed homotopy equivalence, because composing it with the flow for a long time $3T$
in $N'/L'$ gives Salamon's comparison map~\cite{Salamon:1985}. For regular pairs, the projection induces
an isomorphism $H_*(N,L)\cong H_*(N/L,[L])$, natural with respect to inclusions, so
$H_*(N_1,N_0)\to H_*(N_2,M_1')$ is an isomorphism. Hence $H_*(N_1,N_0)\to H_*(N_2,N_0)$ is injective, and
$\partial=0$ by exactness. This holds for every such index triple, so (ii) does not depend on the
triple used to define $\Psi$. The connecting orbits project to solutions of~\eqref{eq:nonaut} by
Lemma~\ref{lem:isolation}(c).

For (iii), (i) and rank--nullity give
\[
\dim\CH_k(S^\osx)=\big(\dim\CH_k(K_+)-\operatorname{rk}\Psi_k\big)+\big(\dim\CH_{k-1}(K_-)-\operatorname{rk}\Psi_{k-1}\big)
\]
for $k\geq1$ and $\dim\CH_0(S^\osx)=\dim\CH_0(K_+)-\operatorname{rk}\Psi_0$, which is the recursion. A linear map between finite dimensional vector spaces
is determined up to isomorphism by its rank, and the indices of $K_\pm$ and $S^\osx$ are invariant under
continuation. Over $\Z$, the invariance of $\Psi$ itself follows from the continuation invariance of
connecting homomorphisms~\cite{Franzosa:1989,Salamon:1985}, which is not used below.
\end{proof}

Equivalently,
one may keep the grading of $S^\osx$ and regard the transport map as the connecting homomorphism
$\partial_k:\CH_k(K_-\times\{\pi\})\to\CH_{k-1}(K_+\times\{0\})$ itself, of degree $-1$ as
in~\eqref{eq:AR}. The two conventions are related by $\Psi_{k-1}=a_{k-1}^{-1}\circ\partial_k\circ s_{k-1}$. We
use the grading of the frozen systems because it compares the UV and IR indices directly.

\begin{remark}[The transport map as a connecting homomorphism]\label{rem:status}
The map $\Psi$ is the connecting homomorphism of the attractor--repeller pair $(K_+\times\{0\},
K_-\times\{\pi\})$ of an autonomous flow on $X\times A_\eta$, composed with the boundary isomorphisms $s_k$
and $a_k$ of Remark~\ref{rem:boundary-iso}. Its content beyond the classical theory of connecting
homomorphisms~\cite{Conley:1978,Franzosa:1989,Salamon:1985} lies in the explicit construction of this pair
from the non-autonomous equation and in the grading by the indices of the UV and IR limit systems. J\"anig's
nonautonomous index also has a connecting homomorphism for attractor--repeller
decompositions~\cite{Janig:2019hom,Janig:2019conn}, defined over a cycle that contains a reversed
history~\cite{Janig:2020}. We do not know whether the two maps agree under hypotheses common to both
settings, and nothing below depends on such an identification.
\end{remark}

\begin{figure}[t]
\centering
\begin{tikzpicture}[>=Stealth,scale=1.0]
 \fill[gray!25] (-1,-1.7) rectangle (0,1.7);
 \fill[gray!25] (7,-1.7) rectangle (8,1.7);
 \draw[dashed,line width=1pt] (-1,-1.7) -- (-1,1.7);
 \draw[dashed,line width=1pt] (8,-1.7) -- (8,1.7);
 \draw[line width=1.6pt] (0,-1.7) -- (0,1.7);
 \draw[line width=1.6pt] (7,-1.7) -- (7,1.7);
 \node[above,align=center,font=\small] at (0,1.7) {$\theta=\pi$\\ UV fiber, flow $f_-$};
 \node[above,align=center,font=\small] at (7,1.7) {$\theta=0$\\ IR fiber, flow $f_+$};
 \node[above,align=center,font=\small] at (3.5,1.7) {upper sheet $0<\theta<\pi$\\ solutions of~\eqref{eq:nonaut}};
 \draw[red!70!black,line width=1.3pt,fill=red!30] (0,0) ellipse (0.4 and 1.0);
 \draw[blue!70!black,line width=1.3pt,fill=blue!25] (7,0) ellipse (0.4 and 0.8);
 \node[fill=red!30,inner sep=1pt,font=\small] at (0,0) {$K_-$};
 \node[fill=blue!25,inner sep=1pt,font=\small] at (7,0) {$K_+$};
 \draw[->,line width=1.4pt] (0.4,0.6) .. controls (2.5,1.35) and (4.5,1.1) .. (6.62,0.42);
 \draw[->,line width=1.4pt] (0.42,0.1) .. controls (2.5,0.25) and (4.5,-0.1) .. (6.6,0.0);
 \draw[->,line width=1.4pt] (0.4,-0.6) .. controls (2.5,-1.3) and (4.5,-1.05) .. (6.62,-0.42);
 \foreach \y in {-1.2,1.2} {
   \draw[->,black!75,line width=1.2pt] (-0.25,\y) -- (-0.8,\y);
   \draw[->,black!75,line width=1.2pt] (7.8,\y) -- (7.3,\y);
 }
 \node[left,font=\small,align=right] at (-1.05,0) {exit\\ $\theta=\pi+\eta$};
 \node[right,font=\small,align=left] at (8.05,0) {entrance\\ $\theta=-\eta$};
 \node[below,font=\small,align=center] at (0,-1.75) {repeller $K_-\times\{\pi\}$\\ index $\Sigma h_{f_-}(K_-)$};
 \node[below,font=\small,align=center] at (7,-1.75) {attractor $K_+\times\{0\}$\\ index $h_{f_+}(K_+)$};
 \node[font=\small] at (3.5,-3.25)
   {$\cdots\to\CH_k(K_-;f_-)\xrightarrow{\ \Psi_k\ }\CH_k(K_+;f_+)\to\CH_k(S^\osx)\to\CH_{k-1}(K_-;f_-)\to\cdots$};
\end{tikzpicture}
\caption{The one-sided extension $S^\osx$ and the transport map. The shaded strips are the parts of the
lower sheet contained in the arc $A_\eta$, where the base flow leaves through $\theta=\pi+\eta$ and enters
through $\theta=-\eta$. They make the UV fiber two-sided, so that $K_-\times\{\pi\}$ acquires the
suspended index, while the IR index is unchanged. The connecting homomorphism of this
attractor--repeller pair is the transport map $\Psi$, and $\Psi\neq0$ forces a connecting solution of the
non-autonomous equation (Theorem~\ref{thm:transport}).}
\label{fig:onesided}
\end{figure}

\begin{remark}[$\Psi$ in terms of the extended phase space]\label{rem:slabform}
Attaching to an index pair $(N,L)$ of $S$ in $\Ss$ a collar beyond the UV fiber produces an index
pair for $S^\osx$ whose quotient is the mapping cone of the inclusion $h_{f_-}(K_-)\hookrightarrow
h_{\Ss}(S)$. Combined with Theorem~\ref{thm:slab}, this identifies $\Psi$ (up to sign) with the map
induced by including the UV boundary index into the index in $\Ss$, followed by the isomorphism of
Theorem~\ref{thm:slab}. In other words, $\Psi$ measures how much of the UV index survives transport
along the physical time direction. We only sketch this identification, which is not used below.
\end{remark}

\begin{remark}[Transition matrices and continuation maps]\label{rem:transition}
When $F(\tau,\cdot)=F_{\lambda(\tau)}$ is a parameter shift, $\Dd$ restricted to a sheet is, up to
reparametrization, the extended system used by Reineck~\cite{Reineck:1988} and by
Franzosa--Mischaikow and McCord--Mischaikow~\cite{Franzosa:1998,McCord:1992tm} to define topological
transition matrices, with one difference, namely that here the rate of the shift is fixed by the physics rather than
sent to zero. If $K_\pm$ carry Morse decompositions, the connection matrix of $S^\osx$ with respect to
the combined decomposition has block form, i.e.\ its diagonal blocks are connection matrices for $f_+$ and
for $f_-$ (the latter shifted up by one degree), and its off-diagonal block is a transition matrix
refining $\Psi$. For gradient families $F_\lambda=-\nabla V_\lambda$, $\Psi$ is the Conley-index
analog of the continuation map of Morse homology, which counts index-zero solutions of the
non-autonomous gradient flow~\cite{Salamon:1990,Schwarz:1993}.
\end{remark}

\subsection{Rate independence}

\begin{theorem}[Common strict entrance/exit blocks]\label{thm:rate}
Let $N\subset X$ be a compact $n$-dimensional submanifold with boundary $\partial N=\partial^{\rm in}
\sqcup\partial^{\rm out}$ (either piece possibly empty). Suppose that this decomposition is the same for
all frozen systems, i.e.\ that, for every $\tau\in[-1,1]$, $F(\tau,\cdot)$ points strictly into $N$ on
$\partial^{\rm in}$ and strictly out of $N$ on $\partial^{\rm out}$. Then, for every $g$ as in
Hypothesis~\ref{hyp:H}, the following hold.
\begin{enumerate}
\item[(i)] $S=\Inv(N\times[-1,1])$ is isolated in $\Ss$, $K_\pm=\Inv(N,f_\pm)$, and
$\CH_*(K_\pm;f_\pm)\cong H_*(N,\partial^{\rm out})$,
\item[(ii)] $\CH_*(S^\osx)=0$, and $\Psi$ is an isomorphism,
\item[(iii)] if $H_*(N,\partial^{\rm out})\neq0$, then for every rate function $g$
equation~\eqref{eq:nonaut} has a solution defined for all $t$ that stays in $N$ and satisfies
$\operatorname{dist}(x(t),K_\mp)\to0$ as $t\to\mp\infty$.
\end{enumerate}
\end{theorem}

\begin{proof}
Points of $\partial N\times[-1,1]$ are strict entrance or exit points and cannot lie on full orbits in
$N\times[-1,1]$, which gives (i), and $(N,\partial^{\rm out})$ is an isolating block for each $f_\pm$. For
(ii), let $N^\osx=N\times A_\eta$ and $L^\osx=\partial^{\rm out}\times A_\eta\cup N\times\{\pi+\eta\}$.
On $N\times\{-\eta\}$ the base flow points into $A_\eta$, and on $N\times\{\pi+\eta\}$ it points out, since
$G>0$ on $(-\pi,0)$. The fiber component is strictly transverse on $\partial N\times A_\eta$, and at the
corners the exit condition holds. Hence $(N^\osx,L^\osx)$ is an index pair for $S^\osx$, and
\begin{equation}
N^\osx/L^\osx\cong\big(N/\partial^{\rm out}\big)\wedge\big(A_\eta/\{\pi+\eta\}\big)\simeq\bar0,
\end{equation}
because $A_\eta$ pointed at an end point is contractible. Exactness of~\eqref{eq:transport-les} then
forces $\Psi$ to be an isomorphism. Part (iii) follows from Theorem~\ref{thm:transport}(ii).
\end{proof}

Theorem~\ref{thm:rate} makes precise the robustness question raised in Sec.~\ref{subsec:questions}. As long as a single
block with a fixed strict entrance/exit decomposition of its boundary serves all frozen systems and
$H_*(N,\partial^{\rm out})\neq0$, UV$\to$IR connections at the level of $K_\pm$ cannot be destroyed by
driving the system faster or slower. The nonvanishing condition is essential. If
$H_*(N,\partial^{\rm out})=0$, then $\Psi$ is an isomorphism between zero groups and forces no
connection. The hypothesis is stronger than the existence of a common isolating neighborhood, because a common
isolating neighborhood alone gives continuation of $S^\osx$ in $g$, and hence invariance of $\Psi$ over a
field (Theorem~\ref{thm:transport}(iii)), but not the vanishing of $\CH_*(S^\osx)$.

The conclusion concerns the isolated invariant sets $K_\pm=\Inv(N,f_\pm)$ and not a particular branch of
equilibria. Theorem~\ref{thm:rate} says nothing about the Morse sets inside $N$, and rate-induced
tipping~\cite{Ashwin:2017} between them is compatible with $\Psi$ being an isomorphism. If
$\partial^{\rm out}=\emptyset$, the block is forward invariant for the non-autonomous system, so no
solution that enters $N$ leaves it at any rate. This excludes escape from $N$. It does not show that a
solution stays near a given branch of attractors of the frozen systems at intermediate times, which is
the tracking question of rate-induced tipping. Tracking needs additional information, for instance a
small block that moves with the branch (Corollary~\ref{cor:moving}).

Theorem~\ref{thm:rate} requires a block that is fixed in $X$. In the applications the natural blocks move
with the time, and the following extension of Theorem~\ref{thm:rate} is the form in which it is used.

\begin{corollary}[Moving blocks]\label{cor:moving}
Let $(x,\tau)\mapsto(\phi_\tau(x),\tau)$ be a homeomorphism of $\Ss$ onto $Y\times[-1,1]$, and let $N\subset Y$
be compact, with a closed subset $\partial^{\rm out}N\subset N$. Suppose that, in the coordinates
$(\xi,\tau)=(\phi_\tau(x),\tau)$ and for every $\tau\in[-1,1]$, points of $\partial^{\rm out}N$ leave $N$
immediately along the compactified flow, and all other boundary points of $N$ enter $N$ immediately and lie
outside $N$ just before. If $N/\partial^{\rm out}N$ is pointed homotopy equivalent to $S^k$, then
$S=\Inv\{(x,\tau):\phi_\tau(x)\in N\}$ is an isolated invariant set of $\Ss$,
$\CH_*(K_\pm;f_\pm;\Lambda)\cong\Lambda[k]$, $\CH_*(S^\osx)=0$, and $\Psi$ is an isomorphism, so $\Psi=\pm1$
with integer coefficients. Moreover, $S$ contains a solution of~\eqref{eq:nonaut} that runs from $K_-$ to
$K_+$.
\end{corollary}

\begin{proof}
The proof of Theorem~\ref{thm:rate} uses only that $N\times[-1,1]$ is an isolating block of the
compactified flow with exit set $\partial^{\rm out}N\times[-1,1]$, uniformly in $\tau$, and it applies
verbatim in the coordinates $(\xi,\tau)$. It gives that $S$ is isolated, that $(N,\partial^{\rm out}N)$
is an index pair for $K_\pm$, so that $\CH_*(K_\pm)\cong\tilde H_*(S^k;\Lambda)=\Lambda[k]$, that
$\CH_*(S^\osx)=0$ and that $\Psi$ is an isomorphism. An automorphism of $\Z$ is $\pm1$. The connecting
solution is Theorem~\ref{thm:rate}(iii).
\end{proof}

\subsection{Hyperbolic connecting solutions and the Fredholm index}\label{subsec:fredholm}

Let $A:\R\to\operatorname{End}(\R^n)$ be continuous and bounded. Let $C^0_b=C^0_b(\R,\R^n)$ be the space
of bounded continuous functions with $\|h\|_{C^0}=\sup_t|h(t)|$, and $C^1_b=C^1_b(\R,\R^n)$ the space of
$C^1$ functions $y$ with $y$ and $\dot y$ bounded, with $\|y\|_{C^1}=\max\big(\|y\|_{C^0},\|\dot
y\|_{C^0}\big)$. Then $Ly\coloneqq\dot y-A(t)y$ is a bounded operator $L:C^1_b\to C^0_b$, the
\emph{linearized operator} of $A$. A constant $A_0\in\operatorname{End}(\R^n)$ is \emph{hyperbolic} if
$\dot y=A_0y$ has an exponential dichotomy on $\R$, i.e.\ there are a projection $P$ that commutes with
$A_0$ and constants $K,\alpha>0$ with
\begin{equation}\label{eq:dichotomy}
\|e^{tA_0}P\|\le Ke^{-\alpha t},\qquad\|e^{-tA_0}(1-P)\|\le Ke^{-\alpha t}\qquad(t\ge0),
\end{equation}
and its \emph{unstable dimension} is $\mu=\operatorname{rank}(1-P)$. This holds if and only if $A_0$ has no
eigenvalue on the imaginary axis, and $\mu$ is then the number of eigenvalues with positive real part,
counted with multiplicity. In all examples below, $A_0$ has a basis of real eigenvectors with nonzero
eigenvalues, and $P$ is the projection onto the span of those with negative eigenvalue. The
\emph{unstable} and \emph{stable subspaces} of $A$ are
\begin{equation}\label{eq:EuEs}
\begin{aligned}
E^u&\coloneqq\{y(0):\dot y=A y,\ y\text{ bounded on }(-\infty,0]\}\subset\R^n,\\
E^s&\coloneqq\{y(0):\dot y=A y,\ y\text{ bounded on }[0,\infty)\}\subset\R^n,
\end{aligned}
\end{equation}
and we call $A$ \emph{transversal} if $E^u+E^s=\R^n$. A connecting solution $x^*$ of~\eqref{eq:nonaut}
is transversal if its linearization $A(t)=D_xf(t,x^*(t))$ is, and its linearized operator is that of
$A$. Remark~\ref{rem:palmergeom} in Appendix~\ref{app:theory} relates this to the transversality of invariant manifolds in the
double.

\begin{proposition}[Fredholm index and transversality]\label{prop:palmer}
\begin{enumerate}
\item[(i)] Assume $g'(-1)>0>g'(1)$. Let $e_\pm$ be hyperbolic equilibria of $f_\pm$ with fiber indices
$\mu_\pm$. Then $p_-=(e_-,\pi)$ and $p_+=(e_+,0)$ are hyperbolic equilibria of $\Dd$ with Morse indices
$\mu_-+1$ and $\mu_+$.
\item[(ii)] Let $A:\R\to\operatorname{End}(\R^n)$ be continuous and bounded, with $A(t)\to A_\pm$ as
$t\to\pm\infty$, where $A_\pm$ are hyperbolic with unstable dimensions $\mu_\pm$. Then $L:C^1_b\to C^0_b$ is
Fredholm~\cite{Palmer:1984,Palmer:1988} of index $\mu_--\mu_+$. Moreover, $\ker L\cong E^u\cap E^s$
via $y\mapsto y(0)$, and $\operatorname{coker}L\cong\R^n/(E^u+E^s)$.
\item[(iii)] In the situation of (ii), $A$ is transversal if and only if $L$ is onto. If $\mu_-=\mu_+$,
this holds if and only if $\ker L=0$, and then $L:C^1_b\to C^0_b$ is an isomorphism.
\end{enumerate}
\end{proposition}

Part (ii) is due to Palmer~\cite{Palmer:1984,Palmer:1988}. For completeness, a proof of the
proposition is given in Appendix~\ref{app:theory}.

With the shift $\mu_-\mapsto\mu_-+1$ at the UV end, the doubled Morse indices reproduce the Fredholm
index of $L$. Namely, the dimension $\mu_--\mu_+$ of the family of connecting solutions of the
non-autonomous equation equals the difference of doubled indices minus the one-dimensional orbit, which
compensates for the absence of time-translation symmetry. Families with $\mu_--\mu_+>0$ are the generic
UV$\to$IR solutions. Connections with $\mu_--\mu_+<0$ are non-generic and can only be protected by
additional structure.

\subsection{Equivariant structure of the full double}

\begin{theorem}[Equivariant structure]\label{thm:equivariant}
Let $S\subset\Ss$ be an isolated invariant set with index pair $(N,L)$, and $Y\coloneqq q^{-1}N/q^{-1}L$.
\begin{enumerate}
\item[(i)] $(q^{-1}N,q^{-1}L)$ is an $R$-invariant index pair of $\tilde S$, and
\begin{equation}
Y^R\cong h_{f_+}(K_+)\vee h_{f_-}(K_-),\qquad Y/R\cong h_{\Ss}(S).
\end{equation}
\item[(ii)] With coefficients in any ring $\Lambda$, the connecting homomorphism $\partial^{\Dd}$ of the
attractor--repeller pair $(K_+\times\{0\},K_-\times\{\pi\})$ of $\tilde S$ vanishes, and
\begin{equation}
0\to\CH_k(K_+;f_+)\to\CH_k(\tilde S)\to\CH_{k-1}(K_-;f_-)\to0
\end{equation}
is split exact. In particular $\CH_k(\tilde S;\Lambda)\cong\CH_k(K_+;f_+;\Lambda)\oplus
\CH_{k-1}(K_-;f_-;\Lambda)$, and $R_*$ acts by $+1$ on the first summand and by $-1$ on the quotient
$\CH_{k-1}(K_-;f_-;\Lambda)$.
\item[(iii)] If $\CH_*(K_\pm;f_\pm)$ are of finite type, then, with coefficients in any field,
$\tilde\chi(h_{\Dd}(\tilde S))=\tilde\chi(h_{f_+}(K_+))-\tilde\chi(h_{f_-}(K_-))
=2\,\tilde\chi(h_{\Ss}(S))-\tilde\chi(h_{f_+}(K_+))-\tilde\chi(h_{f_-}(K_-))
=2\,\tilde\chi(h_{\Ss}(S))-\tilde\chi(Y^R)$.
\end{enumerate}
\end{theorem}

The proof of (i) is a direct check of the index-pair axioms. Part (ii) follows from the projection
$q$, which gives a left inverse of the inclusion of the IR attractor. Both proofs are given in
Appendix~\ref{app:theory}.

Theorem~\ref{thm:equivariant} clarifies the role of the doubling. The double makes all boundary
sets two-sided, and its $\Z_2$-equivariant structure encodes the UV and IR autonomous indices as the
fixed-point data and the extended phase space $\Ss$ as the orbit space. Its connecting homomorphism,
however, vanishes over every coefficient ring and is therefore insensitive to UV$\to$IR connections.
Heuristically, each connection appears twice, once in each sheet, and the two contributions cancel.
The reflection alone gives only $2\partial^{\Dd}=0$, since $R_*$ acts by opposite signs on the UV and IR
indices. The vanishing of $\partial^{\Dd}$ itself uses the projection to $\Ss$ and
Theorem~\ref{thm:slab}. The connection information is carried by the one-sided extension, through
$\Psi$. Likewise, the homology of $Y/R\cong h_{\Ss}(S)$ relative to the fixed-point
set is pure UV data. By (i), (iii) and Theorem~\ref{thm:slab}, its Euler characteristic is
$\tilde\chi(h_{\Ss}(S))-\tilde\chi(Y^R)=-\tilde\chi(h_{f_-}(K_-))$, and in all examples below it is the
suspended UV index, not evidence of a connecting orbit.

\subsection{Reversibility and duality}

Many truncations of kinetic theory are reversible in a generalized sense. The equations are not
invariant under $t\mapsto-t$ alone, since dissipation picks a direction of time, but they are invariant
under time reversal combined with a reflection of the state variables. For the Gubser system of
Sec.~\ref{sec:gubser} the reflection reverses the sign of the temperature, and it exchanges each UV
fixed point with an IR fixed point. Such a symmetry relates the index of a set to that of its partner. In
RG language, a relevant direction at a UV point becomes an irrelevant direction at its partner, and the
indices are complementary.

\begin{theorem}[Duality]\label{thm:duality}
Let $\psi$ be a flow on an $n$-manifold without boundary with a time-reversal symmetry $\mathcal S$. For every isolated
invariant set
$K$, $\mathcal S(K)$ is isolated and, with $\F$ coefficients,
\begin{equation}
\dim\CH_k(\mathcal S(K))=\dim\CH_{n-k}(K).
\end{equation}
If $\mathcal S(K)=K$, then $\tilde\chi(h(K))=(-1)^n\tilde\chi(h(K))$, and in particular $\tilde\chi(h(K))=0$
when $n$ is odd.
\end{theorem}

\begin{proof}
$\mathcal S$ is a homeomorphism that conjugates $\psi$ to $\psi^{-1}$. It therefore maps isolating
neighborhoods of $K$ to isolating neighborhoods of $\mathcal S(K)$, so $\mathcal S(K)$ is isolated, and
index pairs of $K$ for $\psi^{-1}$ to index pairs of $\mathcal S(K)$ for $\psi$, so
$h(\mathcal S(K),\psi)\cong h(K,\psi^{-1})$. By
McCord duality~\eqref{eq:duality} applied to $\psi^{-1}$, $\CH_k(K,\psi^{-1})\cong\CH^{n-k}(K,\psi)$,
whose dimension equals that of $\CH_{n-k}(K,\psi)$ over a field.
\end{proof}

\begin{remark}\label{rem:duality-boundary}
Only the last step uses the absence of boundary. On a manifold with invariant boundary, such as the
Poincar\'e compactifications of Sec.~\ref{sec:full3d}, the identity $h(\mathcal S(K),\psi)\cong
h(K,\psi^{-1})$ still holds, but for sets meeting the boundary the index is one-sided and McCord duality
does not apply as stated. A boundary version of the duality, or a further doubling across the boundary,
would be required. We therefore do not invoke Theorem~\ref{thm:duality} for such sets.
\end{remark}

%======================================================================
\section{Gubser flow in Israel--Stewart theory}\label{sec:gubser}
%======================================================================

\subsection{Set-up and symmetries}

Gubser flow~\cite{Gubser:2010ze} describes a conformal fluid that expands boost-invariantly along the
beam axis and, in addition, radially, as in the transverse expansion of the fireball produced in a
heavy-ion collision. Its symmetry group $SO(3)_q\times SO(1,1)\times\Z_2$ is large enough that, after a
Weyl rescaling to $dS_3\times\R$, the dynamics depends on a single time, the de Sitter time $\rho$. In
Israel--Stewart (IS) theory, which supplements ideal hydrodynamics by a relaxation equation for the shear
stress, the state is described by the temperature $T$ in the rescaled frame and the normalized shear
stress $\pib$, and the relaxation time is $c/T$, as in Ref.~\onlinecite{Behtash:2017wqg}. The global flow
structure of this system, including a classification of its invariant submanifolds and of the flow
lines between UV and IR fixed points, was studied in Ref.~\onlinecite{Behtash:2020jhep} with
skew-product flows and attractor--repeller pairs. The expansion is not stationary in
$\rho$, and this is the source of the explicit time dependence. With $\tau=\tanh\rho$ the Gubser flow of IS
theory is the three-dimensional autonomous system
\begin{align}
\frac{dT}{d\rho}&=\frac{T}{3}\,(\pib-2)\,\tau\eqqcolon\beta_T,\label{eq:gT}\\
\frac{d\pib}{d\rho}&=-\frac43\Big(\pib^2-\frac15\Big)\tau-\frac{T\pib}{c}\eqqcolon\beta_{\pib},\label{eq:gpi}\\
\frac{d\tau}{d\rho}&=1-\tau^2\eqqcolon\beta_\tau,\label{eq:gtau}
\end{align}
where $c>0$ is the relaxation scale and $\rho\to\mp\infty$ is the UV/IR limit. It satisfies
Hypothesis~\ref{hyp:H} with $t=\rho$, $\sigma=\tanh$, $g(\tau)=1-\tau^2$, fiber $X=\R^2\ni(T,\pib)$ and
$F$ polynomial, and the physical region is $T\ge0$. The Jacobian is
\begin{equation}\label{eq:gubserJ}
J=\begin{pmatrix}
\frac{(\pib-2)\tau}{3} & \frac{T\tau}{3} & \frac{T(\pib-2)}{3}\\[0.5ex]
-\frac{\pib}{c} & -\frac{8\pib\tau}{3}-\frac{T}{c} & -\frac43\big(\pib^2-\frac15\big)\\[0.5ex]
0&0&-2\tau
\end{pmatrix}.
\end{equation}
With $a=1/\sqrt5$, the equilibria, the fiber eigenvalues, the fiber indices $\mu$ and the Morse indices
in the double are listed in Table~\ref{tab:gubser}. The normal eigenvalue is $\pm2$ in $\Ss$ and
$\pm1$ in the double.

\begin{table}[t]
\centering
\begin{tabular}{lcccc}
\toprule
point & $(T,\pib,\tau)$ & fiber eigenvalues & $\mu$ & index in $\Dd$\\
\midrule
$\Sig_U(A)$ & $(0,a,-1)$ & $\frac{2-a}{3},\ \frac{8a}{3}$ & 2 & 3\\
$\Sig_I(A)$ & $(0,a,+1)$ & $-\frac{2-a}{3},\ -\frac{8a}{3}$ & 0 & 0\\
$\Sig_U(B)$ & $(0,-a,-1)$ & $\frac{2+a}{3},\ -\frac{8a}{3}$ & 1 & 2\\
$\Sig_I(B)$ & $(0,-a,+1)$ & $-\frac{2+a}{3},\ \frac{8a}{3}$ & 1 & 1\\
$\Sig_U(C)$ & $(+\frac{38c}{15},2,-1)$ & $\frac{21\pm\sqrt{821}}{15}$ & 1 & 2\\
$\Sig_I(C)$ & $(-\frac{38c}{15},2,+1)$ & $-\frac{21\pm\sqrt{821}}{15}$ & 1 & 1\\
\bottomrule
\end{tabular}
\caption{Equilibria of the Gubser system~\eqref{eq:gT}--\eqref{eq:gtau}, with $a=1/\sqrt5$.}
\label{tab:gubser}
\end{table}

The first terms on the right-hand sides of~\eqref{eq:gT} and~\eqref{eq:gpi} describe the effect of the
expansion, and the last term of~\eqref{eq:gpi} describes the relaxation of the shear stress by
collisions, at the rate $T/c$. The UV points $\Sig_U$ are thus the early-time regimes of the expansion, the
IR points $\Sig_I$ the late-time ones, and the $C$-points, at which $T\neq0$, are the only equilibria at
which the collision term matters.

\paragraph{Time reversal and reflection.}\label{subsec:symmetries}
The system~\eqref{eq:gT}--\eqref{eq:gtau} is invariant under the time reversal
\begin{equation}\label{eq:reversor}
\mathcal S:(T,\pib,\tau)\mapsto(-T,\pib,-\tau),\qquad\rho\mapsto-\rho,
\end{equation}
which maps solutions to solutions with the direction of the flow reversed. $\mathcal S$ exchanges the UV
and IR members of each pair in Table~\ref{tab:gubser} and complements their indices, $3-k$. Since the
right-hand side of~\eqref{eq:gT} is proportional to $T$, the plane $T=0$ is invariant and the sign of $T$
is preserved, and the two $C$-points therefore lie in dynamically disconnected half-spaces.

In the double, with $\tau=\cos\theta$ and $\dot\theta=-\sin\theta$, two different involutions act, namely
\begin{equation}
R:(T,\pib,\theta)\mapsto(T,\pib,-\theta),\qquad
\tilde{\mathcal S}:(T,\pib,\theta)\mapsto(-T,\pib,\pi-\theta),\ \ t\mapsto-t.
\end{equation}
The reflection $R$ commutes with the flow and fixes the UV and IR fibers. The lift $\tilde{\mathcal S}$
of the time reversal is a time-reversal symmetry that exchanges the UV and IR fibers, and restricted to $T=0$ its
fixed-point set is the plane $\tau=0$. They generate a Klein four-group $\{1,R,\tilde{\mathcal S},
R\tilde{\mathcal S}\}$, in which $R$ preserves the direction of time while $\tilde{\mathcal S}$ and
$R\tilde{\mathcal S}$ reverse it. In particular the reflection $R$ is \emph{not} the time reversal. The involution $R$ is responsible for the fixed-point/orbit-space structure of
Theorem~\ref{thm:equivariant}, and $\tilde{\mathcal S}$ for the duality of Theorem~\ref{thm:duality}.
Since $\Dd$ is three-dimensional, every $\tilde{\mathcal S}$-invariant isolated invariant set of the
doubled Gubser flow has $\tilde\chi=0$, and $\tilde{\mathcal S}$-partners have complementary indices.

\subsection{The invariant plane \texorpdfstring{$T=0$}{T=0} and the limit \texorpdfstring{$c\to\infty$}{c->infinity}}

Since collisions act at the rate $T/c$, both the plane $T=0$ and the limit $c\to\infty$ switch them off,
and the shear stress then evolves under the expansion alone. On $T=0$ the system does not depend on $c$
and reduces to the warm-up system~\eqref{eq:warmup} under $(\pib,\tau)\leftrightarrow(x,p)$. In the limit $c\to\infty$ the $\pib$-equation decouples from $T$ for
all $T$. The two points $\Sig_{U,I}(C)$ run off to infinity, and the system reduces to the
$(\pib,\tau)$-plane. With $\kappa=8/(3\sqrt5)$, the solutions are
\begin{align}
\pib(\rho)&=a\,\frac{\cosh^{\kappa}\rho+K}{\cosh^{\kappa}\rho-K},\label{eq:gpi-sol}\\
T(\rho)&=T_0\,\cosh^{-(2+a)/3}\rho\;\Big|\frac{\cosh^{\kappa}\rho-K}{1-K}\Big|^{1/4},\label{eq:gT-sol}
\end{align}
with $K\in\R\cup\{\infty\}$ and~\eqref{eq:gT-sol} valid while $\pib$ is finite. Both formulas can be
verified by direct differentiation, using $3\kappa/4=2a$. In terms of $z=e^{2\rho}$ one has $\cosh\rho=(1+z)/(2\sqrt z)$.
The solution~\eqref{eq:gT-sol} is real for every real solution, and the sign of $T$ is that of $T_0$.
For bounded solutions $T\to0$ as $\rho\to\pm\infty$, and
\begin{equation}\label{eq:gubser-conn}
\lim_{\rho\to\pm\infty}\pib=\begin{cases}+a,& K<1,\\ -a,& K=\infty,\end{cases}
\qquad\text{i.e.}\qquad
\begin{aligned}
&\Sig_U(A)\to\Sig_I(A) &&(K<1),\\
&\Sig_U(B)\to\Sig_I(B) &&(K=\infty),
\end{aligned}
\end{equation}
while solutions with $K>1$ ($\pib(0)<-a$) blow up in finite time. The line $\pib=-a$ is the separatrix
between the two universality classes of initial data.

This phase portrait resembles that of the $O(N)$ model of Sec.~\ref{sec:ON}, but with an important
difference. In the $O(N)$ model the fixed points are interior points of the coupling space, so their
Conley indices are defined without further construction, and the exchange of stability at $N=4$
proceeds by a collision of fixed points. In the Gubser flow the fixed points lie on the invariant
boundary $\tau=\pm1$, their indices in $\Ss$ are one-sided, and the constructions of
Sec.~\ref{sec:theory} are needed.

\subsection{Conley indices of the reduced system}\label{subsec:reduced-index}

Consider the reduced system on $T=0$, equivalently $c=\infty$, in the double, namely
\begin{equation}\label{eq:reduced-double}
\dot\pib=-\frac43(\pib^2-a^2)\cos\theta,\qquad\dot\theta=-\sin\theta .
\end{equation}

\paragraph{Non-isolation of the $A$-line.}
By~\eqref{eq:warmup-K}, every neighborhood of the line $\pib\equiv a$ contains entire bounded orbits
($|K|$ small), which touch the boundary of any thin annulus around the line from inside. Hence the
$A$-line is not an isolated invariant set, and no Conley index can be attached to it. Similarly, the
union of all bounded orbits is not compact, since $\max\pib=a(1+K)/(1-K)\to\infty$ as $K\to1^-$. In the plane,
the isolated invariant sets built from the four equilibria are the points themselves and the $B$-circle
below, and a global description requires a compactification.

\begin{figure}[t]
\centering
\includegraphics[width=0.78\textwidth]{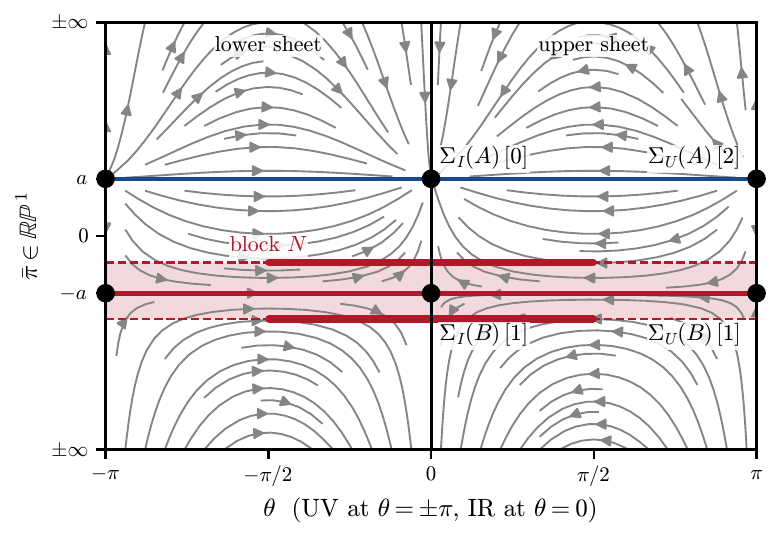}
\caption{The reduced doubled Gubser flow~\eqref{eq:reduced-double} on the torus $T^2=\RP^1\times S^1$,
with $\pib$ plotted through the angle $\arctan\pib$, so that the top and bottom edges ($\pib=\pm\infty$)
are identified, as are the left and right edges ($\theta=\pm\pi$). The four equilibria are shown with
their Morse indices. The reflection $R$ is $\theta\mapsto-\theta$. The red line is the
$B$-circle, consisting of $\Sig_U(B)$, $\Sig_I(B)$ and the separatrix in each sheet. The shaded strip is
the isolating block $N=I_\delta\times S^1$, and its exit set $L$ (thick red segments) is the part of
$\partial N$ with $\cos\theta\ge0$.}
\label{fig:torus}
\end{figure}

\begin{proposition}[Torus compactification]\label{prop:torus}
The $\pib$-equation of~\eqref{eq:reduced-double} is a Riccati equation and extends smoothly to
$\RP^1$. In the coordinate $w=1/\pib$, $\dot w=\frac43(1-a^2w^2)\cos\theta$. The reduced doubled flow is
therefore a smooth flow on $T^2=\RP^1\times S^1$ (Fig.~\ref{fig:torus}), and the following hold for
every $a>0$.
\begin{enumerate}
\item[(i)] It has exactly four equilibria, all hyperbolic, with Morse indices $2,1,1,0$ at
$\Sig_U(A)$, $\Sig_U(B)$, $\Sig_I(B)$, $\Sig_I(A)$. Every other orbit converges to an equilibrium in
forward and in backward time. Off the fixed fibers $\theta=0,\pi$ it runs from $\Sig_U(A)$ to
$\Sig_I(A)$, except for the separatrix $\pib=-a$ in each sheet, which runs from $\Sig_U(B)$ to
$\Sig_I(B)$. The other orbits in the fiber $\theta=\pi$ run from $\Sig_U(A)$ to $\Sig_U(B)$, and those
in the fiber $\theta=0$ run from $\Sig_I(B)$ to $\Sig_I(A)$. Consequently, the four equilibria, ordered
by $\Sig_I(A)<\Sig_I(B)<\Sig_U(B)<\Sig_U(A)$, form a Morse decomposition of $T^2$.
\item[(ii)] $\CH_*(T^2)=H_*(T^2)$. The Morse polynomial $1+2t+t^2$ equals the Poincar\'e polynomial
of $T^2$, so the connection matrix of the full double vanishes, in accordance with
Theorem~\ref{thm:equivariant}(ii).
\item[(iii)] For the one-sided extension of the compactified extended phase space $\RP^1\times[-1,1]$,
Theorem~\ref{thm:rate} with $N=\RP^1$ gives $\CH_*=0$ and $\Psi:H_*(\RP^1)\to H_*(\RP^1)$ an isomorphism.
The one-sided connection matrix (Morse indices $2,1$ for $\Sig_U(A),\Sig_U(B)$ and $1,0$ for
$\Sig_I(B),\Sig_I(A)$) is forced over $\F$, namely
\begin{equation}
\Delta\,\Sig_U(A)=\Sig_I(B),\qquad\Delta\,\Sig_U(B)=\Sig_I(A).
\end{equation}
These entries are realized by the heteroclinic chains $\Sig_U(A)\to\Sig_U(B)\to\Sig_I(B)$ and
$\Sig_U(B)\to\Sig_I(B)\to\Sig_I(A)$ through the separatrix.
\end{enumerate}
\end{proposition}

The proof, which uses the explicit solution of the Riccati equation, is given in Appendix~\ref{app:gubser}.

\paragraph{An isolating block for the separatrix.}
Let $I_\delta=[-a-\delta,-a+\delta]$ with $0<\delta<2a$, and let $N=I_\delta\times S^1$. On $\pib=-a\pm
\delta$, $\dot\pib$ has the sign of $\pm\cos\theta$, so the exit set is
$L=\{-a\pm\delta\}\times\{\cos\theta\ge0\}$ (Fig.~\ref{fig:torus}). At the tangency points $\cos\theta=0$ one finds
$\ddot\pib=-\frac43(\pib^2-a^2)\sin^2\theta$, so tangencies are external and $N$ is an isolating block.
By~\eqref{eq:warmup-K} its invariant set is the \emph{$B$-circle} $\tilde S_B=\{-a\}\times S^1$, which
consists of $\Sig_U(B)$, $\Sig_I(B)$ and the two copies of the separatrix, one in each sheet. The
quotient $N/L$ is an annulus with two disjoint boundary arcs collapsed to a point, so
\begin{equation}\label{eq:fig8}
h(\tilde S_B)\simeq S^1\vee S^1,\qquad \CH_*(\tilde S_B)=\Z^2\ \text{in degree }1 .
\end{equation}
Up to homotopy this is the once-punctured torus. Its fundamental group is free on two generators, and
it is aspherical, being homotopy equivalent to a graph~\cite{Hatcher:2002}. This agrees with
Theorem~\ref{thm:equivariant}(ii), since $\CH_*(\Sig_I(B);f_+)=\Z[1]$ and $\CH_*(\Sig_U(B);f_-)=\Z[0]$
contribute $\Z[1]\oplus\Z[0+1]$. The two degree-one generators can be taken to be the core circle
$\gamma_c$ and the arc $\gamma_t=I_\delta\times\{0\}$ joining the two exit arcs. The reflection
acts by $R_*=\operatorname{diag}(-1,+1)$ in this basis. The fixed-point set and the orbit space of this
index are computed in Appendix~\ref{app:gubser}, in accordance with Theorems~\ref{thm:slab}
and~\ref{thm:equivariant}.

\paragraph{One-sided extension and transport.}
The one-sided extension $S_B^\osx$ has the index pair $I_\delta\times A_\eta$ whose exit set consists of
three disjoint arcs, so again $h(S_B^\osx)\simeq S^1\vee S^1$ and $\CH_*=\Z^2[1]$. Here
$\Psi:\CH_0(\Sig_U(B);f_-)=\Z\to\CH_0(\Sig_I(B);f_+)=0$ vanishes for degree reasons. Consistently, the
linearization along the separatrix is $\delta\pib'=\frac{8a}3\tau\,\delta\pib$, with limits $\mp\frac{8a}3$
of unstable dimensions $0$ and $1$, so Proposition~\ref{prop:palmer}(ii) gives Fredholm index
$\mu_--\mu_+=0-1=-1$, i.e.\ in the reduced system the separatrix is a non-generic connection. It persists only under perturbations that keep $\pib=-a$ an
equilibrium of every frozen vector field $F(\tau,\cdot)$, and the Conley index of the reduced system
does not detect it.

\subsection{Transversality of the separatrix at finite relaxation scale}\label{subsec:transversal}

In the full system at finite $c$, the separatrix $\gamma_B(\rho)=(0,-a,\tanh\rho)$ acquires a
transverse direction $T$, which is coupled to $\pib$ through the term $-T\pib/c$.

\begin{proposition}\label{prop:transversal}
For every $0<c<\infty$, the connecting solution $(T,\pib)\equiv(0,-a)$ from $\Sig_U(B)$ to $\Sig_I(B)$
is transversal, its linearized operator $L$ has Fredholm index $\mu(\Sig_U(B))-\mu(\Sig_I(B))=0$, and
$L:C^1_b\to C^0_b$ is an isomorphism.
\end{proposition}

\begin{proof}
With $k=(2+a)/3$, the linearization along $\gamma_B$ is
\begin{equation}\label{eq:linB}
\delta T'=-k\tau\,\delta T,\qquad\delta\pib'=\frac{8a}{3}\tau\,\delta\pib+\frac{a}{c}\,\delta T,
\qquad\tau=\tanh\rho ,
\end{equation}
i.e.\ $A(\rho)=\tau(\rho)A_1+A_2$ with $A_1=\operatorname{diag}(-k,\frac{8a}3)$ and
$A_2(\delta T,\delta\pib)=(0,\frac ac\delta T)$. Its limits $A_\pm=\pm A_1+A_2$ as $\rho\to\pm\infty$ are
lower triangular with the nonzero eigenvalues $\mp k$ and $\pm\frac{8a}3$, which are distinct, so
both are hyperbolic with unstable dimension $1$, and $L$ has index $1-1=0$ by
Proposition~\ref{prop:palmer}(ii). Next, $\delta T=A\cosh^{-k}\rho$, bounded in both directions. The
homogeneous solutions for $\delta\pib$, $C\cosh^{8a/3}\rho$, are unbounded in both directions. With
$\Phi=\cosh^{8a/3}\rho$, the solutions bounded as $\rho\to-\infty$ and as $\rho\to+\infty$ are,
respectively,
\[
\delta\pib=\Phi(\rho)\int_{-\infty}^{\rho}\Phi^{-1}\,\frac{a}{c}\,\delta T,\qquad
\delta\pib=-\Phi(\rho)\int_\rho^{\infty}\Phi^{-1}\,\frac{a}{c}\,\delta T .
\]
The two agree only if $A\,\frac ac\int_{\R}\cosh^{-8a/3-k}=0$, i.e.\ $A=0$. Hence $\ker L=0$, and the
connecting solution is transversal and $L$ is an isomorphism by Proposition~\ref{prop:palmer}(iii).
\end{proof}

The argument uses only $\tau(\rho)\to\pm1$, so it applies to every rate $g$ as in
Hypothesis~\ref{hyp:H}.

\begin{figure}[t]
\centering
\includegraphics[width=\textwidth]{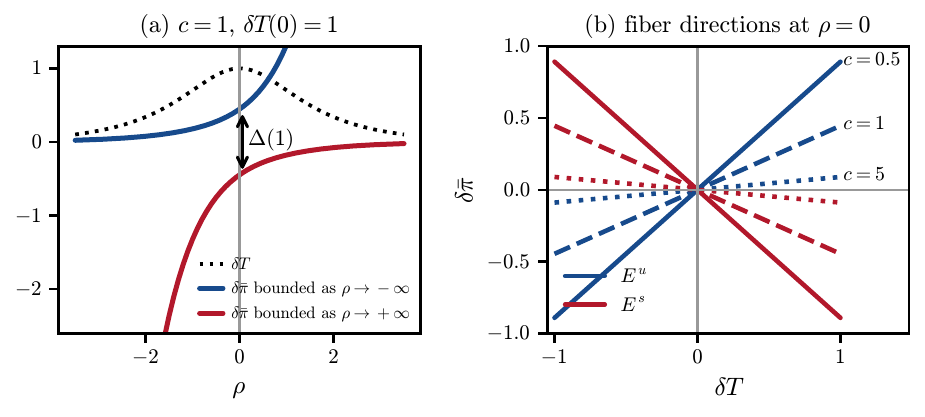}
\caption{Transversality of the Gubser separatrix (Proposition~\ref{prop:transversal}). (a) For $c=1$ and
$\delta T=\cosh^{-k}\rho$, the solution $\delta\pib$ that is bounded as $\rho\to-\infty$ (blue) and the one
that is bounded as $\rho\to+\infty$ (red) differ at $\rho=0$ by $\Delta(1)$, so no nonzero bounded
solution exists and $\ker L=0$. (b) The unstable and stable fiber directions $E^u$ and $E^s$ at $\rho=0$,
of slopes $\pm\Delta(c)/2$, for $c=0.5$, $1$ and $5$. They close up as $c\to\infty$, where transversality
is lost. The slopes refer to the physical normalization of $T$ (Remark~\ref{rem:gap}).}
\label{fig:transversal}
\end{figure}

\begin{remark}[Splitting of $E^u$ and $E^s$, and its coordinate dependence]\label{rem:gap}
With $\nu=\frac23+3a=\frac{8a}3+k$, evaluating the two solutions of the proof for $A=1$ at $\rho=0$ shows
that the subspaces~\eqref{eq:EuEs} of~\eqref{eq:linB} are $E^u=\langle(1,s_u)\rangle$ and
$E^s=\langle(1,s_s)\rangle$ in the $(\delta T,\delta\pib)$ plane, with $s_u=-s_s=\Delta/2$ and
\begin{equation}\label{eq:gap}
\Delta(c)\coloneqq s_u-s_s=\frac{a}{c}\int_{-\infty}^{\infty}\cosh^{-\nu}\rho\,d\rho
=\frac{a}{c}\,\frac{\sqrt\pi\,\Gamma(\nu/2)}{\Gamma\big((\nu+1)/2\big)},
\end{equation}
which is positive and tends to $0$ as $c\to\infty$. Numerically, $\Delta(c)\approx0.892/c$.
Figure~\ref{fig:transversal} illustrates these slopes for several values of $c$.
The invariant content of Proposition~\ref{prop:transversal} is the invertibility of $L$, equivalently
the non-vanishing of the integral $\int_\R\cosh^{-\nu}\rho\,d\rho$ that obstructs a bounded solution with
$A\neq0$. The number $\Delta(c)$ is a \emph{coordinate-normalized slope splitting}. Under a rescaling
$T\mapsto\lambda T$ it becomes $\Delta/\lambda$, and a metric-dependent angle between $E^u$ and $E^s$
changes accordingly. The claims $\Delta(c)\neq0$ and, for the fixed physical normalization of the
temperature $T$, $\Delta(c)\propto1/c$ are meaningful. The numerical prefactor is not an invariant.
\end{remark}

For the isolation of the separatrix we use a block that moves with $\tau$. In the coordinates
$x=T/c$, $y=\pib+a$, the system~\eqref{eq:gT}--\eqref{eq:gpi} reads
\begin{equation}\label{eq:gubser-xy}
x'=\frac{x\tau}{3}\,(y-a-2),\qquad y'=\kappa\,y+a\,x,\qquad\kappa=\frac{8a}{3}\tau-\frac43\tau y-x,
\end{equation}
independently of $c$. With $\varepsilon=\frac1{20}$, let
\begin{equation}\label{eq:gubser-block}
\xi=\frac{a\,x+(1+\tau)\,y}{\varepsilon},\qquad\eta=\frac{(1-\tau)\,y-a\,x}{\varepsilon},\qquad
N_B=\{|\xi|\le1,\ |\eta|\le1\}.
\end{equation}
Up to the constant factor $2a/\varepsilon$, these are the coordinates of $(x,y)$ in the moving frame
$u=(1-\tau,a)$, $s=(-(1+\tau),a)$, whose determinant is $2a$ for every $\tau$.

\begin{corollary}\label{cor:gubserB}
For $0<c<\infty$, let $S_B=\{(0,-a,\tau):\tau\in[-1,1]\}\subset\Ss$.
\begin{enumerate}
\item[(i)] $S_B$ is isolated, with $K_\pm=\{(0,-a)\}$ and $\CH_*(K_\pm;f_\pm)\cong\Z[1]$,
$\CH_*(S_B^\osx)=0$, and $\Psi:\CH_1(\Sig_U(B);f_-)\cong\Z\to\CH_1(\Sig_I(B);f_+)\cong\Z$ equals $\pm1$.
\item[(ii)] $\CH_*(S_B)=\Z[1]$ in $\Ss$ (Theorem~\ref{thm:slab}), and $(K_+\times\{0\},K_-\times\{\pi\})$
is an attractor--repeller pair in $S_B^\osx$ with $\CH_*(K_+\times\{0\})=\Z[1]$ and
$\CH_*(K_-\times\{\pi\})=\Z[2]$.
\item[(iii)] The UV$\to$IR connection persists under every perturbation $(\tilde F,\tilde g)$ of the
compactified vector field, with $\tilde g$ as in Hypothesis~\ref{hyp:H}, such that, on $N_B$ and for
all $\tau\in[-1,1]$, $|\tilde F_T-\beta_T|\le c/200$, $|\tilde F_{\pib}-\beta_{\pib}|\le1/200$ and
$|\tilde g-(1-\tau^2)|\le1/200$. Precisely, the conclusions of (i) hold for the invariant set of $N_B$,
with $K_\pm$ its fibers at $\tau=\pm1$, and it contains a solution from $K_-$ to $K_+$. This includes
perturbations that break the invariance of $T=0$, the $\tanh$ time dependence, or the form of the IS
equations.
\item[(iv)] At $c=\infty$, $S_B$ is not isolated. Indeed, with $k=(2+a)/3$, the curve given by
$T=A\cosh^{-k}\rho$, $\pib=-a$ and $\tau=\tanh\rho$ is a solution for every $A$.
\end{enumerate}
\end{corollary}

\begin{proof}
(i) By Lemma~\ref{lem:gubser-block}(a) of Appendix~\ref{app:gubser}, for every $\tau\in[-1,1]$ one has
$\pm\xi'>\frac12$ on the faces $\xi=\pm1$ of $N_B$ and $\pm\eta'<-\frac12$ on the faces $\eta=\pm1$.
Hence $N_B$ satisfies the hypothesis of Corollary~\ref{cor:moving} with exit set $\{|\xi|=1\}$, and
$N_B/\{|\xi|=1\}\simeq S^1$. The cone estimate of Lemma~\ref{lem:gubser-block}(b) shows that every
solution that stays in $N_B$ has $\xi=\eta=0$, i.e.\ lies on $S_B$. Thus $\Inv(N_B)=S_B$ and
$K_\pm=\{(0,-a)\}$, and (i) follows from Corollary~\ref{cor:moving} with $k=1$.
(ii) follows from (i), Theorem~\ref{thm:slab} and Proposition~\ref{prop:boundary}. (iii) Perturbations
of the stated size change $\xi'$ and $\eta'$ on $N_B$ by less than
$20\big(\frac{a}{200}+\frac{2}{200}+\frac{\varepsilon}{200}\big)<\frac14$, so the face inequalities of
Lemma~\ref{lem:gubser-block}(a) survive. Hence Corollary~\ref{cor:moving} applies to the perturbed system
with the same block. (iv) At
$c=\infty$ the coupling $-T\pib/c$ drops out, $\pib=-a$ remains invariant, and on it~\eqref{eq:gT} is
$T'=-k\tau T$, whose solutions are $A\cosh^{-k}\rho$. For small $A\neq0$ these solutions stay in any
neighborhood of $S_B$ and are not contained in $S_B$.
\end{proof}

Theorem~\ref{thm:equivariant}(ii) also gives $\CH_*(\tilde S_B)
=\Z[1]\oplus\Z[2]$ in the double. Namely, $\CH_1(\tilde S_B)\cong\CH_1(K_+)\cong\Z$, $\CH_2(\tilde S_B)
\cong\CH_1(K_-)\cong\Z$, and all other groups vanish. With integer coefficients, the connection persists, possibly
non-uniquely, under every deformation of $(F,g)$ along which $S_B^\osx$ and $K_-$ continue. Indeed, the
homology index is invariant under continuation, so after the deformation $\CH_*(S^\osx)=0$ and
$\CH_1(K_-)\cong\Z$. By the exact sequence~\eqref{eq:transport-les}, $\Psi$ is then an isomorphism, in
particular $\Psi_1\neq0$, and Theorem~\ref{thm:transport}(ii) gives the connection. Over $\Q$, this is
the assertion $\operatorname{rk}\Psi_1=1$ of Theorem~\ref{thm:transport}(iii). As $c\to\infty$,
$\Delta(c)\to0$ by~\eqref{eq:gap}, and at $c=\infty$ the linearization decouples and $\ker L$ is spanned
by the $T$-mode, which is the linearization of the solutions in (iv). Transversality
(Proposition~\ref{prop:transversal}) and isolation (Corollary~\ref{cor:gubserB}(i)) are thus lost
together.

In the limit $c\to\infty$ and on the invariant plane $T=0$, the variables $(\pib,\tau)$ obey the same
reduced equations. The two systems nevertheless differ. The collisionless system keeps the direction
$T$, and in it $S_B$ lies in the family of solutions $T=A\cosh^{-k}\rho$, $\pib=-a$ of
Corollary~\ref{cor:gubserB}(iv). The index of the separatrix depends on this normal direction. In the plane its Fredholm index is $-1$ and
$\Psi=0$, while in the full system at finite $c$ it is $0$ and $\Psi=\pm1$. The limit $c\to\infty$
therefore discards exactly the coupling that makes the separatrix robust. Since $T\equiv0$ is the edge of
the physical region, $\gamma_B$ is the boundary of the one-parameter family of physical ($T>0$)
solutions emanating from $\Sig_U(B)$ (fiber index $\mu=1$). By isolation, no such solution that stays
close to $\gamma_B$ converges to $\Sig_I(B)$. For the other UV/IR pairs of Table~\ref{tab:gubser}, the
expected dimension $\mu_--\mu_+$ of transversal families of connecting solutions is $2$ for
$\Sig_U(A)\to\Sig_I(A)$, $1$ for $\Sig_U(B)\to\Sig_I(A)$ and $\Sig_U(A)\to\Sig_I(B)$, and $0$ for
$\Sig_U(B)\to\Sig_I(B)$. Connections $\Sig_U(C)\to\Sig_I(C)$ are excluded by the sign of $T$.

\subsection{Physical interpretation}

The three constructions of Sec.~\ref{sec:theory} play distinct roles in this example. The index in
$\Ss$ is the IR index (Theorem~\ref{thm:slab}), and it does not see the separatrix. The full double returns
the UV and IR indices as fixed-point data, but its connecting homomorphism vanishes
(Theorem~\ref{thm:equivariant}). The connection is detected by the transport map of the one-sided
extension, which is $\Psi=0$ in the reduced model and $\Psi=\pm1$ at every finite $c$, with a splitting
$\Delta(c)\propto1/c$ of the stable and unstable directions in the physical normalization
(Remark~\ref{rem:gap}). A change of $\Psi$ along a family can happen only through a loss of isolation,
which here occurs exactly at $c=\infty$. The converse does not hold, since connecting orbits can be
created or annihilated in algebraically canceling pairs without any change of $\Psi$.

In physical terms, the two universality classes are distinguished by the initial shear stress. By
\eqref{eq:gubser-conn}, solutions of the collisionless reduced system that start with $\pib>-a$ approach
the late-time point $\Sig_I(A)$, while those that start with $\pib<-a$ blow up in finite de Sitter time,
which signals the breakdown of the IS description. The separatrix is the sharp threshold between these
behaviors. Its transversality at every finite $c$ means that the threshold is not an artifact of
neglecting collisions. It survives every small modification of the equations that keeps the relevant
sets isolated, such as a change of transport coefficients or the addition of small higher-order
terms, and the topological invariant that guarantees this robustness is the transport map $\Psi=\pm1$. In
the reduced model, where $\Psi=0$, no such protection is available.

%======================================================================
\section{Bjorken flow}\label{sec:bjorken}
%======================================================================

\subsection{Compactification and fixed points}

Bjorken flow models the longitudinally boost-invariant expansion of the matter produced in a heavy-ion
collision. At early times the longitudinal expansion dominates and drives the distribution function far
from equilibrium, while at late times collisions drive it back to local equilibrium. A striking feature,
first seen in hydrodynamic and kinetic models~\cite{Heller:2015dha}, is that solutions with very
different initial data collapse onto a single attractor solution long before equilibrium is reached. In
the moment formulation of Ref.~\onlinecite{Behtash:2019}, truncated at $L=1$, the state is described by a single
normalized moment $c_1$ of the distribution function, with $c_1=0$ corresponding to local equilibrium,
and the dimensionless flow time $w$ runs from the far-from-equilibrium regime $w\to0$ to the
hydrodynamic regime $w\to\infty$. We consider this system in the flow time $0<w<\infty$,
\begin{equation}\label{eq:bj}
\frac{dc_1}{dw}=-\frac{1}{1-c_1/20}\Big[\frac{3}{2\theta_0}\,c_1+\frac1w\,P(c_1)\Big],\qquad
P(c_1)=\frac57c_1-\frac15c_1^2+4,
\end{equation}
with a constant $\theta_0>0$, on $X=(-\infty,20)$, where the vector field is regular. For $w\to0$ the
$P$-term dominates in the time $\log w$, and for $w\to\infty$ the linear term dominates in the time $w$.
The two ends therefore have different natural clocks. A single clock covering both is
\begin{equation}
\frac{dt}{dw}=1+\frac1w,\qquad u=\frac{w}{1+w}\in(0,1),
\end{equation}
in terms of which~\eqref{eq:bj} becomes
\begin{equation}\label{eq:bj-slab}
\frac{dc_1}{dt}=-\frac{\frac{3}{2\theta_0}\,u\,c_1+(1-u)\,P(c_1)}{1-c_1/20},\qquad\frac{du}{dt}=u(1-u)^2 .
\end{equation}
This is of the form~\eqref{eq:slab}. After $\tau=2u-1$, $g(\tau)=\frac14(1+\tau)(1-\tau)^2$ is
polynomial and $F$ is the convex combination of the UV and IR limit fields. The UV end $u=0$ is
normally hyperbolic ($du/dt\simeq u$), while the IR end $u=1$ is normally attracting but degenerate.
Conley theory needs no hyperbolicity, and Proposition~\ref{prop:boundary} applies unchanged, because its
proof uses only the sign of $G$. In particular no regularization of the IR end is needed. In regularized
versions (for instance $dw/d\bar\rho=1/w-\epsilon$), an additional IR equilibrium at
$c_1=O(1/(\epsilon\theta_0))$ appears, which is an artifact of the regulator and runs to infinity as
$\epsilon\to0$. Physically, the regulator replaces the slow approach to equilibrium by a spurious late-time
balance, and the compactification above shows that it is not needed for index-theoretic results.

\paragraph{Fixed points.}
At $u=0$ the equilibria are the zeros of $P$,
\begin{equation}
c_1^\pm=\frac{25\pm3\sqrt{505}}{14}\approx6.60,\ -3.03,
\end{equation}
with fiber eigenvalues $(101\pm17\sqrt{505})/168\approx2.87,\,-1.67$ and normal eigenvalue $1$. Hence
$c_1^+$ has Morse index $2$ and $c_1^-$ Morse index $1$ in the double ($\mu=1,0$). At $u=1$ the only
equilibrium in $X$ is $c_1=0$, with fiber eigenvalue $-3/(2\theta_0)$ and Conley index $S^0$. The two UV
points are far-from-equilibrium fixed points of the expansion-dominated regime. Since $c_1^-$ attracts in
the fiber, its only unstable direction in the double is the time direction, so exactly one solution
emanates from it. This is the attractor solution of Proposition~\ref{prop:bjorken} below. The point
$c_1^+$ repels in the fiber, and a one-parameter family of solutions emanates from it. The IR point $c_1=0$
is local equilibrium. Solutions approach it algebraically, since the term $P(c_1)/w$
in~\eqref{eq:bj} balances the linear term, and the attractor solution of
Proposition~\ref{prop:bjorken} satisfies $c_1^*(w)=-8\theta_0/(3w)+O(w^{-2})$ as $w\to\infty$. The
fiber eigenvalue $-3/(2\theta_0)$ governs the exponential decay, in $w$, of perturbations about this
solution.

\subsection{Existence and uniqueness of the attractor solution}

\begin{proposition}\label{prop:bjorken}
Let $\theta_0>0$, $c_L<c_1^-$ and $0<c_R<c_1^+$. Then $N=[c_L,c_R]$ satisfies the hypotheses of
Theorem~\ref{thm:rate} with $\partial^{\rm out}=\emptyset$. Consequently $K_-=\{c_1^-\}$, $K_+=\{0\}$,
$\Psi:\Z[0]\to\Z[0]$ is an isomorphism, and $\CH_*(S^\osx)=0$. There is a unique solution $c_1^*(w)$
of~\eqref{eq:bj} with $c_1^*\to c_1^-$ as $w\to0$. It satisfies $c_1^*\to0$ as $w\to\infty$, it is
hyperbolic, in the sense that its linearized operator $L:C^1_b\to C^0_b$ is an isomorphism, and its
existence is independent of $\theta_0$ and of the rate at which the
system interpolates between its limits.
\end{proposition}

\begin{proof}
$P<0$ outside $[c_1^-,c_1^+]$ and $P>0$ inside. At $c_L$ both terms of the numerator
in~\eqref{eq:bj-slab} are negative, and at $c_R$ both are positive, for every $u\in[0,1]$, while
$1-c_1/20>0$. Hence $N$ is strictly attracting for every frozen field. On $N$, the zeros of $f_-$ are
the zeros of $P$ in $N$, i.e.\ $c_1^-$ alone, and the only zero of $f_+=-\frac{3}{2\theta_0}c_1/(1-c_1/20)$
is $c_1=0$. Since $N$ is one-dimensional, $\Inv(N,f_\pm)$ consists of these equilibria, so $K_\pm$ are as
stated. Theorem~\ref{thm:rate} gives the isomorphism and existence.

\emph{Uniqueness.} Let $F(c_1,u)$ be the right-hand side of~\eqref{eq:bj-slab}. At $p_-=(c_1^-,0)$ its
fiber derivative is $\partial_{c_1}F=-P'(c_1^-)/(1-c_1^-/20)\approx-1.67$, so there are a neighborhood $U$ of
$c_1^-$ in $N$ and $\delta>0$ with $\partial_{c_1}F<-\frac12$ on $U\times[0,\delta)$. Let $c_1$ and $c_2$ be
two solutions over the same base solution $u(t)$, both tending to $c_1^-$ as $t\to-\infty$. There is $T$
such that $c_1(t),c_2(t)\in U$ and $u(t)<\delta$ for $t\le T$. Since $F(\cdot,u)$ is decreasing on $U$, the
function $z=(c_1-c_2)^2$ satisfies $\dot z=2(c_1-c_2)\big(F(c_1,u)-F(c_2,u)\big)\le0$ on $(-\infty,T]$,
so $z(T)\le z(t)\to0$ as $t\to-\infty$. Hence $c_1(T)=c_2(T)$, and $c_1=c_2$ by uniqueness of solutions.
Thus there is exactly one solution $c_1^*$ with $c_1^*\to c_1^-$ as $w\to0$. The argument uses only
$u\to0$ in backward time, so it holds for every rate.

\emph{IR limit.} Since $c_1^-\in\operatorname{int}N$ and $N\times[0,1]$ is forward invariant, $c_1^*(w)\in N$
for all $w$. The $\omega$-limit set of the orbit lies in $\{u=1\}$ (because $u$ increases to $1$), is
compact and invariant, and is contained in $N$. It is therefore $\{0\}$, the only invariant set of $f_+$
in $N$, so $c_1^*\to0$.

\emph{Invertibility of $L$.} Let $a(t)$ be the derivative of the right-hand side of~\eqref{eq:bj-slab}
with respect to $c_1$ along $(c_1^*(t),u(t))$. Then $a$ is continuous and bounded, and $a(t)\to
a_-\approx-1.67$ as $t\to-\infty$ and $a(t)\to a_+=-3/(2\theta_0)$ as $t\to+\infty$. Both limits are
negative, hence hyperbolic with unstable dimension $0$, and $Ly=\dot y-a(t)y$ is Fredholm of index $0-0=0$ by
Proposition~\ref{prop:palmer}(ii). Every solution $y(t)=y(0)\exp\int_0^ta$ with $y(0)\neq0$ grows
exponentially as $t\to-\infty$, so $\ker L=0$, and $L:C^1_b\to C^0_b$ is an isomorphism by
Proposition~\ref{prop:palmer}(iii). Parts (ii) and (iii) of Proposition~\ref{prop:palmer} concern only
the fiber linearization, so they do not need the normal hyperbolicity $g'(1)<0$, which fails at the
degenerate IR end.
\end{proof}

\begin{figure}[t]
\centering
\includegraphics[width=0.66\textwidth]{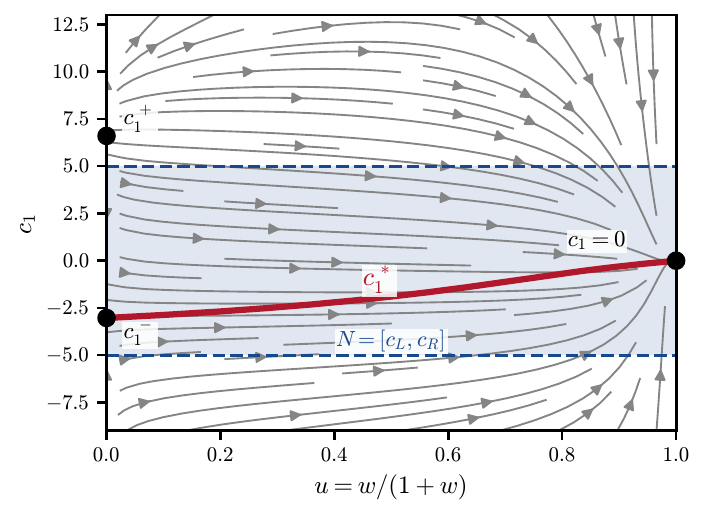}
\caption{The compactified Bjorken system~\eqref{eq:bj-slab} for $\theta_0=1$ in the variables $(u,c_1)$,
$u=w/(1+w)$. The UV fiber $u=0$ carries the equilibria $c_1^\pm$, and the IR fiber $u=1$ carries $c_1=0$.
The shaded band is the trapping region $N=[c_L,c_R]$ of Proposition~\ref{prop:bjorken}, here with
$c_L=-5$ and $c_R=5$. The flow enters it through both edges for every $u$. The red curve is the unique
solution $c_1^*$ leaving $c_1^-$, which is the hydrodynamic attractor. Solutions leaving $c_1^+$ start
outside $N$, and their IR fate is not constrained by $N$.}
\label{fig:bjorken}
\end{figure}

The solution $c_1^*$ (Fig.~\ref{fig:bjorken}) is the analog, in this model, of the hydrodynamic attractor~\cite{Heller:2015dha}.
Proposition~\ref{prop:bjorken} establishes it topologically, in the sense that it is robust under every perturbation for
which a common block exists. It answers, at $L=1$, the question raised in Ref.~\onlinecite{Behtash:2019}
whether a flow line connects the far-from-equilibrium fixed point $c_1^-$, at which time is the only
repelling direction, to the hydrodynamic fixed point. At this truncation it also gives a proof, without
regularization, of the existence of the attractor solution described by the Morse decomposition of
Ref.~\onlinecite{Behtash:2020jhep}. The UV point $c_1^+$ ($\mu=1$) emits a one-parameter family of
solutions, whose IR fate (to $c_1=0$ or to the singular line) is not constrained by $N$.

%======================================================================
\section{The full three-dimensional Gubser flow}\label{sec:full3d}
%======================================================================

Sections~\ref{subsec:reduced-index} and~\ref{subsec:transversal} work with isolated invariant sets in
the invariant plane $T=0$ and its normal direction. We now consider the six-point set of
Table~\ref{tab:gubser}. We prove the compactification, the index of the whole doubled space and the
indices of the invariant circles at infinity (Lemma~\ref{lem:TC}). We then prove the global Morse decomposition,
together with the vanishing of the indices of the Morse sets that contain the $B$- and $C$-points, by
explicit isolating blocks (Theorem~\ref{thm:M}), and we deduce the indices of the attractors
(Corollary~\ref{cor:attractors}).

\subsection{Failure of isolation and Poincar\'e compactification}

In IS theory nothing prevents the shear stress or the temperature from growing without bound in finite
de Sitter time. Such solutions leave the regime in which the truncation is trustworthy, and in the
finite phase space they make the natural invariant sets non-compact. We first show that this already
spoils isolation, and then compactify instead of discarding these solutions, so that their fate becomes part of the index-theoretic approach.

Three hypersurfaces of~\eqref{eq:gT}--\eqref{eq:gtau} are invariant, namely $T=0$ and $\tau=\pm1$. On $T=0$ the
flow reduces to the warm-up system under $(\pib,\tau)\leftrightarrow(x,p)$, with $\Sig_U(A)$,
$\Sig_U(B)$, $\Sig_I(B)$, $\Sig_I(A)$ playing the roles of $A,C,D,B$. Already there, the bounded
invariant set spanned by the connections is not compact. By~\eqref{eq:warmup-sol}, the connections
$\Sig_U(A)\to\Sig_I(A)$ have $\max\pib=a(1+K)/(1-K)\to\infty$ as $K\to1^-$, and the solutions with $K>1$
and $t<-t_*$ lie in $W^u(\Sig_U(A))$ and reach $\pib=+\infty$ at $t=-t_*$. Hence the
connecting orbits between the six points are not contained in any compact subset of the finite phase
space, no compact neighborhood isolates the six-point set together with its connections, and its Conley
index is undefined. Exploratory integration suggests that the unstable manifolds of the other UV points
have escaping branches as well, but this is not needed. The compactification below brings all escaping
orbits into a compact space.

\begin{remark}
Heuristically, the obstruction is also visible algebraically. Along $T=0$ the flow contains the chain
$\Sig_U(A)\to\Sig_U(B)\to\Sig_I(B)\to\Sig_I(A)$ of Morse sets of doubled index $3,2,1,0$, each pair
joined by an orbit on an invariant line, and the middle link is transversal
(Proposition~\ref{prop:transversal}). If the six-point set were isolated with these connections, a
connection matrix with three consecutive nonzero entries would need canceling entries to satisfy
$\Delta^2=0$. The candidates are unavailable, because $\Sig_U(C)\to\Sig_I(C)$ is excluded by the sign of $T$,
and the remaining candidate connections escape.
\end{remark}

\paragraph{Poincar\'e compactification.}
Since $\tau\in[-1,1]$ is already bounded, it suffices to compactify the fiber. Regarding $\tau$ as a
bounded coefficient, the fiber vector field is a polynomial of degree two in $(T,\pib)$, and Poincar\'e
compactification~\cite{Dumortier:2006} replaces the $(T,\pib)$ plane by a closed disc. The extended phase space becomes the solid cylinder
$D^2\times[-1,1]$, whose lateral boundary is an invariant cylinder at infinity. We use the following
coordinates. Near the positive $T$-direction at infinity,
\begin{equation}\label{eq:chartTp}
u=\frac{\pib}{T},\qquad v=\frac1T>0,\qquad ds=|T|\,d\rho,
\end{equation}
in which the flow is polynomial, namely
\begin{equation}\label{eq:flowTp}
\frac{du}{ds}=-\frac uc-\frac53u^2\tau+\frac4{15}v^2\tau+\frac23uv\tau,\qquad
\frac{dv}{ds}=\frac{v\tau}3\,(2v-u),\qquad\frac{d\tau}{ds}=v\,(1-\tau^2).
\end{equation}
Near the negative $T$-direction we use $u=\pib/T$, $\hat v=1/|T|>0$ and $ds=|T|\,d\rho$, obtaining
\begin{equation}\label{eq:flowTm}
\frac{du}{ds}=\frac uc+\frac53u^2\tau-\frac4{15}\hat v^2\tau+\frac23u\hat v\tau,\qquad
\frac{d\hat v}{ds}=\frac{\hat v\tau}3\,(u+2\hat v),\qquad\frac{d\tau}{ds}=\hat v\,(1-\tau^2).
\end{equation}
Near the $\pib$-directions one uses $T/\pib$ and $1/|\pib|$ in the same way. The circle at infinity is
$\{v=0\}$ (resp.\ $\{\hat v=0\}$). It is invariant, and on it the base variable is frozen, $d\tau/ds=0$,
so infinity is foliated by invariant circles and its equilibria form curves over $\tau\in[-1,1]$. At
$v=0$, $du/ds=-u\,(1/c+\frac53u\tau)$, whose zeros are the $T$-axis curve $u=0$ and the diagonal curve
$u=-3/(5c\tau)$. Together with the $\pib$-axis curves found in the coordinates near the $\pib$-directions, these are the three
families of equilibria at infinity. The diagonal curves cross the $\pib$-axis curves transcritically at
$\tau=0$. The $T$-axis curves are not hyperbolic, since the linearization of~\eqref{eq:flowTp} at $u=v=0$ has
eigenvalues $-1/c$ and $0$ in the $(u,v)$ directions. A center-manifold reduction gives
$u=(4c\tau/15)v^2+O(v^3)$, with reduced radial dynamics $dv/ds=(2\tau/3)v^2+O(v^3)$, that is
$dv/d\rho=(2\tau/3)v+O(v^2)$ in physical time. Hence these curves attract radially for $\tau<0$ and
repel for $\tau>0$.

Compactification restores compactness, as the escaping branches acquire $\omega$-limits on the cylinder.
The new invariant sets lie at $|\pib|\to\infty$ and $|T|\to\infty$, i.e.\ in the expansion-dominated and
collision-dominated regimes in which one or the other term of~\eqref{eq:gpi} is neglected entirely.
Whether such limiting regimes deserve standing in the index of the physical theory is a modeling
question, which we return to below.

\subsection{The doubled compactified system and the circles at infinity}

The compactified system is not of the form of Hypothesis~\ref{hyp:H}, because the base equation
$d\tau/ds=v(1-\tau^2)$ depends on the fiber point and vanishes on the cylinder at infinity. We
therefore use the results of Sec.~\ref{sec:theory} only for invariant sets in the finite part
$\{v>0\}$ (respectively $\{\hat v>0\}$). There the compactified vector field is a positive multiple
of~\eqref{eq:gT}--\eqref{eq:gtau}. The two fields have the same oriented orbits, hence the same isolating
neighborhoods, index pairs and Conley indices, and Hypothesis~\ref{hyp:H} holds for the original field.
Sets that meet the cylinder at infinity are treated by explicit blocks (Lemma~\ref{lem:TC} and
Theorem~\ref{thm:M}).

In $\Ss$ the finite equilibria still lie on the invariant boundary $\tau=\pm1$, and by
Theorem~\ref{thm:slab} the index in $\Ss$ of a set in the finite part detects only the IR. We therefore
pass to the double of Sec.~\ref{subsec:double}, replacing $\tau$ by $\cos\theta$ and
$d\tau/ds=v(1-\tau^2)$ by $d\theta/ds=v\,G(\theta)$, and similarly in the other coordinates. The map
$(z,\theta)\mapsto(z,\cos\theta)$ is again a semiconjugacy, since the factor $v$ multiplies both
equations. The result is the solid torus $D^2\times S^1$, whose boundary is the doubled cylinder at
infinity. The whole solid torus is compact and invariant, so it is an isolated invariant set with
empty exit set, and
\begin{equation}\label{eq:solidtorus}
\CH_*(D^2\times S^1;\Z)=H_*(D^2\times S^1;\Z)=\Z[0]\oplus\Z[1].
\end{equation}
Theorem~\ref{thm:equivariant} does not apply to this set. Its fibers over $\theta=0$ and $\theta=\pi$ do
not form an attractor--repeller pair, because the circles at infinity consist of equilibria over every
$\theta$. Equation~\eqref{eq:solidtorus} is a direct computation. The six finite equilibria lie in the
finite part and carry the doubled indices $3,2,2,1,1,0$ of Proposition~\ref{prop:boundary}, and $R$ and
$\tilde{\mathcal S}$ act as in Sec.~\ref{subsec:symmetries}.

In the double the $T$-axis curves close into circles,
\begin{equation}
TC^+\coloneqq\{u=v=0\},\qquad TC^-\coloneqq\{u=\hat v=0\},
\end{equation}
parametrized by $\theta\in S^1$, at $T=+\infty$ and $T=-\infty$ respectively. The time-reversal symmetry $\tilde{\mathcal
S}$ acts by $(u,v,\theta)\mapsto(-u,\hat v=v,\pi-\theta)$ and exchanges the neighborhoods of $T=+\infty$ and
$T=-\infty$.

\begin{lemma}\label{lem:TC}
Let $0<\delta<3/(20c)$, put $\delta'\coloneqq\sqrt{\delta/(4c)}$ and
\begin{equation}
N^+\coloneqq\{|u|\le\delta,\ 0\le v\le\delta'\}\times S^1,\qquad N^-\coloneqq\{|u|\le\delta,\
0\le\hat v\le\delta'\}\times S^1 .
\end{equation}
\begin{enumerate}
\item[(i)] $N^+$ is an isolating block with exit set $L^+=\{v=\delta',\ \cos\theta\ge0\}$ and
$\Inv(N^+)=TC^+$. Hence $TC^+$ is an isolated invariant set, $(N^+,L^+)$ is an index pair for it,
$h(TC^+)=S^1$ and $\CH_*(TC^+;\Lambda)=\Lambda[1]$ for every coefficient ring $\Lambda$.
\item[(ii)] $N^-$ is an isolating block with exit set $L^-=\{|u|=\delta\}\cup\{\hat v=\delta',\
\cos\theta\ge0\}$ and $\Inv(N^-)=TC^-$. Hence $(N^-,L^-)$ is an index pair for $TC^-$, $h(TC^-)=S^2$
and $\CH_*(TC^-;\Lambda)=\Lambda[2]$ for every coefficient ring $\Lambda$.
\item[(iii)] In the compactified extended phase space, i.e.\ with $S^1$ replaced by $\tau\in[-1,1]$, the
sets $N^\pm$ are isolating blocks with exit sets $\{v=\delta',\ \tau\ge0\}$ and $\{|u|=\delta\}\cup\{\hat
v=\delta',\ \tau\ge0\}$, their invariant sets are the $T$-axis curves, and both pointed quotients
$N^\pm/L^\pm$ are contractible. Hence the $T$-axis curves have trivial index $\bar0$ there.
\item[(iv)] For the reversed flow, $N^\pm$ are isolating blocks whose exit sets are the entrance sets
$E^+=\{|u|=\delta\}\cup\{v=\delta',\ \cos\theta\le0\}$ and $E^-=\{\hat v=\delta',\ \cos\theta\le0\}$ of the
forward flow. Hence $h(TC^+)=S^2$ and $h(TC^-)=S^1$ for the reversed flow, and in the setting of (iii)
both $T$-axis curves have trivial index $\bar0$ for the reversed flow as well.
\end{enumerate}
\end{lemma}

The proof, a direct check of the faces of the blocks, is given in Appendix~\ref{app:3d}.

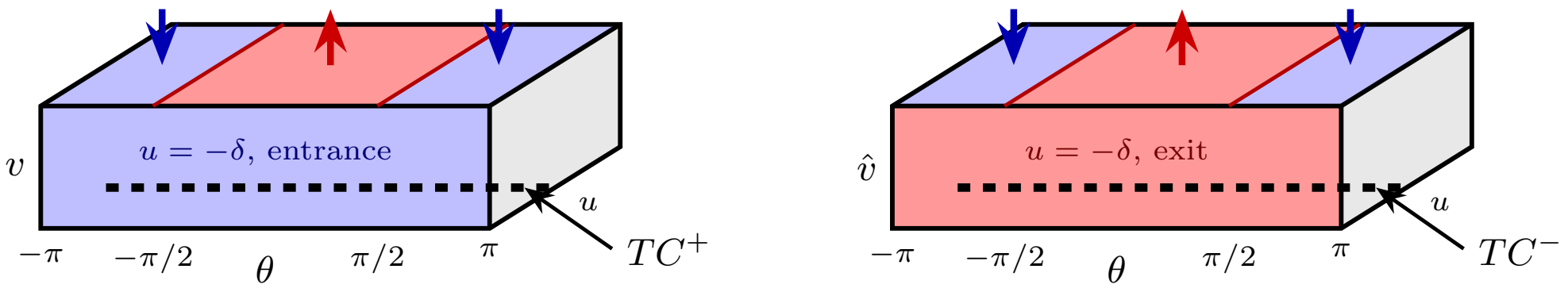
\begin{figure}[t]
\centering
\newcommand{\tcblock}[4]{%
\begin{tikzpicture}[x={(1cm,0cm)},y={(0.8cm,0.5cm)},z={(0cm,1cm)},>=Stealth,font=\small,scale=0.85,line width=1.1pt]
 \draw[black!55,dashed,line width=0.9pt] (0,-0.8,0) -- (0,0.8,0) -- (4.4,0.8,0);
 \draw[black!55,dashed,line width=0.9pt] (0,0.8,0) -- (0,0.8,1.2);
 \fill[#1] (0,-0.8,0) -- (4.4,-0.8,0) -- (4.4,-0.8,1.2) -- (0,-0.8,1.2) -- cycle;
 \fill[blue!25] (0,-0.8,1.2) -- (1.1,-0.8,1.2) -- (1.1,0.8,1.2) -- (0,0.8,1.2) -- cycle;
 \fill[blue!25] (3.3,-0.8,1.2) -- (4.4,-0.8,1.2) -- (4.4,0.8,1.2) -- (3.3,0.8,1.2) -- cycle;
 \fill[red!40] (1.1,-0.8,1.2) -- (3.3,-0.8,1.2) -- (3.3,0.8,1.2) -- (1.1,0.8,1.2) -- cycle;
 \fill[gray!18] (4.4,-0.8,0) -- (4.4,0.8,0) -- (4.4,0.8,1.2) -- (4.4,-0.8,1.2) -- cycle;
 \draw (0,-0.8,0) -- (4.4,-0.8,0) -- (4.4,0.8,0) -- (4.4,0.8,1.2) -- (0,0.8,1.2) -- (0,-0.8,1.2) -- cycle;
 \draw (0,-0.8,1.2) -- (4.4,-0.8,1.2) -- (4.4,-0.8,0);
 \draw (4.4,-0.8,1.2) -- (4.4,0.8,1.2);
 \draw[red!70!black,line width=0.9pt] (1.1,-0.8,1.2) -- (1.1,0.8,1.2) (3.3,-0.8,1.2) -- (3.3,0.8,1.2);
 \draw[line width=2pt,dashed] (0,0,0) -- (4.4,0,0);
 \draw[<-,line width=0.9pt] (4.1,0,0) -- (5.6,-0.8,-0.2) node[right] {#2};
 \draw[->,line width=1.6pt,red!80!black] (2.2,0,1.2) -- (2.2,0,1.75);
 \draw[->,line width=1.6pt,blue!70!black] (0.55,0,1.75) -- (0.55,0,1.2);
 \draw[->,line width=1.6pt,blue!70!black] (3.85,0,1.75) -- (3.85,0,1.2);
 #3
 \foreach \x/\l in {0/{-\pi},1.1/{-\pi/2},3.3/{\pi/2},4.4/{\pi}}
   \node[below,font=\scriptsize] at (\x,-0.8,0) {$\l$};
 \node[below=3pt] at (2.2,-0.8,0) {$\theta$};
 \node[left] at (0,-0.8,0.6) {$#4$};
 \node[below right,font=\scriptsize] at (4.4,0.1,0) {$u$};
\end{tikzpicture}}
\begin{minipage}[b]{0.49\textwidth}
\centering
\tcblock{blue!25}{$TC^+$}{\node[font=\scriptsize,blue!45!black] at (2.2,-0.8,0.75) {$u=-\delta$, entrance};}{v}\\[2pt]
(a) $N^+$, with $\CH_*(TC^+)=\Z[1]$
\end{minipage}\hfill
\begin{minipage}[b]{0.49\textwidth}
\centering
\tcblock{red!40}{$TC^-$}{\node[font=\scriptsize,red!45!black] at (2.2,-0.8,0.75) {$u=-\delta$, exit};}{\hat v}\\[2pt]
(b) $N^-$, with $\CH_*(TC^-)=\Z[2]$
\end{minipage}
\caption{The isolating blocks of Lemma~\ref{lem:TC}, shown in the coordinates $(\theta,u,v)$
or $(\theta,u,\hat v)$, with the faces $\theta=\pm\pi$ identified. The bottom face $v=0$ lies in the
invariant cylinder at infinity and contains the circle $TC^\pm=\{u=0\}$ (thick dashed). Red faces belong
to the exit set and blue faces to the entrance set. On the top face the flow leaves where $\cos\theta\ge0$
and enters where $\cos\theta\le0$, in both blocks. The faces $u=\pm\delta$ (only $u=-\delta$ is visible)
are entrance faces for $N^+$ and exit faces for $N^-$. Hence $L^+$ is a contractible strip, whereas $L^-$ is
two annuli joined by a strip, i.e.\ $L^-\simeq S^1\vee S^1$, and the indices differ by one degree.}
\label{fig:TCblocks}
\end{figure}

The circles $TC^\pm$ are exchanged by $\tilde{\mathcal S}$ (Fig.~\ref{fig:TCblocks}), and their degrees are complementary,
$1+2=3$. Since they lie on the boundary at infinity, this is not an instance of
Theorem~\ref{thm:duality} (Remark~\ref{rem:duality-boundary}) but follows from the explicit blocks. The
doubling has a striking effect on these sets. In $\Ss$ they are Conley-trivial, while on the double
they close into circles and acquire an index.

\subsection{Global Morse decomposition}

We begin by outlining the global picture that motivated the construction. Although it does not enter into the proofs, this picture clarifies the reasoning behind them.

\begin{figure}[t]
\centering
\begin{tikzpicture}[>=Stealth,font=\small,line width=1.1pt,
  ms/.style={draw,line width=1.3pt,rounded corners,align=center,minimum width=1.8cm,minimum height=1.0cm}]
 \node[ms] (UA) at (0,0) {$\Sig_U(A)$\\ $\Z[3]$};
 \node[ms] (TCm) at (2.9,0) {$TC^-$\\ $\Z[2]$};
 \node[ms,dashed] (Mp) at (5.8,0.85) {$M^+$\\ $0$};
 \node[ms,dashed] (Mm) at (5.8,-0.85) {$M^-$\\ $0$};
 \node[ms] (TCp) at (8.7,0) {$TC^+$\\ $\Z[1]$};
 \node[ms] (IA) at (11.6,0) {$\Sig_I(A)$\\ $\Z[0]$};
 \draw[->,line width=2.2pt,red!80!black] (UA) -- node[above] {$\Delta$} (TCm);
 \draw[->] (TCm) -- (Mp);
 \draw[->] (TCm) -- (Mm);
 \draw[->] (Mp) -- (TCp);
 \draw[->] (Mm) -- (TCp);
 \draw[->] (TCp) -- (IA);
 \draw[->,dotted,line width=1.4pt,black!60] (TCm.south) .. controls (3.6,-2.3) and (8.0,-2.3) ..
   node[below,font=\scriptsize] {entry $0$} (TCp.south);
\end{tikzpicture}
\caption{The Morse decomposition of the doubled compactified Gubser flow (Theorem~\ref{thm:M}), ordered
by~\eqref{eq:order3d}, with arrows pointing from higher to lower Morse sets. The indices of $TC^\pm$ are
given by Lemma~\ref{lem:TC}, and the dashed boxes mark the Morse sets $M^\pm$, whose indices vanish by
Theorem~\ref{thm:M}. The only nonzero entry of the connection matrix is $\Delta\,\Sig_U(A)=TC^-$ (thick
red), and the homology of $\Delta$ is $\F[0]\oplus\F[1]$, carried by $\Sig_I(A)$ and $TC^+$, in agreement
with~\eqref{eq:solidtorus}.}
\label{fig:morse3d}
\end{figure}
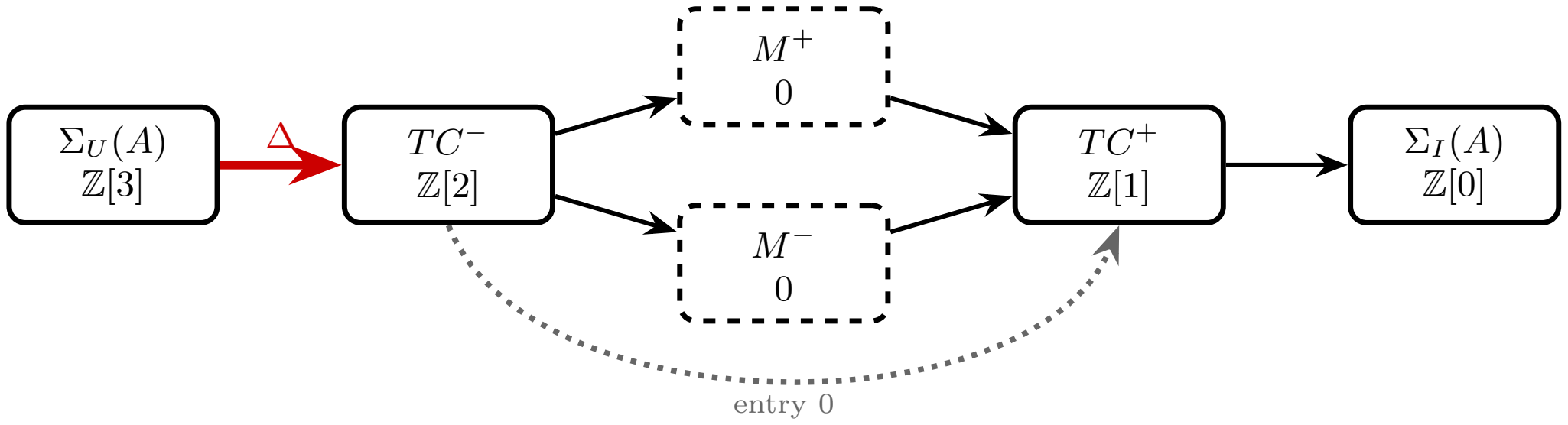

\begin{remark}[Exploratory numerics]\label{rem:numobs}
Exploratory integration of the doubled compactified flow, which is not used in any proof, suggested the
following picture.
\begin{enumerate}
\item[(a)] The flow admits a six-set Morse decomposition into $\Sig_U(A)$, $\Sig_I(A)$, $TC^\pm$, and
\begin{equation}\label{eq:Mpm}
M^-=X^-\cup\Sig_U(B)\cup\Sig_I(B)\cup\{\text{orbits}\},\qquad M^+=X^+\cup\Sig_U(C)\cup\Sig_I(C)\cup
\{\text{orbits}\},
\end{equation}
where $X^\mp$ denote the $\pib^\mp$-axis and diagonal complexes at infinity (defined precisely below). Because the base angle is frozen at infinity, $\epsilon$-chains drift along the
equilibrium curves, which suggests that each family of curves at infinity belongs to the same Morse set as
the finite saddles cyclically related to it.
\item[(b)] There are no connecting orbits between $M^+$ and $M^-$, and the Morse sets are ordered by
\begin{equation}\label{eq:order3d}
\Sig_U(A)>TC^->\{M^+,M^-\}>TC^+>\Sig_I(A),
\end{equation}
with $M^+$ and $M^-$ incomparable (Fig.~\ref{fig:morse3d}).
\end{enumerate}
\end{remark}

We now prove (a) and (b), and that the Morse sets $M^\pm$ have trivial Conley index. The proof uses explicit isolating blocks, and it rests on two elementary properties of
the flow, namely the structure of the equilibria at infinity and two pairs of barrier planes.

On the circle at infinity $C_\theta$ over $\theta\in S^1$, with $\tau=\cos\theta$, the equilibria are
$TC^\pm(\theta)$, the points $P^\pm(\theta)$ in the directions $\pib\to\pm\infty$, and, for $\tau\neq0$, the
diagonal points $D^\pm(\theta)$ on the sides $\pm T>0$, at which $\pib/T=u_*\coloneqq-3/(5c\tau)$. Their
stability and the orbits on $C_\theta$ are determined in Lemma~\ref{lem:barrier} of Appendix~\ref{app:3d}.
The barriers are the planes $\pib=\mp a$, on which $d\pib/d\rho=\pm aT/c$ for every value of $\tau$. Since
the sign of $T$ is preserved, the half-spaces $\{\pib\ge-a\}$ and $\{\pib\le a\}$ are positively invariant
for $T>0$, and $\{\pib\le-a\}$ and $\{\pib\ge a\}$ are positively invariant for $T<0$.

We write $X^-$ for the union over $\theta$ of the points $P^-(\theta)$, of $D^+(\theta)$ with $\tau>0$ and of
$D^-(\theta)$ with $\tau<0$, and $X^+$ for the union of $P^+(\theta)$, of $D^+(\theta)$ with $\tau<0$ and of
$D^-(\theta)$ with $\tau>0$. Since $u_*\to\mp\infty$ as $\tau\to0^\pm$, each diagonal arc ends on a
$\pib$-axis curve of the same complex, and $X^\pm$ are compact. These are the complexes of
Remark~\ref{rem:numobs}(a).

\begin{theorem}[Global Morse decomposition]\label{thm:M}
Let $0<\eta<a$. For every sufficiently small $\delta>0$, let $K^-$ and $K^+$ be the closures of
$\{\pib\le-a+\eta,\ \pib\le-\delta|T|\}$ and of $\{\pib\ge a+\eta,\ \pib\ge\delta|T|\}$ in $D^2\times S^1$,
and let $M^\pm\coloneqq\Inv(K^\pm)$.
\begin{enumerate}
\item[(i)] $K^\mp$ are isolating blocks, and the Conley indices of $M^\pm$ are trivial, i.e.,
\begin{equation}\label{eq:Mzero}
h(M^+)=h(M^-)=\bar0,\qquad\text{so that}\qquad\CH_*(M^+;\Z)=\CH_*(M^-;\Z)=0 .
\end{equation}
\item[(ii)] $M^-$ contains $\Sig_U(B)$, $\Sig_I(B)$ and $X^-$, and $M^+$ contains $\Sig_U(C)$, $\Sig_I(C)$
and $X^+$.
\item[(iii)] The sets $\Sig_U(A)$, $TC^-$, $M^+$, $M^-$, $TC^+$ and $\Sig_I(A)$ form a Morse decomposition
of $D^2\times S^1$ with the order~\eqref{eq:order3d}, and there is no orbit from $M^+$ to $M^-$ or from
$M^-$ to $M^+$.
\end{enumerate}
In particular, the Morse decomposition suggested in Remark~\ref{rem:numobs} exists. The theorem does not
assert that $M^+$ or $M^-$ is chain transitive, so the chain-recurrent set may split further inside $M^\pm$.
\end{theorem}

The blocks $K^\mp$ are bounded by one of the barrier planes, slightly shifted, and by a plane through the
$T$-axis at infinity. Explicit conditions on $\delta$, the exit sets of $K^\mp$ and the proof are given in
Appendix~\ref{app:3d}.

\subsection{Connection matrix, attractors and physical interpretation}

Theorem~\ref{thm:M} reduces the global topology of the doubled compactified flow to four Morse sets with
nontrivial index. We now compute the connection matrix, deduce the indices of the late-time attractors,
and interpret the result.

\paragraph{Connection matrix.}
By Theorem~\ref{thm:M} the sets $M^\pm$ contribute nothing, so the graded module of the Morse
decomposition is generated over $\F$ by $\Sig_U(A)$ in degree 3, $TC^-$ in degree 2, $TC^+$ in degree 1
(Lemma~\ref{lem:TC}) and $\Sig_I(A)$ in degree 0. The connection matrix $\Delta$ has degree $-1$, so its only
possible entries are the coefficients of $TC^-$ in $\Delta\Sig_U(A)$, of $TC^+$ in $\Delta TC^-$ and of
$\Sig_I(A)$ in $\Delta TC^+$. Its homology is $\CH_*(D^2\times S^1;\F)=\F[0]\oplus\F[1]$
by~\eqref{eq:solidtorus}, so $\Delta$ has rank one and must cancel the classes in degrees 3 and 2. Hence
\begin{equation}
\Delta\,\Sig_U(A)=TC^-,\qquad\Delta\,TC^-=\Delta\,TC^+=\Delta\,\Sig_I(A)=0,\qquad
H_*(\Delta)=\F[0]\oplus\F[1].
\end{equation}
Since $\Sig_U(A)$ and $TC^-$ are adjacent in~\eqref{eq:order3d}, the nonzero entry forces~\cite{Franzosa:1989} a connecting orbit
from $\Sig_U(A)$ to $TC^-$. This orbit also follows from two attractor--repeller
sequences, without connection matrix theory, over any nonzero coefficient ring. Let $R_1$ be the Morse set
of $\{TC^-,\Sig_U(A)\}$, the repeller dual to the attractor $A_3$ of Corollary~\ref{cor:attractors} in
$D^2\times S^1$. By~\eqref{eq:solidtorus} and Corollary~\ref{cor:attractors}, $\CH_3(D^2\times S^1)=0$ and
$\CH_2(A_3)=0$, so the sequence of $(A_3,R_1)$ gives $\CH_3(R_1)=0$. In the sequence of $(TC^-,\Sig_U(A))$
in $R_1$, the connecting map $\CH_3(\Sig_U(A))\to\CH_2(TC^-)$ is therefore injective. Here
$\CH_*(\Sig_U(A))=\Z[3]$, because the time-reversal symmetry $\tilde{\mathcal S}$ carries $\Sig_U(A)$ to $\Sig_I(A)$ with the
reversed flow, for which every face of the block $N_A$ of Appendix~\ref{app:3d} is crossed outward, so that $(N_A,\partial N_A)\cong
(D^3,S^2)$ is an index pair. The connecting map is thus nonzero, and $\Sig_U(A)$ is joined to $TC^-$ by an
orbit. The vanishing of $\Delta TC^-$ is consistent with the fact
that no orbit runs from $T=-\infty$ to $T=+\infty$ across the invariant plane $T=0$. In this picture the
topology is carried by two objects. The UV source cancels against the circle at infinity at negative
temperature, and the loop class of the doubled phase space is carried by the circle at infinity at
positive temperature. The indices of the attractors of the Morse decomposition do not need the connection
matrix. They follow from attractor--repeller sequences over every commutative coefficient ring.

\begin{corollary}[Attractor indices]\label{cor:attractors}
Let $A_1\subset A_2\subset A_3$ be the attractors of the Morse decomposition of Theorem~\ref{thm:M} that
correspond to the downward closed subsets
\[
\{\Sig_I(A),TC^+\},\qquad\{\Sig_I(A),TC^+,M^-\},\qquad\{\Sig_I(A),TC^+,M^-,M^+\}
\]
of~\eqref{eq:order3d}, i.e., $A_j$ is the union of these Morse sets and of the connecting orbits between
them. Then, for every commutative ring $\Lambda$ and $j=1,2,3$,
\[
\CH_*(A_j;\Lambda)\cong\Lambda[0]\oplus\Lambda[1],
\]
and the maps $\CH_*(A_1;\Lambda)\to\CH_*(A_2;\Lambda)\to\CH_*(A_3;\Lambda)$ induced by inclusions of index
pairs are isomorphisms. Each $A_j$ has a compact positively invariant isolating neighborhood $B_j$ with
$\Inv(B_j)=A_j$, and every such $B_j$ satisfies $H_*(B_j;\Lambda)\cong\Lambda[0]\oplus\Lambda[1]$.
\end{corollary}

The proof uses only attractor--repeller sequences and is given in Appendix~\ref{app:3d}.

Corollary~\ref{cor:attractors} has a direct physical meaning. A compact positively invariant
neighborhood $B_j$ is a region of states that, once entered, is never left, i.e., a trapping region for
the late-time dynamics. For each of the three attractors, every such region has the homology of a
circle. This loop class is carried by the circle $TC^+$ at infinity at positive temperature, where the
collision rate $T/c$ is unbounded and $\pib/T\to0$, so it is a remnant of the collision-dominated regime. Adding the sectors $M^\pm$, which contain the
separatrix and the $C$-points, changes neither the homology nor, by the isomorphisms of the corollary, the
way it is carried.

We close this section with the physical interpretation of Theorem~\ref{thm:M}. The $C$-points lie in
$M^+$, which carries no index. The sector discarded by the limit $c\to+\infty$ thus contributes no index-theoretic content,
even in the most generous treatment available, namely compactified, doubled, and with every orbit at infinity
retained. The vanishing of $\CH_*(M^-)$ does not conflict with Sec.~\ref{sec:gubser}, because the
results concern different invariant sets. At finite $c$ the separatrix is itself isolated, with doubled
index $\Z[1]\oplus\Z[2]$ and transport map $\Psi=\pm1$ (Corollary~\ref{cor:gubserB}). In the compactified
space $M^-$ also contains the complexes $X^-$ at infinity, whose contributions cancel those of the
separatrix.

%======================================================================
\section{Conley index of the \texorpdfstring{$(4+\epsilon)$}{(4+epsilon)}-dimensional \texorpdfstring{$O(N)$}{O(N)} model with cubic anisotropy}\label{sec:ON}
%======================================================================

The question of Sec.~\ref{subsec:questions}, whether the RG flow changes topology as a discrete parameter is tuned, has
a classic answer in the theory of critical phenomena. In an $N$-component scalar theory with an
isotropic quartic coupling $u$ and a cubic anisotropy $v$, the isotropic $O(N)$ fixed point is stable
against the anisotropy for small $N$ and unstable for large $N$, and the cubic fixed point takes over.
At one loop, as Aharony showed~\cite{Aharony:1973}, the critical value is $N_c=4$. Following Gukov~\cite{Gukov:2016tnp}, we
call this theory the $O(N)$ model, with the understanding that it includes the cubic deformation, which
breaks $O(N)$ to the hypercubic group $S_N\ltimes(\Z_2)^N$ of the cubic fixed point. Gukov uses this
flow to illustrate isolating neighborhoods, and he explains why the exchange of stability at $N_c$ is a
transcritical bifurcation protected by the different symmetries of the two fixed points. The unlabeled
Morse polynomial does not change at $N_c$, since the two fixed points exchange their indices, although
the indices of the labeled fixed points change. The total index cannot detect this transition, and one
must use connection data. In this section we use this model as an
autonomous case, and then let $N$ depend on time to obtain an asymptotically autonomous example.

\subsection{The one-loop flow and its connection data}

\begin{figure}[t]
\centering
\includegraphics[width=\textwidth]{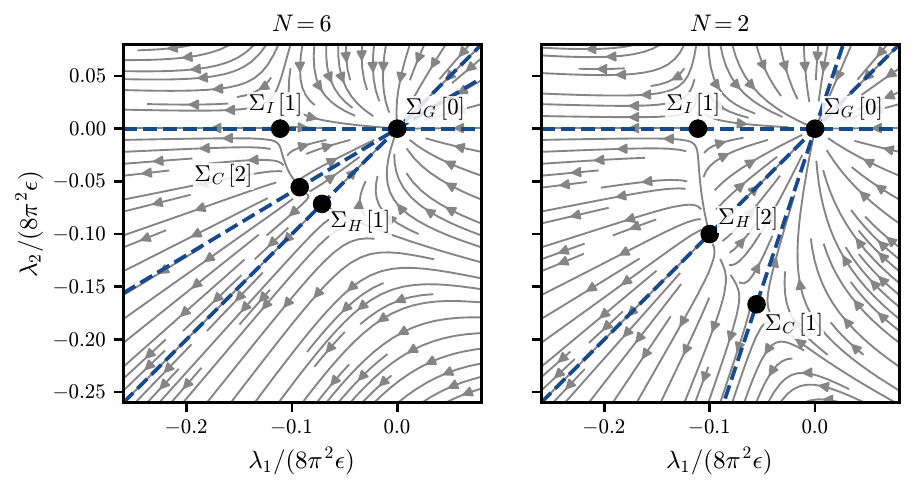}
\caption{Phase portraits of the flow $d\lambda/dt=-\beta(\lambda)$ of~\eqref{eq:ONbeta} in the rescaled
couplings $\lambda/(8\pi^2\epsilon)$, for $N=6$ (left) and $N=2$ (right). The dashed lines are the
three invariant lines $\lambda_2=0$, $\lambda_1=\lambda_2$ and $\lambda_2=3\lambda_1/(N-1)$ of
Lemma~\ref{lem:ONstructure}, and the Morse indices are shown in brackets.
For $N>4$ the source is $\Sig_C$, which lies off the diagonal between $\Sig_I$ and $\Sig_H$. For $N<4$ the
points $\Sig_C$ and $\Sig_H$ have passed through each other on the diagonal at $N=4$ and exchanged
their indices, so that $\Sig_H$ is the source. The multiset of Morse indices, $\{0,1,1,2\}$, is the same
on both sides.}
\label{fig:ON}
\end{figure}

We consider the renormalized flow at leading order of the $(4+\epsilon)$-dimensional $O(N)$ model
with couplings $\lambda=(\lambda_1,\lambda_2)$,
\begin{equation}\label{eq:ONbeta}
\beta_1(\lambda)=\epsilon\lambda_1+\frac{9\lambda_1^2+(N-1)\lambda_2^2}{8\pi^2},\qquad
\beta_2(\lambda)=\epsilon\lambda_2+\frac{6\lambda_1\lambda_2+(N+2)\lambda_2^2}{8\pi^2},
\end{equation}
with flow time directed toward the IR, $d\lambda_i/dt=-\beta_i(\lambda)$, so that the stability index
counts the positive eigenvalues of $J=-\partial\beta/\partial\lambda$. The axis $\lambda_2=0$ and
the diagonal $\lambda_1=\lambda_2$ are invariant lines of the flow, the structural analogs of the
invariant hypersurfaces $T=0$ and $\tau=\pm1$ of the Gubser analysis, and there is a third invariant
line through the origin (Lemma~\ref{lem:ONstructure}). The fixed points are
\begin{equation}
\begin{aligned}
\Sig_G&=(0,0), & \Sig_I&=\Big(-\frac{8\pi^2\epsilon}{9},0\Big),\\
\Sig_C&=\Big(-\frac{8\pi^2\epsilon(N-1)}{9N},-\frac{8\pi^2\epsilon}{3N}\Big), &
\Sig_H&=\Big(-\frac{8\pi^2\epsilon}{N+8},-\frac{8\pi^2\epsilon}{N+8}\Big),
\end{aligned}
\end{equation}
with $\Sig_G,\Sig_I$ on the axis and $\Sig_G,\Sig_H$ on the diagonal. The eigenvalues of $J$ are
\begin{equation}
\Sig_G:(-\epsilon,-\epsilon),\quad\Sig_I:\Big(\epsilon,-\frac\epsilon3\Big),\quad
\Sig_C:\Big(\epsilon,\frac{\epsilon(N-4)}{3N}\Big),\quad\Sig_H:\Big(\epsilon,\frac{\epsilon(4-N)}{N+8}\Big).
\end{equation}
For $\epsilon>0$ and $N>4$ the indices are $(\Sig_G,\Sig_I,\Sig_H,\Sig_C)=(0,1,1,2)$, i.e.\ the Gaussian point
is the IR sink and $\Sig_C$ the source. At $N=4$ the points $\Sig_C$ and $\Sig_H$ collide on the
diagonal, and for $N<4$ the indices become $(0,1,2,1)$, so the source role has passed from $\Sig_C$ to
$\Sig_H$ (Fig.~\ref{fig:ON}).

In terms of $u=\lambda_2$ and $v=\lambda_1-\lambda_2$, \eqref{eq:ONbeta} becomes
\[
\beta_u=\epsilon u+\frac{(N+8)u^2+6uv}{8\pi^2},\qquad\beta_v=\epsilon v+\frac{12uv+9v^2}{8\pi^2},
\]
which is the one-loop beta function of the $N$-vector model with cubic anisotropy~\cite{Aharony:1973},
continued to $d=4+\epsilon$. The labels stand for Gaussian, Ising, Heisenberg and cubic, namely $\Sig_H$,
on the line $v=0$, is the isotropic (Heisenberg) $O(N)$ point, $\Sig_I$, on the line $u=0$, is the
decoupled Ising point, and $\Sig_C$ is the cubic point. The eigenvalue
$\epsilon(4-N)/(N+8)$ of $\Sig_H$ is the crossover exponent of the cubic anisotropy at the isotropic point,
and its sign change at $N=4$ is the exchange of stability described above. For $\epsilon>0$ the
interacting fixed points sit at negative couplings and are ultraviolet fixed points, while the Gaussian
point is infrared stable. Reversing the flow and the signs of the couplings maps the system to the
familiar $d=4-\epsilon$ flow, so the topological results below carry over to that case, with the
Morse indices complemented, $k\mapsto2-k$.

The fixed points and eigenvalues above are exact, and the Morse polynomial $1+2t+t^2=(1+t)^2$ is the
same on both sides of $N=4$, so the unlabeled index data are insensitive to the transition, although
the indices of $\Sig_C$ and $\Sig_H$ are exchanged. We now show
that the connection data are not, and we determine them. Since the flow is polynomial, solutions can
blow up in finite time, so it is a local flow on $\R^2$, to which Conley index theory
applies~\cite{Conley:1978,Rybakowski:1987}. An explicit isolating block makes any analysis at infinity
unnecessary. It is convenient to rescale, $y=\lambda/(8\pi^2\epsilon)$ and $s=\epsilon t$,
which turns the flow into
\begin{equation}\label{eq:ONy}
y_1'=-\big(y_1+9y_1^2+(N-1)y_2^2\big),\qquad y_2'=-\big(y_2+6y_1y_2+(N+2)y_2^2\big),
\end{equation}
with fixed points
\[
\Sig_G=(0,0),\qquad\Sig_I=\big(-\tfrac19,0\big),\qquad\Sig_C=\big(-\tfrac{N-1}{9N},-\tfrac1{3N}\big),\qquad
\Sig_H=\big(-\tfrac1{N+8},-\tfrac1{N+8}\big).
\]
A direct computation shows that these are all the zeros of the
vector field. We assume $N>1$ throughout, so that the weight $N-1$ in the gradient metric of
Lemma~\ref{lem:ONstructure}(i) is positive.

\begin{lemma}[Structure of the one-loop flow]\label{lem:ONstructure}
Let $N>1$.
\begin{enumerate}
\item[(i)] With the constant metric $g=\operatorname{diag}\big(1,\frac{N-1}3\big)$ and
\begin{equation}\label{eq:ONV}
V(y)=\tfrac12y_1^2+\tfrac{N-1}6y_2^2+3y_1^3+(N-1)\,y_1y_2^2+\tfrac{(N-1)(N+2)}9\,y_2^3,
\end{equation}
the flow~\eqref{eq:ONy} is the gradient flow $y'=-g^{-1}\nabla V$. In particular, $V$ strictly decreases
along every non-constant orbit.
\item[(ii)] The lines $y_2=0$, $y_2=y_1$ and $y_2=\frac3{N-1}\,y_1$ are invariant. They pass through
$\Sig_G$ and contain $\Sig_I$, $\Sig_H$ and $\Sig_C$, respectively.
\item[(iii)] The closed quadrant $Q=\{y_1\le0,\ y_2\le0\}$ is positively invariant.
\item[(iv)] Every point with $y_1<-\frac19$, and every point of $Q$ with $y_2<-\frac1{N+2}$, has a
forward orbit that blows up in finite time. More precisely, its forward existence time is at most
$1/(9|y_1(0)|-1)$, respectively $1/((N+2)|y_2(0)|-1)$.
\end{enumerate}
\end{lemma}

The proofs of Lemma~\ref{lem:ONstructure}, Proposition~\ref{prop:ON}, Corollary~\ref{cor:ON} and
Proposition~\ref{prop:ONramp} are elementary but long, and they are given in Appendix~\ref{app:ON}.

The gradient structure~(i) is the familiar one of one-loop beta functions, and the third invariant
line~(ii) carries $\Sig_C$ in the same way as the axis carries $\Sig_I$ and the diagonal carries
$\Sig_H$. At $N=4$ it coincides with the diagonal, which is where $\Sig_C$ and $\Sig_H$ collide.

\begin{proposition}[A uniform isolating block]\label{prop:ON}
Let $N>1$, and let $B_N$ be the union of all bounded full orbits of~\eqref{eq:ONy}. Put
$\delta=\frac1{100}$,
\begin{equation}\label{eq:ONblock}
R=\big[-\tfrac18,\delta\big]\times\big[-\tfrac12,\delta\big],\qquad
L=\big(\{-\tfrac18\}\times[-\tfrac12,\delta]\big)\cup\big([-\tfrac18,\delta]\times\{-\tfrac12\}\big).
\end{equation}
\begin{enumerate}
\item[(a)] For every $N>1$, $R$ is an isolating block with exit set $L$, and
$\Inv(R)=B_N\subset\big[-\frac19,0\big]\times\big[-\frac1{N+2},0\big]$.
\item[(b)] Every orbit in $B_N$ converges to fixed points in both time directions, and every
non-constant orbit in $B_N$ runs from a fixed point $q$ to a fixed point $p$ with $V(p)<V(q)$. The
chain-recurrent set of $B_N$ consists of the fixed points. For $N\neq4$ the four fixed points, ordered
by their values of $V$, form a Morse decomposition of $B_N$.
\item[(c)] $B_N$ has trivial Conley index, $h(B_N)=\bar0$, so that $\CH_*(B_N;\Lambda)=0$ for every
coefficient ring $\Lambda$.
\end{enumerate}
\end{proposition}

In particular, the four fixed points and their connecting orbits form an isolated invariant set. It
continues across $N=4$ with the single isolating block $R$ for all $N>1$, and its index vanishes on
both sides. The index computation~(c) uses only the block, whereas the remaining part of~(a) shows that
$R$ captures every bounded orbit of the flow, so the connection data below describe all of the bounded
dynamics. The connection data follow.

\begin{corollary}[Connection data of the $O(N)$ model]\label{cor:ON}
Let $\Lambda$ be a coefficient ring.
\begin{enumerate}
\item[(i)] For every $N>1$ there are connecting orbits $\Sig_I\to\Sig_G$, $\Sig_H\to\Sig_G$ and
$\Sig_C\to\Sig_G$ along the three invariant lines.
\item[(ii)] Let $N>4$. The four fixed points form a Morse decomposition of $B_N$ with respect to the
order $\Sig_G<\Sig_I,\Sig_H<\Sig_C$, in which the saddles $\Sig_I$ and $\Sig_H$ are unrelated, i.e., no
orbit connects them. The homology indices are
\[
\CH_*(\Sig_G;\Lambda)\cong\Lambda[0],\qquad\CH_*(\Sig_I;\Lambda)\cong\CH_*(\Sig_H;\Lambda)\cong\Lambda[1],
\qquad\CH_*(\Sig_C;\Lambda)\cong\Lambda[2].
\]
Each of $(\Sig_G,\Sig_I)$, $(\Sig_G,\Sig_H)$, $(\Sig_I,\Sig_C)$ and $(\Sig_H,\Sig_C)$ is an
attractor--repeller pair in an isolated invariant set with trivial homology index, so that all connecting
homomorphisms of its sequence~\eqref{eq:AR} are isomorphisms. There are connecting orbits from the source
$\Sig_C$ to both saddles.
\item[(iii)] Let $1<N<4$. Then (ii) holds with $\Sig_C$ and $\Sig_H$ exchanged, i.e., with the order
$\Sig_G<\Sig_I,\Sig_C<\Sig_H$, the indices $\Lambda[0]$, $\Lambda[1]$, $\Lambda[1]$, $\Lambda[2]$ of
$\Sig_G$, $\Sig_I$, $\Sig_C$, $\Sig_H$, the pairs $(\Sig_G,\Sig_I)$, $(\Sig_G,\Sig_C)$, $(\Sig_I,\Sig_H)$,
$(\Sig_C,\Sig_H)$, and connecting orbits from the source $\Sig_H$ to both saddles.
\end{enumerate}
\end{corollary}

\begin{remark}[Connection matrices]\label{rem:ONconn}
Over a field $\F$, a connection matrix $\Delta$ of the Morse decomposition of Corollary~\ref{cor:ON}(ii)
exists~\cite{Franzosa:1989}. Its restriction to each of the intervals $\{\Sig_G,\Sig_I\}$,
$\{\Sig_G,\Sig_H\}$, $\{\Sig_I,\Sig_C\}$ and $\{\Sig_H,\Sig_C\}$ of the order is the connecting
homomorphism of the corresponding attractor--repeller pair, which is an isomorphism by the corollary.
Hence for suitable generators $\Delta\Sig_I=\Delta\Sig_H=\Sig_G$ and $\Delta\Sig_C=a\Sig_I+b\Sig_H$ with
$a,b\neq0$, and $\Delta^2=0$ gives $a+b=0$, so that after rescaling $\Sig_C$ every connection matrix is
given by $\Delta\Sig_I=\Delta\Sig_H=\Sig_G$ and $\Delta\Sig_C=\Sig_I-\Sig_H$. In the same way, for $1<N<4$,
$\Delta\Sig_I=\Delta\Sig_C=\Sig_G$ and $\Delta\Sig_H=\Sig_I-\Sig_C$. For $N>4$ the nonzero
entries, from $\Sig_C$ to $\Sig_I$ and $\Sig_H$ and from these to $\Sig_G$, are those of the
connection matrix given by Gukov~\cite[Eq.~(2.19)]{Gukov:2016tnp}, and only the signs, which depend on
the orientation of the generators, are added here.
Unlike the computation via
transversality~\cite{McCord:1988}, this uses neither the invariant manifolds of the saddles nor the
existence of connecting orbits, which instead follow from the corollary.
\end{remark}

The connection $\Sig_H\to\Sig_G$ along the diagonal is a degree-lowering entry of $\Delta$ for $N>4$, but
for $N<4$ it becomes an index $2\to0$ connection that is invisible to $\Delta$. The connection
$\Sig_C\to\Sig_G$ along the third invariant line undergoes the opposite change. The passage through $N=4$
is therefore detected not by the indices but by the change of connection data, which is quantified by
a transition matrix in the sense of Refs.~\onlinecite{Franzosa:1998,McCord:1992tm}. We compute its analog at a fixed
rate in Proposition~\ref{prop:ONramp} and Remark~\ref{rem:ONrampconn} below. This is the precise form of the robustness question of Sec.~\ref{subsec:questions}, namely whether the RG flow
changes topology as a control parameter (here $N$, in general e.g.\ a fermion number) is tuned. It also
illustrates that connection data can change while the unlabeled Morse polynomial does not.

Two contrasts with the time-dependent examples should be noted. First, the fixed points here are
interior points of the coupling space and hyperbolic, so their Conley indices are defined directly and
no doubling is required. In the non-autonomous cases the fixed points sit on the invariant boundary
$\tau=\pm1$, and Sec.~\ref{sec:theory} must construct these indices. Second, the
stability exchange proceeds by a collision of interior fixed points, whereas the time-dependent phase
transition of Sec.~\ref{subsec:phase} proceeds by a sign change of the control parameter with the fixed
points held in place. By Corollary~\ref{cor:ON}, the two mechanisms realize the same index-level
phenomenon through different geometry, in that the unlabeled Morse polynomial stays the same while
the labeled indices and the connection data jump.

\subsection{A parameter shift through \texorpdfstring{$N=4$}{N=4}}

We now let the control parameter depend on the RG time $t$, i.e.\ on the energy scale. Physically, the
effective number of light components changes along an RG flow whenever some of them acquire a mass and
decouple below a threshold. A smooth shift of $N$ from above $N_c$ to below $N_c$ is the simplest
caricature of this situation. It is the RG analog of a finite-rate quench in the language of Kibble and
Zurek~\cite{Kibble:1976sj,Zurek:1985qw}, in which a control parameter is swept through a critical value in
physical time. It turns the $O(N)$ model into an
asymptotically autonomous system to which Sec.~\ref{sec:theory} applies, and it gives an example in
which the transport map vanishes while the transition matrix of Remark~\ref{rem:transition} does not. Let
$N$ be a Lipschitz function on $[-1,1]$ with values in $(1,\infty)$, let $g$ be a Lipschitz function on
$[-1,1]$ with $g(\pm1)=0$ and $g>0$ on $(-1,1)$, and replace $N$ in~\eqref{eq:ONy} by $N(\sigma(s))$. We
write $N_\mp\coloneqq N(\mp1)$, so that $f_\mp$ is the flow~\eqref{eq:ONy} at $N_\mp$. If $N\in C^1$ and $g$
is as in Hypothesis~\ref{hyp:H}, then Hypothesis~\ref{hyp:H} holds. In general, the compactified
system~\eqref{eq:slab} is Lipschitz and generates a skew-product flow, and we take as the base flow of the
double the flow on $S^1$ that fixes $0$ and $\pi$ and is conjugate by $\theta\mapsto\cos\theta$ to the flow
of $\dot\tau=g(\tau)$ on each open half circle. Then $q$ is again a semiconjugacy, and the proofs of
Lemma~\ref{lem:isolation}, Proposition~\ref{prop:boundary} and Theorem~\ref{thm:rate} apply without change,
since they use only these properties of the two flows. We write $\Sig_\bullet^-$ for the fixed points of
$f_-$ in the UV fiber $\theta=\pi$ of the double, and $\Sig_\bullet^+$ for those of $f_+$ in the IR fiber
$\theta=0$. Here UV and IR refer to the limits $s\to\mp\infty$ of the RG scale, and not to the stability
of the frozen fixed points. For instance, the constant solution at the Gaussian point is a UV$\to$IR
connection from $\Sig_G^-$ to $\Sig_G^+$, although $\Sig_G$ is infrared stable. For $N_->4$ and $N_+\in(1,4)$, by Proposition~\ref{prop:boundary}, the indices in the double
are $1,2,2,3$ for $\Sig_G^-,\Sig_I^-,\Sig_H^-,\Sig_C^-$, and $0,1,2,1$ for $\Sig_G^+,\Sig_I^+,\Sig_H^+,\Sig_C^+$.

\begin{figure}[t]
\centering
\includegraphics[width=\textwidth]{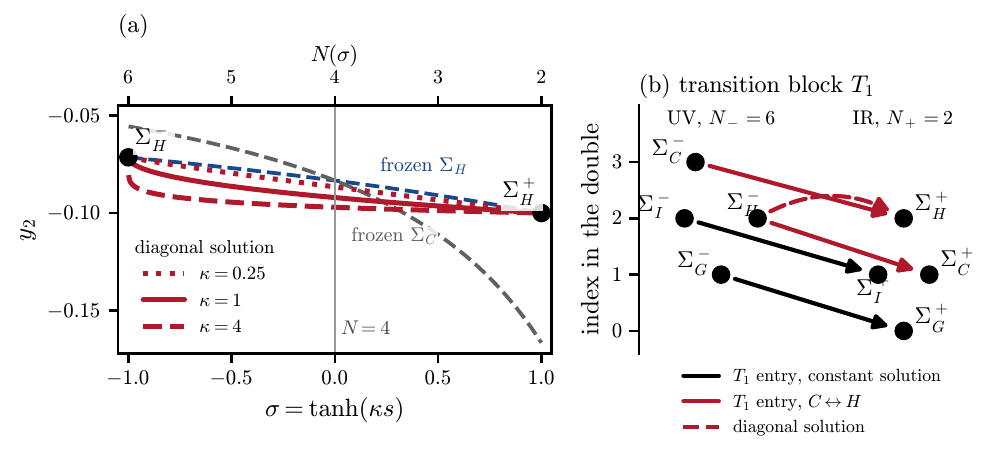}
\caption{The parameter shift of Proposition~\ref{prop:ONramp} for $N(\sigma)=4-2\sigma$, so that $N_-=6$ and
$N_+=2$, with $g(\sigma)=\kappa(1-\sigma^2)$, i.e.\ $\sigma(s)=\tanh(\kappa s)$.
(a)~The diagonal solution~\eqref{eq:ONdiag}, plotted against $\sigma$ for three rates $\kappa$, together
with the $y_2$-coordinates $-1/(N+8)$ and $-1/(3N)$ of the fixed points $\Sig_H$ and $\Sig_C$ of the
frozen flow~\eqref{eq:ONy} at $N=N(\sigma)$. These two curves cross where $N=4$. At every rate the
diagonal solution runs from $\Sig_H^-$ to $\Sig_H^+$, while its shape depends on $\kappa$.
(b)~The eight generators $\Sig_\bullet^\mp$ at the heights given by their indices in the double, and the
nonzero entries of the representative $T_1$ of~\eqref{eq:ONT}, which lower the index by one. The black entries are realized by the constant
solutions $y\equiv\Sig_G$ and $y\equiv\Sig_I$. The red entries exchange the labels $C$ and $H$, and the
entry from $\Sig_H^-$ is realized only through the chain that begins with the diagonal solution (dashed),
of Fredholm index $-1$. The red entries depend on the representative and are absent from $T_0$.}
\label{fig:ONramp}
\end{figure}

\begin{proposition}[Transition data of a parameter shift through $N=4$]\label{prop:ONramp}
Let $S=\Inv(R\times[-1,1])$, with $R$ as in~\eqref{eq:ONblock}, and let $\Lambda$ be a coefficient ring.
\begin{enumerate}
\item[(i)] $S^\osx$ is an isolated invariant set with $\CH_*(S^\osx;\Lambda)=0$, $K_\mp=B_{N_\mp}$ with
$\CH_*(K_\mp;\Lambda)=0$, and $\Psi=0$. Every solution that stays in $R$ converges to a fixed point of
$f_\mp$ as $s\to\mp\infty$. If $N_->4$ and $N_+\in(1,4)$, the eight points $\Sig_\bullet^\mp$ form a Morse
decomposition of $S^\osx$ for the order that is the order of Corollary~\ref{cor:ON} within each fiber and in
which $\Sig_p^+$ lies below $\Sig_q^-$ unless $q=G\neq p$, or $q=I$ and $p\in\{C,H\}$. In particular, no
solution runs from $\Sig_G^-$ to $\Sig_I^+$, $\Sig_C^+$ or $\Sig_H^+$, or from $\Sig_I^-$ to $\Sig_C^+$ or
$\Sig_H^+$.
\item[(ii)] Let $N\in C^1$, let $g$ be as in Hypothesis~\ref{hyp:H} with $g'(-1)>0>g'(1)$, and let $N_->4$
and $N_+\in(1,4)$. The constant solutions $y\equiv\Sig_G$ and $y\equiv\Sig_I$ are transversal connecting
solutions $\Sig_G^-\to\Sig_G^+$ and $\Sig_I^-\to\Sig_I^+$, and their linearized operators are
isomorphisms $C^1_b\to C^0_b$. The only solution that converges to $\Sig_H^+$ as
$s\to\infty$ is the diagonal solution
\begin{equation}\label{eq:ONdiag}
y_1(s)=y_2(s)=-\frac1{p_*(s)},\qquad p_*(s)=\int_s^\infty e^{s-r}\big(N(\sigma(r))+8\big)\,dr,
\end{equation}
which converges to $\Sig_H^-$ as $s\to-\infty$ and whose linearized operator has Fredholm index $-1$. No
solution runs from $\Sig_H^-$ to $\Sig_I^+$ or to $\Sig_C^+$.
\item[(iii)] Let $N_->4$ and $N_+\in(1,4)$, and let $M_A$ be the Morse set of the interval
$\{\Sig_I^+,\Sig_C^+,\Sig_H^+\}$ of the order in (i). Each of $(\Sig_G^+,\Sig_G^-)$, $(\Sig_I^+,\Sig_I^-)$
and $(M_A,\Sig_H^-)$ is an attractor--repeller pair in an isolated invariant set of $S^\osx$ with trivial
homology index. Hence the connecting homomorphisms
\[
\begin{gathered}
\CH_{k+1}(\Sig_G^-;\Lambda)\to\CH_k(\Sig_G^+;\Lambda),\qquad
\CH_{k+1}(\Sig_I^-;\Lambda)\to\CH_k(\Sig_I^+;\Lambda),\\
\CH_{k+1}(\Sig_H^-;\Lambda)\to\CH_k(M_A;\Lambda)
\end{gathered}
\]
are isomorphisms for every $k$, and $\CH_*(M_A;\Lambda)\cong\Lambda[1]$.
\end{enumerate}
\end{proposition}

\begin{remark}[Connection matrices of the shift]\label{rem:ONrampconn}
Let $N_->4$ and $N_+\in(1,4)$, let $\F$ be a field, and let $\Delta_\mp$ be the connection matrices of
Remark~\ref{rem:ONconn} at $N_\mp$. A connection matrix $\Delta$ of the Morse decomposition of
Proposition~\ref{prop:ONramp}(i) over $\F$ exists~\cite{Franzosa:1989}. Its restrictions to the attractor
$K_+\times\{0\}$ and to the repeller $K_-\times\{\pi\}$ are connection matrices of these sets. They are
$\Delta_+$ and, up to sign, by the continuation to a product flow in the proof of
Proposition~\ref{prop:boundary}, $\Delta_-$ with degrees raised by one. We orient the generators so that
both are as in Remark~\ref{rem:ONconn}. The remaining block $T$ has degree $-1$, and its entry from $p$ to
$q$ vanishes unless $q$ lies below $p$. By the order of Proposition~\ref{prop:ONramp}(i) and the degrees,
\[
T\Sig_G^-=t\,\Sig_G^+,\qquad T\Sig_I^-=h\,\Sig_I^+,\qquad T\Sig_H^-=a\,\Sig_I^++b\,\Sig_C^+,\qquad
T\Sig_C^-=\lambda\,\Sig_H^+ .
\]
No point lies strictly between $\Sig_G^+$ and $\Sig_G^-$, so $t$ is the connecting homomorphism of the pair
$(\Sig_G^+,\Sig_G^-)$, and $t\neq0$ by Proposition~\ref{prop:ONramp}(iii). Replacing every IR generator
$\Sig_p^+$ by $t\,\Sig_p^+$, which leaves $\Delta_+$ unchanged, we may assume $t=1$. The condition
$\Delta^2=0$ evaluated on $\Sig_I^-$, $\Sig_H^-$ and $\Sig_C^-$ gives
\[
(1+h)\,\Sig_G^+=0,\qquad(1+a+b)\,\Sig_G^+=0,\qquad(\lambda-1-a)\,\Sig_I^+-(\lambda+b)\,\Sig_C^+=0,
\]
hence $h=-1$, $a=\lambda-1$ and $b=-\lambda$. Thus every connection matrix of $S^\osx$ over $\F$ has the form
$\Delta=\Delta_-\oplus\Delta_++T_\lambda$ for some $\lambda\in\F$, where the transition matrix $T_\lambda$ is
\begin{equation}\label{eq:ONT}
T_\lambda\Sig_G^-=\Sig_G^+,\qquad T_\lambda\Sig_I^-=-\Sig_I^+,\qquad
T_\lambda\Sig_H^-=(\lambda-1)\,\Sig_I^+-\lambda\,\Sig_C^+,\qquad T_\lambda\Sig_C^-=\lambda\,\Sig_H^+ .
\end{equation}
A direct computation shows that the change of basis $\Sig_H^-\mapsto\Sig_H^-+\lambda\Sig_H^+$, which respects
the order because $\Sig_H^+$ lies below $\Sig_H^-$, conjugates $\Delta_-\oplus\Delta_++T_0$ to
$\Delta_-\oplus\Delta_++T_\lambda$. Finally $a+b=-1$, so $T_\lambda\Sig_H^-\neq0$ for every $\lambda$, in
accordance with the third pair of Proposition~\ref{prop:ONramp}(iii). The computation uses neither
transversality~\cite{McCord:1988} nor part (ii) of the proposition.
\end{remark}

The transition matrix is not unique. By Remark~\ref{rem:ONrampconn}, the full connection matrices
$\Delta_\lambda=\Delta_-\oplus\Delta_++T_\lambda$, $\lambda\in\F$, are conjugate by changes of basis that
respect the order, and connection matrices are in general unique only up to such filtered chain
isomorphisms~\cite{Spendlove:2026}. The blocks $T_\lambda$ themselves are not conjugate. The change of
basis mixes $\Sig_H^-$ with $\Sig_H^+$, and $T_0$ has rank $2$, while the representative $\lambda=1$,
\[
T_1\Sig_G^-=\Sig_G^+,\qquad T_1\Sig_I^-=-\Sig_I^+,\qquad T_1\Sig_H^-=-\Sig_C^+,\qquad T_1\Sig_C^-=\Sig_H^+ ,
\]
has rank $4$. Up to signs, $T_1$ exchanges the labels $C$ and $H$. This exchange is not an invariant of
the shift, since $T_0$ has no entry from $\Sig_C^-$ and no entry into $\Sig_C^+$. The data shared by all
$T_\lambda$ are the entries $T\Sig_G^-=\Sig_G^+$ and $T\Sig_I^-=-\Sig_I^+$, which represent the connecting
homomorphisms of the first two pairs of Proposition~\ref{prop:ONramp}(iii), and the relation $a+b=-1$. Since $H_1(M_A;\F)$ is spanned by the common class of $\Sig_I^+$ and $\Sig_C^+$, this
relation says that the grouped connecting homomorphism $\CH_{k+1}(\Sig_H^-)\to\CH_k(M_A)$ of
Proposition~\ref{prop:ONramp}(iii) is an isomorphism. Together with the diagonal
solution~\eqref{eq:ONdiag}, these data detect the passage through $N=4$ at every rate of the shift,
although $\Psi=0$ and the unlabeled Morse polynomials of the two boundary fibers agree
(Fig.~\ref{fig:ONramp}). No representative preserves all labels. Since $T$ lowers the index in the double
by one, a label-preserving $T$ would require $\mu_-(\Sig_\bullet)=\mu_+(\Sig_\bullet)$, and this fails for
$\Sig_C$ and $\Sig_H$.

The geometry behind the entry $T_\lambda\Sig_H^-$ is also instructive. By Proposition~\ref{prop:ONramp}(ii)
no solution runs from $\Sig_H^-$ to an IR saddle, so the nonzero entry is realized only through the chain
$\Sig_H^-\to\Sig_H^+\to\{\Sig_I^+,\Sig_C^+\}$. Its first link is the diagonal solution~\eqref{eq:ONdiag}, of
Fredholm index $-1$. Such connections are non-generic (Sec.~\ref{sec:theory}), and this one is
protected by the invariance of the diagonal. The diagonal is the line of $O(N)$-symmetric couplings,
on which the cubic anisotropy vanishes. Its invariance therefore follows from the $O(N)$ symmetry for
every $N$ and persists beyond one loop, while the explicit formula~\eqref{eq:ONdiag} for $p_*$ is
specific to the one-loop flow. The index
result for $(M_A,\Sig_H^-)$ in Proposition~\ref{prop:ONramp}(iii), on the other hand, uses neither the
diagonal solution nor transversality, only the blocks $R$, $\bar B_G$ and $B_H$ together with the Morse
decomposition of part (i). For perturbations of the flow that break the $O(N)$ symmetry, so that the diagonal is
no longer invariant, one therefore expects the entry to be realized by a direct solution from $\Sig_H^-$ to one of the IR saddles.

%======================================================================
\section{Conclusions and outlook}\label{sec:concl}
%======================================================================

For asymptotically autonomous systems admitting a $C^1$ compactification of time
(Hypothesis~\ref{hyp:H}), we have constructed an explicit UV$\to$IR transport map within the
classical Conley index of an autonomous flow. The extended phase space carries only IR information
(Theorem~\ref{thm:slab}). The double carries the UV and IR indices as the fixed-point data of a $\Z_2$
action, and its connecting homomorphism vanishes over every coefficient ring
(Theorem~\ref{thm:equivariant}). The one-sided extension carries the transport map $\Psi$, graded by
the indices of the UV and IR limit systems, and $\Psi\neq0$ forces connecting solutions
(Theorem~\ref{thm:transport}). Unlike the index of J\"anig over a cycle with a reversed
history~\cite{Janig:2020}, this construction needs no reversed leg, at the price of a prescribed
approach to the limits. If a single block $N$ with a fixed strict entrance/exit decomposition serves
all frozen systems, $\Psi$ is an isomorphism at every admissible rate. If, in addition,
$H_*(N,\partial^{\rm out})\neq0$, the isolated invariant sets $K_\pm$ are connected at every admissible
rate (Theorem~\ref{thm:rate} and Corollary~\ref{cor:moving}). This result does not by itself give
tracking of a particular branch of equilibria. Whether $\Psi$ agrees with a connecting homomorphism
of J\"anig's index under common hypotheses remains open.

In the kinetic-theory models these tools give the following results. The separatrix of the Gubser
flow is a transversal connection at every finite relaxation scale, with a splitting
$\Delta(c)\propto1/c$ in the physical normalization (Proposition~\ref{prop:transversal}), and an
explicit moving block isolates it with $\Psi=\pm1$ (Corollary~\ref{cor:gubserB}). In the Bjorken flow at
$L=1$, the attractor solution exists and is unique at every rate, without regularization
(Proposition~\ref{prop:bjorken}). For the compactified three-dimensional Gubser flow, explicit blocks give
the indices of the invariant circles at infinity and the global Morse decomposition, in which the Morse
sets containing the $B$- and $C$-points have trivial index (Lemma~\ref{lem:TC} and Theorem~\ref{thm:M}).
For the one-loop $O(N)$ model we obtain the connection matrices on both sides of $N=4$
(Remark~\ref{rem:ONconn}), in agreement with Ref.~\onlinecite{Gukov:2016tnp} for $N>4$. For a parameter shift of $N$ through $4$, $\Psi=0$ at every rate, but the
shift has an explicit diagonal connection of Fredholm index $-1$ and an isomorphic grouped connecting
homomorphism from $\Sig_H^-$ to the Morse set of the IR points (Proposition~\ref{prop:ONramp}). Hence the
connection data of the one-sided extension detect a transition that the transport map does not. The full
connection matrix is determined only up to filtered chain isomorphism~\cite{Spendlove:2026}, and its
transition block can change rank under such an isomorphism.

Several extensions are natural. For periodically driven systems the hull is a circle, and the relevant
object is the discrete Conley index of the period map. For random or noisy driving the hull is replaced
by a probability space. Moment hierarchies of the Boltzmann equation with many moments call for the
infinite-dimensional index theory of Rybakowski~\cite{Rybakowski:1987}, and the nonautonomous index of
J\"anig~\cite{Janig:2017,Janig:2020} already covers semilinear parabolic equations. For the compactified Gubser flow
and for the $O(N)$ model, the global results (Theorem~\ref{thm:M}, Proposition~\ref{prop:ON} and
Corollary~\ref{cor:ON}) follow from explicit isolating blocks built from barrier planes, invariant lines
and gradient structure. For systems without such structure, computer-assisted Conley--Morse
methods~\cite{Arai:2009,Bush:2012} are the natural tool, and it would be interesting to see how much of
the $O(N)$ picture survives beyond one loop. Finally, the Fredholm index $\mu_--\mu_+$ counts the free parameters
of a family of UV$\to$IR solutions. Costin~\cite{Costin:1995,Costin:1998} proved that nonlinear systems of
rank one can be solved near a non-resonant fixed point by Borel-summable trans-series, whose free
parameters multiply exponentially small corrections, and resurgent trans-series appear in hydrodynamics
through the factorial divergence of the gradient expansion~\cite{Heller:2015dha,Dorigoni:2014hea}. It would
be interesting to relate the Fredholm index to the number of free trans-series parameters of a family of
UV$\to$IR solutions.

\begin{acknowledgments}
A.~B.\ thanks Sergey Gukov for fruitful and inspiring discussions on Conley index theory during the
program ``Resurgent Asymptotics in Physics and Mathematics'' held at KITP, UC Santa Barbara, in Fall
2017. We also thank Tye Lidman for useful discussions and technical remarks, and Mike Miller for
pointing out the relative Borel construction. S.~K.\ is supported by JSPS KAKENHI Grant Nos.\ 22H05118
and 25K07298. The authors acknowledge using an internally developed agentic workflow in Stacks built on Claude Opus 5.5, Opus 5, Opus 4.8, and GPT-5.6 Sol for developing some of the arguments, proofreading the final manuscript, and verifying some of the mathematical calculations.
The same workflow also helped produce, in part, a formalization of the results of this paper in Lean~4, with partial human input (see the Data Availability statement). The formalization compiles without unproved steps and depends only on the standard axioms of Lean. Two
cited inputs enter as explicit hypotheses, namely McCord's duality theorem~\cite{McCord:1992dual},
used in Theorem~\ref{thm:duality}, and the existence and axioms of connection
matrices~\cite{Franzosa:1989}, used for the connection matrices of Secs.~\ref{sec:gubser},
\ref{sec:full3d} and~\ref{sec:ON}. The stable-manifold part of Remark~\ref{rem:palmergeom}, the
numerical value in Remark~\ref{rem:gap}, and the remarks that are not used in any proof are not
formalized. For Theorem~\ref{thm:equivariant}(ii), the formalization proves $2\partial^{\Dd}=0$ over
every coefficient ring, and $\partial^{\Dd}=0$ when $\CH_*(K_+;f_+;\Lambda)$ has no $2$-torsion, but
not the vanishing of $\partial^{\Dd}$ in general. The asymptotic form of the Bjorken attractor solution
in Sec.~\ref{sec:bjorken} is also not formalized.
\end{acknowledgments}

\section*{Author Declarations}

\subsection*{Conflict of Interest}

The authors have no conflicts to disclose.

\subsection*{Author Contributions}

\textbf{Alireza Behtash}: Conceptualization (lead); Methodology (equal); Formal analysis (equal);
Software (lead); Validation (lead); Writing -- original draft (lead); Writing -- review \& editing (lead).

\textbf{Syo Kamata}: Methodology (equal); Formal analysis (equal); Investigation (supporting);
Validation (supporting); Writing -- review \& editing (supporting).

\textbf{Mauricio M. Martinez}: Validation (supporting); Writing -- review \& editing (supporting).

\section*{Data Availability}

The data that support the findings of this study are available within the article. The Lean formalization
of the results is openly available at
\url{https://github.com/stacks-ai/papers/tree/main/conley-index-dynamical-systems}.

\appendix

%======================================================================
\section{Proofs and complements for Sec.~\ref{sec:theory}}\label{app:theory}
%======================================================================

\subsection{Fredholm index and transversality}

\begin{proof}[Proof of Proposition~\ref{prop:palmer}]
(i) follows from Lemma~\ref{lem:G}. Since $\sin\theta=0$ at $p_\pm$, the $\theta$-derivative of
$F(\cos\theta,x)$ vanishes there, so the linearization of~\eqref{eq:double} at $p_\pm$ is block diagonal,
with blocks $Df_\pm(e_\pm)$ and $G'(0)=g'(1)/2<0$, $G'(\pi)=g'(-1)/2>0$. The dichotomies of the two blocks
combine into a dichotomy of the linearization whose unstable dimension is the sum of theirs. A zero of a $C^1$
vector field at which the linearization has a dichotomy is a hyperbolic equilibrium of its flow, with Morse
index the unstable dimension.

(ii) Let $P_\pm$ be the projections of the dichotomies of $A_\pm$. Since $A(t)\to A_+$, a
Lyapunov--Perron argument on $[T,\infty)$ for large $T$, in which $A(t)-A_+$ is a small perturbation of
the dichotomy of $A_+$, gives two facts. First, $\dim E^s=\operatorname{rank}P_+=n-\mu_+$. Second,
there is a bounded operator $S_+:C^0_b\to\R^n$ such that a solution of $\dot y=A(t)y+h(t)$ is bounded on
$[0,\infty)$ if and only if $y(0)-S_+h\in E^s$. The time reversal $t\mapsto-t$ replaces $A(t)$ by
$-A(-t)$, which tends to $-A_-$ as $t\to\infty$, and $E^u$ by the stable subspace of $-A(-t)$. It gives
in the same way $\dim E^u=\mu_-$ and an operator $S_-$ with the analogous property on $(-\infty,0]$.
Since $\dot y=A(t)y+h(t)$ has a solution with every initial value, $h\in\operatorname{ran}L$ if and only
if $S_+h-S_-h\in E^u+E^s$, so $\operatorname{ran}L$ is closed. A solution of $\dot y=A(t)y$ in $C^1_b$ is
determined by $y(0)\in E^u\cap E^s$, and every such $y(0)$ occurs, so $\ker L\cong E^u\cap E^s$. The map
$h\mapsto[S_+h-S_-h]$ vanishes exactly on $\operatorname{ran}L$, and it is onto $\R^n/(E^u+E^s)$. Indeed, for
$\xi\in\R^n$, let $z$ solve $\dot z=A(t)z$ with $z(0)=\xi$, let $\chi$ be smooth with $\chi=0$ on
$(-\infty,0]$ and $\chi=1$ on $[1,\infty)$, and let $h=\dot\chi z$. Then $\chi z$ and $(\chi-1)z$ solve
$\dot y=A(t)y+h(t)$ and vanish on $(-\infty,0]$ and $[1,\infty)$, respectively, so $S_-h\in E^u$ and
$S_+h+\xi\in E^s$. Hence $\operatorname{coker}L\cong\R^n/(E^u+E^s)$, both are finite-dimensional, and
\[
\operatorname{ind}L=\dim(E^u\cap E^s)-\big(n-\dim(E^u+E^s)\big)=\dim E^u+\dim E^s-n=\mu_--\mu_+ .
\]

(iii) By (ii), $L$ is onto if and only if $E^u+E^s=\R^n$. If $\mu_-=\mu_+$, then $\dim E^u+\dim E^s=n$,
so $E^u+E^s=\R^n$ if and only if $E^u\cap E^s=0$, i.e.\ $\ker L=0$. A bounded bijection between Banach
spaces is an isomorphism.
\end{proof}

\begin{remark}[Geometric transversality]\label{rem:palmergeom}
In the situation of Proposition~\ref{prop:palmer}(i), let $x^*$ be a solution of~\eqref{eq:nonaut} with $x^*(t)\to e_\pm$ as
$t\to\pm\infty$, and let $\gamma=(x^*,\theta^*)$ be the corresponding orbit in the upper sheet of $\Dd$.
Since $F\in C^1$, $A(t)=D_xF(\sigma(t),x^*(t))$ is continuous and bounded, with limits
$A_\pm=Df_\pm(e_\pm)$, so parts (ii) and (iii) of the proposition apply. Now linearize~\eqref{eq:double} along $\gamma$. The
$\theta$-component $\delta\dot\theta=G'(\theta^*)\delta\theta$ decouples, and its solutions are the
multiples of $\dot\theta^*$. The $x$-component of $c\,\dot\gamma$ solves the inhomogeneous equation driven
by $c\,\dot\theta^*$, so every variation is $c\,\dot\gamma+(y,0)$ with $\dot y=A(t)y$. Since $p_\pm$ are
hyperbolic, the stable manifold theorem identifies the tangent space of $W^u(p_-)$ (resp.\ $W^s(p_+)$) at
$\gamma(0)$ with the initial values of the variations that are bounded on $(-\infty,0]$ (resp.\
$[0,\infty)$). The $\theta$-mode $\dot\gamma$ decays at both ends, so
\begin{equation}
T_{\gamma(0)}W^u(p_-)=\langle\dot\gamma(0)\rangle\oplus(E^u\times\{0\}),\qquad T_{\gamma(0)}W^s(p_+)=
\langle\dot\gamma(0)\rangle\oplus(E^s\times\{0\}).
\end{equation}
Since $\dot\gamma(0)$ has a nonzero $\theta$-component, the two tangent spaces span $\R^{n+1}$ if and
only if $E^u+E^s=\R^n$. Thus $x^*$ is transversal if and only if $W^u(p_-)$ and $W^s(p_+)$ intersect
transversally along $\gamma$. If, in addition, $\mu_-=\mu_+$, then $W^u(p_-)\cap W^s(p_+)$ is
one-dimensional near $\gamma(0)$, and a standard argument shows that $S^\osx=\gamma\cup\{p_-,p_+\}$ is
isolated. Indeed, an orbit in a small neighborhood of $S^\osx$ is either $p_\pm$ or has $\alpha$- and
$\omega$-limits $p_-$ and $p_+$, hence lies in $W^u(p_-)\cap W^s(p_+)$, and it crosses the section
$\{\theta=\theta^*(0)\}$ near $\gamma(0)$, so it equals $\gamma$. The connection map of a transversal
heteroclinic orbit between hyperbolic points of indices $m+1$ and $m$ is $\pm1$ times the number of such
orbits~\cite{McCord:1988}, so $\Psi=\pm1$ and $\CH_*(S^\osx)=0$. In the examples below we do not use this
argument, which rests on the invariant manifold theory of $p_\pm$. Instead, isolation and $\Psi$ are
obtained from explicit blocks through Corollary~\ref{cor:moving}, and transversality is proved
through Proposition~\ref{prop:palmer}(iii).
\end{remark}

\subsection{Equivariant structure of the double}

\begin{proof}[Proof of Theorem~\ref{thm:equivariant}]
(i) The axioms of an index pair are orbitwise conditions. Since $q$ maps orbits onto orbits and is
proper, they pull back. For example, if $y\in q^{-1}L$ and $\psi([0,t],y)\subset q^{-1}N$, then
$\varphi([0,t],q y)\subset N$, so $\varphi([0,t],qy)\subset L$ and $\psi([0,t],y)\subset q^{-1}L$. The
exit and isolation conditions are checked in the same way. $R$-invariance is clear. By
Lemma~\ref{lem:double}(b), $q$ identifies two points of $q^{-1}N$ exactly when they lie in the same
$R$-orbit. Since $q^{-1}N$ is compact, $q$ induces a homeomorphism $q^{-1}N/R\cong N$, which carries
$q^{-1}L/R$ onto $L$, so $Y/R\cong(q^{-1}N/R)/(q^{-1}L/R)\cong N/L$. The fixed points of $R$ in $q^{-1}N$
form $N_+\times\{0\}\sqcup N_-\times\{\pi\}$ with $N_\pm\times\{\pm1\}=N\cap X_\pm$, and likewise for $L$,
so $Y^R=(N_+/L_+)\vee(N_-/L_-)$. Both identifications are homeomorphisms of quotient spaces, not only
bijections. Indeed, the induced maps $Y/R\to N/L$ and $(N_+/L_+)\vee(N_-/L_-)\to Y^R$ are continuous
bijections from compact spaces, and their targets are Hausdorff, since $N/L$ is Hausdorff for every
compact $N$ and closed $L$ in a Hausdorff space. Finally, $(N_\pm,L_\pm)$ is an index pair of $K_\pm$ for
the restricted flow $f_\pm$ by the same orbitwise argument.

(ii) Write $A_\Dd=K_+\times\{0\}$ and $U_\Dd=K_-\times\{\pi\}$ for the attractor and repeller in
$\tilde S$, and $A_\Ss=K_+\times\{1\}$ and $U_\Ss=K_-\times\{-1\}$ for those in $S$
(Theorem~\ref{thm:slab}). Then $q^{-1}(A_\Ss)=A_\Dd$ and $q^{-1}(U_\Ss)=U_\Dd$. Choose an index triple
$N_0\subset N_1\subset N_2$ for $(A_\Ss,U_\Ss)$ in $S$ whose pairs are regular in the sense of the proof of
Theorem~\ref{thm:transport}. By the orbitwise argument of (i), applied to the pairs $(N_2,N_0)$,
$(N_1,N_0)$ and $(N_2,N_1)$, the preimages $q^{-1}N_0\subset q^{-1}N_1\subset q^{-1}N_2$ form an
$R$-invariant index triple for $(A_\Dd,U_\Dd)$ in $\tilde S$. Its pairs are regular, because flow collars
pull back under the semiconjugacy $q$. Let $i_\Dd$ and $i_\Ss$ be the maps induced by the inclusions
$(q^{-1}N_1,q^{-1}N_0)\subset(q^{-1}N_2,q^{-1}N_0)$ and $(N_1,N_0)\subset(N_2,N_0)$, and let $q_*$ and
$q_{A*}$ be the maps induced by $q$ on the pairs $(q^{-1}N_2,q^{-1}N_0)$ and $(q^{-1}N_1,q^{-1}N_0)$.
Functoriality gives
\[
q_*\circ i_\Dd=i_\Ss\circ q_{A*}.
\]
Since $q$ commutes with the flows, it commutes with the flow-defined comparison maps between index
pairs, so these maps do not depend on the triple. By Theorem~\ref{thm:slab}, $i_\Ss$ is an isomorphism.

We show that $q_{A*}$ is an isomorphism. Let $(N_K,L_K)$ be an index pair of $K_+$ for $f_+$, and
consider, for $\lambda\in[0,1]$, the flows $\dot x=F(1-\lambda(1-\tau),x)$, $\dot\tau=g(\tau)$ on
$N_K\times[\cos\eta,1]$ and $\dot x=F(1-\lambda(1-\cos\theta),x)$, $\dot\theta=G(\theta)$ on
$N_K\times[-\eta,\eta]$. The map $q$ is a semiconjugacy between them for every $\lambda$. As in the proof
of Proposition~\ref{prop:boundary}, the base equations do not involve $x$, so these sets isolate
$A_\Ss$ and $A_\Dd$ for every $\lambda$, and the flows at $\lambda=1$ are those of $\Ss$ and $\Dd$. The
continuation isomorphisms are induced by inclusions of index pairs of the flows at fixed $\lambda$ into
index pairs of the flows with $\lambda$ as an additional parameter. The preimage under $q\times\mathrm{id}$
of such a parametrized index pair is again one, so $q_{A*}$ at $\lambda=1$ is conjugate to $q_{A*}$ at
$\lambda=0$. At $\lambda=0$ the flows are products, with index pairs $(N_K\times[\cos\eta,1],
L_K\times[\cos\eta,1])$ and $(N_K\times[-\eta,\eta],L_K\times[-\eta,\eta])$, and $q=\mathrm{id}\times\cos$
is a homotopy equivalence of these pairs, since both intervals are contractible. Hence $q_{A*}$ is an
isomorphism.

Consequently $r\coloneqq q_{A*}^{-1}\circ i_\Ss^{-1}\circ q_*$ satisfies $r\circ i_\Dd=\mathrm{id}$, so
$i_\Dd$ is injective. By exactness of~\eqref{eq:AR}, the image of $\partial^{\Dd}$ is the kernel of
$i_\Dd$, so $\partial^{\Dd}=0$, and $r$ splits the resulting short exact sequence. The outer terms are
identified by Proposition~\ref{prop:boundary}, which holds with any coefficients. Finally, $R$ preserves
$\tilde S$, $A_\Dd$ and $U_\Dd$, the maps of the sequence are natural under $R$, and
Proposition~\ref{prop:boundary} gives $R_*=+1$ on $\CH_*(A_\Dd)$ and $R_*=-1$ on $\CH_*(U_\Dd)$.

(iii) The exact sequence~\eqref{eq:AR} with the identifications of Proposition~\ref{prop:boundary} has
finite-dimensional terms, and the alternating sum of dimensions along an exact sequence vanishes. This
gives the first equality over any field. The second follows from
Theorem~\ref{thm:slab}, which gives $\tilde\chi(h_{\Ss}(S))=\tilde\chi(h_{f_+}(K_+))$. For the third,
we show $\tilde H_*(Y^R)\cong\CH_*(K_+;f_+)\oplus\CH_*(K_-;f_-)$ directly, without using homology of wedge
sums. The fixed-point set $\mathrm{Fix}(R)=X\times\{0\}\sqcup X\times\{\pi\}$ is closed and invariant,
and the fixed parts $N_F\supset L_F$ of $q^{-1}N\supset q^{-1}L$ form an index pair of
$\tilde S\cap\mathrm{Fix}(R)$ for the flow on $\mathrm{Fix}(R)$, since all the axioms restrict. The
inclusion induces a continuous bijection $N_F/L_F\to Y^R$ from a compact space to a Hausdorff space, so
$Y^R\cong N_F/L_F$. By uniqueness of the index, the homotopy type of $Y^R$ is therefore the same for all
index pairs $(N,L)$ of $S$, and we may take $(N,L)$ regular in the sense of the proof of
Theorem~\ref{thm:transport}. Then $(N_F,L_F)$ is regular as well, since the collar restricts, and
$\tilde H_*(Y^R)\cong\tilde H_*(N_F/L_F)\cong H_*(N_F,L_F)$. The open sets $\{\cos\theta>0\}$ and
$\{\cos\theta<0\}$ are disjoint and cover $N_F$. Excision and the exact sequence of a triple therefore
give $H_*(N_F,L_F)\cong H_*(N_+,L_+)\oplus H_*(N_-,L_-)$, and the summands are $\CH_*(K_\pm;f_\pm)$ by
(i), because $(N_\pm,L_\pm)$ inherit the collar. The identities are consistent with the Euler
characteristic identity $\chi(Y)=2\chi(Y/R)-\chi(Y^R)$ for involutions.
\end{proof}

\begin{remark}[Borel cohomology]\label{rem:borel}
The same data can be phrased equivariantly. Let $Z_{\Z_2}=Z\times_{\Z_2}E\Z_2$ be Borel's
construction~\cite{Borel:1960,tomDieck:1987}, so that, as shown in Ref.~\onlinecite{Hausmann:2014}, $H^*(B\Z_2;\F)=\F[u]$ with
$|u|=1$. For a pointed $\Z_2$-space $(Z,z_0)$ with fixed base point, the reduced
Borel cohomology is $\tilde H^*_{\Z_2}(Z)\coloneqq H^*(Z_{\Z_2},B\Z_2)$, which removes all of
$H^*(B\Z_2)$ rather than one copy of $\F$. If $(q^{-1}N,q^{-1}L)$ is a finite $\Z_2$-CW pair, the
localization theorem~\cite{tomDieck:1987,Allday:1993} and (i) give
$u^{-1}\tilde H^*_{\Z_2}(Y;\F)\cong\big(\tilde H^*(h_{f_+}(K_+);\F)\oplus\tilde H^*(h_{f_-}(K_-);\F)\big)
\otimes\F[u,u^{-1}]$. Since $R$ acts freely off $Y^R$, also $H^*_{\Z_2}(Y,Y^R;\F)\cong H^*(Y/R,Y^R;\F)$.
Both isomorphisms repackage (i) and (iii) and are not used below.
\end{remark}

%======================================================================
\section{Proofs and complements for Sec.~\ref{sec:gubser}}\label{app:gubser}
%======================================================================

\subsection{The torus compactification}

\begin{proof}[Proof of Proposition~\ref{prop:torus}]
Write $\pib=\tan\varphi$ and $z=(\cos2\varphi,\sin2\varphi)$, which identifies $\RP^1$ with the unit
circle, so that $\pib=z_2/(1+z_1)$ and $w=z_2/(1-z_1)$. With $\cos\theta$ replaced by $1$, the
$\pib$-equation becomes the smooth field $\dot z=k(z)\,(-z_2,z_1)$ on the circle, where
$k(z)=\frac43\big((1+a^2)z_1-(1-a^2)\big)$. Let $\psi_s$ denote its flow. It has exactly the two
equilibria $\pib=\pm a$, i.e., the zeros of $k$. Since $\pib'=-\frac43(\pib^2-a^2)$ and
$w'=\frac43(1-a^2w^2)$, every other orbit of $\psi$ converges to $\pib=a$ as $s\to+\infty$ and to
$\pib=-a$ as $s\to-\infty$. The $\theta$-equation is autonomous. With $\tau=\cos\theta$ its flow is
$\tau(t)=(\tau\cosh t+\sinh t)/D(t,\tau)$, where $D(t,\tau)=\cosh t+\tau\sinh t>0$, and the doubled flow
is
\begin{equation}\label{eq:torus-skew}
(z,\theta)\mapsto\big(\psi_{s(t,\tau)}(z),\,\theta(t)\big),\qquad s(t,\tau)=\int_0^t\tau(r)\,dr=\log D(t,\tau),
\end{equation}
the cocycle property of $s$ being equivalent to $D(t,\tau(r))\,D(r,\tau)=D(t+r,\tau)$.

(i) The equilibria are the points with $\sin\theta=0$ and $\pib=\pm a$. In the coordinates
$(\pib,\theta)$ the linearization at such a point is
$\operatorname{diag}\big(-\frac83\pib\cos\theta,\,-\cos\theta\big)$, which gives the stated Morse
indices. If $|\tau|<1$, then $\theta(t)\to0$ as $t\to+\infty$, $\theta(t)\to\pi$ as $t\to-\infty$, and
$D(t,\tau)\ge\frac{1-|\tau|}2e^{|t|}$, so that $s(t,\tau)\to+\infty$ as $t\to\pm\infty$.
By~\eqref{eq:torus-skew} such an orbit runs from $\Sig_U(A)$ to $\Sig_I(A)$, unless $\pib=-a$, in which
case it runs from $\Sig_U(B)$ to $\Sig_I(B)$. On the fiber $\theta=0$ one has $s=t$, and on the fiber
$\theta=\pi$ one has $s=-t$, which gives the orbits in the fixed fibers. Thus every orbit that is not an
equilibrium runs from an equilibrium to a strictly smaller one. A closed ball around an equilibrium
that contains no other equilibrium then contains no other entire orbit, since the limits of such an
orbit would both equal the center. Hence each equilibrium is an isolated invariant set, and the
connecting orbits give the Morse decomposition.

(ii) The whole torus is an isolating block with empty exit set, so $(T^2,\emptyset)$ is an index pair
and $\CH_*(T^2)=H_*(T^2)$. Over a field this has total dimension four, the number of equilibria. The
homology of the connection matrix complex is isomorphic to $\CH_*(T^2)$, so the kernel of $\Delta$ has
dimension equal to the rank of $\Delta$ plus four, and $\Delta=0$ by the rank-nullity theorem.

(iii) The one-sided extension is the skew product $(z,\tau)\mapsto(\psi_{s(t,\tau)}(z),\tau(t))$ on
$\RP^1\times[-1,1]$, and $N=\RP^1$ is a common block with empty exit set, since $\RP^1$ has no boundary.
Theorem~\ref{thm:rate} gives $\CH_*=0$ and the isomorphism $\Psi$. The UV and IR blocks of the connection
matrix vanish by the argument of (ii), since each fiber index is $H_*(\RP^1)$, of rank two. The condition
$H(\Delta)=0$ then forces both off-diagonal entries to be nonzero, i.e., equal to one over $\F$. The
chains are given by~\eqref{eq:torus-skew}. In the fiber $\tau=-1$ one has $s=-t$, and the orbits between
the two equilibria run from $\Sig_U(A)$ to $\Sig_U(B)$. The separatrix $\pib=-a$ runs from $\Sig_U(B)$
to $\Sig_I(B)$, and in the fiber $\tau=1$ one has $s=t$, so the orbits there run from $\Sig_I(B)$ to
$\Sig_I(A)$.
\end{proof}

\subsection{Equivariant data of the separatrix}

Let $N=I_\delta\times S^1$ and $L$ be the isolating block of the separatrix and its exit set
(Sec.~\ref{subsec:reduced-index}), and $Y=N/L$. Theorem~\ref{thm:equivariant} gives, and a direct computation
confirms, the following.
\begin{itemize}
\item The fixed-point set is $Y^R=(I_\delta/\partial I_\delta)\vee(I_\delta\times\{\pi\})_+\simeq S^1\vee S^0=
h_{f_+}(\Sig_I(B))\vee h_{f_-}(\Sig_U(B))$.
\item The orbit space is $Y/R\simeq S^1=h_{\Ss}(S_B)\cong h_{f_+}(\Sig_I(B))$, in accordance with
Theorem~\ref{thm:slab}.
\item Relative to the fixed points, $H^*(Y/R,Y^R;\F)=\F[1]$. This is the suspended UV index
$\tilde H^0(h_{f_-}(\Sig_U(B)))=\tilde H^0(S^0)$ shifted by one, as noted after
Theorem~\ref{thm:equivariant}. In the language of Remark~\ref{rem:borel}, this group is
$H^*_{\Z_2}(Y,Y^R;\F)$, and the reduced Borel cohomology $\tilde H^*_{\Z_2}(Y;\F)$ is a free
$\F[u]$-module of rank $\dim\tilde H^*(Y^R)=2$ on degree-one generators.
\end{itemize}

\subsection{The moving block of the separatrix}\label{app:gubser-block}

\begin{lemma}\label{lem:gubser-block}
Let $c_1=\frac{7\sqrt5}{30}-\frac13$ and $c_2=\frac{3\sqrt5}{10}-\frac23$, so that $0<c_1<\frac15$ and
$0<c_2<\frac1{200}$. Along~\eqref{eq:gubser-xy} and $\tau'=1-\tau^2$, the coordinates~\eqref{eq:gubser-block}
satisfy
\begin{equation}\label{eq:xieta-dot}
\begin{aligned}
\xi'&=(1+c_1\tau+c_2\tau^2)\,\xi+c_2\tau(1+\tau)\,\eta+R_\xi,\\
\eta'&=-(1-c_1\tau+c_2\tau^2)\,\eta+c_2\tau(1-\tau)\,\xi+R_\eta,
\end{aligned}
\end{equation}
with the quadratic remainders
\begin{align*}
R_\xi&=\frac{\xi+\eta}{240}\Big[\big((3\sqrt5-5)\tau^2+(6\sqrt5-5)\tau+3\sqrt5\big)\eta
+\big((3\sqrt5-5)\tau^2-3\tau-3\sqrt5\big)\xi\Big],\\
R_\eta&=-\frac{\xi+\eta}{240}\Big[\big((3\sqrt5-5)\tau^2+3\tau-3\sqrt5\big)\eta
+\big((3\sqrt5-5)\tau^2-(6\sqrt5-5)\tau+3\sqrt5\big)\xi\Big].
\end{align*}
For all $(\xi,\eta)\in N_B$ and $\tau\in[-1,1]$,
\begin{enumerate}
\item[(a)] $\pm\xi'>\frac12$ on the faces $\xi=\pm1$ and $\pm\eta'<-\frac12$ on the faces $\eta=\pm1$,
\item[(b)] $\xi\xi'-\eta\eta'\ge\frac35(\xi^2+\eta^2)$.
\end{enumerate}
Consequently, the only solutions of the compactified flow that stay in $N_B$ for all times are those
with $\xi=\eta=0$, and the same holds for the frozen flows $f_\pm$.
\end{lemma}

\begin{proof}
The identities~\eqref{eq:xieta-dot} follow by substituting $x=\varepsilon\big((1-\tau)\xi-(1+\tau)\eta\big)/(2a)$
and $y=\varepsilon(\xi+\eta)/2$ into~\eqref{eq:gubser-xy} and differentiating~\eqref{eq:gubser-block},
including the terms that come from $\tau'=1-\tau^2$. For $|\tau|\le1$, the coefficients in the brackets
of $R_\xi$ and $R_\eta$ are bounded in absolute value by $12\sqrt5-10<17$ and $6\sqrt5-2<12$. Since
$|\xi+\eta|\le2$ on $N_B$, this gives $|R_\xi|,|R_\eta|\le\frac{29}{120}<\frac14$.

(a) On the face $\xi=\pm1$, \eqref{eq:xieta-dot} gives $\pm\xi'\ge1-c_1-2c_2-\frac14>\frac12$, and the
faces $\eta=\pm1$ are treated in the same way.

(b) The quadratic part of $\xi\xi'-\eta\eta'$ equals
$\xi^2+\eta^2+c_1\tau(\xi^2-\eta^2)+c_2\tau^2(\xi+\eta)^2\ge(1-c_1)(\xi^2+\eta^2)$. With $A=\xi+\eta$
and $B=\xi-\eta$, the cubic part is
\[
\xi R_\xi-\eta R_\eta=\frac{A}{240}\Big[(3\sqrt5-5)\tau^2A^2-3\tau AB-3\sqrt5B^2\Big].
\]
Since $|A|\le2$, $3|AB|\le\frac32(A^2+B^2)$ and $A^2+B^2=2(\xi^2+\eta^2)$, its absolute value is at most
$\frac{3\sqrt5+3/2}{60}(\xi^2+\eta^2)<\frac17(\xi^2+\eta^2)$. Hence
$\xi\xi'-\eta\eta'\ge\big(1-\frac15-\frac17\big)(\xi^2+\eta^2)>\frac35(\xi^2+\eta^2)$.

For the last assertion, let $V=\xi^2-\eta^2$ along a solution that stays in $N_B$ for all times. By (b),
$V'\ge\frac65(\xi^2+\eta^2)\ge\frac65|V|$. The set $\{\xi=\eta=0\}$ is invariant, so a solution that
meets it at one time lies in it at all times. If $(\xi,\eta)\neq0$ and $V\ge0$ at some time $\rho_0$,
then $V>0$ for $\rho>\rho_0$, and $V'\ge\frac65V$ forces $V\to\infty$ as $\rho\to\infty$. If $V<0$ at
all times, then $V'\ge-\frac65V$ forces $V\to-\infty$ as $\rho\to-\infty$. Both contradict $|V|\le1$ on
$N_B$. The estimates (a) and (b) hold for every $\tau\in[-1,1]$, including the fibers $\tau=\pm1$, where
$\tau'=0$, so the argument also applies to $f_\pm$.
\end{proof}

%======================================================================
\section{Proofs for Sec.~\ref{sec:full3d}}\label{app:3d}
%======================================================================

\subsection{The circles at infinity}

\begin{proof}[Proof of Lemma~\ref{lem:TC}]
Throughout, $|u|\le\delta$, $0\le v\le\delta'$ and $|\tau|\le1$, where $\tau=\cos\theta$. The bound
$\delta<3/(20c)$ gives $\delta<\delta'$, $\delta'<15/(16c)$ and $\frac53\delta+\frac{14}{15}\delta'<\frac1{2c}$.
For $G(\theta)=-\sin\theta$ the function $\tau=\cos\theta$ satisfies $d\tau/ds=v(1-\tau^2)$ in the double.
The arguments below for the block property and for the invariant set use only~\eqref{eq:flowTp}, this
equation and $|\tau|\le1$, and not the topology of the base.

\emph{Block property of $N^+$.} On the faces $|u|=\delta$, by~\eqref{eq:flowTp},
$\operatorname{sign}(u)\,du/ds\le-\delta/c+\frac53\delta^2+\frac4{15}\delta'^2+\frac23\delta\delta'
=-\frac{14\delta}{15c}+\frac53\delta^2+\frac23\delta\delta'<0$, so these faces are strict entrance faces. On
the face $v=\delta'$, $dv/ds=\frac{\delta'\tau}3(2\delta'-u)$ with $2\delta'-u\ge2\delta'-\delta>0$, so points
with $\cos\theta>0$ exit and points with $\cos\theta<0$ enter. At $\cos\theta=0$ one has $dv/ds=0$,
$d\tau/ds=\delta'>0$ and $d^2v/ds^2=\frac{\delta'}3(2\delta'-u)\,d\tau/ds>0$, so $v$ has a strict local
minimum and the tangency is external. The face $v=0$ is invariant. Hence $N^+$ is an isolating block~\cite{Conley:1971}
with exit set $L^+$.

\emph{Corners.} The block $N^+$ is a manifold with corners along the edges $\{|u|=\delta,\ v=\delta'\}$,
where an entrance face meets the top face. The face $v=0$ lies in the invariant boundary of the
compactified phase space and is not part of the boundary of $N^+$ relative to that space. No rounding of
the corners is needed, because the analysis above shows directly that $(N^+,L^+)$ is an index pair in the
sense of Refs.~\onlinecite{Conley:1978,Salamon:1985}, a notion that requires no regularity of $\partial N^+$. Indeed,
$L^+$ is closed and contained in $\partial N^+$, so $N^+=\overline{N^+\setminus L^+}$ isolates
$\Inv(N^+)$, and $\Inv(N^+)=TC^+$ lies in the interior of $N^+\setminus L^+$ relative to the
compactified phase space. Every point of $L^+$ leaves $N^+$ immediately in forward time, including the corner points,
where $u$ moves inward while $v$ either increases strictly ($\cos\theta>0$) or has a strict local
minimum ($\cos\theta=0$). Hence $L^+$ is positively invariant relative to $N^+$. Finally, every other
boundary point, including the corner points with $\cos\theta<0$, where both faces are strict entrance
faces, is a strict entrance point, so an orbit that leaves $N^+$ does so through $L^+$.

\emph{Invariant set of $N^+$.} Let $\zeta$ be a full orbit in $N^+$ with $v>0$ somewhere, hence everywhere.
Consider $P\coloneqq\{|u|\le v\}$. On $u=v$ one finds $d(u-v)/ds=-v/c-\frac{16}{15}v^2\tau<0$, and on $u=-v$,
$d(u+v)/ds=v/c-\frac{16}{15}v^2\tau>0$, so no orbit leaves $P$ in forward time while it stays in $N^+$.
Off $P$, i.e.\ for $|u|>v$, the bound $v^2<|u|v$ gives
$d\ln|u|/ds\le-1/c+\frac53\delta+\frac4{15}\delta'+\frac23\delta'<-1/(2c)$. If $\zeta(s_0)\notin P$, then
$\zeta(s)\notin P$ for all $s\le s_0$, and $|u|$ grows at least like $e^{(s_0-s)/(2c)}$ as $s\to-\infty$,
so $\zeta$ leaves $N^+$. Hence $\zeta\subset P$. In $P$ we have $2v-u\in[v,3v]$, so $dv/ds$ has the sign
of $\tau$, and $d\tau/ds=v(1-\tau^2)\ge0$. If $\zeta$ lies in the fiber $\tau=1$, then $d(1/v)/ds\le-1/3$, and
$v$ reaches $\delta'$ in finite forward time. If $\zeta$ lies in the fiber $\tau=-1$, then $d(1/v)/ds\ge1/3$,
and $v$ reaches $\delta'$ in finite backward time. Otherwise $|\tau|<1$ along $\zeta$, and
\begin{equation}
\Psi\coloneqq v^6(1-\tau^2)>0,\qquad \frac{d\Psi}{ds}=2v^6(1-\tau^2)\,\tau\,(v-u),\qquad
\frac{d\tau}{ds}=v(1-\tau^2)=\frac{\Psi}{v^5}\ge\frac{\Psi}{\delta'^5},
\end{equation}
where $v-u\ge0$ in $P$. Hence $\Psi$ is nondecreasing where $\tau\ge0$ and nonincreasing where $\tau\le0$.
If $\tau(s_1)\ge0$ for some $s_1$, then $\tau\ge0$ on $[s_1,\infty)$, because $\tau$ is nondecreasing, so
$d\tau/ds\ge\Psi(s_1)/\delta'^5>0$ there, which contradicts $\tau\le1$. If $\tau<0$ everywhere, then
$d\tau/ds\ge\Psi(0)/\delta'^5>0$ on $(-\infty,0]$, which contradicts $\tau\ge-1$. Therefore
$\Inv(N^+)\subset\{v=0\}$. There, $du/ds=-u(1/c+\frac53u\tau)$ and $|u|\le\delta<3/(5c)$, so $|u|$ strictly
decreases unless $u=0$, and full orbits have $u\equiv0$. This proves $\Inv(N^+)=TC^+$.

\emph{Index of $TC^+$.} The bounds on $\delta$ and $\delta'$ give $y_1=(1+u^2+v^2)^{-1/2}\ge\frac12\min(1,c)$
on $N^+$, so the coordinates $(u,v,\theta)$ give a homeomorphism from $N^+\subset D^2\times S^1$ onto
$\{|u|\le\delta,\ 0\le v\le\delta'\}\times S^1$, with inverse
$(u,v,\theta)\mapsto\big((1,u,v)/\sqrt{1+u^2+v^2},\theta\big)$, and it carries $L^+$ onto
$\{v=\delta',\ \cos\theta\ge0\}$. We compute in these coordinates. The homotopy $(u,v,\theta)\mapsto((1-t)u,(1-t)v+t\delta',\theta)$, $t\in[0,1]$,
maps $N^+$ into itself and $L^+$ into $L^+$, and at $t=1$ it is a retraction of $N^+$ onto the circle
$C\coloneqq\{u=0,\ v=\delta'\}$. Hence $N^+/L^+$ is pointed homotopy equivalent to $C/(C\cap L^+)$, where
$C\cap L^+=\{\cos\theta\ge0\}$ is a closed arc. Put $m\coloneqq\max(-\cos\theta,0)$. The map
$\theta\mapsto(1-2m^2,\,2m\sin\theta)$ takes values in $S^1$, sends the arc to $(1,0)$, and is injective on
its complement $\{\cos\theta<0\}$ with image $S^1\setminus\{(1,0)\}$. It therefore induces a continuous
bijection $C/(C\cap L^+)\to S^1$ of pointed spaces, which is a homeomorphism because the source is compact
and the target is Hausdorff. Hence $h(TC^+)=S^1$ and $\CH_*(TC^+;\Lambda)=\tilde H_*(S^1;\Lambda)=\Lambda[1]$.

(ii) The time-reversal symmetry $\tilde{\mathcal S}$ maps $N^+$ onto $N^-$, reverses time, and therefore exchanges
entrance and exit. The entrance set of $N^+$ is $\{|u|=\delta\}\cup\{v=\delta',\cos\theta\le0\}$, and its
image is $L^-$, since $\cos(\pi-\theta)=-\cos\theta$. Hence $N^-$ is an isolating block with exit set $L^-$,
and $\Inv(N^-)=\tilde{\mathcal S}(TC^+)=TC^-$. The corner argument for $N^+$ carries over with entrance
and exit exchanged. All corner points now lie in $L^-$, because the faces $u=\pm\delta$ belong to
$L^-$, and they leave immediately through these strict exit faces. Hence $(N^-,L^-)$ is again an index
pair. The homotopy $(u,\hat v,\theta)\mapsto(u,(1-t)\hat v+t\delta',\theta)$ maps $N^-$ into itself and $L^-$
into $L^-$, and at $t=1$ it is a retraction of $N^-$ onto the annulus $A\coloneqq\{|u|\le\delta,\ \hat
v=\delta'\}$. Hence $N^-/L^-$ is pointed homotopy equivalent to $A/(A\cap L^-)$, where $A\cap
L^-=\{|u|=\delta\}\cup\{\cos\theta\ge0\}$. Put $s\coloneqq u/\delta$, $r\coloneqq\sqrt{1-s^2}$ and
$w\coloneqq\max(-\cos\theta,0)$. The map
\begin{equation}
(u,\theta)\mapsto\big(1-2r^2w^2,\ 2srw^2,\ -2rw\sin\theta\big)
\end{equation}
takes values in $S^2$, because $w^2\sin^2\theta=w^2-w^4$. It sends $A\cap L^-$, where $rw=0$, to $(1,0,0)$.
On the complement, where $rw>0$, the first coordinate determines $rw$, the third then determines
$\sin\theta$, hence $\cos\theta<0$ and $w$, and the second determines $s$, so the map is injective there.
Its image is $S^2\setminus\{(1,0,0)\}$. As for $TC^+$, it induces a homeomorphism $A/(A\cap L^-)\to S^2$ of
pointed spaces. Hence $h(TC^-)=S^2$ and $\CH_*(TC^-;\Lambda)=\tilde H_*(S^2;\Lambda)=\Lambda[2]$.

(iii) With $S^1$ replaced by $[-1,1]$, the proofs of the block property and of $\Inv(N^\pm)=TC^\pm$ apply
verbatim, since they use only~\eqref{eq:flowTp}, $d\tau/ds=v(1-\tau^2)$ and $|\tau|\le1$, and in the coordinates
near $T=-\infty$ the time-reversal symmetry $\tilde{\mathcal S}$ acts by $\tau\mapsto-\tau$. The homotopy
$(u,v,\tau)\mapsto((1-t)u,(1-t)v+t\delta',(1-t)\tau+t)$ maps $N^+$ into itself and $L^+$ into $L^+$, and it
ends at the point $(0,\delta',1)\in L^+$. Hence $N^+/L^+$ is contractible relative to its base point.
Similarly, $(u,\hat v,\tau)\mapsto(u,(1-t)\hat v+t\delta',(1-t)\tau+t)$ maps $N^-$ into itself and $L^-$ into
$L^-$, and it ends in $\{\hat v=\delta',\ \tau=1\}\subset L^-$, so $N^-/L^-$ is contractible as well.

(iv) Reversing time exchanges entrance and exit sets. The time-reversal symmetry $\tilde{\mathcal S}$ reverses time and
maps $(N^+,E^+)$ onto $(N^-,L^-)$ and $(N^-,E^-)$ onto $(N^+,L^+)$. Hence the reversed flow near
$TC^\pm$ is conjugate to the forward flow near $TC^\mp$, and (iv) follows from (i), (ii) and (iii).
\end{proof}

\subsection{Equilibria at infinity, barriers and the fiber flows}

\begin{lemma}[Equilibria at infinity and barriers]\label{lem:barrier}
\begin{enumerate}
\item[(i)] On the circle at infinity $C_\theta$ over $\theta\in S^1$, with $\tau=\cos\theta$, the equilibria
are $TC^\pm(\theta)$, the points $P^\pm(\theta)$ in the directions $\pib\to\pm\infty$, and, for $\tau\neq0$,
the diagonal points $D^\pm(\theta)$, at which $u=u_*\coloneqq-3/(5c\tau)$ on the side $\pm T>0$. Within
$C_\theta$, $TC^+(\theta)$ is attracting and $TC^-(\theta)$ is repelling. In the fiber directions,
$D^+(\theta)$ is a source and $D^-(\theta)$ is a sink, $P^+(\theta)$ is a source for $\tau>0$ and a sink
for $\tau<0$, and $P^-(\theta)$ is a sink for $\tau>0$ and a source for $\tau<0$.
\item[(ii)] Let $\phi$ be the polar angle on $C_\theta$, with $T=r\cos\phi$ and $\pib=r\sin\phi$, so that
$TC^\pm$ are $\phi=0,\pi$, $P^\pm$ are $\phi=\pm\pi/2$, and $D^\pm$ are $\phi_+$ and $\phi_++\pi$, where
$\tan\phi_+=u_*$ and $\phi_+\in(-\pi/2,\pi/2)$. Up to a positive time change, the flow on $C_\theta$ is
\begin{equation}\label{eq:polar}
\phi'=-\sin\phi\cos\phi\,\Big(\frac1c\cos\phi+\frac53\tau\sin\phi\Big).
\end{equation}
For $\tau=0$ this is $\phi'=-\sin\phi\cos^2\phi/c$, so that $P^\pm$ are double zeros, and the two arcs of
$C_\theta\setminus\{TC^\pm\}$ consist of the orbits $TC^-\to P^\pm\to TC^+$. For $\tau>0$ one has
$\phi_+\in(-\pi/2,0)$, and the six arcs of $C_\theta$ are the orbits
\begin{equation}\label{eq:arcs}
D^+\to P^-,\ \ D^+\to TC^+,\ \ P^+\to TC^+,\ \ P^+\to D^-,\ \ TC^-\to D^-,\ \ TC^-\to P^- .
\end{equation}
For $\tau<0$ the arcs are the images of~\eqref{eq:arcs} under $\tilde{\mathcal S}$, namely $P^-\to D^-$,
$TC^-\to D^-$, $TC^-\to P^+$, $D^+\to P^+$, $D^+\to TC^+$ and $P^-\to TC^+$. Here $X\to Y$ means that every
orbit that starts in the open arc between $X$ and $Y$ stays in it, converges to $X$ in backward time and
to $Y$ in forward time. In particular each open arc is a single orbit.
\item[(iii)] Let $\tau(\rho)$ be any continuous function, for instance a solution of~\eqref{eq:gtau} or the
base coordinate $\cos\theta$ of the double, and let $(T,\pib)$ solve~\eqref{eq:gT}--\eqref{eq:gpi} with
this coefficient. The sign of $T$ is preserved, and on the planes $\pib=\mp a$ one has $d\pib/d\rho=\pm
aT/c$. Hence, for $T>0$, the sets $\{\pib\ge-a\}$ and $\{\pib\le a\}$ are positively invariant, and for
$T<0$ the sets $\{\pib\le-a\}$ and $\{\pib\ge a\}$ are positively invariant, with strict crossing of the
planes $\pib=\pm a$ in each case.
\end{enumerate}
\end{lemma}

\begin{proof}
(i) At $v=0$, \eqref{eq:flowTp} gives $du/ds=-u\,(1/c+\frac53u\tau)$, with zeros $u=0$ and $u=u_*$. The
derivative of the right-hand side is $-1/c$ at $u=0$ and $1/c$ at $u_*$, and at $u_*$ one has
$dv/ds=-\frac{\tau u_*}3v+O(v^2)=\frac v{5c}+O(v^2)$. Hence $TC^+$ attracts within $C_\theta$ and $D^+$ is a
source. In~\eqref{eq:flowTm} all signs at $\hat v=0$ are reversed, which gives the corresponding claims for $TC^-$
and $D^-$. Near $P^+$ we use $w=T/\pib$, $z=1/\pib$ and $ds=\pib\,d\rho$, in which
\begin{equation}
\frac{dw}{ds}=w\Big(\frac53\tau-\frac23z\tau-\frac43a^2z^2\tau+\frac wc\Big),\qquad
\frac{dz}{ds}=z\Big(\frac43(1-a^2z^2)\tau+\frac wc\Big),
\end{equation}
so that $P^+$ has the eigenvalues $\frac53\tau$ and $\frac43\tau$. Near $P^-$ one uses $ds=|\pib|\,d\rho$,
which reverses the signs.
(ii) The quadratic part of~\eqref{eq:gT}--\eqref{eq:gpi} is $\dot T=T\pib\tau/3$,
$\dot\pib=-\frac43\tau\pib^2-T\pib/c$, and $r^2\dot\phi=T\dot\pib-\pib\dot T$. On the circle at infinity
this gives~\eqref{eq:polar} after division by $r>0$. Put $g(\phi)=\frac1c\cos\phi+\frac53\tau\sin\phi$.
Its zeros are $\phi_+$ and $\phi_++\pi$, and $g(0)=1/c>0$, so $g>0$ exactly on $(\phi_+,\phi_++\pi)$. For
$\tau>0$ the signs of $-\sin\phi\cos\phi$ and $g$ on the six arcs are
\begin{center}
\begin{tabular}{lcccccc}
arc & $(\phi_+,0)$ & $(0,\frac\pi2)$ & $(\frac\pi2,\phi_++\pi)$ & $(\phi_++\pi,\pi)$ & $(-\pi,-\frac\pi2)$ &
$(-\frac\pi2,\phi_+)$\\
$-\sin\phi\cos\phi$ & $+$ & $-$ & $+$ & $+$ & $-$ & $+$\\
$g$ & $+$ & $+$ & $+$ & $-$ & $-$ & $-$\\
$\phi'$ & $+$ & $-$ & $+$ & $-$ & $+$ & $-$
\end{tabular}
\end{center}
which gives~\eqref{eq:arcs}, since a solution of a Lipschitz scalar equation that starts between two
consecutive zeros of the right-hand side stays between them, is strictly monotone, and converges to them.
For $\tau=0$, $\phi'$ is negative on $(0,\pi)$ and positive on $(-\pi,0)$. For
$\tau<0$ we use $\tilde{\mathcal S}$, which maps $C_\theta$ to $C_{\pi-\theta}$, reverses time, fixes $P^\pm$,
and exchanges $TC^+$ with $TC^-$ and $D^+$ with $D^-$.
(iii) By~\eqref{eq:gT}, $T$ solves the linear equation $dT/d\rho=kT$ with the continuous coefficient
$k=(\pib-2)\tau/3$, so its sign is preserved. On $\pib=\mp a$, \eqref{eq:gpi} gives $d\pib/d\rho=\pm aT/c$
because $a^2=1/5$, for every value of $\tau$. A comparison argument then gives the invariance of the four
half-spaces.
\end{proof}

We denote by $D_0$ and $D_\pi$ the IR and UV fibers $\theta=0$ and $\theta=\pi$,
which are invariant closed discs.

\begin{lemma}[The fiber flows]\label{lem:fiber}
Let $\delta$ and the blocks $N^\pm$ be as in Lemma~\ref{lem:TC}.
\begin{enumerate}
\item[(i)] The equilibria of the flow on $D_0$ are the sink $\Sig_I(A)$, the saddles $\Sig_I(B)$ and
$\Sig_I(C)=(T,\pib)=(-38c/15,2)$, the repellers $TC^-$, $P^+$ and $D^+$, the attractors $P^-$ and $D^-$,
and the point $TC^+$. The unstable manifold of $\Sig_I(B)$ lies in $\{T=0\}$ and consists of two orbits, which
end at $\Sig_I(A)$ and at $P^-$. The orbits of $D_0$ that converge to $TC^+$ lie on the circle at infinity,
and they come from $D^+$ and $P^+$.
\item[(ii)] $D_0$ contains no periodic orbit.
\item[(iii)] Every orbit in $D_0$ converges to an equilibrium in forward and in backward time, i.e.,
every $\alpha$- and $\omega$-limit set of an orbit in $D_0$ is a single equilibrium.
\end{enumerate}
The same conclusions hold for $D_\pi$, with $\Sig_I$ replaced by $\Sig_U$, $TC^\pm$ by $TC^\mp$, $D^\pm$ by
$D^\mp$ and repellers by attractors, and with $\Sig_U(C)=(38c/15,2)$. Parts (ii) and (iii) hold more
generally for the compactified fiber flow with any frozen value $\tau\neq0$ and any $c\neq0$.
\end{lemma}

\begin{proof}
The result for $D_\pi$ follows from that for $D_0$ by the time-reversal symmetry $\tilde{\mathcal S}$, which maps
$D_0$ onto $D_\pi$ and reverses time. (i) The finite equilibria with $\tau=1$ and their types are
those of Table~\ref{tab:gubser}. The points $P^\pm$ and $D^\pm$ are hyperbolic sources or sinks of the fiber
flow by Lemma~\ref{lem:barrier}(i). The set $N^+\cap D_\pi$ is a compact positively
invariant neighborhood of $TC^+(\pi)$ in $D_\pi$ with $\Inv(N^+)=TC^+$, so $TC^+(\pi)$ attracts in $D_\pi$,
and by $\tilde{\mathcal S}$ the point $TC^-(0)$ repels in $D_0$. At $\Sig_I(B)$ the unstable eigenvalue
$8a/3$ belongs to the $\pib$-direction, which is tangent to the invariant line $T=0$. There
$d\pib/d\rho=-\frac43(\pib^2-a^2)$, so the two branches run to $\pib=a$ and to $\pib=-\infty$. For the last
claim of (i), let $\gamma$ be an orbit of $D_0$ with $T>0$ that converges to $TC^+$. Then $\gamma$
eventually lies in $N^+\cap D_0$. There an orbit in the cone $P=\{|u|\le v\}$ has $d(1/v)/ds\le-\frac13$,
and outside $P$ the coordinate $|u|$ decays exponentially while $d\ln v/ds\ge-|u|/3$, so $\gamma$ enters
$P$ and then reaches the top face of $N^+$, which is impossible. On the circle at infinity the orbits
that end at $TC^+$ come from $D^+$ and $P^+$ by~\eqref{eq:arcs}.

(ii) A periodic orbit is its own $\omega$-limit set, so by (iii) it is an equilibrium.

(iii) We prove the general version for a frozen value $\tau\neq0$ and any $c\neq0$. We represent the
compactified fiber by the closed upper hemisphere of unit vectors $y=(y_1,y_2,y_3)$ with $y_3\ge0$, where
$(T,\pib)=(y_1,y_2)/y_3$ on the finite part $y_3>0$, and $y_3=0$ is the circle at infinity. Up to a
positive continuous time change, which does not affect limit sets, the fiber flow is generated by the
tangential projection $Y-\langle Y,y\rangle y$ of the field
\begin{equation}
Y(y)=\Big(\frac\tau3\,y_1(y_2-2y_3),\ -\frac43\,\tau\Big(y_2^2-\frac{y_3^2}5\Big)-\frac{y_1y_2}c,\ 0\Big),
\end{equation}
and we write $s$ for its time. Along an orbit in the finite part, $dT/ds=y_3\beta_T$ and
$d\pib/ds=y_3\beta_{\pib}$. Along an orbit with $y_1>0$, the coordinates~\eqref{eq:chartTp} give
$du/ds=y_1F(u,v)$, where $F(u,v)$ denotes the right-hand side of the $u$-equation in~\eqref{eq:flowTp}.
Both $y_1$ and $y_3$ satisfy linear equations $dy_i/ds=k_iy_i$ with continuous coefficients $k_i$, so
their signs are constant along orbits. The reflection $T\mapsto-T$ conjugates the flow for $(c,\tau)$ to
the flow for $(-c,\tau)$, and the time reversal $s\mapsto-s$ turns the flow for $(c,\tau)$ into the flow
for $(-c,-\tau)$. It therefore suffices to show that every orbit $\gamma$ with $y_1\ge0$ converges to an
equilibrium as $s\to\infty$, and we may assume that $\gamma$ is not an equilibrium. An orbit that meets
an equilibrium is that equilibrium, so $\gamma$ never meets one.

We use two scalar facts about a $C^1$ function $x$ on $\R$. (a) Suppose that for every level $\lambda$
outside a finite set there is a time after which $x'$ has one fixed strict sign at all times with
$x=\lambda$. Then $x$ converges in $[-\infty,\infty]$, since otherwise $x$ would cross an admissible level
between $\liminf x$ and $\limsup x$ in both directions at arbitrarily late times. (b) If $x$ converges to a
finite limit and $x'$ converges, then $x'\to0$.

Suppose first that $\gamma$ lies in the finite part, and put $g=\pib-2$, $T_C=-38c\tau/15$ and $K=T-T_C$.
The point $(T_C,2)$ is the finite equilibrium $\Sig(C)$ of the fiber, and by~\eqref{eq:gT}
and~\eqref{eq:gpi},
\begin{equation}
\frac{dg}{ds}=-\frac{2y_3}{c}\,K\quad\text{where }g=0,\qquad
\frac{dK}{ds}=-\frac{38c\tau^2y_3}{45}\,g\quad\text{where }K=0.
\end{equation}
We claim that the sign of $g$ is eventually constant. If $g$ never vanishes, this is clear. Otherwise let
$g(s_0)=0$. Then $k_0=K(s_0)\neq0$, since $\gamma$ does not meet $\Sig(C)$. The functions $G=-ck_0g$ and
$\tilde K=k_0K$ satisfy $dG/ds=2y_3\tilde K$ where $G=0$ and $d\tilde K/ds=\frac{38}{45}\tau^2y_3G$ where
$\tilde K=0$. At $s_0$ one has $G=0$, $\tilde K=k_0^2>0$ and $dG/ds>0$, so $G$ and $\tilde K$ are positive
on an interval $(s_0,s_1]$. The open quadrant $\{G>0,\ \tilde K>0\}$ cannot be left in forward time,
because at a first exit time either $G=0<\tilde K$ and $dG/ds>0$, or $\tilde K=0<G$ and $d\tilde K/ds>0$,
or $G=\tilde K=0$ and $\gamma$ meets $\Sig(C)$. Hence $g$ has the sign of $-ck_0$ for $s\ge s_1$. Since
$dT/ds=\frac13y_3\,T\,g\,\tau$, the coordinate $T$ is eventually monotone, and as $T\ge0$, it converges to
some $\ell\in[0,\infty]$.

If $\ell<\infty$, then at a level $\pib=\lambda$ one has $d\pib/ds=y_3\beta_{\pib}(T,\lambda)$ with
$y_3>0$, and $\beta_{\pib}(T,\lambda)\to\beta_{\pib}(\ell,\lambda)$. This limit is a quadratic polynomial
in $\lambda$ with leading coefficient $-\frac43\tau\neq0$, so it vanishes for at most two values of
$\lambda$, and by (a) $\pib$ converges in $[-\infty,\infty]$. If $\pib\to\pm\infty$, then $\gamma$
converges to $P^\pm$. If $\pib\to L\in\R$, then $\gamma$ converges to the finite point $(\ell,L)$, $y_3$
has a positive limit, and (b) applied to $T$ and to $\pib$ gives $\beta_T(\ell,L)=\beta_{\pib}(\ell,L)=0$,
so the limit is an equilibrium.

If $\ell=\infty$, then $y_1>0$ and $v=1/T\to0$ in the coordinates~\eqref{eq:chartTp}. At a level $u=\lambda$
one has $du/ds=y_1F(\lambda,v)$ with $y_1>0$, and $F(\lambda,v)\to F(\lambda,0)=-\lambda(1/c+\frac53\lambda
\tau)$, which vanishes only for $\lambda\in\{0,u_*\}$. By (a), $u$ converges in $[-\infty,\infty]$. If
$u\to\pm\infty$, then $\gamma$ converges to $P^\pm$. If $u\to L\in\R$, then $\gamma$ converges to the
point $(u,v)=(L,0)$ at infinity, $y_1$ has a positive limit, and (b) gives $F(L,0)=0$, so the limit is
$TC^+$ or $D^+$.

Suppose finally that $\gamma$ lies on the circle at infinity. If $y_1>0$, then $v=0$ along $\gamma$, and
the previous paragraph applies verbatim. If $y_1=0$, then $\gamma$ is one of the equilibria $P^\pm$, which
was excluded.
\end{proof}

\subsection{The global Morse decomposition}

\paragraph{Homogeneous coordinates.}
As in the proof of Lemma~\ref{lem:fiber}(iii), we represent the compactified fiber $D^2$ by the closed
upper hemisphere of unit vectors $y=(y_1,y_2,y_3)$ with $y_3\ge0$, where $(T,\pib)=(y_1,y_2)/y_3$ on the
finite part and $y_3=0$ is the circle at infinity, so that $u=y_2/y_1$ and $v=y_3/y_1$ in the
coordinates~\eqref{eq:chartTp}. Up to a positive continuous time change, the doubled compactified flow on
$D^2\times S^1$ is generated by
\begin{equation}\label{eq:homflow}
y'=Y-Sy,\qquad\theta'=-y_3\sin\theta,\qquad S\coloneqq\langle Y,y\rangle,
\end{equation}
with $Y$ as in that proof and $\tau=\cos\theta$, so that $\tau'=y_3(1-\tau^2)$. Since $Y_3=0$, every linear
function $\ell=l_1y_1+l_2y_2+l_3y_3$ satisfies $\ell'=l_1Y_1+l_2Y_2-S\ell$. In particular $y_1$ and $y_3$
satisfy linear equations, so their signs are constant along orbits, and $\tau$ is nondecreasing, strictly
increasing where $y_3>0$ and $|\tau|<1$. For $\sigma\neq0$ put
\[
H_\sigma\coloneqq y_2+\sigma y_3,\qquad k_\sigma\coloneqq\frac{4c(a^2-\sigma^2)}{3\sigma},\qquad
Q_\sigma\coloneqq y_1+k_\sigma\tau y_3 .
\]
Because $a^2=1/5$, on the plane $H_\sigma=0$ one has $Y_2=\sigma y_3Q_\sigma/c$, and therefore
\begin{equation}\label{eq:Hsigma}
H_\sigma'=\kappa H_\sigma+\frac{\sigma y_3}c\,Q_\sigma
\end{equation}
with a continuous coefficient $\kappa$. Likewise $B_1\coloneqq y_2+ay_3=H_a$ and $B_2\coloneqq ay_3-y_2=-H_{-a}$
satisfy $B_i'=\kappa_iB_i+ay_1y_3/c$, which is the homogeneous form of Lemma~\ref{lem:barrier}(iii). We
use repeatedly the comparison principle for $g'=\kappa g+m$ with continuous $\kappa$, namely, if $m\ge0$ on
$[s_0,s_1]$ and $g(s_0)\ge0$, then $g(s_1)\ge0$, and $g(s_1)>0$ if moreover $g(s_0)>0$ or $m(s_1)>0$. For
instance, along an orbit with $y_1\ge0$ the conditions $B_i\ge0$ and $B_i>0$ persist in forward time.

\paragraph{Precise form of Theorem~\ref{thm:M}.}
Fix $\eta\in(0,a)$, put $k_\mp\coloneqq4c\eta(2a\mp\eta)/\big(3(a\mp\eta)\big)$, so that $k_-=k_{a-\eta}$ and
$k_+=k_{-(a+\eta)}$, and let $\delta>0$ satisfy
\begin{equation}\label{eq:kcond}
\begin{gathered}
\text{(k1)}\ \ \delta\Big(\frac53+\frac4{15(a-\eta)^2}+\frac2{3(a-\eta)}\Big)<\frac1c,\\
\text{(k2)}\ \ \delta k_-<a-\eta,\ \ \delta k_+<a+\eta,\qquad
\text{(k3)}\ \ 5c\delta\le3 .
\end{gathered}
\end{equation}
These conditions hold for all sufficiently small $\delta>0$. In the homogeneous coordinates the
blocks of Theorem~\ref{thm:M} are
\begin{equation}\label{eq:Kpm}
K^-\coloneqq\{H_{a-\eta}\le0,\ y_2+\delta|y_1|\le0\},\qquad
K^+\coloneqq\{H_{-(a+\eta)}\ge0,\ y_2-\delta|y_1|\ge0\},
\end{equation}
i.e., $K^-$ is the closure of $\{\pib\le-a+\eta,\ \pib\le-\delta|T|\}$ and $K^+$ is the closure of
$\{\pib\ge a+\eta,\ \pib\ge\delta|T|\}$ in $D^2\times S^1$.
We prove that, under these conditions, $K^\mp$ are isolating blocks with exit sets
\begin{equation}\label{eq:Lpm}
L^\mp=K^\mp\cap\big(\{H_{\pm(a\mp\eta)}=0,\ y_1+k_\mp\tau y_3\ge0\}\cup\{y_2\pm\delta y_1=0\}\big),
\end{equation}
that every equilibrium with $y_2<0$ lies in $M^-$, that every equilibrium with $y_2>ay_3$ lies in $M^+$,
and that the conclusions of Theorem~\ref{thm:M} hold. Since the equilibria listed in
Theorem~\ref{thm:M}(ii) satisfy these inequalities, this proves the theorem.

\begin{proof}[Proof of Theorem~\ref{thm:M}]
\emph{Faces of $K^-$.} On $K^-$ one has $-y_2\ge\delta|y_1|$ and $-y_2\ge(a-\eta)y_3$, so $y_2<0$, since
otherwise $y=0$. On the plane face $\{H_{a-\eta}=0\}\cap K^-$ one has $y_3>0$, and by~\eqref{eq:Hsigma}
the orbit leaves $K^-$ where $Q_{a-\eta}>0$ and enters where $Q_{a-\eta}<0$. Where $Q_{a-\eta}=0$, the
function $F\coloneqq Q_{a-\eta}/y_3=T+k_-\tau$ vanishes, i.e., $T=-k_-\tau$, and by~\eqref{eq:gT}
\begin{equation}\label{eq:Fprime}
F'=y_3\big(\beta_T+k_-(1-\tau^2)\big)=y_3k_-\Big(1-\tau^2+\frac{2+a-\eta}3\,\tau^2\Big)>0 ,
\end{equation}
so $Q_{a-\eta}>0$ immediately afterwards, and the comparison principle applied to~\eqref{eq:Hsigma} shows
that the orbit leaves $K^-$ immediately. On the side face $\{y_2+\delta y_1=0\}\cap K^-$, which lies near
$TC^+$ where $u=-\delta$, one has $y_1>0$ and $(a-\eta)y_3\le\delta y_1$, so that
\[
(y_2+\delta y_1)'=\frac{\delta y_1^2}c-\frac53\delta^2\tau y_1^2+\frac4{15}\tau y_3^2-\frac23\delta\tau y_1y_3
\ge\delta y_1^2\Big(\frac1c-\delta\Big(\frac53+\frac4{15(a-\eta)^2}+\frac2{3(a-\eta)}\Big)\Big)>0
\]
by (k1). Hence this face is a strict exit face. In the same way the side face
$\{y_2-\delta y_1=0\}\cap K^-$, which lies near $TC^-$, has $y_1<0$ and $(y_2-\delta y_1)'<0$, so it is a
strict entrance face. Every other boundary point of $K^-$ lies on a face that is crossed strictly inward,
and on no face that is crossed outward, so it is a strict entrance point. Hence $K^-$ is an isolating block
with exit set $L^-$. For $K^+$ the same computations apply with $a+\eta$ in place of $a-\eta$, since (k1)
implies the corresponding inequality for $a+\eta$. The plane face is crossed outward where
$Q_{-(a+\eta)}=y_1+k_+\tau y_3>0$, with $F'>0$ on the tangency curve because $a+\eta<2$, the face
$\{y_2-\delta y_1=0\}$ near $TC^+$ is a strict exit face, and the face $\{y_2+\delta y_1=0\}$ near $TC^-$ is
a strict entrance face. Hence $K^+$ is an isolating block with exit set $L^+$.

\emph{Indices.} Write $K=K^\mp$ and $L=L^\mp$, and put $e=\mp1$, $\sigma=a\mp\eta$ and $k=k_\mp$, so that
$K=\{\sigma y_3\le ey_2,\ \delta|y_1|\le ey_2\}$ and $ey_2>0$ on $K$. The functions $U\coloneqq\delta
y_1/(ey_2)$ and $Z\coloneqq\sigma y_3/(ey_2)$ are continuous on $K$, and for each $\theta$ the map
$y\mapsto(U,Z)$ is a homeomorphism of the slice of $K$ over $\theta$ onto the square $[-1,1]\times[0,1]$,
with inverse $(U,Z)\mapsto(U/\delta,e,Z/\sigma)/|(U/\delta,e,Z/\sigma)|$. Since $Q_\sigma\ge0$ if and only if
$U\ge-\delta k\tau Z/\sigma$, it maps the slice of $L$ onto the arc
\[
J_\theta\coloneqq\{Z=1,\ U\ge u_0\}\cup\{U=1\},\qquad u_0\coloneqq-\delta k\tau/\sigma\in(-1,1),
\]
where $|u_0|<1$ by (k2). Put $\lambda\coloneqq\min\big(2,\max\big(0,(U-u_0)/(1-u_0)+1-Z\big)\big)$ and
$r(U,Z)\coloneqq\big(u_0+\min(\lambda,1)(1-u_0),\,2-\max(\lambda,1)\big)$. Then $r$ is a retraction of the
square onto $J_\theta$, which runs along $J_\theta$ from $(u_0,1)$ at $\lambda=0$ to $(1,0)$ at $\lambda=2$,
and it depends continuously on $\theta$. The straight-line homotopy from the identity to $r$ stays in the
square, fixes $J_\theta$ pointwise and ends in $J_\theta$. Transported back to $K$, it deforms $K$ into $L$
through maps that fix $L$, so $K/L$ is contractible and $h(M^\mp)=[K/L]=\bar0$, which proves (i).

\emph{Equilibria.} In the finite part an equilibrium has $S=0$, $\sin\theta=0$ and $Y_1=Y_2=0$, i.e.,
$\tau=\pm1$ and $\beta_T=\beta_{\pib}=0$, so it is one of the points of Table~\ref{tab:gubser}. The points
$\Sig_{U,I}(B)$, with $T=0$ and $\pib=-a$, satisfy $H_{a-\eta}=-\eta y_3$ and $y_2+\delta|y_1|=-ay_3$, so they
lie in $K^-$. The points $\Sig_{U,I}(C)$, with $\pib=2$ and $|T|=38c/15$, satisfy $\pib>a+\eta$ and
$\delta|y_1|=\frac{38c\delta}{15}y_3\le\frac{38}{25}y_3<y_2$ by (k3), so they lie in $K^+$. At infinity
$y_3=0$, and $(Y_1,Y_2)$ is parallel to $(y_1,y_2)$, i.e.,
\[
y_2Y_1-y_1Y_2=y_1y_2\Big(\frac53\tau y_2+\frac{y_1}c\Big)=0 .
\]
The solutions are $P^\pm$ with $y_1=0$, $TC^\pm$ with $y_2=0$, and the diagonal points with
$y_1=-\frac53c\tau y_2$, which satisfy $\delta|y_1|\le\frac53c\delta|y_2|\le|y_2|$ by (k3). Hence $P^-$ and
the diagonal points with $y_2<0$ lie in $K^-$, and $P^+$ and those with $y_2>0$ lie in $K^+$. An
equilibrium in $K^\pm$ lies in $\Inv(K^\pm)=M^\pm$, and the list shows that the equilibria with $y_2<0$ lie
in $M^-$ and those with $y_2>ay_3$ lie in $M^+$. This proves (ii), and also that every equilibrium lies in
one of the six sets of (iii), since the remaining ones are $\Sig_{U,I}(A)$ and $TC^\pm$.

\emph{Limit sets contain equilibria.} Let $x\in D^2\times S^1$. Along its orbit $\tau$ is nondecreasing and
the sign of $\sin\theta$ is constant, since $(\sin\theta)'=-y_3\tau\sin\theta$, so $\theta$ converges and is
constant on every orbit $\zeta$ in $\omega(x)$. Along $\zeta$ the coordinate $y$ is therefore an orbit of the
fiber flow with frozen $\tau$, and $y_3(1-\tau^2)=y_3\tau\sin\theta=0$. If $\tau\neq0$, then $\zeta$ converges
to an equilibrium of the fiber flow by Lemma~\ref{lem:fiber}(iii). If $\tau=0$, then $y_3=0$ along $\zeta$,
and $\zeta$ converges by Lemma~\ref{lem:barrier}(ii). In both cases the limit lies in the closed invariant
set $\omega(x)$, and it is an equilibrium of~\eqref{eq:homflow}, because the $\theta$-component of the field
vanishes along $\zeta$ and hence at the limit. Thus every $\omega$-limit set contains an equilibrium, and by
the time-reversal symmetry $\tilde{\mathcal S}$, which acts by $(y_1,y_2,y_3,\theta)\mapsto(-y_1,y_2,y_3,\pi-\theta)$, so does
every $\alpha$-limit set.

\emph{Convex neighborhoods.} We call a closed set $N$ a convex neighborhood of a set $M$ if $\Inv(N)=M$,
$M\subset\operatorname{int}N$ and every orbit meets $N$ in an interval of times. Then $M$ is an isolated
invariant set. If $\omega(x)$ meets $M$, then the orbit of $x$ lies in $\operatorname{int}N$ at arbitrarily
late times, hence in $N$ for all large times, and $\omega(x)\subset\Inv(N)=M$. Similarly $\alpha(x)\subset M$
if $\alpha(x)$ meets $M$, and $x\in M$ if both meet $M$. Each of the six sets has a convex neighborhood.

(a) $K^\pm$ for $M^\pm$. Since $K^\pm$ are isolating blocks, $M^\pm\subset\operatorname{int}K^\pm$. Suppose
that an orbit leaves $\{H_{a-\eta}\le0\}$ and returns, and let $[s_1,s_2]$ be an excursion, i.e.,
$H_{a-\eta}=0$ at $s_1$ and $s_2$ and $H_{a-\eta}>0$ in between. If $y_3=0$ along the orbit, then
$H_{a-\eta}'=\kappa H_{a-\eta}$ and $H_{a-\eta}\equiv0$, which is impossible. Otherwise $y_3>0$,
and~\eqref{eq:Hsigma} gives $Q_{a-\eta}\ge0$ at $s_1$ and $Q_{a-\eta}\le0$ at $s_2$, i.e., $F(s_1)\ge0\ge
F(s_2)$, where $F'=y_3(\beta_T+k_-(1-\tau^2))$ and $\beta_T=T(\pib-2)\tau/3$. During the excursion
$\pib\le a<2$. Indeed $B_2>0$ at $s_1$ and $s_2$, where $\pib=-a+\eta$. If $y_1\ge0$, then $B_2\ge0$
persists in forward time from $s_1$. If $y_1<0$, then $B_2\le0$ persists in forward time, so $B_2$ cannot
become negative before $s_2$. If $T\ge0$, then $F(s_2)\le0$ forces $\tau(s_2)\le0$,
hence $\tau\le0$ on $[s_1,s_2]$, so $\beta_T\ge0$ and $F$ is nondecreasing. Then $F\equiv0$ on $[s_1,s_2]$,
and $F'=0$ forces $\tau=-1$ and $T=0$, whence $F=-k_-\neq0$, which is a contradiction. If $T<0$, then
$F(s_1)\ge0$ forces $\tau(s_1)>0$, so $\tau>0$ on $[s_1,s_2]$, $\beta_T>0$ and $F$ is strictly increasing,
which is again a contradiction. The same argument applies to $\{H_{-(a+\eta)}\ge0\}$, along whose
excursions $\pib\le a+\eta<2$. Within an interval of times on which the orbit stays in the half-space, an
excursion out of $\{y_2+\delta y_1\le0\}$ would end at a point of the strict exit face where the derivative
is nonpositive, and an excursion out of $\{y_2-\delta y_1\le0\}$ would begin at a point of the strict
entrance face where the derivative is nonnegative, which is impossible. Hence every orbit meets $K^-$, and
likewise $K^+$, in an interval of times.

(b) $N_A\coloneqq\{\tau\ge\frac12,\ |y_1|\le\frac{ac}2y_3,\ \frac a2y_3\le y_2\le\frac{3a}2y_3\}$ for
$\Sig_I(A)$. On $N_A$ one has $y_3>0$. On the faces $y_1=\pm\frac{ac}2y_3$ one has
$(\pm y_1-\frac{ac}2y_3)'=\pm Y_1=\pm\frac\tau3y_1(y_2-2y_3)<0$, since $y_2\le\frac{3a}2y_3<2y_3$. With
$\tau\ge\frac12$ and $|y_1|/c\le\frac a2y_3$, the field $Y_2=-\frac43\tau(y_2^2-a^2y_3^2)-y_1y_2/c$ satisfies
$Y_2\le-\frac{a^2}{12}y_3^2$ on the face $y_2=\frac{3a}2y_3$ and $Y_2\ge\frac{a^2}4y_3^2$ on the face
$y_2=\frac a2y_3$, and $\tau$ is nondecreasing. Hence $N_A$ is positively invariant, and therefore every orbit
meets it in an interval of times. On the compact invariant set $\Inv(N_A)$ we have
\[
\begin{gathered}
(T^2)'=\frac23y_3T^2(\pib-2)\tau,\qquad(1-\tau)'=-y_3(1-\tau^2),\\
\big((\pib-a)^2\big)'=-\frac83y_3(\pib-a)^2(\pib+a)\quad\text{where }T=0\text{ and }\tau=1 .
\end{gathered}
\]
Since $y_3>0$, $\tau\ge\frac12$ and $|\pib|<2$ on $N_A$, each of these functions has negative derivative
wherever it is positive. Such a function cannot attain a positive maximum on a compact invariant set,
because its value slightly before the maximum point would be larger. Hence $T\equiv0$ on $\Inv(N_A)$, then
$\tau\equiv1$, and then $\pib\equiv a$. Thus $\Inv(N_A)=\Sig_I(A)$, which is an interior point of $N_A$.

(c) $N^+$ for $TC^+$. Choose $\delta_0\in(0,3/(20c))$ with $4c\delta_0<(a-\eta)^2$, put
$\delta_0'=\sqrt{\delta_0/(4c)}$, so that $\delta_0<(a-\eta)\delta_0'$, and let $N^+$ be the block of
Lemma~\ref{lem:TC} with parameter $\delta_0$. Then $\Inv(N^+)=TC^+\subset\operatorname{int}N^+$. An orbit that
leaves $N^+$ crosses its exit set at a point $p$ with $v=\delta_0'$, $|u|\le\delta_0$ and $\tau\ge0$, i.e.,
$y_3=\delta_0'y_1$ and $|y_2|\le\delta_0y_1$ with $y_1>0$. After $p$ one has $y_1>0$, $y_3>0$ and $\tau>0$.
Moreover $B_2(p)\ge(a\delta_0'-\delta_0)y_1>0$, so $B_2\ge0$ after $p$, and $\psi\coloneqq y_3-\delta_0'y_1$
satisfies
\[
\psi'=-S\psi+\frac{\delta_0'}3\,\tau y_1(2y_3-y_2),
\]
where the last term is nonnegative, because $y_2\le ay_3$, and positive after $p$. Since $\psi(p)=0$, the
comparison principle gives $\psi>0$ at all later times, whereas $\psi\le0$ on $N^+$. Hence the orbit never
returns, and $N^+$ is a convex neighborhood of $TC^+$. We also note that $H_{a-\eta}(p)\ge((a-\eta)\delta_0'
-\delta_0)y_1>0$ and that $Q_{a-\eta}=y_1+k_-\tau y_3>0$ after $p$, so $H_{a-\eta}>0$ at all later times
by~\eqref{eq:Hsigma}. Thus after leaving $N^+$ an orbit has
\begin{equation}\label{eq:afterTC}
y_1>0,\qquad B_2\ge0,\qquad H_{a-\eta}>0 .
\end{equation}

(d) The time-reversal symmetry $\tilde{\mathcal S}$ reverses time, maps $K^\pm$ onto themselves, and exchanges $TC^+$ with
$TC^-$ and $\Sig_I(A)$ with $\Sig_U(A)$. It therefore maps convex neighborhoods of $TC^+$ and $\Sig_I(A)$ to
convex neighborhoods of $TC^-$ and $\Sig_U(A)$.

\emph{Disjointness.} The points $\Sig_{I,U}(A)$ have $y_1=0$, $y_2=ay_3>0$ and $\tau=\pm1$, the circles
$TC^\pm$ have $y=(\pm1,0,0)$, $K^-$ lies in $\{y_2<0\}$, and $K^+$ lies in $\{y_2>0,\ y_2\ge(a+\eta)y_3\}$.
Hence the six sets are pairwise disjoint.

\emph{Exclusions.} Let $x$ lie outside the six sets. By the previous steps $\omega(x)$ contains an
equilibrium, which lies in one of the six sets, so $\omega(x)$ lies in one of them, and similarly
$\alpha(x)$. They are different sets, since otherwise $x$ would lie in that set. We exclude the pairs in
which $\alpha(x)$ is not above $\omega(x)$ in~\eqref{eq:order3d}.
\begin{itemize}
\item $\alpha(x)\subset\Sig_I(A)$ is impossible, because then the orbit lies in $N_A$ at arbitrarily early
times, hence at all times, and $x\in\Inv(N_A)$. By $\tilde{\mathcal S}$, $\omega(x)\subset\Sig_U(A)$ is
impossible as well.
\item If $\alpha(x)\subset TC^+$, then the orbit leaves $N^+$, since $x\notin TC^+$, and~\eqref{eq:afterTC}
holds afterwards. Hence $\omega(x)\subset\{y_1\ge0\}$ is disjoint from $TC^-$. The orbit does not lie in
$K^-\subset\{H_{a-\eta}\le0\}$ for large times, so $\omega(x)\not\subset M^-$, and it does not lie in $K^+$ for
large times, since $B_2\le-\eta y_3$ on $K^+$ and $y_3=0$ would force $y=0$, so $\omega(x)\not\subset M^+$.
By $\tilde{\mathcal S}$, if $\omega(x)\subset TC^-$, then $\alpha(x)$ is not contained in $TC^+$ or $M^\pm$.
\item If $y_1\ge0$, $\alpha(x)\subset M^-$ and $\omega(x)\subset M^+$, then the orbit lies in $K^-$ at some
time, where $B_2\ge0$ because $y_2<0$. Hence $B_2\ge0$ afterwards, and the orbit does not lie in $K^+$ for
large times, as in the previous item.
\item If $y_1>0$, $\alpha(x)\subset M^+$ and $\omega(x)\subset M^-$, then the orbit lies in $K^+$ at some time,
where $B_1>0$ because $y_2>0$. Hence $B_1>0$ afterwards. For large times the orbit lies in $K^-$, where
$y_2<0<y_2+ay_3$ gives $y_3>0$ and $-a<\pib\le-a+\eta$, and on $\omega(x)\subset K^-\cap\{B_1\ge0\}$ likewise
$y_3>0$. Since $\theta$ is constant on the orbits in $\omega(x)$, $\tau'=y_3(1-\tau^2)=0$ there, so $\tau$
converges to $\pm1$ along the orbit of $x$. If $\tau\to-1$, then $\tau\equiv-1$, and for large times $T>0$
satisfies $T'=\frac13y_3T(2-\pib)>0$ and $T\le a/\delta$, because $\delta y_1\le-y_2<ay_3$. Hence $T$
converges to some $\ell>0$, and since $y_1-\ell y_3=y_3(T-\ell)\to0$, the set $\omega(x)$ lies in the
invariant set $\{y_1=\ell y_3\}$, on which $T\equiv\ell$, whereas $T'>0$ there. If $\tau\to1$, then $\tau>0$
for large times, and since $-a<\pib<0$ and $T>0$,
\[
\pib'=y_3\beta_{\pib}=y_3\Big(\frac43(a^2-\pib^2)\tau-\frac{T\pib}c\Big)>0 .
\]
Hence $\pib$ converges to some $\ell\in(-a,0]$, and $\omega(x)$ lies in $\{y_2=\ell y_3,\ \tau=1,\
y_1\ge0\}$, on which $\pib\equiv\ell$, whereas $\beta_{\pib}=\frac43(a^2-\ell^2)-T\ell/c>0$ there.
\end{itemize}
The time-reversal symmetry $\tilde{\mathcal S}$ maps an orbit with $y_1<0$ from $M^-$ to $M^+$ to an orbit with $y_1>0$
from $M^+$ to $M^-$, and an orbit with $y_1\le0$ from $M^+$ to $M^-$ to an orbit with $y_1\ge0$ from $M^-$ to
$M^+$. Hence no orbit outside the six sets runs between $M^+$ and $M^-$. The excluded pairs are exactly those
in which $\alpha(x)$ is not above $\omega(x)$, so $\alpha(x)$ lies strictly above $\omega(x)$
in~\eqref{eq:order3d}.

\emph{Conclusion.} The space $D^2\times S^1$ is compact and invariant, hence isolated, and the six sets are
isolated, pairwise disjoint and ordered as required, so they form a Morse decomposition with the
order~\eqref{eq:order3d}. Finally, an orbit inside one of the six sets has both limit sets in that set,
which does not meet both $M^-$ and $M^+$. Hence there is no orbit from $M^+$ to $M^-$ or from $M^-$ to
$M^+$, which proves (iii).
\end{proof}

\subsection{The attractors}

\begin{proof}[Proof of Corollary~\ref{cor:attractors}]
\emph{The two Morse sets in $A_1$.} The set $N_A$ of the proof of Theorem~\ref{thm:M}(b) is compact and
positively invariant, and $\Inv(N_A)=\Sig_I(A)$ lies in its interior, so $(N_A,\emptyset)$ is an index pair
for $\Sig_I(A)$. Consider the homotopy that moves $y$ along the segment to $(0,a,1)$ and
$(\cos\theta,\sin\theta)$ along the segment to $(1,0)$, and then projects both radially back to the unit
sphere and the unit circle. Along the segments $y_3>0$ and $\cos\theta\ge\frac12$, so the projections are
defined. Apart from $\tau\ge\frac12$, the inequalities defining $N_A$ are linear and homogeneous in $y$,
and $(0,a,1)$ satisfies them, so they hold along the segment and after projection. The segment from a
point of the circle to $(1,0)$ lies in the unit disc, so the projection does not decrease
$\cos\theta$. Hence the homotopy maps $N_A$ into itself, and at its end every point is at $A$. Thus $N_A$
is contractible, $h(\Sig_I(A))=S^0$ and $\CH_*(\Sig_I(A);\Lambda)=\Lambda[0]$. By Lemma~\ref{lem:TC}(i), $\CH_*(TC^+;\Lambda)=\Lambda[1]$.

\emph{Trapping regions.} The solid torus $D^2\times S^1$ is compact and invariant, so
$(D^2\times S^1,\emptyset)$ is an index pair for it. Each $A_j$ is an attractor in $D^2\times S^1$, and an
index triple refining this index pair~\cite{Salamon:1985} contains an index pair $(B_j,\emptyset)$ for
$A_j$. Since its exit set is empty, $B_j$ is positively invariant, $\Inv(B_j)=A_j$ and $A_j$ lies in the
interior of $B_j$. Conversely, if $B_j$ is any compact positively invariant isolating neighborhood with
$\Inv(B_j)=A_j$, then $(B_j,\emptyset)$ is an index pair for $A_j$, so
$\CH_*(A_j;\Lambda)=\tilde H_*(B_j\sqcup\{*\};\Lambda)=H_*(B_j;\Lambda)$. It remains to compute
$\CH_*(A_j;\Lambda)$.

\emph{The attractor $A_1$.} The attractor--repeller pair $(\Sig_I(A),TC^+)$ in $A_1$ gives the exact
sequence~\cite{Conley:1978}
\begin{multline*}
0=\CH_1(\Sig_I(A))\to\CH_1(A_1)\to\CH_1(TC^+)\\
\xrightarrow{\ \partial\ }\CH_0(\Sig_I(A))\xrightarrow{\ i\ }\CH_0(A_1)\to\CH_0(TC^+)=0,
\end{multline*}
with coefficients in $\Lambda$, and $\CH_k(A_1;\Lambda)=0$ for $k\ge2$, because $\CH_k$ of both
$\Sig_I(A)$ and $TC^+$ vanishes in these degrees. In particular $i$ is onto. The set $B_1$ is nonempty,
since it contains $TC^+$. Pick $b\in B_1$. The map $B_1\sqcup\{*\}\to S^0$ that sends $B_1$ to the point
other than the base point has the pointed section that sends this point to $b$. Hence it induces a
surjection $\CH_0(A_1;\Lambda)=\tilde H_0(B_1\sqcup\{*\};\Lambda)\to\tilde H_0(S^0;\Lambda)=\Lambda$.
Composing it with $i$ and with an isomorphism $\Lambda\cong\CH_0(\Sig_I(A);\Lambda)$ gives a surjective
$\Lambda$-linear map $f\colon\Lambda\to\Lambda$. Then $f(x)=xf(1)$, and $yf(1)=1$ for some $y$, so $f$ is
injective, because $\Lambda$ is commutative. Hence $i$ is injective, $\partial=0$,
$\CH_0(A_1;\Lambda)\cong\Lambda$ and $\CH_1(A_1;\Lambda)\cong\CH_1(TC^+;\Lambda)\cong\Lambda$.

\emph{The attractors $A_2$ and $A_3$.} The attractor--repeller pairs $(A_1,M^-)$ in $A_2$ and $(A_2,M^+)$ in
$A_3$ give exact sequences
\[
\cdots\to\CH_k(A_1;\Lambda)\to\CH_k(A_2;\Lambda)\to\CH_k(M^-;\Lambda)\to\CH_{k-1}(A_1;\Lambda)\to\cdots,
\]
and similarly for $(A_2,M^+)$. In these sequences the maps
$\CH_k(A_1;\Lambda)\to\CH_k(A_2;\Lambda)$ and $\CH_k(A_2;\Lambda)\to\CH_k(A_3;\Lambda)$ are induced by
inclusions of index pairs. Since
$h(M^\pm)=\bar0$ by~\eqref{eq:Mzero}, $\CH_*(M^\pm;\Lambda)=0$, and these maps are isomorphisms.
\end{proof}

%======================================================================
\section{Proofs for Sec.~\ref{sec:ON}}\label{app:ON}
%======================================================================

\begin{proof}[Proof of Lemma~\ref{lem:ONstructure}]
(i) One checks that
\[
\partial_{y_1}V=y_1+9y_1^2+(N-1)y_2^2,\qquad\partial_{y_2}V=\frac{N-1}3\big(y_2+6y_1y_2+(N+2)y_2^2\big).
\] The metric $g$ is positive definite for
$N>1$, and $dV/ds=-\langle\nabla V,g^{-1}\nabla V\rangle<0$ away from the zeros of the vector field.

(ii) On the line $y_2=ky_1$ one finds $y_2'-ky_1'=k(k-1)\big((N-1)k-3\big)y_1^2$, which vanishes
identically exactly for $k\in\{0,1,3/(N-1)\}$. The coordinates of the fixed points give the second
claim.

(iii) The boundary of $Q$ consists of $\Sig_G$, the half-line $\{y_2=0,\ y_1<0\}$, which is invariant by
(ii), and the half-line $\{y_1=0,\ y_2<0\}$, on which $y_1'=-(N-1)y_2^2<0$, so the flow enters $Q$ there.
By uniqueness of solutions no orbit leaves $Q$ in forward time.

(iv) Put $x=-y_1$ and $w=-y_2$. Then $x'=-x+9x^2+(N-1)y_2^2\ge x(9x-1)$. If $x(0)>\frac19$, then $x$ is
nondecreasing, hence $x\ge x(0)$ along the forward orbit, and
\[
\Big(\frac1x\Big)'=-\frac{x'}{x^2}\le-\Big(9-\frac1x\Big)\le-\Big(9-\frac1{x(0)}\Big),
\]
so the positive function $1/x$ would vanish at time $1/(9x(0)-1)$, and the forward existence time is at
most this value. On $Q$, which the forward orbit does not leave by (iii), one has $y_1\le0$ and hence
$w'=w\big((N+2)w-1-6y_1\big)\ge w\big((N+2)w-1\big)$, and the same argument with $N+2$ in place of $9$
applies to $w$.
\end{proof}

\begin{proof}[Proof of Proposition~\ref{prop:ON}]
(a) We first check the faces of $R$, for every $N>1$. On $y_1=\delta$ one has
$y_1'=-\big(\delta+9\delta^2+(N-1)y_2^2\big)<0$, and on $y_1=-\frac18$ one has
$y_1'=-\big(\frac1{64}+(N-1)y_2^2\big)<0$. On $y_2=\delta$, since $1+6y_1\ge\frac14$ on $R$, one has
$y_2'=-\delta\big(1+6y_1+(N+2)\delta\big)\le-\frac\delta4$, and on $y_2=-\frac12$, since $y_1\le\delta$ and
$N+2>3$, one has $y_2'=\frac12\big(1+6y_1-\frac{N+2}2\big)<\frac12\big(1+6\delta-\frac32\big)<0$. Hence the
faces $y_1=\delta$ and $y_2=\delta$ are strict entrance faces, and the faces in $L$ are strict exit
faces. At the corner $(\delta,\delta)$ both faces are strict entrance faces, and the other three
corners lie in $L$ and leave $R$ immediately through a strict exit face. As in the proof of
Lemma~\ref{lem:TC}, $(R,L)$ is therefore an index pair, and no full orbit in $R$ meets $\partial R$, so
$R$ is an isolating block with exit set $L$.

Every full orbit in $R$ is bounded, so $\Inv(R)\subset B_N$. Conversely, let $\gamma$ be a bounded full
orbit. Each of the four bounds below follows from one growth argument, since a coordinate that grows at
least linearly in one time direction contradicts boundedness.
\begin{itemize}
\item $y_1\ge-\frac19$. If $x=-y_1$ satisfies $x(s_0)>\frac19$, then $x'\ge x(9x-1)$, as in the proof of
Lemma~\ref{lem:ONstructure}(iv), so $x\ge x(s_0)$ and $x'\ge x(s_0)(9x(s_0)-1)>0$ for $s\ge s_0$.
\item $y_1\le0$. If $y_1(s_0)>0$, then $y_1'\le-y_1$, so $y_1\ge y_1(s_0)$ and $y_1'\le-y_1(s_0)$ for
$s\le s_0$.
\item $y_2\le0$. If $y_2(s_0)>0$, then, as long as $y_2\ge y_2(s_0)$, the first bound gives $1+6y_1\ge\frac13$
and hence $y_2'=-y_2\big(1+6y_1+(N+2)y_2\big)\le-\frac13y_2(s_0)$, so $y_2$ grows at least linearly for
$s\le s_0$.
\item $y_2\ge-\frac1{N+2}$. By the previous two bounds $\gamma\subset Q$, so $w=-y_2$ satisfies
$w'\ge w((N+2)w-1)$, and the first argument applies.
\end{itemize}
Hence $\gamma\subset\big[-\frac19,0\big]\times\big[-\frac1{N+2},0\big]$. Since $\frac1{N+2}<\frac13$, this
rectangle lies in the interior of $R$, and therefore $B_N\subset\Inv(R)$. The bounds hold in particular
for the fixed points.

(b) By Lemma~\ref{lem:ONstructure}(i), along every orbit $dV/ds=-D$ with
$D\coloneqq\langle\nabla V,g^{-1}\nabla V\rangle\ge0$, and $D$ vanishes exactly at the fixed points. Let
$\gamma\subset B_N$ be an orbit, which lies in the compact set $R$ by (a), and let $p$ be a cluster point
of $\gamma(s)$ as $s\to\infty$. If $p$ were not a fixed point, then $V$ would drop by some $\kappa>0$ along
the orbit of $p$ in time one, by continuity of the flow it would drop by at least $\kappa/2$ in time one
along the orbits of all points near $p$, and $\gamma$ passes near $p$ at arbitrarily late times. Then
$V(\gamma(s))\to-\infty$, which contradicts the boundedness of $V$ on $R$. Hence all cluster points are
fixed points. Since the set of cluster points of a curve in a compact set is nonempty and connected, and
there are finitely many fixed points, $\gamma$ converges to a fixed point $p$, and likewise to a fixed
point $q$ in backward time. Then $V(p)\le V(\gamma(s))\le V(q)$ for all $s$. If $V(q)\le V(p)$, then $V$ is
constant along $\gamma$, so $D=0$ along $\gamma$ and $\gamma$ is a fixed point. Hence every non-constant
orbit runs from $q$ to $p$ with $V(p)<V(q)$.

Fixed points are chain recurrent. Let $x\in B_N$ be a non-fixed point, and choose $s_0>0$ with
$V(\varphi_{s_0}x)<V(x)$. Since $V$ takes finitely many values at fixed points, there are
$V(\varphi_{s_0}x)<a<b<V(x)$ such that $[a,b]$ contains none of them. On the compact set $B_N\cap\{a\le
V\le b\}$ one has $D\ge d$ for some $d>0$, so every $y\in B_N$ with $V(y)\le b$ satisfies $V(\varphi_ty)<a$
for $t\ge T_0\coloneqq(b-a)/d+1$. Since $V$ is uniformly continuous on $B_N$, there is $\varepsilon>0$ such
that $|V(y)-V(z)|<b-a$ whenever $|y-z|<\varepsilon$. Consider an $(\varepsilon,T)$-chain
$x=x_0,x_1,\dots,x_{n+1}$ in $B_N$ with jump times $t_i\ge T\coloneqq\max(s_0,T_0)$. Then
$V(\varphi_{t_0}x_0)\le V(\varphi_{s_0}x)<a$, hence $V(x_1)<b$, and inductively $V(\varphi_{t_i}x_i)<a$ and
$V(x_{i+1})<b$ for all $i$. Since $V(x)>b$, the chain does not return to $x$, so $x$ is not chain
recurrent.

For $N\neq4$ the four fixed points are distinct. Each fixed point $p$ is an isolated invariant set,
since a small closed ball $U\subset\operatorname{int}R$ around $p$ contains no other fixed point, and a
full orbit in $U$ converges in both time directions to fixed points in $U$, i.e., to $p$, so $V$ is
constant along it, and it equals $p$. By the previous paragraph, every point of $B_N$ that is not a fixed
point has its $\omega$-limit at a fixed point with a strictly smaller value of $V$ than its
$\alpha$-limit. Hence the four points, ordered by $V$, form a Morse decomposition of $B_N$. The
flow-defined order, which is generated by the connecting orbits, is contained in this order, so it is a
strict partial order and the four points also form a Morse decomposition with respect to it.

(c) By (a), $(R,L)$ is an index pair for $B_N$, so $h(B_N)$ is the homotopy type of $R/L$. The exit set
$L$ consists of two adjacent sides of the rectangle $R$, so it is an arc, and $R/L$ is contractible. Hence
$h(B_N)=\bar0$ and $\CH_*(B_N;\Lambda)=\tilde H_*(R/L;\Lambda)=0$.
\end{proof}

\begin{proof}[Proof of Corollary~\ref{cor:ON}]
(i) On the line $y=\varsigma\,(1,k)$ with $k\in\{0,1,3/(N-1)\}$, the flow~\eqref{eq:ONy} reduces to
$\varsigma'=-\varsigma(1+\alpha\varsigma)$ with $\alpha=9$, $N+8$ and $9N/(N-1)$, respectively. This moves every
point strictly between the fixed point $\varsigma=-1/\alpha$ and $0$ toward the origin.

(ii) and (iii). Write $\Sig_s$ for the saddle other than $\Sig_I$ and $\Sig_o$ for the source, so that
$(\Sig_s,\Sig_o)=(\Sig_H,\Sig_C)$ for $N>4$ and $(\Sig_s,\Sig_o)=(\Sig_C,\Sig_H)$ for $1<N<4$.

\emph{The Morse decomposition.} At the fixed points
\[
V(\Sig_G)=0,\qquad V(\Sig_I)=\tfrac1{486},\qquad V(\Sig_C)=\tfrac{(N-1)(N+2)}{486N^2},\qquad
V(\Sig_H)=\tfrac{N+2}{18(N+8)^2},
\]
so that
\[
\begin{gathered}
V(\Sig_C)-V(\Sig_H)=\tfrac{(N+2)(N-4)^3}{486N^2(N+8)^2},\qquad V(\Sig_C)-V(\Sig_I)=\tfrac{N-2}{486N^2},\\
V(\Sig_H)-V(\Sig_I)=\tfrac{(N-1)(10-N)}{486(N+8)^2}.
\end{gathered}
\]
Hence $\Sig_G$ has the smallest and $\Sig_o$ the largest value of $V$ among the four fixed points. By
Proposition~\ref{prop:ON}(b), every non-constant orbit of $B_N$ runs from a fixed point $q$ to a fixed
point $p$ with $V(p)<V(q)$, so no such orbit starts at $\Sig_G$ or ends at $\Sig_o$. It remains to exclude
orbits between $\Sig_I$ and $\Sig_s$. Each invariant line $y_2=ky_1$ of
Lemma~\ref{lem:ONstructure}(ii) satisfies $y_2'-ky_1'=-(y_2-ky_1)\,\kappa(y)$ with
\[
\kappa=1+6y_1+(N+2)y_2,\qquad\kappa=1+9y_1+3y_2,\qquad\kappa=1+9y_1+(N-1)y_2
\]
for $k=0$, $1$ and $3/(N-1)$, respectively. Let $q$ be a fixed point on such a line with $\kappa(q)>0$,
and let $y(s)$ be an orbit of $B_N$ with $\alpha$-limit $q$. Then $W=y_2-ky_1$ satisfies $W'=-\kappa W$
along it, and $\kappa>0$ for $s\le s_0$, so $W^2$ is nonincreasing on $(-\infty,s_0]$. Since $W(s)\to0$ as
$s\to-\infty$, $W$ vanishes on $(-\infty,s_0]$, and by uniqueness for the linear equation $W'=-\kappa W$
it vanishes identically, i.e., the orbit lies on the line. Now $\kappa(\Sig_I)=\frac13$ on the axis,
$\kappa(\Sig_H)=\frac{N-4}{N+8}$ on the diagonal and $\kappa(\Sig_C)=\frac{4-N}{3N}$ on the third line, which
are positive at the saddles. An orbit leaving $\Sig_I$ therefore lies on the axis, which does not contain
$\Sig_s$, and an orbit leaving $\Sig_s$ lies on the line through $\Sig_s$, which does not contain $\Sig_I$.
This gives the Morse decomposition with the stated order.

\emph{Blocks for $\Sig_G$ and for two attractors.} Put $e=\frac1{100(N+8)}$. On the square
$B_G=[-e,e]^2$ the flow enters strictly through all four faces, so $B_G$ is an isolating block with empty
exit set. Its invariant set is $\Sig_G$, since a full orbit in $B_G$ converges in both time directions to
$\Sig_G$, the only fixed point in $B_G$, so that $V$ is constant along it. As $B_G$ is convex,
$h(\Sig_G)=\big(B_G\sqcup\{*\},*\big)$ is the pointed $0$-sphere and $\CH_*(\Sig_G;\Lambda)\cong\Lambda[0]$.

For a saddle $\Sig$, let $A(\Sig)$ be the set of points of $B_N$ whose $\alpha$-limit is $\Sig_G$ or
$\Sig$, which is the Morse set of the attractor $\{\Sig_G,\Sig\}$ of the order. Consider the thin boxes
\[
\begin{aligned}
B_I&=\big[-\tfrac18,e\big]\times[-e,e],\\
B_H&=\big\{-\tfrac1{N+8}-e_H\le y_2\le e_H,\ |y_1-y_2|\le e_H\big\},& e_H&=\tfrac{N-4}{100(N+8)(N+4)},
\quad N>4,\\
B_C&=\big\{-\tfrac{N-1}{9N}-e_C\le y_1\le e_C,\ \big|y_2-\tfrac3{N-1}y_1\big|\le e_C\big\},& e_C&
=\tfrac{(4-N)(N-1)}{100N(N+11)},\quad 1<N<4,
\end{aligned}
\]
around the segments from $\Sig_I$, $\Sig_H$, $\Sig_C$ to $\Sig_G$, and let $L_\bullet$ be the face of
$B_\bullet$ farthest from $\Sig_G$, namely $y_1=-\frac18$, $y_2=-\frac1{N+8}-e_H$ and
$y_1=-\frac{N-1}{9N}-e_C$, respectively. A direct computation on the four faces shows that the
flow leaves $B_\bullet$ strictly through $L_\bullet$ and enters strictly through the other three faces, so
$B_\bullet$ is an isolating block with exit set $L_\bullet$. Its invariant set is $A(\Sig_\bullet)$.
Indeed, every full orbit in $B_\bullet$ is bounded, so it lies in $B_N$ and converges to fixed points in
$B_\bullet$, which are $\Sig_G$ and $\Sig_\bullet$. Conversely, an orbit of $B_N$ with $\alpha$-limit
$\Sig_\bullet$ lies on the invariant line through $\Sig_\bullet$, its $\omega$-limit is the only fixed point
of that line with a smaller value of $V$, namely $\Sig_G$, and it stays in the segment between them, which
lies in $B_\bullet$. Since $B_\bullet/L_\bullet$ is a box modulo one of its faces, it is contractible, and
$\CH_*(A(\Sig_I);\Lambda)=\CH_*(A(\Sig_s);\Lambda)=0$.

\emph{Indices and pairs.} By Proposition~\ref{prop:ON}(c), $\CH_*(B_N;\Lambda)=0$. For an attractor--repeller
pair $(A,R)$ in a set $S$ with $\CH_*(S;\Lambda)=0$, exactness of~\eqref{eq:AR} shows that every
connecting homomorphism $\partial\colon\CH_{k+1}(R;\Lambda)\to\CH_k(A;\Lambda)$ is an isomorphism. The
pairs $(\Sig_G,\Sig_I)$ in $A(\Sig_I)$ and $(\Sig_G,\Sig_s)$ in $A(\Sig_s)$ thus give
$\CH_*(\Sig_I;\Lambda)\cong\CH_*(\Sig_s;\Lambda)\cong\Lambda[1]$. Let $A'$ be the set of points of $B_N$ whose
$\alpha$-limit is not $\Sig_o$, the Morse set of the attractor $\{\Sig_G,\Sig_I,\Sig_s\}$. Since
$\CH_*(A(\Sig_I);\Lambda)=0$, the sequence of the pair $(A(\Sig_I),\Sig_s)$ in $A'$ gives
$\CH_*(A';\Lambda)\cong\CH_*(\Sig_s;\Lambda)\cong\Lambda[1]$, and the pair $(A',\Sig_o)$ in $B_N$ gives
$\CH_*(\Sig_o;\Lambda)\cong\Lambda[2]$. For a saddle $\Sig$ with other saddle $\Sig'$, let $M(\Sig,\Sig_o)$ be
the set of points of $B_N$ whose $\omega$-limit is $\Sig$ or $\Sig_o$. Then $(A(\Sig'),M(\Sig,\Sig_o))$ is an
attractor--repeller pair in $B_N$, and $\CH_*(A(\Sig');\Lambda)=0$ gives
$\CH_*(M(\Sig,\Sig_o);\Lambda)\cong\CH_*(B_N;\Lambda)=0$. Within $M(\Sig,\Sig_o)$, $(\Sig,\Sig_o)$ is an
attractor--repeller pair. Together with $A(\Sig_I)$ and $A(\Sig_s)$ this gives the four pairs.

\emph{Connecting orbits.} If an attractor--repeller pair $(A,R)$ in $S$ has no connecting orbit from $R$ to
$A$, then $S=A\sqcup R$ and $\CH_*(S)\cong\CH_*(A)\oplus\CH_*(R)$. For $S=M(\Sig,\Sig_o)$, $A=\Sig$ and
$\Lambda=\Z$ one has $\CH_*(S;\Z)=0$ but $\CH_*(\Sig;\Z)\cong\Z[1]$, so there is an orbit from $\Sig_o$ to
$\Sig$ for both saddles $\Sig$.
\end{proof}

\begin{proof}[Proof of Proposition~\ref{prop:ONramp}]
(i) The face inequalities in the proof of Proposition~\ref{prop:ON}(a) hold for every $N>1$, so $R$ is a
block with the same entrance and exit faces for every frozen system. Theorem~\ref{thm:rate}, with the
corners of $R$ treated as in Lemma~\ref{lem:TC}, shows that $S^\osx$ is isolated with $\CH_*(S^\osx)=0$,
and that $K_\mp=B_{N_\mp}$. Since $\CH_*(K_\mp)=H_*(R,L)=0$, the map $\Psi$ is zero.

For the convergence, let $y$ be a solution in $R$, and let $V_+$ and $D_+$ be the function~\eqref{eq:ONV}
and the dissipation $D$ of the proof of Proposition~\ref{prop:ON}(b) at $N=N_+$. Since the vector
field~\eqref{eq:ONy} depends on $N$ only through the terms $-y_2^2$ in both components, along $y$
\[
\frac{d}{ds}V_+(y)=-D_+(y)+\big(N(\sigma(s))-N_+\big)E(y),\qquad
E(y)=-y_2^2\big(\partial_{y_1}V_++\partial_{y_2}V_+\big)(y),
\]
and the last term tends to $0$ as $s\to\infty$. Suppose that $V_+$ takes two values $a<b$ on the
$\omega$-limit set $\Omega$ of $y$. Since $V_+$ has finitely many critical values, there is $m\in(a,b)$
that is not a critical value, and on the compact set $R\cap\{V_+=m\}$ one has $D_+\ge d$ for some $d>0$.
Once the last term is smaller than $d$, the function $V_+(y(s))$ crosses the level $m$ only downward, which
contradicts the fact that it takes values below and above $m$ at arbitrarily late times. Hence $V_+$ is
constant on $\Omega$. By Lemma~\ref{lem:isolation}(c), $\Omega$ is an invariant set of $f_+$, so $V_+$ is
constant along the orbits of $f_+$ in $\Omega$, which are therefore fixed points by
Lemma~\ref{lem:ONstructure}(i). As there are finitely many fixed points and $\Omega$ is connected, $y$
converges. The argument as $s\to-\infty$ is the same, with $V_-$ in place of $V_+$.

Now let $N_->4$ and $N_+\in(1,4)$. By Corollary~\ref{cor:ON}, the four fixed points form Morse
decompositions of $K_\mp$ with the orders of the corollary. An orbit of $S^\osx$ in the open upper sheet
projects to a solution in $R$, so by the convergence just shown it runs from some $\Sig_q^-$ to some
$\Sig_p^+$. Each of the eight points is isolated, because every other orbit of $S^\osx$ has a limit point
among the remaining seven. It remains to show the two exceptions. Let $y$ be a solution in $R$ with
$y(s)\to\Sig_G$ as $s\to-\infty$. Since $\frac{d}{ds}|y|^2=-2|y|^2+O(|y|^3)$, with a constant that is uniform
in $N\le\max N$, the function $|y|^2$ is nonincreasing for $s\le s_0$, and it tends to $0$ as
$s\to-\infty$. Hence $y=0$ for $s\le s_0$, and $y\equiv0$ by uniqueness, which ends at $\Sig_G^+$. Let now
$y(s)\to\Sig_I$ as $s\to-\infty$. Then $y_2'=-\kappa\,y_2$ with $\kappa=1+6y_1+\big(N(\sigma(s))+2\big)y_2$,
and $\kappa\to\frac13$ as $s\to-\infty$, so the argument for the invariant lines in the proof of
Corollary~\ref{cor:ON} shows that $y_2\equiv0$. The solution therefore lies on the axis, and it ends at
$\Sig_G^+$ or $\Sig_I^+$, since $\Sig_C^+$ and $\Sig_H^+$ do not lie on the axis.

(ii) The points $\Sig_G=(0,0)$ and $\Sig_I=(-\frac19,0)$ do not depend on $N$, and neither do the
linearizations there, which are $-I$ and $\operatorname{diag}(1,-\frac13)$. The linearized operator $L$
therefore has constant hyperbolic coefficients $A_0$, so $\mu_-=\mu_+$, and $\ker L=0$, since by the
dichotomy~\eqref{eq:dichotomy} every nonzero solution of $\dot y=A_0y$ is unbounded. By
Proposition~\ref{prop:palmer}(iii), the two connections are transversal and $L$ is an isomorphism. The diagonal is invariant for every $N$
(Lemma~\ref{lem:ONstructure}(ii)), and on it $\varsigma=y_1=y_2$ satisfies $\varsigma'=-\varsigma\big(1+(N+8)\varsigma\big)$,
so that $p=-1/\varsigma$ satisfies the linear equation $p'=p-\big(N(\sigma(s))+8\big)$. Its only solution that is
bounded as $s\to\infty$ is $p_*$. As a weighted average of $N+8$, $p_*$ takes values in $(9,\infty)$, and
$p_*(s)\to N_\mp+8$ as $s\to\mp\infty$. Hence~\eqref{eq:ONdiag} lies in $(-\frac19,0)^2\subset\operatorname{int}R$
and runs from $\Sig_H^-$ to $\Sig_H^+$. Its linearization is continuous and bounded and tends to the
Jacobians at $\Sig_H$ for $N=N_\mp$, whose eigenvectors $(1,1)$ and $(N-1,-3)$ have the eigenvalues $1$
and $(4-N)/(N+8)$. Hence $\mu(\Sig_H^-)=1$ and $\mu(\Sig_H^+)=2$, and the Fredholm index is $1-2=-1$ by
Proposition~\ref{prop:palmer}(ii).

The remaining claims follow from the off-diagonal coordinate $w=y_1-y_2$. The terms of~\eqref{eq:ONy}
that involve $N$ cancel in the difference of the two components, and along every solution
\[
w'=-k\,w,\qquad k=1+9y_1+3y_2 ,
\]
for every value of $N$. Hence a solution that meets the diagonal once lies on it. At
$\Sig_H(N)=\big(-\frac1{N+8},-\frac1{N+8}\big)$ one has $k=\frac{N-4}{N+8}$. Let $y$ be a solution with
$y(s)\to\Sig_H^+$ as $s\to\infty$. Since $N_+<4$, there is $T$ with $k\le0$ on $[T,\infty)$, so
$(w^2)'=-2kw^2\ge0$ there. As $w(s)\to0$, this forces $w(T)=0$, so $y$ lies on the diagonal. Then
$p=-1/y_1$ solves $p'=p-\big(N(\sigma(s))+8\big)$ and is bounded on $[T,\infty)$. Since $p-p_*$ is a
multiple of $e^s$, we obtain $p=p_*$, i.e., $y$ is the diagonal solution. Similarly, let $y(s)\to\Sig_H^-$ as
$s\to-\infty$. Since $N_->4$, $k\ge0$ on some $(-\infty,T]$, so $w^2$ is nonincreasing there and tends to $0$
as $s\to-\infty$. Hence $w(T)=0$, and $y$ lies on the diagonal for all $s$, as do its limits as
$s\to\infty$. The diagonal does not contain $\Sig_I^+=(-\frac19,0)$, and it contains $\Sig_C^+$ only if
$N_+=4$. This argument uses only that $N$ and $g$ are Lipschitz.

(iii) For a set $J$ of the eight points that is an interval of the order of (i), write $\mathcal M(J)$ for
its Morse set, i.e., the set of points of $S^\osx$ whose $\alpha$- and $\omega$-limits lie in $J$. If $J$ is
closed downward, this is the set of points whose $\alpha$-limit lies in $J$. Let $\bar N=\max N$,
$\bar e=\frac1{100(\bar N+8)}$, and
\[
\bar B_G=[-\bar e,\bar e]^2,\qquad\bar B_I=\big[-\tfrac18,\delta\big]\times[-\bar e,\bar e]\subset R.
\]
The face computations for $B_G$ and $B_I$ in the proof of Corollary~\ref{cor:ON} and for $R$ in the proof of
Proposition~\ref{prop:ON}(a) hold for every $N\in(1,\bar N]$. Hence $\bar B_G$ is a common block with empty
exit set and $\bar B_I$ is a common block with exit face $y_1=-\frac18$, and by Theorem~\ref{thm:rate}(ii)
the one-sided extensions $S_G^\osx$ and $S_I^\osx$ of $S_G=\Inv(\bar B_G\times[-1,1])$ and
$S_I=\Inv(\bar B_I\times[-1,1])$ have trivial homology index. By the two exceptions in the proof of (i), an
orbit of $S^\osx$ whose $\alpha$-limit is $\Sig_G^\mp$ is $y\equiv0$, and one whose $\alpha$-limit is
$\Sig_I^\mp$ lies on the axis, so both stay in $\bar B_I$, and the former also stays in $\bar B_G$.
Conversely, the $\alpha$-limit of an orbit in $\bar B_G$ or $\bar B_I$ is a fixed point in that box, and
the only such points are $\Sig_G$, respectively $\Sig_G$ and $\Sig_I$. Hence
\[
S_G^\osx=\mathcal M(\Sig_G^\pm),\qquad S_I^\osx=\mathcal M(\Sig_G^\pm,\Sig_I^\pm).
\]
In particular the fibers of $S_G$ are $\{\Sig_G\}$, so $(\Sig_G^+,\Sig_G^-)$ is an attractor--repeller
pair in $S_G^\osx$ by Lemma~\ref{lem:isolation}(c). Next, $\big(\mathcal M(\Sig_G^\pm),\mathcal M(\Sig_I^+,\Sig_I^-)\big)$
is an attractor--repeller pair in $S_I^\osx$, and both $S_I^\osx$ and $\mathcal M(\Sig_G^\pm)$ have trivial
homology index, so the exact sequence~\eqref{eq:AR} gives $\CH_*(\mathcal M(\Sig_I^+,\Sig_I^-))=0$. Within
this set, $(\Sig_I^+,\Sig_I^-)$ is an attractor--repeller pair.

For the last pair, let $P_1$ consist of the four IR points together with $\Sig_G^-$ and $\Sig_H^-$, which
is closed downward. An orbit that ends at a UV point lies in the UV fiber, so the Morse set of the
complement $\{\Sig_I^-,\Sig_C^-\}$ is $U\times\{\pi\}$, where $U=M(\Sig_I,\Sig_C)\subset B_{N_-}$ is the set
of the proof of Corollary~\ref{cor:ON}, which shows $\CH_*(U)=0$. By Proposition~\ref{prop:boundary},
$\CH_*(U\times\{\pi\})\cong\CH_{*-1}(U)=0$. The pair $\big(\mathcal M(P_1),U\times\{\pi\}\big)$ in $S^\osx$ thus
gives $\CH_*(\mathcal M(P_1))=0$, and the pair $\big(\mathcal M(\Sig_G^\pm),M_1\big)$ in $\mathcal M(P_1)$,
where $M_1$ is the Morse set of $\{\Sig_I^+,\Sig_C^+,\Sig_H^+,\Sig_H^-\}$, gives $\CH_*(M_1)=0$. Within $M_1$,
$(M_A,\Sig_H^-)$ is an attractor--repeller pair. Finally
$\CH_*(\Sig_H^-)\cong\CH_{*-1}(\Sig_H;f_-)\cong\Lambda[2]$ by Proposition~\ref{prop:boundary} and
Corollary~\ref{cor:ON}(ii), so $\CH_*(M_A)\cong\Lambda[1]$.
\end{proof}

%======================================================================
\section{Notation}\label{app:appendix}
%======================================================================

We collect notions for processes $\varphi(t,s)=\varphi(t,s,\cdot)$ on a metric space $(X,d)$,
see Refs.~\onlinecite{Kloeden:2011,Rasmussen:2007lnm,Rasmussen:2007tams}. A \emph{non-autonomous set} is a subset
$M\subset\R\times X$ with fibers $M(t)=\{x:(t,x)\in M\}$. It is \emph{invariant} if
$\varphi(t,s)M(s)=M(t)$ for all $t\ge s$, and \emph{forward} (\emph{backward}) \emph{invariant} if
$\varphi(t,s)M(s)\subset M(t)$ ($\supset M(t)$) for $t\ge s$. For a non-autonomous set $N$,
\begin{equation}
\Inv(N)(t)\coloneqq\{\xi(t):\ \xi:\R\to X\text{ a solution with }\xi(r)\in N(r)\text{ for all }r\}.
\end{equation}
$N$ with compact fibers is a \emph{non-autonomous isolating neighborhood} if $\Inv(N)(t)\subset
\operatorname{int}N(t)$ for all $t$, uniformly in the sense that the time-extended set
$\{(\sigma(t),x):x\in N(t)\}$ has compact closure in the extended phase space $\Ss$ and the closure is an isolating
neighborhood there. For asymptotically autonomous systems this reduces the non-autonomous notion to
the autonomous one on $\Ss$ used in the main text.

Let $d(A|B)\coloneqq\sup_{x\in A}\inf_{y\in B}d(x,y)$ be the Hausdorff semi-distance and $U_\eta(A)$ the
$\eta$-neighborhood of $A$. An invariant non-autonomous set $A$ with compact fibers is a
\emph{pullback attractor} if $\lim_{s\to-\infty}d(\varphi(t,s)D\,|\,A(t))=0$ for every bounded $D$ and every
$t$, and a \emph{forward attractor} if $\lim_{t\to\infty}d(\varphi(t+s,s)D\,|\,A(t+s))=0$ for every $s$.
Following Rasmussen~\cite{Rasmussen:2007tams}, $A$ is a \emph{past attractor} if there is $\eta>0$ with
\begin{equation}
\lim_{t\to\infty}\sup_{s\le0}d\big(\varphi(s,s-t)U_\eta(A(s-t))\,\big|\,A(s)\big)=0,
\end{equation}
and a \emph{future attractor} if
\begin{equation}
\lim_{t\to\infty}\sup_{s\ge0}d\big(\varphi(s+t,s)U_\eta(A(s))\,\big|\,A(s+t)\big)=0.
\end{equation}
Past and future repellers are defined with time reversed. For the extended phase space of an asymptotically autonomous
system, past (future) notions probe only the UV (IR) limit flow, in accordance with
Theorem~\ref{thm:slab}.

For semiflows $\pi$ on a metric space $Z$, as arise for skew products over the forward hull
$\mathcal H^+(p)=\cl\{\theta_t p:t\ge0\}$, Rybakowski~\cite{Rybakowski:1987} uses the following notion.
A pair $(N_1,N_2)$ of closed subsets of $Z$ is an \emph{FM-index pair} for an isolated invariant set $K$
if (1) $N_2\subset N_1$ and $N_2$ is $N_1$-positively invariant, (2) $N_2$ is an exit ramp for $N_1$, i.e.\ if
$x\in N_1$ and $\pi(t,x)\notin N_1$ for some $t>0$, then there is $t'\in[0,t]$ with
$\pi([0,t'],x)\subset N_1$ and $\pi(t',x)\in N_2$, and (3) $K\subset\operatorname{int}(N_1\setminus N_2)$ and
$K$ is the largest invariant set in $\cl(N_1\setminus N_2)$. For the compact extended phase spaces of the main text these
reduce to the index pairs of Sec.~\ref{sec:background}.

For completeness, we list below the main symbols and the place where each is introduced. A few symbols
are used with different meanings in different sections, and the context always fixes the meaning. In
particular, $R$ is the reflection of the double in Sec.~\ref{sec:theory} and the rectangle~\eqref{eq:ONblock}
in Sec.~\ref{sec:ON}, $L$ is the exit set of an index pair or the linearized operator, $\Delta$ is a
connection matrix or the splitting $\Delta(c)$, $\Sig_I$ is an IR point of the Gubser flow or the Ising
point of the $O(N)$ model, and $\Sigma$ without a subscript is the reduced suspension.

\begin{notationlist}
\ntrow{\textbf{Symbol}}{\textbf{Meaning}}{\textbf{Introduced in}}
\ntrule
\ntgroup{General}
\ntrow{$\R,\Z,\Q$}{real numbers, integers, rationals}{}
\ntrow{$\F$, $\mathbb F_p$, $\mathbb K$}{field with two elements, field with $p$ elements, general field}{}
\ntrow{$\Lambda$}{coefficient ring}{Sec.~\ref{sec:background}}
\ntrow{$\Lambda[k]$}{$\Lambda$ in degree $k$ and $0$ in all other degrees}{Sec.~\ref{sec:background}}
\ntrow{$\RP^1$, $S^k$, $D^2$}{real projective line, $k$-sphere, closed disc}{}
\ntrow{$\cl$, $\operatorname{int}$}{closure, interior}{}
\ntgroup{Non-autonomous systems and the extended phase space}
\ntrow{$X$}{phase space (fiber) of the non-autonomous system, open in $\R^n$}{Eq.~\eqref{eq:nonaut}}
\ntrow{$f(t,x)$}{time-dependent vector field}{Eq.~\eqref{eq:nonaut}}
\ntrow{$\varphi(t,s,x)$}{process, value at time $t$ of the solution with $x(s)=x$}{Eq.~\eqref{eq:process}}
\ntrow{$\theta_t$, $\phi$, $\Pi_t$}{base flow, cocycle, skew-product flow}{Sec.~\ref{sec:nonaut}}
\ntrow{$\mathcal H(f)$, $f_s$}{hull of $f$, translate $f(s+\cdot,\cdot)$}{Lemma~\ref{lem:hull}}
\ntrow{$f_-$, $f_+$}{UV and IR limit vector fields}{Sec.~\ref{sec:nonaut}}
\ntrow{$g$, $\sigma$}{rate function on $[-1,1]$, time map with $\sigma'=g(\sigma)$}{Hypothesis~\ref{hyp:H}}
\ntrow{$F(\tau,x)$}{vector field in the compactified time $\tau$, $f(t,x)=F(\sigma(t),x)$}{Hypothesis~\ref{hyp:H}}
\ntrow{$\tau$}{compactified time, $\tau\in[-1,1]$}{Eq.~\eqref{eq:slab}}
\ntrow{$\Ss$}{compactified extended phase space $X\times[-1,1]$}{Eq.~\eqref{eq:slab}}
\ntrow{$X_\pm$}{IR and UV boundary fibers $X\times\{\pm1\}$}{Sec.~\ref{subsec:slab}}
\ntrow{$\mu(e)$}{fiber index, number of eigenvalues of $Df_\pm(e)$ with positive real part}{Sec.~\ref{subsec:slab}}
\ntgroup{Conley index}
\ntrow{$\psi$, $Y$}{(local) flow on a locally compact metric space $Y$}{Sec.~\ref{sec:background}}
\ntrow{$\Inv(N)$}{maximal invariant subset of $N$}{Sec.~\ref{sec:background}}
\ntrow{$(N,L)$}{index pair, $L$ the exit set}{Sec.~\ref{sec:background}}
\ntrow{$h(S)$}{Conley index (pointed homotopy type of $N/L$)}{Sec.~\ref{sec:background}}
\ntrow{$h_{\Ss}$, $h_{\Dd}$, $h_{f_\pm}$}{Conley index computed in $\Ss$, in $\Dd$, for the flow of $f_\pm$}{Prop.~\ref{prop:boundary}}
\ntrow{$\CH_*(S)$, $\CH^*(S)$}{homology and cohomology Conley index}{Sec.~\ref{sec:background}}
\ntrow{$\bar0$}{homotopy type of a pointed one-point space (trivial index)}{Sec.~\ref{sec:background}}
\ntrow{$\tilde\chi$}{reduced Euler characteristic}{Sec.~\ref{sec:background}}
\ntrow{$\partial^{\rm in}$, $\partial^{\rm out}$}{entrance and exit parts of the boundary of a block}{Theorem~\ref{thm:rate}}
\ntrow{$\alpha(y)$, $\omega(y)$}{$\alpha$- and $\omega$-limit sets}{Sec.~\ref{sec:background}}
\ntrow{$\{M_i\}$}{Morse decomposition}{Sec.~\ref{sec:background}}
\ntrow{$(A,R)$}{attractor--repeller pair}{Eq.~\eqref{eq:AR}}
\ntrow{$\partial$}{connecting homomorphism}{Eq.~\eqref{eq:AR}}
\ntrow{$\Delta$}{connection matrix}{Sec.~\ref{sec:background}}
\ntrow{$\Sigma$, $\vee$, $\wedge$}{reduced suspension, wedge sum, smash product}{Sec.~\ref{sec:background}}
\ntrow{$\mathcal S$}{time-reversal symmetry, $\mathcal S\circ\psi_t=\psi_{-t}\circ\mathcal S$}{Sec.~\ref{subsec:equiv}}
\ntgroup{Double, one-sided extension and transport map}
\ntrow{$\Dd$}{double $X\times S^1$}{Eq.~\eqref{eq:double}}
\ntrow{$\theta$, $G$}{base coordinate on $S^1$ with $\tau=\cos\theta$, base vector field $\dot\theta=G(\theta)$}{Lemma~\ref{lem:G}}
\ntrow{$R$}{reflection $(x,\theta)\mapsto(x,-\theta)$, exchanges the two sheets}{Eq.~\eqref{eq:double}}
\ntrow{$q$}{projection $\Dd\to\Ss$, $(x,\theta)\mapsto(x,\cos\theta)$}{Lemma~\ref{lem:double}}
\ntrow{$S$}{isolated invariant set of $\Ss$}{Sec.~\ref{subsec:double}}
\ntrow{$K_\pm$}{IR and UV parts of $S$, $S\cap X_\pm=K_\pm\times\{\pm1\}$}{Sec.~\ref{subsec:double}}
\ntrow{$\tilde S$}{lift $q^{-1}(S)$ to the double}{Sec.~\ref{subsec:double}}
\ntrow{$S^\osx$}{one-sided extension $\tilde S\cap(X\times[0,\pi])$}{Sec.~\ref{subsec:double}}
\ntrow{$A_\eta$}{arc $[-\eta,\pi+\eta]\subset S^1$ that isolates $S^\osx$}{Lemma~\ref{lem:isolation}}
\ntrow{$\Psi$, $\Psi_k$}{UV$\to$IR transport map $\CH_k(K_-;f_-)\to\CH_k(K_+;f_+)$}{Theorem~\ref{thm:transport}}
\ntrow{$\partial^{\Dd}$}{connecting homomorphism of the double}{Theorem~\ref{thm:equivariant}}
\ntrow{$Y$, $Y^R$, $Y/R$}{quotient $q^{-1}N/q^{-1}L$, its $R$-fixed points, its orbit space}{Theorem~\ref{thm:equivariant}}
\ntrow{$Z_{\Z_2}$, $\tilde H^*_{\Z_2}$}{Borel construction, reduced Borel cohomology}{Remark~\ref{rem:borel}}
\ntgroup{Linearization and Fredholm index}
\ntrow{$A(t)$, $A_\pm$}{linearization along a solution and its limits}{Sec.~\ref{subsec:fredholm}}
\ntrow{$C^0_b$, $C^1_b$}{bounded continuous and bounded $C^1$ functions $\R\to\R^n$}{Sec.~\ref{subsec:fredholm}}
\ntrow{$L$}{linearized operator $Ly=\dot y-A(t)y$}{Sec.~\ref{subsec:fredholm}}
\ntrow{$P$, $K$, $\alpha$}{projection and constants of an exponential dichotomy}{Eq.~\eqref{eq:dichotomy}}
\ntrow{$\mu_\pm$}{unstable dimensions of $A_\pm$}{Prop.~\ref{prop:palmer}}
\ntrow{$E^u$, $E^s$}{unstable and stable subspaces}{Eq.~\eqref{eq:EuEs}}
\ntrow{$W^u$, $W^s$}{unstable and stable manifolds}{Remark~\ref{rem:palmergeom}}
\ntgroup{Gubser flow}
\ntrow{$\rho$}{de Sitter time, $\tau=\tanh\rho$}{Sec.~\ref{sec:gubser}}
\ntrow{$T$, $\pib$}{temperature in the rescaled frame, normalized shear stress}{Sec.~\ref{sec:gubser}}
\ntrow{$c$}{relaxation scale, the relaxation time is $c/T$}{Eq.~\eqref{eq:gpi}}
\ntrow{$a$}{$1/\sqrt5$}{Table~\ref{tab:gubser}}
\ntrow{$\beta_T$, $\beta_{\pib}$, $\beta_\tau$}{right-hand sides of the Gubser system}{Eqs.~\eqref{eq:gT}--\eqref{eq:gtau}}
\ntrow{$\Sig_U(\cdot)$, $\Sig_I(\cdot)$}{UV and IR equilibria, labeled $A$, $B$, $C$}{Table~\ref{tab:gubser}}
\ntrow{$\tilde{\mathcal S}$}{lift of the time reversal $\mathcal S$ to the double}{Sec.~\ref{subsec:symmetries}}
\ntrow{$K$, $\kappa$}{constant of motion of the warm-up system, exponent $8/(3\sqrt5)$}{Eq.~\eqref{eq:warmup-K}}
\ntrow{$\Delta(c)$}{splitting of $E^u$ and $E^s$ along the separatrix}{Eq.~\eqref{eq:gap}}
\ntrow{$u$, $v$, $\hat v$, $s$}{Poincar\'e coordinates near $T=\pm\infty$ and the associated time}{Eq.~\eqref{eq:chartTp}}
\ntrow{$TC^\pm$, $P^\pm$, $D^\pm$}{equilibria at infinity in the $T$-, $\pib$- and diagonal directions}{Sec.~\ref{sec:full3d}}
\ntrow{$K^\pm$, $M^\pm$}{isolating blocks and Morse sets of the global decomposition}{Theorem~\ref{thm:M}}
\ntgroup{Bjorken flow}
\ntrow{$w$}{dimensionless flow time}{Eq.~\eqref{eq:bj}}
\ntrow{$c_1$}{normalized moment of the distribution function}{Eq.~\eqref{eq:bj}}
\ntrow{$\theta_0$}{relaxation parameter}{Eq.~\eqref{eq:bj}}
\ntrow{$P(c_1)$}{polynomial governing the early-time regime}{Eq.~\eqref{eq:bj}}
\ntrow{$u$}{compactified time $w/(1+w)$, $\tau=2u-1$}{Eq.~\eqref{eq:bj-slab}}
\ntrow{$c_1^\pm$}{UV fixed points (zeros of $P$)}{Sec.~\ref{sec:bjorken}}
\ntgroup{$O(N)$ model with cubic anisotropy}
\ntrow{$N$, $\epsilon$}{number of components, $d=4+\epsilon$}{Sec.~\ref{sec:ON}}
\ntrow{$\lambda_1$, $\lambda_2$, $\beta_1$, $\beta_2$}{couplings and their one-loop beta functions}{Eq.~\eqref{eq:ONbeta}}
\ntrow{$y$, $s$}{rescaled couplings $\lambda/(8\pi^2\epsilon)$, rescaled time $\epsilon t$}{Eq.~\eqref{eq:ONy}}
\ntrow{$\Sig_G$, $\Sig_I$, $\Sig_H$, $\Sig_C$}{Gaussian, Ising, Heisenberg and cubic fixed points}{Sec.~\ref{sec:ON}}
\ntrow{$V$}{Lyapunov (gradient) function of the one-loop flow}{Lemma~\ref{lem:ONstructure}}
\ntrow{$R$, $L$}{uniform isolating block and its exit set}{Eq.~\eqref{eq:ONblock}}
\ntrow{$B_N$}{union of all bounded full orbits}{Prop.~\ref{prop:ON}}
\ntrow{$N_\mp$}{UV and IR values of $N$ in the parameter shift}{Sec.~\ref{sec:ON}}
\ntrow{$\Sig_\bullet^\mp$}{fixed points of $f_\mp$ in the UV and IR fibers of the double}{Sec.~\ref{sec:ON}}
\ntrow{$T_\lambda$, $T_1$}{transition matrix of the parameter shift, and its value at $\lambda=1$}{Eq.~\eqref{eq:ONT}}
\end{notationlist}

\bibliography{conley_refs}

%aipnum4-2.bst 2019-01-14 (MD) hand-edited version of apsrev4-1.bst
\begin{thebibliography}{77}%
\makeatletter
\providecommand \@ifxundefined [1]{%
 \@ifx{#1\undefined}
}%
\providecommand \@ifnum [1]{%
 \ifnum #1\expandafter \@firstoftwo
 \else \expandafter \@secondoftwo
 \fi
}%
\providecommand \@ifx [1]{%
 \ifx #1\expandafter \@firstoftwo
 \else \expandafter \@secondoftwo
 \fi
}%
\providecommand \natexlab [1]{#1}%
\providecommand \enquote  [1]{``#1''}%
\providecommand \bibnamefont  [1]{#1}%
\providecommand \bibfnamefont [1]{#1}%
\providecommand \citenamefont [1]{#1}%
\providecommand \href@noop [0]{\@secondoftwo}%
\providecommand \href [0]{\begingroup \@sanitize@url \@href}%
\providecommand \@href[1]{\@@startlink{#1}\@@href}%
\providecommand \@@href[1]{\endgroup#1\@@endlink}%
\providecommand \@sanitize@url [0]{\catcode `\\12\catcode `\$12\catcode
  `\&12\catcode `\#12\catcode `\^12\catcode `\_12\catcode `\%12\relax}%
\providecommand \@@startlink[1]{}%
\providecommand \@@endlink[0]{}%
\providecommand \url  [0]{\begingroup\@sanitize@url \@url }%
\providecommand \@url [1]{\endgroup\@href {#1}{\urlprefix }}%
\providecommand \urlprefix  [0]{URL }%
\providecommand \Eprint [0]{\href }%
\providecommand \doibase [0]{https://doi.org/}%
\providecommand \selectlanguage [0]{\@gobble}%
\providecommand \bibinfo  [0]{\@secondoftwo}%
\providecommand \bibfield  [0]{\@secondoftwo}%
\providecommand \translation [1]{[#1]}%
\providecommand \BibitemOpen [0]{}%
\providecommand \bibitemStop [0]{}%
\providecommand \bibitemNoStop [0]{.\EOS\space}%
\providecommand \EOS [0]{\spacefactor3000\relax}%
\providecommand \BibitemShut  [1]{\csname bibitem#1\endcsname}%
\let\auto@bib@innerbib\@empty
%</preamble>
\bibitem [{\citenamefont {Bjorken}(1983)}]{Bjorken:1982qr}%
  \BibitemOpen
  \bibfield  {author} {\bibinfo {author} {\bibfnamefont {J.~D.}\ \bibnamefont
  {Bjorken}},\ }\bibfield  {title} {\enquote {\bibinfo {title} {{Highly
  relativistic nucleus-nucleus collisions: the central rapidity region}},}\
  }\href@noop {} {\bibfield  {journal} {\bibinfo  {journal} {Phys. Rev. D}\
  }\textbf {\bibinfo {volume} {27}},\ \bibinfo {pages} {140} (\bibinfo {year}
  {1983})}\BibitemShut {NoStop}%
\bibitem [{\citenamefont {Gubser}(2010)}]{Gubser:2010ze}%
  \BibitemOpen
  \bibfield  {author} {\bibinfo {author} {\bibfnamefont {S.~S.}\ \bibnamefont
  {Gubser}},\ }\bibfield  {title} {\enquote {\bibinfo {title} {{Symmetry
  constraints on generalizations of Bjorken flow}},}\ }\href@noop {} {\bibfield
   {journal} {\bibinfo  {journal} {Phys. Rev. D}\ }\textbf {\bibinfo {volume}
  {82}},\ \bibinfo {pages} {085027} (\bibinfo {year} {2010})},\ \Eprint
  {https://arxiv.org/abs/1006.0006} {arXiv:1006.0006} \BibitemShut {NoStop}%
\bibitem [{\citenamefont {Gubser}\ and\ \citenamefont
  {Yarom}(2011)}]{Gubser:2010ui}%
  \BibitemOpen
  \bibfield  {author} {\bibinfo {author} {\bibfnamefont {S.~S.}\ \bibnamefont
  {Gubser}}\ and\ \bibinfo {author} {\bibfnamefont {A.}~\bibnamefont {Yarom}},\
  }\bibfield  {title} {\enquote {\bibinfo {title} {{Conformal hydrodynamics in
  Minkowski and de Sitter spacetimes}},}\ }\href@noop {} {\bibfield  {journal}
  {\bibinfo  {journal} {Nucl. Phys. B}\ }\textbf {\bibinfo {volume} {846}},\
  \bibinfo {pages} {469} (\bibinfo {year} {2011})}\BibitemShut {NoStop}%
\bibitem [{\citenamefont {Behtash}, \citenamefont {Cruz-Camacho},\ and\
  \citenamefont {Martinez}(2018)}]{Behtash:2017wqg}%
  \BibitemOpen
  \bibfield  {author} {\bibinfo {author} {\bibfnamefont {A.}~\bibnamefont
  {Behtash}}, \bibinfo {author} {\bibfnamefont {C.~N.}\ \bibnamefont
  {Cruz-Camacho}},\ and\ \bibinfo {author} {\bibfnamefont {M.}~\bibnamefont
  {Martinez}},\ }\bibfield  {title} {\enquote {\bibinfo {title}
  {{Far-from-equilibrium attractors and nonlinear dynamical systems approach to
  the Gubser flow}},}\ }\href@noop {} {\bibfield  {journal} {\bibinfo
  {journal} {Phys. Rev. D}\ }\textbf {\bibinfo {volume} {97}},\ \bibinfo
  {pages} {044041} (\bibinfo {year} {2018})},\ \Eprint
  {https://arxiv.org/abs/1711.01745} {arXiv:1711.01745} \BibitemShut {NoStop}%
\bibitem [{\citenamefont {Behtash}\ \emph
  {et~al.}(2019{\natexlab{a}})\citenamefont {Behtash}, \citenamefont {Kamata},
  \citenamefont {Martinez},\ and\ \citenamefont
  {Cruz-Camacho}}]{Behtash:2018moe}%
  \BibitemOpen
  \bibfield  {author} {\bibinfo {author} {\bibfnamefont {A.}~\bibnamefont
  {Behtash}}, \bibinfo {author} {\bibfnamefont {S.}~\bibnamefont {Kamata}},
  \bibinfo {author} {\bibfnamefont {M.}~\bibnamefont {Martinez}},\ and\
  \bibinfo {author} {\bibfnamefont {C.~N.}\ \bibnamefont {Cruz-Camacho}},\
  }\bibfield  {title} {\enquote {\bibinfo {title} {{Non-perturbative
  rheological behavior of a far-from-equilibrium expanding plasma}},}\
  }\href@noop {} {\bibfield  {journal} {\bibinfo  {journal} {Phys. Lett. B}\
  }\textbf {\bibinfo {volume} {797}},\ \bibinfo {pages} {134914} (\bibinfo
  {year} {2019}{\natexlab{a}})},\ \Eprint {https://arxiv.org/abs/1805.07881}
  {arXiv:1805.07881} \BibitemShut {NoStop}%
\bibitem [{\citenamefont {Israel}\ and\ \citenamefont
  {Stewart}(1979)}]{Israel:1979wp}%
  \BibitemOpen
  \bibfield  {author} {\bibinfo {author} {\bibfnamefont {W.}~\bibnamefont
  {Israel}}\ and\ \bibinfo {author} {\bibfnamefont {J.~M.}\ \bibnamefont
  {Stewart}},\ }\bibfield  {title} {\enquote {\bibinfo {title} {{Transient
  relativistic thermodynamics and kinetic theory}},}\ }\href@noop {} {\bibfield
   {journal} {\bibinfo  {journal} {Ann. Phys.}\ }\textbf {\bibinfo {volume}
  {118}},\ \bibinfo {pages} {341} (\bibinfo {year} {1979})}\BibitemShut
  {NoStop}%
\bibitem [{\citenamefont {Behtash}\ \emph
  {et~al.}(2019{\natexlab{b}})\citenamefont {Behtash}, \citenamefont {Kamata},
  \citenamefont {Martinez},\ and\ \citenamefont {Shi}}]{Behtash:2019}%
  \BibitemOpen
  \bibfield  {author} {\bibinfo {author} {\bibfnamefont {A.}~\bibnamefont
  {Behtash}}, \bibinfo {author} {\bibfnamefont {S.}~\bibnamefont {Kamata}},
  \bibinfo {author} {\bibfnamefont {M.}~\bibnamefont {Martinez}},\ and\
  \bibinfo {author} {\bibfnamefont {H.}~\bibnamefont {Shi}},\ }\bibfield
  {title} {\enquote {\bibinfo {title} {{Dynamical systems and nonlinear
  transient rheology of the far-from-equilibrium Bjorken flow}},}\ }\href@noop
  {} {\bibfield  {journal} {\bibinfo  {journal} {Phys. Rev. D}\ }\textbf
  {\bibinfo {volume} {99}},\ \bibinfo {pages} {116012} (\bibinfo {year}
  {2019}{\natexlab{b}})},\ \Eprint {https://arxiv.org/abs/1901.08632}
  {arXiv:1901.08632} \BibitemShut {NoStop}%
\bibitem [{\citenamefont {Wallace}\ and\ \citenamefont
  {Zia}(1974)}]{Wallace:1974dx}%
  \BibitemOpen
  \bibfield  {author} {\bibinfo {author} {\bibfnamefont {D.~J.}\ \bibnamefont
  {Wallace}}\ and\ \bibinfo {author} {\bibfnamefont {R.~K.~P.}\ \bibnamefont
  {Zia}},\ }\bibfield  {title} {\enquote {\bibinfo {title} {{Gradient flow and
  the renormalization group}},}\ }\href@noop {} {\bibfield  {journal} {\bibinfo
   {journal} {Phys. Lett. A}\ }\textbf {\bibinfo {volume} {48}},\ \bibinfo
  {pages} {325} (\bibinfo {year} {1974})}\BibitemShut {NoStop}%
\bibitem [{\citenamefont {Friedan}\ and\ \citenamefont
  {Konechny}(2010)}]{Friedan:2009ik}%
  \BibitemOpen
  \bibfield  {author} {\bibinfo {author} {\bibfnamefont {D.}~\bibnamefont
  {Friedan}}\ and\ \bibinfo {author} {\bibfnamefont {A.}~\bibnamefont
  {Konechny}},\ }\bibfield  {title} {\enquote {\bibinfo {title} {{Gradient
  formula for the beta-function of 2d quantum field theory}},}\ }\href@noop {}
  {\bibfield  {journal} {\bibinfo  {journal} {J. Phys. A}\ }\textbf {\bibinfo
  {volume} {43}},\ \bibinfo {pages} {215401} (\bibinfo {year} {2010})},\
  \Eprint {https://arxiv.org/abs/0910.3109} {arXiv:0910.3109} \BibitemShut
  {NoStop}%
\bibitem [{\citenamefont {Zamolodchikov}(1986)}]{Zamolodchikov:1986gt}%
  \BibitemOpen
  \bibfield  {author} {\bibinfo {author} {\bibfnamefont {A.~B.}\ \bibnamefont
  {Zamolodchikov}},\ }\bibfield  {title} {\enquote {\bibinfo {title}
  {{Irreversibility of the flux of the renormalization group in a 2D field
  theory}},}\ }\href@noop {} {\bibfield  {journal} {\bibinfo  {journal} {JETP
  Lett.}\ }\textbf {\bibinfo {volume} {43}},\ \bibinfo {pages} {730} (\bibinfo
  {year} {1986})}\BibitemShut {NoStop}%
\bibitem [{\citenamefont {Jafferis}\ \emph {et~al.}(2011)\citenamefont
  {Jafferis}, \citenamefont {Klebanov}, \citenamefont {Pufu},\ and\
  \citenamefont {Safdi}}]{Jafferis:2011zi}%
  \BibitemOpen
  \bibfield  {author} {\bibinfo {author} {\bibfnamefont {D.~L.}\ \bibnamefont
  {Jafferis}}, \bibinfo {author} {\bibfnamefont {I.~R.}\ \bibnamefont
  {Klebanov}}, \bibinfo {author} {\bibfnamefont {S.~S.}\ \bibnamefont {Pufu}},\
  and\ \bibinfo {author} {\bibfnamefont {B.~R.}\ \bibnamefont {Safdi}},\
  }\bibfield  {title} {\enquote {\bibinfo {title} {{Towards the $F$-theorem:
  $\mathcal N=2$ field theories on the three-sphere}},}\ }\href@noop {}
  {\bibfield  {journal} {\bibinfo  {journal} {JHEP}\ }\textbf {\bibinfo
  {volume} {06}},\ \bibinfo {pages} {102} (\bibinfo {year} {2011})}\BibitemShut
  {NoStop}%
\bibitem [{\citenamefont {Casini}\ and\ \citenamefont
  {Huerta}(2012)}]{Casini:2012ei}%
  \BibitemOpen
  \bibfield  {author} {\bibinfo {author} {\bibfnamefont {H.}~\bibnamefont
  {Casini}}\ and\ \bibinfo {author} {\bibfnamefont {M.}~\bibnamefont
  {Huerta}},\ }\bibfield  {title} {\enquote {\bibinfo {title} {{On the RG
  running of the entanglement entropy of a circle}},}\ }\href@noop {}
  {\bibfield  {journal} {\bibinfo  {journal} {Phys. Rev. D}\ }\textbf {\bibinfo
  {volume} {85}},\ \bibinfo {pages} {125016} (\bibinfo {year}
  {2012})}\BibitemShut {NoStop}%
\bibitem [{\citenamefont {Jack}\ and\ \citenamefont
  {Osborn}(1990)}]{Jack:1990eb}%
  \BibitemOpen
  \bibfield  {author} {\bibinfo {author} {\bibfnamefont {I.}~\bibnamefont
  {Jack}}\ and\ \bibinfo {author} {\bibfnamefont {H.}~\bibnamefont {Osborn}},\
  }\bibfield  {title} {\enquote {\bibinfo {title} {{Analogs for the $c$ theorem
  for four-dimensional renormalizable field theories}},}\ }\href@noop {}
  {\bibfield  {journal} {\bibinfo  {journal} {Nucl. Phys. B}\ }\textbf
  {\bibinfo {volume} {343}},\ \bibinfo {pages} {647} (\bibinfo {year}
  {1990})}\BibitemShut {NoStop}%
\bibitem [{\citenamefont {Komargodski}\ and\ \citenamefont
  {Schwimmer}(2011)}]{Komargodski:2011vj}%
  \BibitemOpen
  \bibfield  {author} {\bibinfo {author} {\bibfnamefont {Z.}~\bibnamefont
  {Komargodski}}\ and\ \bibinfo {author} {\bibfnamefont {A.}~\bibnamefont
  {Schwimmer}},\ }\bibfield  {title} {\enquote {\bibinfo {title} {{On
  renormalization group flows in four dimensions}},}\ }\href@noop {} {\bibfield
   {journal} {\bibinfo  {journal} {JHEP}\ }\textbf {\bibinfo {volume} {12}},\
  \bibinfo {pages} {099} (\bibinfo {year} {2011})}\BibitemShut {NoStop}%
\bibitem [{\citenamefont {Banks}\ and\ \citenamefont
  {Zaks}(1982)}]{Banks:1981nn}%
  \BibitemOpen
  \bibfield  {author} {\bibinfo {author} {\bibfnamefont {T.}~\bibnamefont
  {Banks}}\ and\ \bibinfo {author} {\bibfnamefont {A.}~\bibnamefont {Zaks}},\
  }\bibfield  {title} {\enquote {\bibinfo {title} {{On the phase structure of
  vector-like gauge theories with massless fermions}},}\ }\href@noop {}
  {\bibfield  {journal} {\bibinfo  {journal} {Nucl. Phys. B}\ }\textbf
  {\bibinfo {volume} {196}},\ \bibinfo {pages} {189} (\bibinfo {year}
  {1982})}\BibitemShut {NoStop}%
\bibitem [{\citenamefont {Kaplan}\ \emph {et~al.}(2009)\citenamefont {Kaplan},
  \citenamefont {Lee}, \citenamefont {Son},\ and\ \citenamefont
  {Stephanov}}]{Kaplan:2009kr}%
  \BibitemOpen
  \bibfield  {author} {\bibinfo {author} {\bibfnamefont {D.~B.}\ \bibnamefont
  {Kaplan}}, \bibinfo {author} {\bibfnamefont {J.-W.}\ \bibnamefont {Lee}},
  \bibinfo {author} {\bibfnamefont {D.~T.}\ \bibnamefont {Son}},\ and\ \bibinfo
  {author} {\bibfnamefont {M.~A.}\ \bibnamefont {Stephanov}},\ }\bibfield
  {title} {\enquote {\bibinfo {title} {{Conformality lost}},}\ }\href@noop {}
  {\bibfield  {journal} {\bibinfo  {journal} {Phys. Rev. D}\ }\textbf {\bibinfo
  {volume} {80}},\ \bibinfo {pages} {125005} (\bibinfo {year}
  {2009})}\BibitemShut {NoStop}%
\bibitem [{\citenamefont {Ashwin}\ \emph {et~al.}(2012)\citenamefont {Ashwin},
  \citenamefont {Wieczorek}, \citenamefont {Vitolo},\ and\ \citenamefont
  {Cox}}]{Ashwin:2012}%
  \BibitemOpen
  \bibfield  {author} {\bibinfo {author} {\bibfnamefont {P.}~\bibnamefont
  {Ashwin}}, \bibinfo {author} {\bibfnamefont {S.}~\bibnamefont {Wieczorek}},
  \bibinfo {author} {\bibfnamefont {R.}~\bibnamefont {Vitolo}},\ and\ \bibinfo
  {author} {\bibfnamefont {P.}~\bibnamefont {Cox}},\ }\bibfield  {title}
  {\enquote {\bibinfo {title} {{Tipping points in open systems: bifurcation,
  noise-induced and rate-dependent examples in the climate system}},}\
  }\href@noop {} {\bibfield  {journal} {\bibinfo  {journal} {Phil. Trans. R.
  Soc. A}\ }\textbf {\bibinfo {volume} {370}},\ \bibinfo {pages} {1166}
  (\bibinfo {year} {2012})}\BibitemShut {NoStop}%
\bibitem [{\citenamefont {Ashwin}, \citenamefont {Perryman},\ and\
  \citenamefont {Wieczorek}(2017)}]{Ashwin:2017}%
  \BibitemOpen
  \bibfield  {author} {\bibinfo {author} {\bibfnamefont {P.}~\bibnamefont
  {Ashwin}}, \bibinfo {author} {\bibfnamefont {C.}~\bibnamefont {Perryman}},\
  and\ \bibinfo {author} {\bibfnamefont {S.}~\bibnamefont {Wieczorek}},\
  }\bibfield  {title} {\enquote {\bibinfo {title} {{Parameter shifts for
  nonautonomous systems in low dimension: bifurcation- and rate-induced
  tipping}},}\ }\href@noop {} {\bibfield  {journal} {\bibinfo  {journal}
  {Nonlinearity}\ }\textbf {\bibinfo {volume} {30}},\ \bibinfo {pages} {2185}
  (\bibinfo {year} {2017})}\BibitemShut {NoStop}%
\bibitem [{\citenamefont {Kibble}(1976)}]{Kibble:1976sj}%
  \BibitemOpen
  \bibfield  {author} {\bibinfo {author} {\bibfnamefont {T.~W.~B.}\
  \bibnamefont {Kibble}},\ }\bibfield  {title} {\enquote {\bibinfo {title}
  {{Topology of cosmic domains and strings}},}\ }\href@noop {} {\bibfield
  {journal} {\bibinfo  {journal} {J. Phys. A}\ }\textbf {\bibinfo {volume}
  {9}},\ \bibinfo {pages} {1387} (\bibinfo {year} {1976})}\BibitemShut
  {NoStop}%
\bibitem [{\citenamefont {Zurek}(1985)}]{Zurek:1985qw}%
  \BibitemOpen
  \bibfield  {author} {\bibinfo {author} {\bibfnamefont {W.~H.}\ \bibnamefont
  {Zurek}},\ }\bibfield  {title} {\enquote {\bibinfo {title} {{Cosmological
  experiments in superfluid helium?}}}\ }\href@noop {} {\bibfield  {journal}
  {\bibinfo  {journal} {Nature}\ }\textbf {\bibinfo {volume} {317}},\ \bibinfo
  {pages} {505} (\bibinfo {year} {1985})}\BibitemShut {NoStop}%
\bibitem [{\citenamefont {Conley}(1978)}]{Conley:1978}%
  \BibitemOpen
  \bibfield  {author} {\bibinfo {author} {\bibfnamefont {C.}~\bibnamefont
  {Conley}},\ }\href@noop {} {\emph {\bibinfo {title} {{Isolated Invariant Sets
  and the Morse Index}}}},\ \bibinfo {series} {CBMS Regional Conference Series
  in Mathematics}, Vol.~\bibinfo {volume} {38}\ (\bibinfo  {publisher}
  {American Mathematical Society},\ \bibinfo {address} {Providence, RI},\
  \bibinfo {year} {1978})\BibitemShut {NoStop}%
\bibitem [{\citenamefont {Mischaikow}\ and\ \citenamefont
  {Mrozek}(2002)}]{Mischaikow:2002}%
  \BibitemOpen
  \bibfield  {author} {\bibinfo {author} {\bibfnamefont {K.}~\bibnamefont
  {Mischaikow}}\ and\ \bibinfo {author} {\bibfnamefont {M.}~\bibnamefont
  {Mrozek}},\ }\bibfield  {title} {\enquote {\bibinfo {title} {{Conley
  index}},}\ }in\ \href@noop {} {\emph {\bibinfo {booktitle} {Handbook of
  Dynamical Systems}}},\ Vol.~\bibinfo {volume} {2},\ \bibinfo {editor} {edited
  by\ \bibinfo {editor} {\bibfnamefont {B.}~\bibnamefont {Fiedler}}}\ (\bibinfo
   {publisher} {North-Holland},\ \bibinfo {address} {Amsterdam},\ \bibinfo
  {year} {2002})\ pp.\ \bibinfo {pages} {393--460}\BibitemShut {NoStop}%
\bibitem [{\citenamefont {Arai}\ \emph {et~al.}(2009)\citenamefont {Arai},
  \citenamefont {Kalies}, \citenamefont {Kokubu}, \citenamefont {Mischaikow},
  \citenamefont {Oka},\ and\ \citenamefont {Pilarczyk}}]{Arai:2009}%
  \BibitemOpen
  \bibfield  {author} {\bibinfo {author} {\bibfnamefont {Z.}~\bibnamefont
  {Arai}}, \bibinfo {author} {\bibfnamefont {W.}~\bibnamefont {Kalies}},
  \bibinfo {author} {\bibfnamefont {H.}~\bibnamefont {Kokubu}}, \bibinfo
  {author} {\bibfnamefont {K.}~\bibnamefont {Mischaikow}}, \bibinfo {author}
  {\bibfnamefont {H.}~\bibnamefont {Oka}},\ and\ \bibinfo {author}
  {\bibfnamefont {P.}~\bibnamefont {Pilarczyk}},\ }\bibfield  {title} {\enquote
  {\bibinfo {title} {{A database schema for the analysis of global dynamics of
  multiparameter systems}},}\ }\href@noop {} {\bibfield  {journal} {\bibinfo
  {journal} {SIAM J. Appl. Dyn. Syst.}\ }\textbf {\bibinfo {volume} {8}},\
  \bibinfo {pages} {757} (\bibinfo {year} {2009})}\BibitemShut {NoStop}%
\bibitem [{\citenamefont {Bush}\ \emph {et~al.}(2012)\citenamefont {Bush},
  \citenamefont {Gameiro}, \citenamefont {Harker}, \citenamefont {Kokubu},
  \citenamefont {Mischaikow}, \citenamefont {Obayashi},\ and\ \citenamefont
  {Pilarczyk}}]{Bush:2012}%
  \BibitemOpen
  \bibfield  {author} {\bibinfo {author} {\bibfnamefont {J.}~\bibnamefont
  {Bush}}, \bibinfo {author} {\bibfnamefont {M.}~\bibnamefont {Gameiro}},
  \bibinfo {author} {\bibfnamefont {S.}~\bibnamefont {Harker}}, \bibinfo
  {author} {\bibfnamefont {H.}~\bibnamefont {Kokubu}}, \bibinfo {author}
  {\bibfnamefont {K.}~\bibnamefont {Mischaikow}}, \bibinfo {author}
  {\bibfnamefont {I.}~\bibnamefont {Obayashi}},\ and\ \bibinfo {author}
  {\bibfnamefont {P.}~\bibnamefont {Pilarczyk}},\ }\bibfield  {title} {\enquote
  {\bibinfo {title} {{Combinatorial-topological framework for the analysis of
  global dynamics}},}\ }\href@noop {} {\bibfield  {journal} {\bibinfo
  {journal} {Chaos}\ }\textbf {\bibinfo {volume} {22}},\ \bibinfo {pages}
  {047508} (\bibinfo {year} {2012})}\BibitemShut {NoStop}%
\bibitem [{\citenamefont {Smale}(1960)}]{Smale:1960}%
  \BibitemOpen
  \bibfield  {author} {\bibinfo {author} {\bibfnamefont {S.}~\bibnamefont
  {Smale}},\ }\bibfield  {title} {\enquote {\bibinfo {title} {{Morse
  inequalities for a dynamical system}},}\ }\href@noop {} {\bibfield  {journal}
  {\bibinfo  {journal} {Bull. Amer. Math. Soc.}\ }\textbf {\bibinfo {volume}
  {66}},\ \bibinfo {pages} {43} (\bibinfo {year} {1960})}\BibitemShut {NoStop}%
\bibitem [{\citenamefont {Smale}(1961)}]{Smale:1961}%
  \BibitemOpen
  \bibfield  {author} {\bibinfo {author} {\bibfnamefont {S.}~\bibnamefont
  {Smale}},\ }\bibfield  {title} {\enquote {\bibinfo {title} {{On gradient
  dynamical systems}},}\ }\href@noop {} {\bibfield  {journal} {\bibinfo
  {journal} {Ann. of Math.}\ }\textbf {\bibinfo {volume} {74}},\ \bibinfo
  {pages} {199} (\bibinfo {year} {1961})}\BibitemShut {NoStop}%
\bibitem [{\citenamefont {Witten}(1982)}]{Witten:1982im}%
  \BibitemOpen
  \bibfield  {author} {\bibinfo {author} {\bibfnamefont {E.}~\bibnamefont
  {Witten}},\ }\bibfield  {title} {\enquote {\bibinfo {title} {{Supersymmetry
  and Morse theory}},}\ }\href@noop {} {\bibfield  {journal} {\bibinfo
  {journal} {J. Differential Geom.}\ }\textbf {\bibinfo {volume} {17}},\
  \bibinfo {pages} {661} (\bibinfo {year} {1982})}\BibitemShut {NoStop}%
\bibitem [{\citenamefont {Schwarz}(1993)}]{Schwarz:1993}%
  \BibitemOpen
  \bibfield  {author} {\bibinfo {author} {\bibfnamefont {M.}~\bibnamefont
  {Schwarz}},\ }\href@noop {} {\emph {\bibinfo {title} {{Morse Homology}}}},\
  \bibinfo {series} {Progress in Mathematics}, Vol.\ \bibinfo {volume} {111}\
  (\bibinfo  {publisher} {Birkh{\"a}user},\ \bibinfo {address} {Basel},\
  \bibinfo {year} {1993})\BibitemShut {NoStop}%
\bibitem [{\citenamefont {Fujimori}\ \emph {et~al.}(2018)\citenamefont
  {Fujimori}, \citenamefont {Honda}, \citenamefont {Kamata}, \citenamefont
  {Misumi},\ and\ \citenamefont {Sakai}}]{Fujimori:2018kqp}%
  \BibitemOpen
  \bibfield  {author} {\bibinfo {author} {\bibfnamefont {T.}~\bibnamefont
  {Fujimori}}, \bibinfo {author} {\bibfnamefont {M.}~\bibnamefont {Honda}},
  \bibinfo {author} {\bibfnamefont {S.}~\bibnamefont {Kamata}}, \bibinfo
  {author} {\bibfnamefont {T.}~\bibnamefont {Misumi}},\ and\ \bibinfo {author}
  {\bibfnamefont {N.}~\bibnamefont {Sakai}},\ }\bibfield  {title} {\enquote
  {\bibinfo {title} {{Resurgence and Lefschetz thimble in 3d $\mathcal N=2$
  supersymmetric Chern--Simons matter theories}},}\ }\href@noop {} {\bibfield
  {journal} {\bibinfo  {journal} {Prog. Theor. Exp. Phys.}\ }\textbf {\bibinfo
  {volume} {2018}},\ \bibinfo {pages} {123B03} (\bibinfo {year} {2018})},\
  \Eprint {https://arxiv.org/abs/1805.12137} {arXiv:1805.12137} \BibitemShut
  {NoStop}%
\bibitem [{\citenamefont {Conley}\ and\ \citenamefont
  {Smoller}(1971)}]{Conley:1971cs}%
  \BibitemOpen
  \bibfield  {author} {\bibinfo {author} {\bibfnamefont {C.}~\bibnamefont
  {Conley}}\ and\ \bibinfo {author} {\bibfnamefont {J.}~\bibnamefont
  {Smoller}},\ }\bibfield  {title} {\enquote {\bibinfo {title} {{Shock waves as
  limits of progressive wave solutions of higher order equations}},}\
  }\href@noop {} {\bibfield  {journal} {\bibinfo  {journal} {Comm. Pure Appl.
  Math.}\ }\textbf {\bibinfo {volume} {24}},\ \bibinfo {pages} {459} (\bibinfo
  {year} {1971})}\BibitemShut {NoStop}%
\bibitem [{\citenamefont {Smoller}(1983)}]{Smoller:1983}%
  \BibitemOpen
  \bibfield  {author} {\bibinfo {author} {\bibfnamefont {J.}~\bibnamefont
  {Smoller}},\ }\href@noop {} {\emph {\bibinfo {title} {{Shock Waves and
  Reaction--Diffusion Equations}}}},\ \bibinfo {series} {Grundlehren der
  Mathematischen Wissenschaften}, Vol.\ \bibinfo {volume} {258}\ (\bibinfo
  {publisher} {Springer},\ \bibinfo {address} {New York},\ \bibinfo {year}
  {1983})\BibitemShut {NoStop}%
\bibitem [{\citenamefont {Gukov}(2017)}]{Gukov:2016tnp}%
  \BibitemOpen
  \bibfield  {author} {\bibinfo {author} {\bibfnamefont {S.}~\bibnamefont
  {Gukov}},\ }\bibfield  {title} {\enquote {\bibinfo {title} {{RG flows and
  bifurcations}},}\ }\href@noop {} {\bibfield  {journal} {\bibinfo  {journal}
  {Nucl. Phys. B}\ }\textbf {\bibinfo {volume} {919}},\ \bibinfo {pages} {583}
  (\bibinfo {year} {2017})},\ \Eprint {https://arxiv.org/abs/1608.06638}
  {arXiv:1608.06638} \BibitemShut {NoStop}%
\bibitem [{\citenamefont {Markus}(1956)}]{Markus:1956}%
  \BibitemOpen
  \bibfield  {author} {\bibinfo {author} {\bibfnamefont {L.}~\bibnamefont
  {Markus}},\ }\bibfield  {title} {\enquote {\bibinfo {title} {{Asymptotically
  autonomous differential systems}},}\ }in\ \href@noop {} {\emph {\bibinfo
  {booktitle} {Contributions to the Theory of Nonlinear Oscillations III}}},\
  \bibinfo {series} {Annals of Mathematics Studies}, Vol.~\bibinfo {volume}
  {36}\ (\bibinfo  {publisher} {Princeton University Press},\ \bibinfo
  {address} {Princeton, NJ},\ \bibinfo {year} {1956})\ pp.\ \bibinfo {pages}
  {17--29}\BibitemShut {NoStop}%
\bibitem [{\citenamefont {Thieme}(1992)}]{Thieme:1992}%
  \BibitemOpen
  \bibfield  {author} {\bibinfo {author} {\bibfnamefont {H.~R.}\ \bibnamefont
  {Thieme}},\ }\bibfield  {title} {\enquote {\bibinfo {title} {{Convergence
  results and a Poincar{\'e}--Bendixson trichotomy for asymptotically
  autonomous differential equations}},}\ }\href@noop {} {\bibfield  {journal}
  {\bibinfo  {journal} {J. Math. Biol.}\ }\textbf {\bibinfo {volume} {30}},\
  \bibinfo {pages} {755} (\bibinfo {year} {1992})}\BibitemShut {NoStop}%
\bibitem [{\citenamefont {Mischaikow}, \citenamefont {Smith},\ and\
  \citenamefont {Thieme}(1995)}]{Mischaikow:1995}%
  \BibitemOpen
  \bibfield  {author} {\bibinfo {author} {\bibfnamefont {K.}~\bibnamefont
  {Mischaikow}}, \bibinfo {author} {\bibfnamefont {H.}~\bibnamefont {Smith}},\
  and\ \bibinfo {author} {\bibfnamefont {H.~R.}\ \bibnamefont {Thieme}},\
  }\bibfield  {title} {\enquote {\bibinfo {title} {{Asymptotically autonomous
  semiflows: chain recurrence and Lyapunov functions}},}\ }\href@noop {}
  {\bibfield  {journal} {\bibinfo  {journal} {Trans. Amer. Math. Soc.}\
  }\textbf {\bibinfo {volume} {347}},\ \bibinfo {pages} {1669} (\bibinfo {year}
  {1995})}\BibitemShut {NoStop}%
\bibitem [{\citenamefont {Kloeden}\ and\ \citenamefont
  {Rasmussen}(2011)}]{Kloeden:2011}%
  \BibitemOpen
  \bibfield  {author} {\bibinfo {author} {\bibfnamefont {P.~E.}\ \bibnamefont
  {Kloeden}}\ and\ \bibinfo {author} {\bibfnamefont {M.}~\bibnamefont
  {Rasmussen}},\ }\href@noop {} {\emph {\bibinfo {title} {{Nonautonomous
  Dynamical Systems}}}},\ \bibinfo {series} {Mathematical Surveys and
  Monographs}, Vol.\ \bibinfo {volume} {176}\ (\bibinfo  {publisher} {American
  Mathematical Society},\ \bibinfo {address} {Providence, RI},\ \bibinfo {year}
  {2011})\BibitemShut {NoStop}%
\bibitem [{\citenamefont {Rasmussen}(2007{\natexlab{a}})}]{Rasmussen:2007tams}%
  \BibitemOpen
  \bibfield  {author} {\bibinfo {author} {\bibfnamefont {M.}~\bibnamefont
  {Rasmussen}},\ }\bibfield  {title} {\enquote {\bibinfo {title} {{Morse
  decompositions of nonautonomous dynamical systems}},}\ }\href@noop {}
  {\bibfield  {journal} {\bibinfo  {journal} {Trans. Amer. Math. Soc.}\
  }\textbf {\bibinfo {volume} {359}},\ \bibinfo {pages} {5091} (\bibinfo {year}
  {2007}{\natexlab{a}})}\BibitemShut {NoStop}%
\bibitem [{\citenamefont {Rasmussen}(2007{\natexlab{b}})}]{Rasmussen:2007lnm}%
  \BibitemOpen
  \bibfield  {author} {\bibinfo {author} {\bibfnamefont {M.}~\bibnamefont
  {Rasmussen}},\ }\href@noop {} {\emph {\bibinfo {title} {{Attractivity and
  Bifurcation for Nonautonomous Dynamical Systems}}}},\ \bibinfo {series}
  {Lecture Notes in Mathematics}, Vol.\ \bibinfo {volume} {1907}\ (\bibinfo
  {publisher} {Springer},\ \bibinfo {address} {Berlin},\ \bibinfo {year}
  {2007})\BibitemShut {NoStop}%
\bibitem [{\citenamefont {J{\"a}nig}(2017)}]{Janig:2017}%
  \BibitemOpen
  \bibfield  {author} {\bibinfo {author} {\bibfnamefont {A.}~\bibnamefont
  {J{\"a}nig}},\ }\bibfield  {title} {\enquote {\bibinfo {title} {{A
  non-autonomous Conley index}},}\ }\href@noop {} {\bibfield  {journal}
  {\bibinfo  {journal} {J. Fixed Point Theory Appl.}\ }\textbf {\bibinfo
  {volume} {19}},\ \bibinfo {pages} {1825} (\bibinfo {year}
  {2017})}\BibitemShut {NoStop}%
\bibitem [{\citenamefont {J{\"a}nig}(2019{\natexlab{a}})}]{Janig:2019hom}%
  \BibitemOpen
  \bibfield  {author} {\bibinfo {author} {\bibfnamefont {A.}~\bibnamefont
  {J{\"a}nig}},\ }\bibfield  {title} {\enquote {\bibinfo {title}
  {{Nonautonomous Conley index theory. The homology index and
  attractor-repeller decompositions}},}\ }\href@noop {} {\bibfield  {journal}
  {\bibinfo  {journal} {Topol. Methods Nonlinear Anal.}\ }\textbf {\bibinfo
  {volume} {53}},\ \bibinfo {pages} {57} (\bibinfo {year}
  {2019}{\natexlab{a}})}\BibitemShut {NoStop}%
\bibitem [{\citenamefont {J{\"a}nig}(2019{\natexlab{b}})}]{Janig:2019conn}%
  \BibitemOpen
  \bibfield  {author} {\bibinfo {author} {\bibfnamefont {A.}~\bibnamefont
  {J{\"a}nig}},\ }\bibfield  {title} {\enquote {\bibinfo {title}
  {{Nonautonomous Conley index theory: the connecting homomorphism}},}\
  }\href@noop {} {\bibfield  {journal} {\bibinfo  {journal} {Topol. Methods
  Nonlinear Anal.}\ }\textbf {\bibinfo {volume} {53}},\ \bibinfo {pages} {427}
  (\bibinfo {year} {2019}{\natexlab{b}})}\BibitemShut {NoStop}%
\bibitem [{\citenamefont {J{\"a}nig}(2020)}]{Janig:2020}%
  \BibitemOpen
  \bibfield  {author} {\bibinfo {author} {\bibfnamefont {A.}~\bibnamefont
  {J{\"a}nig}},\ }\bibfield  {title} {\enquote {\bibinfo {title}
  {{Equilibrium-like solutions of asymptotically autonomous differential
  equations}},}\ }\href@noop {} {\bibfield  {journal} {\bibinfo  {journal} {J.
  Dyn. Differ. Equ.}\ }\textbf {\bibinfo {volume} {32}},\ \bibinfo {pages}
  {1249} (\bibinfo {year} {2020})}\BibitemShut {NoStop}%
\bibitem [{\citenamefont {Wieczorek}, \citenamefont {Xie},\ and\ \citenamefont
  {Jones}(2021)}]{Wieczorek:2021}%
  \BibitemOpen
  \bibfield  {author} {\bibinfo {author} {\bibfnamefont {S.}~\bibnamefont
  {Wieczorek}}, \bibinfo {author} {\bibfnamefont {C.}~\bibnamefont {Xie}},\
  and\ \bibinfo {author} {\bibfnamefont {C.~K. R.~T.}\ \bibnamefont {Jones}},\
  }\bibfield  {title} {\enquote {\bibinfo {title} {{Compactification for
  asymptotically autonomous dynamical systems: theory, applications and
  invariant manifolds}},}\ }\href@noop {} {\bibfield  {journal} {\bibinfo
  {journal} {Nonlinearity}\ }\textbf {\bibinfo {volume} {34}},\ \bibinfo
  {pages} {2970} (\bibinfo {year} {2021})}\BibitemShut {NoStop}%
\bibitem [{\citenamefont {Reineck}(1988)}]{Reineck:1988}%
  \BibitemOpen
  \bibfield  {author} {\bibinfo {author} {\bibfnamefont {J.~F.}\ \bibnamefont
  {Reineck}},\ }\bibfield  {title} {\enquote {\bibinfo {title} {{Connecting
  orbits in one-parameter families of flows}},}\ }\href@noop {} {\bibfield
  {journal} {\bibinfo  {journal} {Ergodic Theory Dynam. Systems}\ }\textbf
  {\bibinfo {volume} {8$^*$}},\ \bibinfo {pages} {359} (\bibinfo {year}
  {1988})}\BibitemShut {NoStop}%
\bibitem [{\citenamefont {Franzosa}\ and\ \citenamefont
  {Mischaikow}(1998)}]{Franzosa:1998}%
  \BibitemOpen
  \bibfield  {author} {\bibinfo {author} {\bibfnamefont {R.}~\bibnamefont
  {Franzosa}}\ and\ \bibinfo {author} {\bibfnamefont {K.}~\bibnamefont
  {Mischaikow}},\ }\bibfield  {title} {\enquote {\bibinfo {title} {{Algebraic
  transition matrices in the Conley index theory}},}\ }\href@noop {} {\bibfield
   {journal} {\bibinfo  {journal} {Trans. Amer. Math. Soc.}\ }\textbf {\bibinfo
  {volume} {350}},\ \bibinfo {pages} {889} (\bibinfo {year}
  {1998})}\BibitemShut {NoStop}%
\bibitem [{\citenamefont {McCord}\ and\ \citenamefont
  {Mischaikow}(1992)}]{McCord:1992tm}%
  \BibitemOpen
  \bibfield  {author} {\bibinfo {author} {\bibfnamefont {C.}~\bibnamefont
  {McCord}}\ and\ \bibinfo {author} {\bibfnamefont {K.}~\bibnamefont
  {Mischaikow}},\ }\bibfield  {title} {\enquote {\bibinfo {title} {{Connected
  simple systems, transition matrices, and heteroclinic bifurcations}},}\
  }\href@noop {} {\bibfield  {journal} {\bibinfo  {journal} {Trans. Amer. Math.
  Soc.}\ }\textbf {\bibinfo {volume} {333}},\ \bibinfo {pages} {397} (\bibinfo
  {year} {1992})}\BibitemShut {NoStop}%
\bibitem [{\citenamefont {Salamon}(1990)}]{Salamon:1990}%
  \BibitemOpen
  \bibfield  {author} {\bibinfo {author} {\bibfnamefont {D.}~\bibnamefont
  {Salamon}},\ }\bibfield  {title} {\enquote {\bibinfo {title} {{Morse theory,
  the Conley index and Floer homology}},}\ }\href@noop {} {\bibfield  {journal}
  {\bibinfo  {journal} {Bull. London Math. Soc.}\ }\textbf {\bibinfo {volume}
  {22}},\ \bibinfo {pages} {113} (\bibinfo {year} {1990})}\BibitemShut
  {NoStop}%
\bibitem [{\citenamefont {Robbin}\ and\ \citenamefont
  {Salamon}(1995)}]{Robbin:1995}%
  \BibitemOpen
  \bibfield  {author} {\bibinfo {author} {\bibfnamefont {J.}~\bibnamefont
  {Robbin}}\ and\ \bibinfo {author} {\bibfnamefont {D.}~\bibnamefont
  {Salamon}},\ }\bibfield  {title} {\enquote {\bibinfo {title} {{The spectral
  flow and the Maslov index}},}\ }\href@noop {} {\bibfield  {journal} {\bibinfo
   {journal} {Bull. London Math. Soc.}\ }\textbf {\bibinfo {volume} {27}},\
  \bibinfo {pages} {1} (\bibinfo {year} {1995})}\BibitemShut {NoStop}%
\bibitem [{\citenamefont {Heller}\ and\ \citenamefont
  {Spali{\'n}ski}(2015)}]{Heller:2015dha}%
  \BibitemOpen
  \bibfield  {author} {\bibinfo {author} {\bibfnamefont {M.~P.}\ \bibnamefont
  {Heller}}\ and\ \bibinfo {author} {\bibfnamefont {M.}~\bibnamefont
  {Spali{\'n}ski}},\ }\bibfield  {title} {\enquote {\bibinfo {title}
  {{Hydrodynamics beyond the gradient expansion: resurgence and
  resummation}},}\ }\href@noop {} {\bibfield  {journal} {\bibinfo  {journal}
  {Phys. Rev. Lett.}\ }\textbf {\bibinfo {volume} {115}},\ \bibinfo {pages}
  {072501} (\bibinfo {year} {2015})},\ \Eprint
  {https://arxiv.org/abs/1503.07514} {arXiv:1503.07514} \BibitemShut {NoStop}%
\bibitem [{\citenamefont {Romatschke}(2018)}]{Romatschke:2017vte}%
  \BibitemOpen
  \bibfield  {author} {\bibinfo {author} {\bibfnamefont {P.}~\bibnamefont
  {Romatschke}},\ }\bibfield  {title} {\enquote {\bibinfo {title}
  {{Relativistic fluid dynamics far from local equilibrium}},}\ }\href@noop {}
  {\bibfield  {journal} {\bibinfo  {journal} {Phys. Rev. Lett.}\ }\textbf
  {\bibinfo {volume} {120}},\ \bibinfo {pages} {012301} (\bibinfo {year}
  {2018})}\BibitemShut {NoStop}%
\bibitem [{\citenamefont {Blaizot}\ and\ \citenamefont
  {Yan}(2018)}]{Blaizot:2017ucy}%
  \BibitemOpen
  \bibfield  {author} {\bibinfo {author} {\bibfnamefont {J.-P.}\ \bibnamefont
  {Blaizot}}\ and\ \bibinfo {author} {\bibfnamefont {L.}~\bibnamefont {Yan}},\
  }\bibfield  {title} {\enquote {\bibinfo {title} {{Fluid dynamics of out of
  equilibrium boost invariant plasmas}},}\ }\href@noop {} {\bibfield  {journal}
  {\bibinfo  {journal} {Phys. Lett. B}\ }\textbf {\bibinfo {volume} {780}},\
  \bibinfo {pages} {283} (\bibinfo {year} {2018})}\BibitemShut {NoStop}%
\bibitem [{\citenamefont {Florkowski}, \citenamefont {Heller},\ and\
  \citenamefont {Spali{\'n}ski}(2018)}]{Florkowski:2017olj}%
  \BibitemOpen
  \bibfield  {author} {\bibinfo {author} {\bibfnamefont {W.}~\bibnamefont
  {Florkowski}}, \bibinfo {author} {\bibfnamefont {M.~P.}\ \bibnamefont
  {Heller}},\ and\ \bibinfo {author} {\bibfnamefont {M.}~\bibnamefont
  {Spali{\'n}ski}},\ }\bibfield  {title} {\enquote {\bibinfo {title} {{New
  theories of relativistic hydrodynamics in the LHC era}},}\ }\href@noop {}
  {\bibfield  {journal} {\bibinfo  {journal} {Rep. Prog. Phys.}\ }\textbf
  {\bibinfo {volume} {81}},\ \bibinfo {pages} {046001} (\bibinfo {year}
  {2018})}\BibitemShut {NoStop}%
\bibitem [{\citenamefont {Jankowski}\ and\ \citenamefont
  {Spali{\'n}ski}(2023)}]{Jankowski:2023fdz}%
  \BibitemOpen
  \bibfield  {author} {\bibinfo {author} {\bibfnamefont {J.}~\bibnamefont
  {Jankowski}}\ and\ \bibinfo {author} {\bibfnamefont {M.}~\bibnamefont
  {Spali{\'n}ski}},\ }\bibfield  {title} {\enquote {\bibinfo {title}
  {{Hydrodynamic attractors in ultrarelativistic nuclear collisions}},}\
  }\href@noop {} {\bibfield  {journal} {\bibinfo  {journal} {Prog. Part. Nucl.
  Phys.}\ }\textbf {\bibinfo {volume} {132}},\ \bibinfo {pages} {104048}
  (\bibinfo {year} {2023})}\BibitemShut {NoStop}%
\bibitem [{\citenamefont {Behtash}\ \emph {et~al.}(2020)\citenamefont
  {Behtash}, \citenamefont {Kamata}, \citenamefont {Martinez},\ and\
  \citenamefont {Shi}}]{Behtash:2020jhep}%
  \BibitemOpen
  \bibfield  {author} {\bibinfo {author} {\bibfnamefont {A.}~\bibnamefont
  {Behtash}}, \bibinfo {author} {\bibfnamefont {S.}~\bibnamefont {Kamata}},
  \bibinfo {author} {\bibfnamefont {M.}~\bibnamefont {Martinez}},\ and\
  \bibinfo {author} {\bibfnamefont {H.}~\bibnamefont {Shi}},\ }\bibfield
  {title} {\enquote {\bibinfo {title} {{Global flow structure and exact formal
  transseries of the Gubser flow in kinetic theory}},}\ }\href@noop {}
  {\bibfield  {journal} {\bibinfo  {journal} {J. High Energy Phys.}\ }\textbf
  {\bibinfo {volume} {07}},\ \bibinfo {pages} {226} (\bibinfo {year} {2020})},\
  \Eprint {https://arxiv.org/abs/1911.06406} {arXiv:1911.06406} \BibitemShut
  {NoStop}%
\bibitem [{\citenamefont {Aharony}(1973)}]{Aharony:1973}%
  \BibitemOpen
  \bibfield  {author} {\bibinfo {author} {\bibfnamefont {A.}~\bibnamefont
  {Aharony}},\ }\bibfield  {title} {\enquote {\bibinfo {title} {{Critical
  behavior of anisotropic cubic systems}},}\ }\href@noop {} {\bibfield
  {journal} {\bibinfo  {journal} {Phys. Rev. B}\ }\textbf {\bibinfo {volume}
  {8}},\ \bibinfo {pages} {4270} (\bibinfo {year} {1973})}\BibitemShut
  {NoStop}%
\bibitem [{\citenamefont {Sell}(1971)}]{Sell:1971}%
  \BibitemOpen
  \bibfield  {author} {\bibinfo {author} {\bibfnamefont {G.~R.}\ \bibnamefont
  {Sell}},\ }\href@noop {} {\emph {\bibinfo {title} {{Topological Dynamics and
  Ordinary Differential Equations}}}}\ (\bibinfo  {publisher} {Van Nostrand
  Reinhold},\ \bibinfo {address} {London},\ \bibinfo {year} {1971})\BibitemShut
  {NoStop}%
\bibitem [{\citenamefont {Salamon}(1985)}]{Salamon:1985}%
  \BibitemOpen
  \bibfield  {author} {\bibinfo {author} {\bibfnamefont {D.}~\bibnamefont
  {Salamon}},\ }\bibfield  {title} {\enquote {\bibinfo {title} {{Connected
  simple systems and the Conley index of isolated invariant sets}},}\
  }\href@noop {} {\bibfield  {journal} {\bibinfo  {journal} {Trans. Amer. Math.
  Soc.}\ }\textbf {\bibinfo {volume} {291}},\ \bibinfo {pages} {1} (\bibinfo
  {year} {1985})}\BibitemShut {NoStop}%
\bibitem [{\citenamefont {Conley}\ and\ \citenamefont
  {Easton}(1971)}]{Conley:1971}%
  \BibitemOpen
  \bibfield  {author} {\bibinfo {author} {\bibfnamefont {C.}~\bibnamefont
  {Conley}}\ and\ \bibinfo {author} {\bibfnamefont {R.}~\bibnamefont
  {Easton}},\ }\bibfield  {title} {\enquote {\bibinfo {title} {{Isolated
  invariant sets and isolating blocks}},}\ }\href@noop {} {\bibfield  {journal}
  {\bibinfo  {journal} {Trans. Amer. Math. Soc.}\ }\textbf {\bibinfo {volume}
  {158}},\ \bibinfo {pages} {35} (\bibinfo {year} {1971})}\BibitemShut
  {NoStop}%
\bibitem [{\citenamefont {Franzosa}(1989)}]{Franzosa:1989}%
  \BibitemOpen
  \bibfield  {author} {\bibinfo {author} {\bibfnamefont {R.}~\bibnamefont
  {Franzosa}},\ }\bibfield  {title} {\enquote {\bibinfo {title} {{The
  connection matrix theory for Morse decompositions}},}\ }\href@noop {}
  {\bibfield  {journal} {\bibinfo  {journal} {Trans. Amer. Math. Soc.}\
  }\textbf {\bibinfo {volume} {311}},\ \bibinfo {pages} {561} (\bibinfo {year}
  {1989})}\BibitemShut {NoStop}%
\bibitem [{\citenamefont {McCord}(1988)}]{McCord:1988}%
  \BibitemOpen
  \bibfield  {author} {\bibinfo {author} {\bibfnamefont {C.}~\bibnamefont
  {McCord}},\ }\bibfield  {title} {\enquote {\bibinfo {title} {{The connection
  map for attractor-repeller pairs}},}\ }\href@noop {} {\bibfield  {journal}
  {\bibinfo  {journal} {Trans. Amer. Math. Soc.}\ }\textbf {\bibinfo {volume}
  {307}},\ \bibinfo {pages} {195} (\bibinfo {year} {1988})}\BibitemShut
  {NoStop}%
\bibitem [{\citenamefont {McCord}(1992)}]{McCord:1992dual}%
  \BibitemOpen
  \bibfield  {author} {\bibinfo {author} {\bibfnamefont {C.}~\bibnamefont
  {McCord}},\ }\bibfield  {title} {\enquote {\bibinfo {title}
  {{Poincar{\'e}--Lefschetz duality for the homology Conley index}},}\
  }\href@noop {} {\bibfield  {journal} {\bibinfo  {journal} {Trans. Amer. Math.
  Soc.}\ }\textbf {\bibinfo {volume} {329}},\ \bibinfo {pages} {233} (\bibinfo
  {year} {1992})}\BibitemShut {NoStop}%
\bibitem [{\citenamefont {Floer}(1987)}]{Floer:1987}%
  \BibitemOpen
  \bibfield  {author} {\bibinfo {author} {\bibfnamefont {A.}~\bibnamefont
  {Floer}},\ }\bibfield  {title} {\enquote {\bibinfo {title} {{A refinement of
  the Conley index and an application to the stability of hyperbolic invariant
  sets}},}\ }\href@noop {} {\bibfield  {journal} {\bibinfo  {journal} {Ergodic
  Theory Dynam. Systems}\ }\textbf {\bibinfo {volume} {7}},\ \bibinfo {pages}
  {93} (\bibinfo {year} {1987})}\BibitemShut {NoStop}%
\bibitem [{\citenamefont {Bartsch}(1993)}]{Bartsch:1993}%
  \BibitemOpen
  \bibfield  {author} {\bibinfo {author} {\bibfnamefont {T.}~\bibnamefont
  {Bartsch}},\ }\href@noop {} {\emph {\bibinfo {title} {{Topological Methods
  for Variational Problems with Symmetries}}}},\ \bibinfo {series} {Lecture
  Notes in Mathematics}, Vol.\ \bibinfo {volume} {1560}\ (\bibinfo  {publisher}
  {Springer},\ \bibinfo {address} {Berlin},\ \bibinfo {year}
  {1993})\BibitemShut {NoStop}%
\bibitem [{\citenamefont {Lamb}\ and\ \citenamefont
  {Roberts}(1998)}]{Lamb:1998}%
  \BibitemOpen
  \bibfield  {author} {\bibinfo {author} {\bibfnamefont {J.~S.~W.}\
  \bibnamefont {Lamb}}\ and\ \bibinfo {author} {\bibfnamefont {J.~A.~G.}\
  \bibnamefont {Roberts}},\ }\bibfield  {title} {\enquote {\bibinfo {title}
  {{Time-reversal symmetry in dynamical systems: a survey}},}\ }\href@noop {}
  {\bibfield  {journal} {\bibinfo  {journal} {Physica D}\ }\textbf {\bibinfo
  {volume} {112}},\ \bibinfo {pages} {1} (\bibinfo {year} {1998})}\BibitemShut
  {NoStop}%
\bibitem [{\citenamefont {Palmer}(1984)}]{Palmer:1984}%
  \BibitemOpen
  \bibfield  {author} {\bibinfo {author} {\bibfnamefont {K.~J.}\ \bibnamefont
  {Palmer}},\ }\bibfield  {title} {\enquote {\bibinfo {title} {{Exponential
  dichotomies and transversal homoclinic points}},}\ }\href@noop {} {\bibfield
  {journal} {\bibinfo  {journal} {J. Differential Equations}\ }\textbf
  {\bibinfo {volume} {55}},\ \bibinfo {pages} {225} (\bibinfo {year}
  {1984})}\BibitemShut {NoStop}%
\bibitem [{\citenamefont {Palmer}(1988)}]{Palmer:1988}%
  \BibitemOpen
  \bibfield  {author} {\bibinfo {author} {\bibfnamefont {K.~J.}\ \bibnamefont
  {Palmer}},\ }\bibfield  {title} {\enquote {\bibinfo {title} {{Exponential
  dichotomies and Fredholm operators}},}\ }\href@noop {} {\bibfield  {journal}
  {\bibinfo  {journal} {Proc. Amer. Math. Soc.}\ }\textbf {\bibinfo {volume}
  {104}},\ \bibinfo {pages} {149} (\bibinfo {year} {1988})}\BibitemShut
  {NoStop}%
\bibitem [{\citenamefont {Hatcher}(2002)}]{Hatcher:2002}%
  \BibitemOpen
  \bibfield  {author} {\bibinfo {author} {\bibfnamefont {A.}~\bibnamefont
  {Hatcher}},\ }\href@noop {} {\emph {\bibinfo {title} {{Algebraic
  Topology}}}}\ (\bibinfo  {publisher} {Cambridge University Press},\ \bibinfo
  {address} {Cambridge},\ \bibinfo {year} {2002})\BibitemShut {NoStop}%
\bibitem [{\citenamefont {Dumortier}, \citenamefont {Llibre},\ and\
  \citenamefont {Art{\'e}s}(2006)}]{Dumortier:2006}%
  \BibitemOpen
  \bibfield  {author} {\bibinfo {author} {\bibfnamefont {F.}~\bibnamefont
  {Dumortier}}, \bibinfo {author} {\bibfnamefont {J.}~\bibnamefont {Llibre}},\
  and\ \bibinfo {author} {\bibfnamefont {J.~C.}\ \bibnamefont {Art{\'e}s}},\
  }\href@noop {} {\emph {\bibinfo {title} {{Qualitative Theory of Planar
  Differential Systems}}}},\ Universitext\ (\bibinfo  {publisher} {Springer},\
  \bibinfo {address} {Berlin},\ \bibinfo {year} {2006})\BibitemShut {NoStop}%
\bibitem [{\citenamefont {Rybakowski}(1987)}]{Rybakowski:1987}%
  \BibitemOpen
  \bibfield  {author} {\bibinfo {author} {\bibfnamefont {K.~P.}\ \bibnamefont
  {Rybakowski}},\ }\href@noop {} {\emph {\bibinfo {title} {{The Homotopy Index
  and Partial Differential Equations}}}},\ Universitext\ (\bibinfo  {publisher}
  {Springer},\ \bibinfo {address} {Berlin},\ \bibinfo {year}
  {1987})\BibitemShut {NoStop}%
\bibitem [{\citenamefont {Spendlove}\ and\ \citenamefont
  {Vandervorst}(2026)}]{Spendlove:2026}%
  \BibitemOpen
  \bibfield  {author} {\bibinfo {author} {\bibfnamefont {K.}~\bibnamefont
  {Spendlove}}\ and\ \bibinfo {author} {\bibfnamefont {R.}~\bibnamefont
  {Vandervorst}},\ }\bibfield  {title} {\enquote {\bibinfo {title} {{Graded
  differential vector spaces, Cartan--Eilenberg systems and conjectures in
  Conley index theory}},}\ }\href@noop {} {\bibfield  {journal} {\bibinfo
  {journal} {J. Dyn. Differ. Equ.}\ }\textbf {\bibinfo {volume} {38}},\
  \bibinfo {pages} {695} (\bibinfo {year} {2026})}\BibitemShut {NoStop}%
\bibitem [{\citenamefont {Costin}(1995)}]{Costin:1995}%
  \BibitemOpen
  \bibfield  {author} {\bibinfo {author} {\bibfnamefont {O.}~\bibnamefont
  {Costin}},\ }\bibfield  {title} {\enquote {\bibinfo {title} {{Exponential
  asymptotics, transseries, and generalized Borel summation for analytic,
  nonlinear, rank-one systems of ordinary differential equations}},}\
  }\href@noop {} {\bibfield  {journal} {\bibinfo  {journal} {Int. Math. Res.
  Not.}\ }\textbf {\bibinfo {volume} {1995}},\ \bibinfo {pages} {377} (\bibinfo
  {year} {1995})}\BibitemShut {NoStop}%
\bibitem [{\citenamefont {Costin}(1998)}]{Costin:1998}%
  \BibitemOpen
  \bibfield  {author} {\bibinfo {author} {\bibfnamefont {O.}~\bibnamefont
  {Costin}},\ }\bibfield  {title} {\enquote {\bibinfo {title} {{On Borel
  summation and Stokes phenomena for rank-1 nonlinear systems of ordinary
  differential equations}},}\ }\href@noop {} {\bibfield  {journal} {\bibinfo
  {journal} {Duke Math. J.}\ }\textbf {\bibinfo {volume} {93}},\ \bibinfo
  {pages} {289} (\bibinfo {year} {1998})}\BibitemShut {NoStop}%
\bibitem [{\citenamefont {Dorigoni}(2019)}]{Dorigoni:2014hea}%
  \BibitemOpen
  \bibfield  {author} {\bibinfo {author} {\bibfnamefont {D.}~\bibnamefont
  {Dorigoni}},\ }\bibfield  {title} {\enquote {\bibinfo {title} {{An
  introduction to resurgence, trans-series and alien calculus}},}\ }\href@noop
  {} {\bibfield  {journal} {\bibinfo  {journal} {Ann. Phys.}\ }\textbf
  {\bibinfo {volume} {409}},\ \bibinfo {pages} {167914} (\bibinfo {year}
  {2019})},\ \Eprint {https://arxiv.org/abs/1411.3585} {arXiv:1411.3585}
  \BibitemShut {NoStop}%
\bibitem [{\citenamefont {Borel}(1960)}]{Borel:1960}%
  \BibitemOpen
  \bibfield  {author} {\bibinfo {author} {\bibfnamefont {A.}~\bibnamefont
  {Borel}},\ }\href@noop {} {\emph {\bibinfo {title} {{Seminar on
  Transformation Groups}}}},\ \bibinfo {series} {Annals of Mathematics
  Studies}, Vol.~\bibinfo {volume} {46}\ (\bibinfo  {publisher} {Princeton
  University Press},\ \bibinfo {address} {Princeton, NJ},\ \bibinfo {year}
  {1960})\BibitemShut {NoStop}%
\bibitem [{\citenamefont {tom Dieck}(1987)}]{tomDieck:1987}%
  \BibitemOpen
  \bibfield  {author} {\bibinfo {author} {\bibfnamefont {T.}~\bibnamefont {tom
  Dieck}},\ }\href@noop {} {\emph {\bibinfo {title} {{Transformation
  Groups}}}},\ \bibinfo {series} {De Gruyter Studies in Mathematics},
  Vol.~\bibinfo {volume} {8}\ (\bibinfo  {publisher} {Walter de Gruyter},\
  \bibinfo {address} {Berlin},\ \bibinfo {year} {1987})\BibitemShut {NoStop}%
\bibitem [{\citenamefont {Hausmann}(2014)}]{Hausmann:2014}%
  \BibitemOpen
  \bibfield  {author} {\bibinfo {author} {\bibfnamefont {J.-C.}\ \bibnamefont
  {Hausmann}},\ }\href@noop {} {\emph {\bibinfo {title} {{Mod Two Homology and
  Cohomology}}}},\ Universitext\ (\bibinfo  {publisher} {Springer},\ \bibinfo
  {address} {Cham},\ \bibinfo {year} {2014})\BibitemShut {NoStop}%
\bibitem [{\citenamefont {Allday}\ and\ \citenamefont
  {Puppe}(1993)}]{Allday:1993}%
  \BibitemOpen
  \bibfield  {author} {\bibinfo {author} {\bibfnamefont {C.}~\bibnamefont
  {Allday}}\ and\ \bibinfo {author} {\bibfnamefont {V.}~\bibnamefont {Puppe}},\
  }\href@noop {} {\emph {\bibinfo {title} {{Cohomological Methods in
  Transformation Groups}}}}\ (\bibinfo  {publisher} {Cambridge University
  Press},\ \bibinfo {address} {Cambridge},\ \bibinfo {year} {1993})\BibitemShut
  {NoStop}%
\end{thebibliography}%

\end{document}